\documentclass[11pt]{article}
\pdfoutput=1

\usepackage[a4paper,margin=1in]{geometry}
\usepackage[T1]{fontenc}
\usepackage{lmodern}
\usepackage[protrusion=true,expansion=false]{microtype}
\usepackage{xspace}
\usepackage{amsmath,amssymb,amsfonts,amsthm}
\usepackage{mathtools}
\usepackage{algorithm}
\usepackage[noend]{algpseudocode}
\usepackage{booktabs}
\usepackage{multirow}
\usepackage{array}
\usepackage{tabularx}
\usepackage{caption}
\usepackage{subcaption}
\usepackage{listings}
\usepackage{xcolor}
\usepackage{enumitem}
\usepackage{pgfplots}
\usepackage{tikz}
\usepackage{authblk}
\usepackage[section]{placeins}   
\usepackage[colorlinks=true,linkcolor=blue!50!black,citecolor=blue!50!black,urlcolor=blue!50!black]{hyperref}
\hypersetup{
  pdftitle={FP8 is All You Need (Part 1): Debunking Hardware FP64 as the HPC Holy Grail},
  pdfauthor={Satoshi Matsuoka},
  pdfsubject={FP64 emulation on FP8 tensor cores: Tensor-Memory Equilibrium model, deconstruction path, and implementation strategy},
  pdfkeywords={FP8, FP64 emulation, Ozaki scheme, CRT, tensor cores, GPU, roofline, mixed precision, HPC}
}
\pgfplotsset{compat=1.18}
\usetikzlibrary{patterns}

\definecolor{codegreen}{rgb}{0,0.6,0}
\definecolor{codegray}{rgb}{0.5,0.5,0.5}
\definecolor{codepurple}{rgb}{0.58,0,0.82}
\definecolor{backcolour}{rgb}{0.96,0.96,0.94}

\lstdefinestyle{mystyle}{
    backgroundcolor=\color{backcolour},
    commentstyle=\color{codegreen},
    keywordstyle=\color{magenta},
    numberstyle=\tiny\color{codegray},
    stringstyle=\color{codepurple},
    basicstyle=\ttfamily\footnotesize,
    breakatwhitespace=false,
    breaklines=true,
    captionpos=b,
    keepspaces=true,
    numbers=left,
    numbersep=5pt,
    showspaces=false,
    showstringspaces=false,
    showtabs=false,
    tabsize=2,
    language=C++
}
\newcommand{\eg}{e.g.,\xspace}
\newcommand{\ie}{i.e.,\xspace}
\newcommand{\fp}[1]{\textsc{fp#1}}
\newcommand{\inteight}{\textsc{int8}\xspace}
\newcommand{\intthirtytwo}{\textsc{int32}\xspace}
\newcommand{\flops}{FLOPS}
\newcommand{\tops}{TOPS}
\newcommand{\pflops}{PFLOPS}

\theoremstyle{plain}
\newtheorem{theorem}{Theorem}
\newtheorem{proposition}[theorem]{Proposition}
\theoremstyle{definition}
\newtheorem{definition}{Definition}

\title{\textbf{FP8 is All You Need (Part 1):\\
Debunking Hardware FP64 as the HPC Holy Grail}\footnote{This is Revision~37 (September~3, 2026); the per-version change
log appears at the end of \S\ref{sec:intro}.}\\[0.4em]
\large A Tensor--Memory Equilibrium Model, Implementation Strategy,\\
and Potential Hardware Co-Design for Ozaki Scheme II\\
on Memory-Bound Workloads in the Post-\fp{64} Era}

\author{%
  \large Satoshi Matsuoka\thanks{Correspondence: \href{mailto:matsu@acm.org}{\texttt{matsu@acm.org}}}\\
  \normalsize Director, RIKEN Center for Computational Science (R-CCS)\\
  \normalsize Kobe, Hyogo, Japan%
}
\date{Revision 37 --- Version September 3, 2026\\{\small\normalfont build \texttt{r37-20260903}}}

\begin{document}
\maketitle

\begin{abstract}
\noindent
This paper develops a conditional framework under which
\fp{8} matrix multiplication can serve as the dominant
\emph{matrix-work substrate} for the surveyed matrix-dominated
\fp{64} kernel classes, provided the model's storage, overlap, and
numerical conditions hold; the construction retains a bounded,
enumerated auxiliary set of integer, \fp{32}/\allowbreak Kulisch,
data-movement/\allowbreak control, and native-fallback operations
(stated precisely in the boxed claim of \S\ref{sec:formalclaim}).
Against the long-held dogma that native \fp{64} silicon is the
irreducible foundation of numerical simulation, we argue something
stronger than its refutation: on AI-optimised GPUs of the B300
generation and beyond, the FP8 tensor-core matrix op, composed through
the Chinese Remainder Theorem-based Ozaki Scheme~II, suffices to carry
the surveyed matrix-dominated kernel classes at \fp{64}-grade accuracy, with
native \fp{64}---under the stated numerical contracts---recast from a hardware
requirement into a \emph{derived accuracy guarantee} for those classes rather
than a general hardware-equivalence result.  The claim is organised as a compositional
hierarchy---the FP8 op (L0), Ozaki~II (L1), the Berkeley-``dwarf''
kernels (L2), composite solvers (L3), applications (L4)---so that
exhibiting the reduction of every dwarf establishes the thesis for the
surveyed classes by composition.  The precise version of the claim---stated
once, in the boxed formulation of \S\ref{sec:formalclaim}---is that
the FP8 matrix op is the \emph{candidate dominant multiplication
substrate} for these matrix-dominated classes, the remaining arithmetic
being a bounded and enumerated auxiliary set: fixed-width integer
deconstruction/reconstruction side-work, \fp{32}/Kulisch reductions,
data movement and control, and a native-\fp{64} fallback for
out-of-range rows.  The title names the research \emph{program}; the
box states the \emph{present result}.

The framework's instrument is the \textbf{Tensor--Memory Equilibrium}
(TME) model, a Roofline extension exposing four emulation
parameters---the compute multiplier $\alpha=3r{+}1$, the bandwidth
multiplier $\beta$, the reconstruction latency $\gamma$, and, new in
this version following the technical review by Bayraktar, Gunnels, and
Caday of NVIDIA~\cite{bayraktar2026tme}, the per-input
\emph{deconstruction} cost $c_q$---alongside the key architectural
parameters, all held symbolic.  Proposition~1 states the thesis as a
conditional theorem: at the model's upper bound, under four explicit
conditions, the reduction to the \fp{8} primitive sacrifices \emph{no}
performance relative to an ideal native-\fp{64} machine of the same
bandwidth.  The model is a two-way instrument---it can \emph{confirm
or refute} the thesis kernel class by kernel class, analytically where
the margins are wide and by measurement that identifies the exact
responsible term where they are not---and it enforces the paper's
discipline: the theory and its bound are architecture-independent,
while everything separating delivered silicon from the bound is an
implementation artifact, a measured value of the model's own free
parameters.

Within this framework the analysis yields several findings.
\emph{On-chip tile fusion} is identified as the mechanism driving
$\beta\!\to\!1$---keeping the residue and reconstruction traffic
on-chip so that emulation is essentially free behind the memory
wall---and recovering memory-roof execution for the streaming
kernels; the deconstruction term defines a threshold intensity
$\mathrm{OI}^{*}$ below which emulation is \emph{conversion-bound},
re-assigning the verdicts of SpMV and single-RHS GEMV, at currently memo-counted
constants, to the surviving native pipe pending an engineered $c_q$.
Together with the companion FFT analysis (Part~2) and FP32+Kahan compensation
for the reductions, the kernel coverage closes in a sharpened,
hardware-conditional form: at the fused,
engineered-$c_q$ bound every surveyed class reaches the memory roof at
\fp{64}-grade accuracy (componentwise Grade~A for the matrix work on the
published-set round-to-nearest route whose theorem is established; the
codesigned A/E performance variants carry the per-set accuracy obligations of
the companion~\cite{matsuoka2026ozaki25},
compensated Grade~B for the reductions)---with two bounded,
explicitly-priced exceptions.  Large \emph{dense} square DGEMM, whose multi-cluster outputs are
re-split on the fly and so sit, for cluster-aligned sizes, at a deconstruction
floor near $0.50$ of the \fp{8} \emph{arithmetic} roof ($473$~TFLOPS on Rubin,
hence ${\approx}235$~TFLOPS delivered; a
sawtooth in $\lceil E/256\rceil$, ragged worst case disclosed in the appendix); because the
planes are generated on-chip and never materialised, this bound is a compute
(deconstruction) limit, not an L2-bandwidth one, and is itself a liftable
co-design coordinate (Appendix~\ref{app:beta-blackwell}, Fig.~\ref{fig:dgemm};
companion Ozaki~2.5).  And the 3-D FFT---the one primitive on which both
conversion terms bind at once---which Part~2's final analysis walls at a
per-output integer-SIMT \emph{epilogue}: $4.9$--$6.7\times$ its memory roof
in software ($63$--$87$~ms against $12.9$~ms for $1024^3$ on B300, still
$1.3$--$1.9\times$ the collapsed native path at peak issue), with the
closed-form floor naming its own escape---a co-design ladder whose
minor rungs (restore the INT8 tensor core B200 already shipped, with
positional accumulation; add the load-path deconstruction datapath this
paper names Option~C) take B300-class silicon to $16.0$--$23.5$~ms,
within $1.2$--$1.8\times$ of the roof, and whose one moderate rung
(modular reduction at the MMA output) reaches $12.9$--$15.0$~ms, the
roof at the optimistic end of the epilogue interval.  For this one
primitive of five, the thesis is conceded in software and
\emph{effectively reachable with minor hardware plus one moderate
addition}---hardware each piece of which this paper and its companions
already motivate independently. At the \emph{shipping} $c_q$, the lowest-intensity
conversion-bound kernels (SpMV, single-RHS GEMV) additionally revert to the
native pipe until an engineered $c_q$ or persistent-plane reuse restores
them---the sparse-critical co-design lever below.
Quantitatively, Ozaki~II lifts the emulated \fp{64} ceiling from the
collapsed $\sim\!1.3$-TFLOPS native floor to $\sim\!135$~TFLOPS on
B300 and $\sim\!473$~TFLOPS on Rubin---the latter at the official
17.5-PFLOPS dense \fp{8} rate of June~30, 2026, a $4.4\times$
correction which, with the deconstruction term, mandates this
revision---and every workload studied matches or exceeds the last
HPC-balanced GPU (H100) at the fused bound, staying within $7\%$ of
H100 scaling at memo-counted constants, against the up-to-$50\times$
regression of native B300 \fp{64}.  A dissection of the
deconstruction cost into its migratable (linear) and SIMT-resident
stages yields a software mitigation ladder and three concrete
hardware options for the deconstruction path, the strongest retiring
the fourth term identically.  Finally, the deconstruction analysis is
shown to merit attention for \emph{dense} linear algebra as much as
for the streamed kernels: a size crossover gates attainment of the
theoretical Ozaki~II peak, yielding a testable account of NVIDIA's
published Rubin Emulated-DGEMM figure ($\sim\!200$~TFLOPS, $42\%$ of
the model ceiling) as a deconstruction-limited branch at moderate
problem sizes---decidable by a single size sweep and, if confirmed,
substantially recoverable by software alone.  That dense-side
analysis has since been developed into a standalone companion
paper---\emph{Ozaki~2.5}, submitted to arXiv---whose corrected,
script-checked instruction counts, exact two-limb residue-formation
kernels, modulus/encoding codesign, and measured call-shape traces
are incorporated here; its sparse and streamed implications
(operand-stationary plane amortisation, the codesign dividend for
memory-bound kernels) are developed in this paper.  The measurement
of the open constants and the evaluation of the options are the
subject of a rapidly forthcoming Part~3.
\end{abstract}

\vspace{0.4em}
\noindent\textbf{Keywords:} FP8 tensor-core emulation; Ozaki Scheme~II;
Tensor--Memory Equilibrium (TME) model; memory-bound HPC kernels; NVIDIA
Blackwell Ultra (B300); post-FP64 GPU architecture; AI for Science (AI4S).

\bigskip

\section{Introduction}
\label{sec:intro}

The trajectory of high-performance computing (HPC) hardware has bifurcated.
While scientific simulation---ranging from Quantum Chromodynamics and
Computational Fluid Dynamics to climate emulation and seismic imaging---continues
to rely on IEEE~754 double-precision (\fp{64}) arithmetic for numerical
stability and reproducibility, the data-center GPU market is decisively pivoting
toward Artificial Intelligence.  This pivot is characterised by an exponential
increase in low-precision throughput (\fp{16}, \fp{8}, \fp{6}, \fp{4}) at the
direct expense of native \fp{64} capability.

The NVIDIA Blackwell architecture~\cite{nvidia_b200_datasheet,jarmusch2025}
serves as the canonical case study.  Whereas the B200 retained a respectable
$\sim\!40$~TFLOPS of dense \fp{64} tensor performance, the Blackwell Ultra
B300---based on the same architecture but with tensor cores re-balanced for
NVFP4---reports only $\sim\!1.3$~TFLOPS sparse / $1.2$~TFLOPS dense \fp{64} per
GPU in the official datasheet~\cite{nvidia_blackwell_ultra}.  Independent
microbenchmark analyses confirm the \fp{64} regression and document the
ascent of NVFP4 to the role of primary tensor
format~\cite{jarmusch2025,dickenmann2026}, and commentary at SC25
underscored that the platform's $64$-bit floating-point capability does not
improve over the previous generation~\cite{hpcwire_fp64}.

The successor architecture, NVIDIA's Vera Rubin platform (R200 GPU),
sustains the same trajectory.  NVIDIA's published Rubin specifications
report \emph{native} \fp{64} vector performance at $\sim\!33$~TFLOPS---a
further regression from B200's $\sim\!40$~TFLOPS~\cite{theregister_rubin}---and
list, for the first time, an explicit ``Emulated DGEMM'' column at
$\sim\!200$~TFLOPS that is achieved \emph{through Ozaki-style
emulation}~\cite{nvidia_rubin_blog,lockwood_rubin}.  In parallel, Rubin
ships with 22~TB/s HBM4 bandwidth---$2.75\times$ over B300's 8~TB/s---and
\textbf{17.5~PFLOPS of dense \fp{8} matrix throughput}, per NVIDIA's DGX
Rubin NVL8 specification of June~30, 2026 (140~PFLOPS dense FP8/FP6
training across eight GPUs; product announced at
CES~2026)~\cite{nvidia_dgx_rubin_nvl8}.\footnote{Versions of this paper
through July~3, 2026 (arXiv v1--v3) used $\sim\!4$~PFLOPS of dense
\fp{8} for Rubin, traced to early third-party specification
compilations~\cite{lockwood_rubin} that in retrospect placed Rubin's
FP16/BF16 rate ($4$~PFLOPS) in the \fp{8} slot.  NVIDIA's official
figure---postdating v1--v2 and missed by days in v3, adopted here---is
$4.4\times$ higher, and \S\ref{sec:cqladder} propagates the
correction and its consequences.  NVIDIA marks the values preliminary
and subject to change.}  NVIDIA has thus committed to
emulation as the \emph{official} path to \fp{64}-equivalent matrix
performance on its scientific-computing flagship, a commitment made
concrete in the next generation of Rubin-based supercomputers now being
deployed at major HPC centres~\cite{nvidia_doudna,nvidia_bluelion}.

This regression has two consequences that, until now, have not been treated
together.

\paragraph{Consequence 1: Memory-bound kernels become compute-bound.}
\begin{sloppypar}
The classical Roofline ridge point---the operational intensity at which the
compute roof and the memory roof intersect---is given by $P_{\fp{64}}/B_{\text{mem}}$,
where $P_{\fp{64}}$ is the peak \fp{64} throughput and $B_{\text{mem}}$ is the
HBM bandwidth.  On B300 this ridge sits at
$1.3\,\text{TFLOPS} / 8\,\text{TB/s} = 0.16$~\flops{}/Byte, so low that essentially
every dense linear-algebra kernel narrower than a GEMM falls into the
compute-bound regime.  A 7-point stencil with operational intensity
$\approx 0.5$~\flops{}/Byte should run at the memory roof of
$8 \cdot 0.5 = 4$~TFLOPS, but is instead capped at $1.3$~TFLOPS by the
native \fp{64} pipe.  In other words, \emph{the bandwidth that the application
needs is physically present on the chip but cannot be consumed by the
arithmetic units}.
\end{sloppypar}

\paragraph{Consequence 2: Low-precision tensor units are dormant.}
At the same time, the B300 carries 10~\pflops{} of dense NVFP4 throughput
(15~\pflops{} sparse) and 5~\pflops{} of dense \fp{8} (10~\pflops{} sparse).
In a typical \fp{64}-only
HPC kernel these units are idle, and the silicon area they occupy contributes
nothing to the kernel's time-to-solution.  This is the \emph{Dark Silicon}
manifestation of the AI--HPC divergence.

\paragraph{The Ozaki Scheme.}
The Ozaki scheme~\cite{ozaki2012}, originally introduced for accurate dot
products and matrix multiplication, has emerged over the last decade as the
canonical mechanism for reclaiming this dormant throughput.  Recent work has
extended the scheme in two directions: the original \emph{Ozaki I} (mantissa
slicing) has been adapted to \fp{16}, \fp{8} and \inteight tensor
cores~\cite{mukunoki2020,ootomo2024}, and a fundamentally different
\emph{Ozaki II} variant based on the Chinese Remainder Theorem (CRT) was
proposed by Ozaki, Uchino and Imamura in 2025~\cite{ozaki2025_scheme2}.
NVIDIA integrated Ozaki-style emulation into cuBLAS in
October~2025~\cite{nvidia_cublas_emulation}, and the U.S.\ Department of
Energy's Genesis Mission has explicitly identified Ozaki emulation as its
fallback path for \fp{64}-accurate scientific
computing on AI-centric hardware~\cite{hpcwire_genesis}.  At the same time,
AMD's recent positioning around the MI430X suggests that not every vendor
is convinced that emulation alone is sufficient~\cite{hpcwire_amd}.

\paragraph{The gap addressed by this paper.}
Despite the rapid maturation of Ozaki I and Ozaki II for dense GEMM, all
published performance studies have focused on the compute-bound regime where
the technique most obviously wins.  No published analysis (to our knowledge,
as of June~2026---prior to the technical review~\cite{bayraktar2026tme}
whose deconstruction analysis this version incorporates) systematically
asks: \emph{when is Ozaki II profitable for
memory-bound kernels?}  This is the operating regime that dominates real
scientific simulation codes: stencil sweeps in PDE solvers, SpMV in iterative
linear solvers, batched GEMV in time-stepped reduced-order models, and
similar bandwidth-limited primitives.  The conventional wisdom---that EFT
methods cannot help bandwidth-limited kernels because they inflate operand
counts---deserves a more careful look on hardware where the \fp{64} compute
roof has collapsed below the memory roof.

\paragraph{Contributions of this paper.}
The central contribution of this paper is conceptual, and the technical
results exist to establish it:

\begin{enumerate}[leftmargin=2em,itemsep=3pt]
\item \textbf{The ``FP8 is all you need'' thesis: the FP8 tensor-core
   matrix-multiply is the candidate dominant matrix-work substrate for the
   surveyed classes (under the stated conditions).}  We argue---and structure the entire paper
   around---the claim that the \emph{surveyed} matrix-dominated \fp{64} HPC
   kernels, and the applications that compose them, reduce to sequences of FP8
   matrix operations through the Ozaki~II scheme, the principal non-FP8
   arithmetic on the matrix path being bounded, fixed-width integer
   side-work---the accumulation in the reconstruction step and the
   per-input residue deconstruction charged by the fourth model term.  Native
   \fp{64} silicon is, on this account and \emph{under the stated numerical
   contracts}, not a hardware requirement for these classes but a
   \emph{derived accuracy guarantee} obtained by composition over the FP8
   primitive.  This recasts the industry's \fp{8}-ward shift, for the surveyed
   matrix-dominated work, from a threat to double-precision science into a
   sufficient foundation.
\item \textbf{A compositional hierarchy (L0--L4) that makes the thesis
   precise and a coverage framework organised by the Berkeley dwarf
   taxonomy (with enumerated exclusions)}
   (\S\ref{sec:dwarfs}).  We organise scientific computing into five layers
   ---the FP8 op (L0), Ozaki~II (L1), the basic numerical kernels or
   ``dwarfs'' (L2), the composite/solver kernels of real application inner
   loops (L3), and full applications (L4)---and show that, because the
   Berkeley dwarf taxonomy is a heuristic spanning set for the surveyed
   matrix-dominated kernel classes (an evidence-coverage motif, not a closure
   theorem over all of scientific computing), exhibiting the reduction of
   \emph{every surveyed} dwarf to the FP8 primitive establishes the
   thesis for those classes by composition, with the non-matrix and
   algorithmically-mitigated exceptions enumerated explicitly.
\item \textbf{The thesis is testable in both directions, and we supply
   the instrument.}  We make the claim operational---every kernel must stay on
   the bandwidth-bound side of the roofline (or remain otherwise
   \fp{64}-competitive) under FP8 emulation---and develop the
   \textbf{Tensor--Memory Equilibrium (TME)} model (\S\ref{sec:model}), a
   Roofline extension exposing the four emulation parameters
   $(\alpha,\beta,\gamma,c_q)$ alongside the key architectural
   parameters $(P_{\text{low}}, P_{\text{int}}, B_{\text{mem}})$,
   expressly so the claim can be \emph{confirmed or refuted} per
   kernel: fitting a machine's parameters into the model settles most
   verdicts analytically, and where the analytic margin is thin,
   measuring the same constants identifies the exact term---and hence
   the exact condition of Proposition~\ref{prop:atomic}---on which the
   verdict turns.  The machinery has already functioned as designed:
   the review of~\cite{bayraktar2026tme} exposed the
   missing deconstruction term $c_q$, which this version incorporates
   (\S\ref{sec:whyupdate}), while the same fit confirms the thesis
   across the dwarfs (L2, \S\ref{sec:impl}--\ref{sec:eval}) and their
   compositions (L3, \S\ref{sec:l3})---and that confirmation, with the
   conversion-bound regime now explicitly bounded, is the paper's
   theoretical content, stated formally as
   Proposition~\ref{prop:atomic}.
\item \textbf{The mechanism that makes the reduction efficient, not merely
   possible.}  We identify \textbf{on-chip tile fusion} (SMEM/TMEM-resident
   residue state; the register file holds only addresses and control,
   Appendix~\ref{app:beta-blackwell}) as the structural
   reason the bandwidth multiplier $\beta\!\to\!1$ for streaming kernels,
   keeping the residue/reconstruction traffic on-chip so that emulation is
   essentially free behind the memory wall, and we give the kernel-design
   discipline this requires (\S\ref{sec:impl}, Appendix~\ref{app:sparse}).
   This is what turns ``\fp{8} can emulate \fp{64}'' into ``\fp{8} keeps the
   memory-bound kernels at the memory roof.''
\item \textbf{Quantitative substantiation across architectures and a
   generational baseline.}  We project modeled upper-bound emulated \fp{64}
   throughput across the full operational-intensity spectrum for B300 and
   Rubin~R200 (\S\ref{sec:arch},~\S\ref{sec:eval}), re-baseline against the
   last HPC-balanced GPU (H100) to show emulation matches or exceeds it on
   every workload studied at the fused bound---and stays within $7\%$ of
   H100 scaling at the memo-counted reference deconstruction constants via
   the hybrid native/emulated path---while native B300 regresses by up to
   $\sim\!50\times$ (\S\ref{sec:h100baseline}), settle the INT8-versus-FP8
   substrate question as a division of labour
   (\S\ref{sec:int8vsfp8}), and---together with a companion FFT
   analysis~\cite{matsuoka2026fft} and FP32+Kahan compensation for BLAS-1
   reductions---close the kernel coverage so that, at the fused
   engineered-$c_q$ bound, every surveyed class
   reaches the memory roof at \fp{64}-grade accuracy (Grade~A matrix
   work, Grade-B compensated reductions)---with two bounded exceptions:
   large dense square DGEMM, whose on-the-fly-generated multi-cluster
   outputs sit at a cluster-aligned deconstruction floor near $0.50$ of
   the \emph{arithmetic} roof $P_{\fp{8}}/\alpha$ ($473$~TFLOPS on Rubin),
   not of the memory roof just named (a compute limit, L2-insensitive;
   Fig.~\ref{fig:dgemm}); and the 3-D FFT, whose per-output
   integer-SIMT epilogue holds software at $4.9$--$6.7\times$ its memory
   roof (still $1.3$--$1.9\times$ B300's collapsed native path) and whose
   roof is approached under the companion's minor-hardware rungs and
   reached with one moderate addition~\cite{matsuoka2026fft}---subject to the
   deconstruction condition of \S\ref{sec:model} (\S\ref{sec:l3}).
\item \textbf{The deconstruction analysis reaches dense linear
   algebra, not only the streamed kernels}
   (\S\ref{sec:rubinuplift}).  Although the fourth term binds the
   memory-bound kernels first, we show it also gates \emph{dense}
   Ozaki~II: attainment of the theoretical $P_{\fp{8}}/(3r{+}1)$ peak
   requires the outer-dimension harmonic mean to clear a crossover
   $n^{*}$ (Eq.~\eqref{eq:densecrossover}) that grows with the
   generation---${\approx}350$ on GB300, ${\approx}1{,}200$ on Rubin
   at memo-counted constants---so that moderate-size, tall-skinny, and
   batched GEMMs are conversion-bound even though nominally
   compute-bound.  Optimising the deconstruction path---the software
   ladder of \S\ref{sec:cqladder}, ultimately the hardware options of
   \S\ref{sec:hwproposal}---therefore extends near-peak dense
   emulation toward all sizes, and yields a testable account of
   NVIDIA's published Rubin Emulated-DGEMM figure as the fourth-term
   branch (Eq.~\eqref{eq:pdense}), conditionally worth
   $1.8$--$2.2\times$ by software alone at the announced operating
   region if a size sweep confirms that origin (floor
   ${\approx}1.76\times$ under FP8/INT8 tensor serialisation via the
   INT8-tensor-contention-free \texttt{dp4a} routes, and
   ${\approx}1.3$--$1.5\times$ if SIMT--\fp{8} overlap is also denied
   to both sides).  The dense-side
   development is now a standalone companion
   paper~\cite{matsuoka2026ozaki25}---exact two-limb residue
   formation, script-checked instruction counts, modulus/encoding
   codesign, shape-switched dispatch, and measured call-shape
   traces---whose corrected constants this version incorporates
   throughout.
\item \textbf{Potential hardware co-design, and the sparse/streamed
   Ozaki-2.5 discipline} (\S\ref{sec:hwproposal}).  Dividing the theory
   from the fourth term that limits it localises the co-design coordinates
   precisely.  \emph{On the hardware side}, we set out three concrete options
   for the deconstruction datapath---the strongest, \textbf{Option~C}, a
   stream-side residue-decompose transform on the copy path, retires the
   fourth term ($c_q{\to}0$) identically and un-binds the conversion-bound
   sparse kernels at once (Fig.~\ref{fig:optionc}); this is the $c_q$ half of
   the co-design, with the companion Ozaki~2.5 supplying the dense-DGEMM half
   (cluster reach, TMEM tile, and L2 bandwidth for the re-split floor), the
   two cross-referenced as one design space, none yet present in silicon.
   \emph{On the software side}, new to this version, we
   port the companion's findings to this paper's own territory:
   \emph{operand-stationary plane conversion} for fixed sparse
   operators, stated honestly as a bandwidth-for-instructions trade
   (convert the value array once at setup at ${\approx}3.8\times$ the
   nnz footprint; the planes are re-read every iteration, capping
   delivered throughput near $0.35$ of the raw-value \emph{memory} roof---worse
   than fused conversion on GB300, comparable on Rubin, leaving the
   streamed-vector share charged identically); dispatch of the
   width-bounded dense fraction of sparse solvers by $\bar{n}$ at
   the BLAS boundary, grounded in the companion's measured traces;
   and the \emph{sparse-specific codesign dividend}---the all-byte
   modulus set's pass-count surcharge is free for memory-bound
   kernels while its one-pass reduction legitimately restores a
   $c_q^{\text{res}} \approx 5.3$ per modulus, making the codesigned
   set the better default for the streamed class in its \texttt{dp4a}
   and lazy realisations (the inline tensor-migrated residual still
   favours the published set at the sets' own modulus counts).
\item \textbf{A clear theory/implementation division}
   (\S\ref{sec:formalclaim}).  This paper establishes the \emph{theory
   and its upper bound}: Proposition~\ref{prop:atomic} states the thesis
   as a conditional theorem, and the analytic sections verify its
   conditions (C1)--(C4) satisfiable in the model, class by class.
   Everything separating delivered silicon from the bound is an
   \emph{implementation artifact}---a measured value of the model's own
   free parameters $(\beta, c_q, P^{\text{eff}}_{\text{int}},
   \gamma)$---quantified for 2026 silicon in \S\ref{sec:cqladder}
   and measured end-to-end in Part~3
   (\S\ref{sec:futurework}).  The questions the structural claim leaves
   open are exactly those conditions, concentrated in
   \S\ref{sec:falsification} rather than
   threaded through the theory.
\end{enumerate}

We claim none of the underlying numerical machinery as our own: the Ozaki
scheme, the CRT-based Ozaki~II, the \fp{8} adaptation, the error
analysis, the cuBLAS integration, and the ADP-style accuracy guarantees
are due to the authors cited throughout
(Ozaki~II~\cite{ozaki2025_scheme2}, the \fp{8}
variant~\cite{uchino2026_fp8,mukunoki2025_fp8},
error analysis~\cite{ozaki_error_analysis_2026}, the cuBLAS
integration~\cite{nvidia_cublas_emulation}, and ADP~\cite{schwarz2025}).
Our contribution is to assemble them into a single structural
claim about scientific computing---FP8 as the candidate dominant
matrix-work substrate for the surveyed classes---and to
build the model, the kernel discipline, the cross-architecture projections,
and the completeness audit needed to establish it.

\paragraph{Why this is an opportunity, not a threat.}  The thesis above
inverts a widely held fear.  The AI market's decisive shift toward
low-precision \emph{floating-point}---\fp{8} for inference, then for
training~\cite{deepseekv3}, and now four-bit NVFP4 pretraining matching an
\fp{8} baseline~\cite{nvidia_nvfp4_2025}---was widely expected to
\emph{compromise} double-precision science and to fork the product line
into AI and \fp{64}-preserving SKUs.  We argue the opposite, and more
strongly than a compatibility claim: the very \fp{8} silicon that the AI
workloads are paying for is not merely \emph{adequate} for \fp{64} science
but is its natural and sufficient foundation.  The mechanism, stated once
for the whole paper, is that emulation \emph{extends the achievable compute
ceiling of the roofline}: on a \fp{64}-collapsed part the native ceiling
has fallen \emph{below} the memory roof, forcing physically memory-bound
kernels to become compute-bound on the scant \fp{64} pipe; Ozaki~II lifts
that ceiling back above the memory roof so the kernels are \emph{recovered}
to their natural bandwidth-bound regime.  What remains is kernel-engineering
investment to realise the FP8 building blocks---a well-scoped, tractable
implementation programme (\S\ref{sec:futurework}), motivated by the imminent
ubiquity of FP8-rich silicon.

\subsection{What This Paper Establishes: Theory as Upper Bound, Implementation as Artifact}
\label{sec:formalclaim}

This is, first and deliberately, a \emph{formalisation} paper, and the
point deserves its own subsection because everything else in the paper
is organised around it.  What we establish is a \textbf{theoretical
framework}---the Tensor--Memory Equilibrium model, with its four
emulation parameters $(\alpha,\beta,\gamma,c_q)$ and its architecture
parameters $(P_{\text{low}}, P_{\text{int}}, B_{\text{mem}})$ held
symbolic---within which the following can be stated and checked as a
mathematical claim rather than a benchmark: \emph{low-precision
matrix-multiply---as low as \fp{8}---can serve as the dominant
matrix-work substrate for the surveyed matrix-dominated \fp{64} kernel
classes, under the model's conditions and without performance sacrifice
at its upper bound}.  ``Without performance sacrifice'' has a precise meaning at
the theory level: at the model's upper bound, every surveyed kernel
class attains exactly the performance an ideal machine would deliver
---the memory roof where the kernel is bandwidth-bound, the emulation
ceiling where it is compute-bound---so the reduction from \fp{64}
semantics to the \fp{8} primitive costs \emph{nothing} at the bound.
Proposition~\ref{prop:atomic} in \S\ref{sec:model} states this formally
as a conditional theorem, with four explicit conditions (C1)--(C4); the
analytic content of \S\ref{sec:impl}--\S\ref{sec:l3} is the
verification, class by class, that those conditions are
\emph{satisfiable in the model} with stated margins.  The framework is
deliberately a two-way instrument: because every architectural quantity
enters as a fitted parameter, the same model that confirms the
conditions on one machine can refute them on another, the verdict is
reached analytically wherever the margins are wide, and---whichever way
it falls---it arrives with the responsible term attached, so that
implementation measurement, when needed, tests a named constant rather
than a vague hypothesis.

\begin{center}
\fbox{\parbox{0.92\linewidth}{%
\textbf{The defensible core claim, stated once.}  \emph{For the
surveyed matrix-dominated kernel classes, and under the stated
storage, overlap, and numerical conditions, FP8 Ozaki-II is a
candidate dominant multiplication substrate: fixed-width integer
operations, \fp{32}/Kulisch reductions, data movement and control,
and native fallback remain explicit, enumerated auxiliary
mechanisms.}}}
\end{center}

\noindent
The title names the research \emph{program}---the conditional theorem
at the model's bound, and the engineering agenda for closing its
conditions; the boxed statement is the \emph{present result}, which
every projection in this paper supports and no claim exceeds.

The second half of the discipline is an equally explicit division:
\textbf{everything that separates delivered silicon from the bound is
an implementation artifact, and every artifact is a parameter of the
same model}---the delivered $\beta$, the engineered deconstruction cost
$c_q$, the effective integer budget $P^{\text{eff}}_{\text{int}}$, the
delivered tile efficiency, the reconstruction throughput behind
$\gamma$.  Artifacts are not refutations of the theory; they are
measured values of its free parameters on particular hardware, and they
change with engineering effort and with hardware generation (as this
version's Rubin correction and the companion's audited deconstruction
counts~\cite{matsuoka2026ozaki25} both illustrate).  The paper therefore keeps the
two layers physically separate: the theory and its bound live in
\S\ref{sec:dwarfs} and \S\ref{sec:model} and are architecture-independent;
today's artifacts are quantified in one section, \S\ref{sec:eval}
(the constants in \S\ref{sec:cqladder}, their consequences in
\S\ref{sec:rubinuplift}), against 2026 silicon; the hardware options that would
eliminate the largest of them are \emph{set out} here
(\S\ref{sec:hwproposal})---dividing the theory from its limiting term is what
localises them---with their systematic end-to-end measurement and evaluation
the declared subject of the Part~3 companion paper.  A reader who accepts the theory but doubts the constants and a
reader who accepts today's constants but doubts the theory are, by
construction, disagreeing with different sections.

\subsection{Why This Revision: the Deconstruction Term and the Rubin FP8 Correction}
\label{sec:whyupdate}

This version of the paper differs from its predecessors in two
substantive respects---one theoretical, one observational---and both
reasons deserve a plain statement up front.  The first is a first-order
correction to the paper's own model.
The TME model as previously published charged the tensor-core work
($\alpha$), the HBM traffic ($\beta$), and the \emph{output}-side
reconstruction ($\gamma$ per output element)---but not the
\emph{input}-side \textbf{deconstruction}: before a single MMA can issue,
every streamed \fp{64} operand element must be scaled, rounded, and
reduced modulo each of the $r$ moduli, elementwise and nonlinear work
that runs on the SIMT integer pipe, not on the tensor cores.  As the
technical review by Bayraktar, Gunnels, and Caday of
NVIDIA~\cite{bayraktar2026tme} observed, this term is the exact dual of
$\gamma$---reconstruction is charged per output, deconstruction per
input---and for memory-bound kernels the number of inputs exceeds the
number of outputs by orders of magnitude, so the omission was
\emph{first-order}, not a refinement.  A model that omits a first-order,
architecture-independent term is not an architecture-independent model;
it is an incorrect one.  The term therefore belongs in this paper, in
symbolic form, even though its \emph{constants} are
architecture-specific.

The implications are material and are carried throughout this version.
The deconstruction cost defines a threshold operational intensity
$\mathrm{OI}^{*}$ (\S\ref{sec:model}) below which an emulated kernel is
\emph{conversion-bound}: the integer pipe, not the memory system or the
tensor cores, is the binding resource, and emulation cannot pay for
itself against even a collapsed native \fp{64} pipe.  At the reference
constants currently counted on the shipping cuBLAS emulation
path~\cite{bayraktar2026tme}, the lowest-intensity kernels---SpMV and
single-right-hand-side GEMV---fall below $\mathrm{OI}^{*}$ and are
accordingly re-assigned in the claims tables to the surviving native
\fp{64} pipe, and the batched-GEMV and stencil headlines are reduced.
This verdict is not the last word, because $\mathrm{OI}^{*}$ is itself a
co-design target from two sides. In software, the dominant sparse pattern
is \emph{operand-stationary}: a fixed matrix applied repeatedly (Krylov and
iterative-refinement solvers, the same $A$ across dozens of SpMVs), for
which \emph{deferred/persistent} deconstruction---convert $A$'s residue
planes \emph{once} and reuse them across every application (Mode~P,
\S\ref{sec:hwproposal})---amortises the conversion \emph{instructions}
toward zero across the iteration. This is not a free rescue but a
bandwidth-for-instructions trade: the persistent planes stream at $p$
bytes/nonzero against $8$ ($\times2.83$ for set~S), so it lifts these
kernels off the conversion-bound floor only when the plane array is
cache/L2-resident (or when SIMT issue, not memory, is the bottleneck). For
an $A$ too large to keep resident the planes cost $p/8\times$ the native HBM
traffic and the verdict can revert; and it does nothing for a single SpMV or
a changing $A$ (each new pattern reconverts)---there the fused,
engineered-$c_q$ kernel is the mitigation. Where it does apply, the plane
array's residency, not the per-call conversion, governs. In hardware,
lowering $c_q$ directly (the deconstruction-path options of
\S\ref{sec:hwproposal}) moves the threshold for the single-shot case
too---the same $c_q$ lever the companion Ozaki~2.5 uses to lift the dense
floor. Whether the shipping
constants are a floor or an artifact of a code path never
engineered to minimise deconstruction is a measurable question---the
review itself states the two-sided decision criterion (a conversion kernel
cheaper than $c_q\!\approx\!7$ integer instructions per modulus per
element restores single-RHS GEMV parity; one at or above it confirms
the conversion-bound verdict as delivered)---and it is the first
experiment of the measurement programme.  The engineering of that
conversion path has since been developed into a standalone companion
paper, \emph{Ozaki~2.5} (submitted to
arXiv~\cite{matsuoka2026ozaki25}); this version incorporates its
corrected, script-checked instruction counts throughout
(\S\ref{sec:cqladder}) and develops their sparse and streamed
implications.  With the companion in place, the performance-projection
section has been restructured accordingly: the realistic constants are
established \emph{first} (\S\ref{sec:cqladder}), imported rather than
re-derived, and the projections, substrate discussion, and hardware
proposal then follow from them; the provenance of the memo-counted
shipping constants is summarised in Appendix~\ref{app:memo}.  A
subsequent external review prompted, and this revision adopts, the
storage-mode and tensor-concurrency analyses of
\S\ref{sec:hwproposal}, the algorithm-attribution care of
\S\ref{sec:model} (Scheme~I versus~II on the shipping anchors), the
conditional restatement of Proposition~\ref{prop:atomic}, and
rebuilt kernel byte ledgers throughout.  Two further corrections from
the same review are adopted: the tensor-core tile-granularity charge
$W_{\text{mma}}=(8/B)\,W$ for kernels with $B<8$ right-hand sides, and
corrected batched-GEMV operational intensities ($B/4$~\flops{}/B for
\fp{64}-streamed operands).

The second mandate is observational, and it is drastic.  On June~30,
2026---after the June~8 and June~16 versions of this paper, and days
before the July~3 one---NVIDIA published the DGX Rubin NVL8 product
specification, whose figures fix Rubin's dense \fp{8} tensor rate at
\textbf{17.5~PFLOPS per GPU}~\cite{nvidia_dgx_rubin_nvl8}:
\emph{$4.4\times$ the $\sim\!4$-PFLOPS figure all earlier versions of
this paper carried} from third-party compilations.  A projection paper
whose subject architecture's headline rate moves by $4.4\times$ is
obligated to re-issue: the Rubin Ozaki~II ceiling rises from
$\approx\!108$ to $\approx\!473$~TFLOPS, NVIDIA's published
Emulated-DGEMM figure moves from apparently \emph{exceeding} the model
bound to sitting comfortably inside it at $\approx\!42\%$ (a
cross-check of the
$(3r{+}1)$ cost structure---and, as \S\ref{sec:rubinuplift} now shows,
a gap of exactly the magnitude the deconstruction term itself
generates at moderate problem sizes, raising the testable possibility
that the fourth term, not the tensor cores, is what the published
figure is measuring), and the generational narrative
inverts---emulation margin now grows with the generation
(\S\ref{sec:cqladder}).  The two mandates interact, and
\S\ref{sec:cqladder} analyses the interaction in detail: the \fp{8}
uplift supplies headroom that can absorb much of the deconstruction
\emph{arithmetic} onto the tensor pipes---while the accompanying
bandwidth growth makes the SIMT-side residual of the fourth term
\emph{more} binding, not less, sharpening rather than dissolving the
hardware question below.

What does \emph{not} change is the structure of the thesis---but it
sharpens.  If deconstruction is the one place where the FP8-primitive
stack genuinely needs help on low-intensity kernels, then the hardware
question stops being ``how much \fp{64} silicon must survive?''\ and
becomes ``how cheaply can the deconstruction path be provided?''---by a
larger integer-issue budget, by a fixed-function
\fp{64}$\to$residue conversion datapath (a cvt-class unit,
hypothesized to cost far less area than an \fp{64} ALU array; PPA is
a Part-3 measurement), or by the small surviving native pipe that
B300 in fact retains.  $\mathrm{OI}^{*}$ is precisely the equation that
sizes these alternatives.  The architecture-specific measurement of the
new constants ($c_q$, the integer-issue budget $P_{\text{int}}$, the
converted fraction $\varphi$, delivered tile efficiency), and the
evaluation of these hardware remedies, are deliberately \emph{deferred}
from this paper: they are the subject of a \textbf{Part~3} companion
paper, now in preparation for rapid publication,
with substantially expanded
experiments and scrutiny on GB200-class systems.  This paper retains the
architecture-independent formulation---the four-term model with $c_q$
and $P_{\text{int}}$ as symbolic parameters---and states every claim at
both the fused upper bound and the memo-counted reference constants.

\subsection{The FP8 Primitive Hierarchy and the Berkeley Dwarfs}
\label{sec:dwarfs}

\begin{figure}[htbp]
\centering
\includegraphics[width=\linewidth]{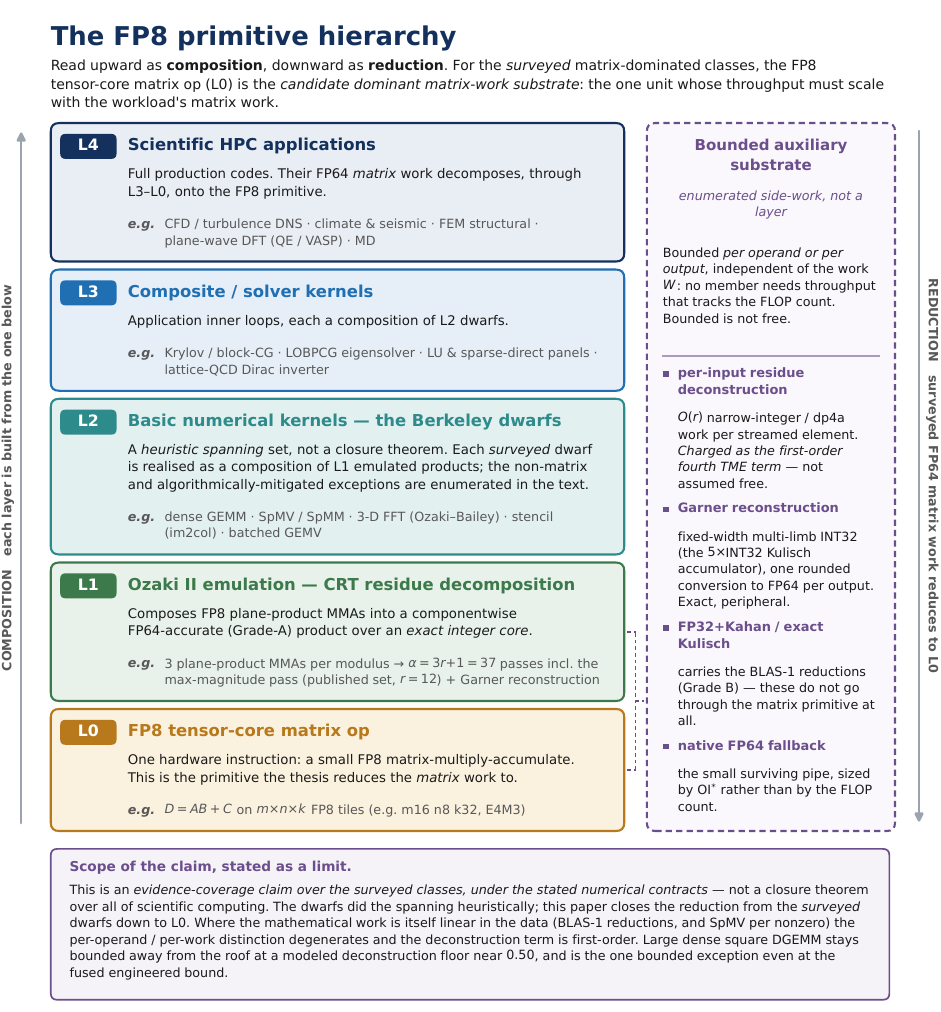}
\caption{The FP8 primitive hierarchy.  Read upward as composition and
downward as reduction.  L0, the FP8 tensor-core matrix-multiply, is the
\emph{candidate dominant matrix-work substrate} for the surveyed classes; higher
layers also require the explicitly enumerated auxiliary operations and fallback
paths described in the text.  The Berkeley dwarfs are layer~L2; this paper
establishes that the \emph{surveyed} L2 dwarfs---and hence the surveyed
composite kernels (L3) and the applications (L4) built from them---reduce
through the Ozaki~II layer (L1) to the L0 primitive at \fp{64}-grade accuracy
\emph{under the stated numerical contracts}, with the non-matrix and
algorithmically-mitigated exceptions enumerated explicitly (this is an
evidence-coverage claim over the surveyed classes, not a closure theorem over
all of scientific computing).  The integer part of that
enumerated auxiliary set is bounded, fixed-width side-work: the
fixed-width \emph{multi-limb} integer accumulation of Garner
reconstruction (narrow per-digit arithmetic; the reconstructed value
spans multiple 32-bit limbs, e.g.\ the $5{\times}$INT32 Kulisch
accumulator of the companion paper, with one rounded conversion to
\fp{64}) and the per-input residue
deconstruction of \S\ref{sec:model}, both exact and peripheral in
arithmetic---though the latter's throughput cost is charged as a
first-order model term.}
\label{fig:hierarchy}
\end{figure}

The structure of the thesis is a strict compositional hierarchy, shown in
Figure~\ref{fig:hierarchy}, and it is most naturally understood as the
completion of a programme begun by the Berkeley
``dwarfs''~\cite{asanovic2006berkeley,colella2004dwarfs}.  That work
identified the recurring patterns of computation in scientific and
parallel computing---a small, \emph{heuristic spanning} set of equivalence classes
(dense and sparse linear algebra, spectral methods, structured and
unstructured grids, particle methods, and so on)---deliberately at a
rough, architecture-agnostic level, so that one could reason about
scientific computing at the right altitude: above the particulars of any
one machine, below the particulars of any one application.  The dwarfs
answered \emph{``what are the recurring kernels?''}  They left open a
question that the post-\fp{64} era makes urgent: \emph{what low-precision
matrix primitive is sufficiently general to carry the matrix work of the
surveyed classes at \fp{64}-grade accuracy under the stated contracts
(\S\ref{sec:background})?}  This paper proposes the FP8 tensor-core matrix
op as the candidate dominant substrate, while retaining the enumerated
auxiliary operations and fallback paths, and in doing so reduces the
matrix work of each class to the FP8 primitive.

\paragraph{What ``primitive'' means---the definition the thesis rests
on.}  Before the hierarchy, the load-bearing definition.  We call an
operation the \emph{primitive} of a computation if its executed count
scales proportionally with the mathematical work $W$; we call
operations whose count is \emph{bounded per operand or per output},
independent of $W$---here the $O(r)$ narrow-integer deconstruction per
streamed element, the $O(r)$ reconstruction per output, addressing, and
control---the \emph{bounded substrate}.  One scope limit must be
stated at the definition, not discovered later: for kernels whose
mathematical work is itself linear in the data---BLAS-1 reductions
and, per nonzero, SpMV---bounded-per-operand work is $\Theta(W)$, and
the per-operand/per-work distinction degenerates.  That is precisely
why the fourth TME term is first-order for the streamed class, and
why the thesis for that class is not ``the substrate is
asymptotically free'' but ``the substrate's demand is bounded per
element and providable by minor, non-scaling hardware,'' with the
BLAS-1 reductions carried by an explicitly enumerated auxiliary
substrate (\fp{32}+Kahan or the exact Kulisch route) rather than by
the matrix primitive at all.  The distinction is otherwise the same
one always applied, silently, to classical machines: a computation
whose numerical work reduces to \fp{64} FMAs was never said to have
``multiple primitives'' because it also performs address arithmetic,
loads, and stores.  The thesis of this paper is that, \emph{for the surveyed matrix-dominated
classes}, \emph{the \fp{8} tensor-core matrix op is the sole matrix-work
primitive in this defined sense}: the
one unit whose throughput must scale with the workload's matrix work,
supported by a fixed, enumerated auxiliary substrate (narrow
integer/\texttt{dp4a} deconstruction, \fp{32}/Kulisch reductions,
conversion data movement, control, and a native fallback) none of
whose members needs throughput that tracks the FLOP count.  The bounded
substrate is not free---the fourth term of \S\ref{sec:model} is
precisely its price list, and \S\ref{sec:cqladder} and
\S\ref{sec:rubinuplift} are the study of
how to pay it---but ``charged'' and ``co-primitive'' are different
claims: the substrate's demand is bounded per element, providable by
minor pipes, narrow dot-product instructions, fixed function, or the
primitive itself (\S\ref{sec:cqladder} and \S\ref{sec:hwproposal}
instantiate these on
current silicon), and no unit carrying it needs throughput that tracks
the FLOP count.

The hierarchy has five layers.  At the bottom, \textbf{L0} is the dominant
hardware matrix primitive: a single FP8 matrix-multiply-accumulate
$D = AB + C$ on small tiles (e.g.\ $m16\,n8\,k32$, E4M3).  Directly above
it, \textbf{L1} is the Ozaki~II scheme, which composes three \fp{8}
plane-product MMAs per modulus---$3r{+}1$ MMA instructions in total
including the max-magnitude pass---followed by Garner reconstruction, over an
exact integer core, into a componentwise \fp{64}-accurate (Grade-A)
matrix product (\S\ref{sec:background}).  \textbf{L2} is
the set of basic numerical kernels---\emph{the dwarfs themselves}: dense
GEMM, SpMV and SpMM, the three-dimensional FFT, structured stencils,
batched GEMV---each realised as a composition of L1 emulated products.
\textbf{L3} is the layer of composite and solver kernels that constitute
the inner loops of real applications---Krylov and block-CG iterations,
LOBPCG eigensolvers, dense and sparse-direct factorisation panels, the
lattice-QCD Dirac inverter---each a composition of L2 dwarfs;
\S\ref{sec:l3} substantiates this layer directly.  Finally, \textbf{L4} is
the full scientific application---computational fluid dynamics and
turbulence DNS, climate and seismic codes, finite-element structural
analysis, plane-wave density-functional theory, molecular dynamics---whose
\fp{64} work decomposes, through L3 and L2 and L1, onto the L0 primitive.

The picture \emph{is} the thesis, and its force comes from the
exhaustiveness the dwarfs already established.  Their taxonomy was
proposed as spanning scientific computing to first order; to the extent
it does, if every dwarf (L2) reduces to the FP8 primitive through
Ozaki~II, then by composition every L3 kernel and L4 application built
from those dwarfs reduces to it as well.  We are not claiming to have
inspected every code in existence---we are observing that the dwarfs did
the spanning, and closing the reduction from the dwarfs down to a single
operation one layer below them.  In the conventional view, L2 would rest
on a hardware \fp{64} multiplier; the contribution of this paper is to
show that it rests instead on the FP8 tensor-core op, with \fp{64}
demoted from a hardware primitive to a derived guarantee.  We believe such
a top-down, primitive-level treatment has been conspicuously absent from
recent debates about reduced precision and AI--HPC convergence, which tend
to begin and end at the level of individual kernels or individual
architectures.

This is a falsifiable claim, not a slogan, and the Tensor--Memory
Equilibrium model of \S\ref{sec:model} is the instrument built to test
it.  The claim is operational: \emph{every dwarf, and every composition of
dwarfs, must stay on the bandwidth-bound side of the roofline (or remain
otherwise \fp{64}-competitive) when built on FP8 Ozaki~II}.  That is a
statement about performance characteristics---operational intensity, the
emulated compute ceiling $P_{\fp{8}}/\alpha$, the bandwidth multiplier
$\beta$, the reconstruction cost $\gamma$, the deconstruction cost
$c_q$---and a performance model is
exactly the right tool to confirm or refute it, provided it captures those
quantities.  The test proceeds in two stages: the fitted model decides
most kernel classes analytically, and where the margin is thin, the
Part~3 measurements identify the term on which the class stands or
falls.  Were the model to show any canonical kernel driven
compute-bound on the FP8 ceiling, or any composition forced persistently
past the memory-bound threshold with no architectural escape, or any
kernel demanding a primitive that is neither an FP8 matrix op nor the
peripheral INT32 reduction, the thesis would be false and the model would
say so.  That it instead returns a positive verdict across the dwarfs is
the theoretical content of this paper, formalised as
Proposition~\ref{prop:atomic}.  We are explicit about the division
of labour (\S\ref{sec:formalclaim}): \emph{this paper establishes the
theory and its upper bound; what separates any given silicon from the
bound is an implementation artifact}, quantified in
\S\ref{sec:cqladder} and measured on production
silicon in Part~3 (\S\ref{sec:futurework}).  The questions that this
structural claim leaves genuinely open---chiefly the efficiency with which
each reduction is realised, captured by $\beta$ under composition---are
second-stage matters of architecture and engineering layered on top of the
bound, and we treat them as such, concentrating them in
\S\ref{sec:falsification} rather than threading caution through the
analysis.

\paragraph{Roadmap.}
\S\ref{sec:background} reviews the two Ozaki schemes (the L1 layer) and
related FP64-emulation work.  \S\ref{sec:arch} establishes the
architectural baselines.  \S\ref{sec:model} develops the TME performance
model.  \S\ref{sec:impl} gives the design of the basic memory-bound
kernels (L2).  \S\ref{sec:eval} reports the projections.  \S\ref{sec:l3}
substantiates the composite/solver layer (L3), showing that the kernels
representative of real applications reduce to the same primitive.
\S\ref{sec:discussion} discusses the broader landscape and implications,
and \S\ref{sec:falsification} states the conditions that decide the
thesis---confirming or falsifying it---and confronts the principal
objections to it.
\S\ref{sec:conclusion} concludes.  Appendices contain the error-analysis
sketch, corrected pseudocode, Garner's derivation, and a detailed analysis
of the sparse primitive establishing that the bandwidth multiplier remains
near unity for sparse kernels (Appendix~\ref{app:sparse}).

\begin{table}[h]
\caption{Notation systems used in this paper (and the companion),
gathered in one place.  The hierarchy-layer and ladder-rung namespaces
are distinct: hierarchy layers L0--L4 appear throughout; ladder rungs
L1--L4 appear only in the deconstruction context of
\S\ref{sec:cqladder}--\S\ref{sec:rubinuplift}.}
\label{tab:glossary}
\centering
\scriptsize
\setlength{\tabcolsep}{4pt}
\begin{tabular}{p{3.2cm}p{10.6cm}}
\toprule
\textbf{system} & \textbf{members} \\
\midrule
Hierarchy layers (\S\ref{sec:dwarfs}) & L0 FP8 MMA $\cdot$ L1 Ozaki~II $\cdot$ L2 kernels (dwarfs) $\cdot$ L3 composite solvers $\cdot$ L4 applications \\
Software ladder (\S\ref{sec:cqladder}) & S0 shipping ($c_q{=}16$, memo-counted) $\cdot$ L1 direct \texttt{dp4a} ($10.3$) $\cdot$ L2 two-limb tensor migration ($6.3/5.8$) $\cdot$ codesign ($5.3$) $\cdot$ L3 lazy reduction ($4.3$) $\cdot$ L4 Kulisch-carrier reconstruction \\
Modulus sets (companion) & S published ($r{=}12$, $30$~B/scalar) $\cdot$ A all-byte ($r{=}15$, $44$~B) $\cdot$ E hybrid ($r{=}13$, $33$~B) $\cdot$ D signed-byte ($r{=}17$) \\
Storage modes (\S\ref{sec:hwproposal}) & F fused (planes in CTA/cluster, multiplicity $\lambda$) $\cdot$ M materialised (global workspace) $\cdot$ P persistent (across calls) \\
Hardware options (\S\ref{sec:hwproposal}) & A narrow-integer issue rebalance $\cdot$ B \texttt{cvt.fp64.residues} instruction $\cdot$ C stream-side (TMA) conversion \\
Accuracy grades (\S\ref{sec:background}) & Grade A exact-product $\cdot$ Grade B compensated $\cdot$ IEEE-behaviour excluded $\cdot$ native fallback \\
Platforms & H100, B200 (measured anchors) $\cdot$ B300/GB300 (``B300'' columns: pre-freeze illustrative constants; the companion freezes the GB300 SKU as \emph{NVL72-derived per-GPU} values---$5.0$~PF \fp{8}, $166.7$~\tops{} \inteight{}, $1.4$~TF \fp{64}, $8$~TB/s; the HGX~B300 SKU differs---and all of \S\ref{sec:cqladder} onward uses the frozen values, deltas $\le 7\%$) $\cdot$ R200/Rubin (preliminary announced specifications, ``up to'') \\
\bottomrule
\end{tabular}
\end{table}

\subsection{What Has Changed Since the arXiv Public Release}
\label{sec:since-arxiv}

The arXiv public release of this paper (arXiv:2606.06510v3, July~3,
2026) predates the NVIDIA technical review and everything that
followed from it.  A reader returning from that version should treat
the present revision as a substantially deeper paper making a more
precisely scoped claim, and the distance between the two is itself a
result worth recording.  Five arcs summarise it.  \emph{(i)~The model
gained its fourth term}: the per-input deconstruction cost $c_q$,
identified by the NVIDIA technical review of Bayraktar, Gunnels and
Caday~\cite{bayraktar2026tme}, now prices the conversion path on both
sides of the tensor cores, defines the conversion-bound regime below
$\mathrm{OI}^{*}$, and turned out to be the single most productive
correction in the paper's history---nearly everything below traces
back to it.  \emph{(ii)~The dense-side analysis became a companion
paper}: \emph{Ozaki~2.5}~\cite{matsuoka2026ozaki25} develops the
deconstruction path as an engineering object in its own right---an
exact carry-corrected residue-formation layout (its Lemma~1) closing
what were previously open obligations, a route menu with per-route
schedules and accuracy contracts, an honest cluster-aligned
dense-DGEMM floor at $0.50$ of the arithmetic roof with an HPL worked
example, and the preferred co-design target (Option~C) that lifts the
floor to the roof.  \emph{(iii)~The FFT question was answered, in
both directions}: the spectral companion (Part~2,
arXiv:2606.23698~\cite{matsuoka2026fft}) constructs a full-\fp{64}
$1024^3$ FFT with no \fp{64} arithmetic anywhere in its kernel, and
finds---against this paper's earlier optimism---that on shipping
silicon its binding resource is a per-output integer-SIMT epilogue
holding software at $4.9$--$6.7\times$ the memory roof (still
$1.3$--$1.9\times$ B300's collapsed native path); the same
closed-form floor names the hardware that removes the wall---minor
rungs to within $1.2$--$1.8\times$ of the roof, one moderate rung to
reach it---and its load-path rung is this paper's own Option~C
converter.  The thesis accordingly sharpened rather than weakened: four
of the five canonical primitives reach the memory roof in software, and
the fifth is \emph{effectively reachable with minor,
independently-motivated hardware plus one moderate addition}.  \emph{(iv)~Every load-bearing constant acquired a
machine-checked owner}: a shared cost engine and a consistency suite
of several hundred checks (the artifact's \texttt{README} states the
exact count for the shipped revision) now re-derive every printed number in this paper
and the Ozaki~2.5 companion from one source, sweep the companion
document set for retracted wordings, and gate the released bundle on
a one-command reproduction (Part~2 ships the parallel
\texttt{ledger.py}\,/\,\texttt{verify.py} apparatus).
\emph{(v)~The revision history itself became part of the method}:
successive review rounds by independent reviewers are dispositioned
item-by-item in a released response letter, and revision identifiers
now change whenever content does.  The paragraphs below record the
per-revision detail.

\paragraph{Changes in this version (Revision~32, July~25, 2026).}
This revision upgrades the TME model
from three emulation parameters to four, incorporating the input-side
\emph{deconstruction} (conversion) term identified in the technical review
of the July~3 version (arXiv:2606.06510v3) by Bayraktar, Gunnels, and
Caday of NVIDIA~\cite{bayraktar2026tme}: every streamed operand element
must be scaled, rounded, and reduced modulo each of the $r$ moduli on the
SIMT integer pipe before any tensor-core MMA can issue, a cost that is
first-order for memory-bound kernels and that earlier versions omitted
(\S\ref{sec:whyupdate}, \S\ref{sec:model}).  The same review's
tile-granularity correction, $W_{\text{mma}}=(8/B)\,W$ for kernels with
fewer than eight right-hand sides, is also adopted.  The model now
carries an explicit conversion-bound regime below the threshold intensity
$\mathrm{OI}^{*}$, the claims tables are re-stated at both the fused
upper bound and the currently \emph{memo-counted} reference deconstruction
constants, and the batched-GEMV operational intensities are corrected to
$B/4$.  A second, independent mandate for this version: the Rubin
dense-\fp{8} rate is corrected from the
superseded $\sim\!4$~PFLOPS of earlier versions to the official
17.5~PFLOPS (NVIDIA DGX Rubin NVL8 specification, June~30,
2026~\cite{nvidia_dgx_rubin_nvl8}, postdating v1--v2)---a $4.4\times$
revision that raises the Rubin
Ozaki~II ceiling to $\approx\!473$~TFLOPS, reconciles the published
$\sim\!200$-TFLOPS Emulated-DGEMM figure with the model bound, and
opens tensor-side headroom whose interaction with the deconstruction
term---what the uplift can absorb, and what it fundamentally
cannot---is analysed in \S\ref{sec:cqladder}.  The engineering of the
deconstruction path is now a standalone companion paper
(\emph{Ozaki~2.5}~\cite{matsuoka2026ozaki25}); this version imports
its script-checked constants, restructures the projection section
around them, and summarises the memo provenance in
Appendix~\ref{app:memo}.  A same-day external review prompted further
repairs, all adopted: explicit \emph{storage modes} for the residue
planes (fused/materialised/persistent, with plane-traffic bounds), an
FP8/INT8 tensor-service \emph{concurrency bracket} with its
INT8-tensor-contention-free \texttt{dp4a} floor, algorithm-attribution care for
the shipping-path anchors (Scheme~I versus~II), a conditional
restatement of Proposition~1, and rebuilt kernel byte
ledgers.  A second round (July~22) corrected the storage model's
plane-traffic factor (write \emph{and} read), added L2-resident and
split-$k$ schedule regimes with per-shape eligibility predicates,
bracketed SIMT--\fp{8} overlap with consistent-worlds floors, and
recomputed the trace composites under per-record schedule
filters.  A third round (July~22) gave the reconstruction term a
route-specific rate bound (\textsc{int8}-hosted mixed radix versus
SIMT Garner, with corrected $k$-thresholds), made the additive
full-traffic memory ledger explicit (with an audit that, on a later
re-check, turned out to have been reported wrongly---see the seventh
round below), replaced the coverage figure
with a boundary-enumerated switched minimum, corrected the
compute-balance band to $56$--$333{:}1$, added per-kernel residency and
split-$k$ resource ledgers, and hardened the CRT bound with an explicit
$\mu$-form.  A fourth round (July~23) corrected that coverage figure
again: an intermediate version had charged a plane-\emph{feeding} L2
bandwidth to a \emph{materialised} large-output schedule (yielding an
``$0.79$, L2-bandwidth-conditional'' minimum), but materialisation is
feed-starved and the design generates residues on the fly, so the
operative bound is the on-the-fly re-split (deconstruction-$\lambda$)
load---a cluster-aligned ${\approx}0.50$-of-roof floor for large dense
DGEMM (against Rubin's $473$-TFLOPS \emph{arithmetic} roof),
\emph{insensitive} to L2 bandwidth---and the cluster reach
governing that re-split was fixed from $1024$ to the correct $256$
(Appendix~\ref{app:beta-blackwell}, Fig.~\ref{fig:dgemm}).  A fifth round
(July~24, Rev~30) unified that floor with the crossover as a single
deconstruction curve (the $\lambda{=}1$ crossover is the convert-once upper
envelope, the floor its multi-cluster continuation), scoped the
falsifiable-prediction knee to its storage mode, added the differentiated
hardware co-design---$c_q$ here as the sparse-critical lever
(\S\ref{sec:hwproposal}), reach/L2 in the companion Ozaki~2.5 for dense
DGEMM---and reframed the Mode-P sparse mitigation as a
bandwidth-for-instructions trade contingent on plane residency. A sixth
round (Rev~31) states the deconstruction ceiling \emph{up front}---the honest
large-DGEMM floor ($0.50$ of the $473$-TFLOPS arithmetic roof) before the
constructions that raise
it---mirroring the companion Ozaki~2.5's reframe into a design-space map
(route menu, limitations table, an HPL worked example, and the
preferred co-design Plan~A/Plan~B to reach the roof).  A seventh round
(Rev~32) corrected two omissions in the companion engine's storage
ledger that the third round's audit had reported as absent: the
HBM-materialisation branch never charged the $c{=}2p$ plane traffic it
claimed to, and the L2-resident branch charged a streaming feed
smaller than one pass over the workspace below $\bar n = 64$.  Both are
now charged, which retires the HBM-materialised branch from the trace
set entirely and moves one composite (CCSD on Rubin,
${\approx}1.73\times \to {\approx}1.69\times$); the audit is now a test
in the companion artifact rather than a claim in the text.
Prior-version history: the June~8 preprint (v1) used $\alpha=r$;
the June~16 (v2) and July~3 (v3) revisions corrected the \fp{8}
compute-multiplier to $\alpha=3r{+}1$ (dense B300 ceiling
$\sim\!135$~TFLOPS, $\sim\!104\times$ native, not $\sim\!380\times$),
repositioned the contribution in the Berkeley-``dwarfs'' tradition, and
added the per-kernel $\beta$-budget analysis of the Blackwell-native
dataflow (Appendix~\ref{app:beta-blackwell}).  Architecture-specific
measurement of the new constants, and the hardware remedies they
motivate, are deferred to a rapidly forthcoming Part~3
(\S\ref{sec:whyupdate}, \S\ref{sec:futurework}).

\paragraph{Changes in this version (Revision~33, July~28, 2026).}
An eighth round retires a withdrawn per-copy-engine hardware-width claim
that had resurfaced in paraphrase ---
``a few hundred narrow MACs in a copy engine'' --- after the exact phrase
``${\approx}300$ narrow MACs'' was already retired in an earlier
revision; the discussion now states only the derivable aggregate
transform rate ($166$--$325$~TMAC/s) and leaves per-engine width to
Part~3.  Three PPA-adjacent phrasings that read as established results
rather than hypotheses under the model --- the conversion datapath
called ``far smaller than an \fp{64} ALU array,'' Plan~A/Plan~B called
``minimal-hardware,'' and the thesis phrase
``a cheap deconstruction path'' --- are qualified or reworded so that
area and power stay explicitly Part-3, matching the companion
Ozaki~2.5's parallel fixes in its Draft~22.  This revision number itself
is part of the fix: two independent final-phase reviews of Revision~32
found it no longer identified a single document, since further wording
corrections had been applied under an unchanged revision label between
one reviewed PDF and the next. The retired-wording regression test
(\texttt{verify/paper\_numbers.py}) is extended to cover the paraphrases
this round retracts, not only their first-quoted forms.

\paragraph{Changes in this version (Revision~34, August~4, 2026).}
A ninth round synchronises this paper with the final (v3) revision of
the spectral companion Part~2~\cite{matsuoka2026fft}, whose own
five-round correction cycle superseded the FFT projection this paper
had been quoting.  The earlier summary---that B300 meets a Kulisch
INT32 sub-floor with margin and projects
${\sim}18$~ms against the
$12.9$~ms memory roof---derived from the companion's v2-era
formulation, whose integer-issue rate was since re-audited (it had
been charged at an FMA-counted figure, among other repairs) and whose
final form finds a per-output integer-SIMT epilogue of $203$--$281$
instructions holding software at $63$--$87$~ms.  Every FFT-coverage
statement in this paper is updated accordingly (abstract,
contributions, \S\ref{sec:h100baseline}, \S\ref{sec:l3}, and the
coverage tables), the four-floor rule is replaced by the companion's
final per-instruction-class floors, and the kernel-coverage claim now
carries the FFT explicitly as its second bounded,
hardware-conditional exception.  We record the direction of this
change candidly---it is the one place in the paper's history where a
companion's deeper analysis \emph{lowered} a coverage claim---and
equally that the landing is constructive: the wall is closed-form,
its removal is the co-design ladder (minor rungs, one moderate) whose
load-path rung this paper already asks for as Option~C, and the sharpened thesis is the
stronger scientific statement.  The dedicated summary subsection
\S\ref{sec:since-arxiv} is new in this revision.

\paragraph{Changes in this version (Revision~35, August~4, 2026, later the same day).}
Statements that presented the Part-3 measurement campaign as a
committed joint undertaking with named external partners are reworded
throughout: Part~3 is affirmed and its scope is unchanged, but this
paper no longer names collaboration partners for work not yet formally
agreed.  The acknowledgements retain thanks for past reviews and
discussions, which are matters of record.  No technical content
changes.

\paragraph{Changes in this version (Revision~36, August~19, 2026).}
A final adversarial cross-paper audit of the three-paper bundle found
one seam the per-paper verification suites could not see, and this
revision closes it.  The spectral companion~\cite{matsuoka2026fft}
normalises the SIMT integer pipe at $41.7$~T~inst/s (spec-derived)
where this paper and Ozaki~2.5 use the memo's $75$~T~inst/s; the two
are now explicitly reconciled (\S\ref{sec:model}, reconciliation
note), with the $41.7$ case carried as an engine-verified sensitivity
($\mathrm{OI}^{*}$, the keep-up budgets) and the Part-3 pipe census
named as the arbiter.  The FFT co-design ladder is described at the
companion's own two cost tiers throughout---minor rungs to within
$1.2$--$1.8\times$ of the roof, one moderate rung to reach it---where
earlier revisions had compressed it to ``minor''.  Cross-citations no
longer name arXiv version numbers that do not yet exist; this revision
is dated on its actual date.  No numerical result changes.

\paragraph{Changes in this version (Revision~37, September~3, 2026).}
A final pre-submission read of the three-paper bundle by a fresh
reviewer model found no error that changes a result, and three
statements that the companion engine or the companion papers
contradicted, all repaired here.  The reconciliation note of
\S\ref{sec:model} no longer claims that the dense floor is independent
of $P_{\text{int}}$: the tensor-side terms are, but the floor is a
minimum over terms, and the companion's engine at $41.7$~T~inst/s puts
the Rubin large-square floor at ${\approx}148$~TFLOPS ($0.31$ of $473$)
rather than $0.50$, with GB300 at route~E's $125$~TFLOPS roof rather
than the \texttt{dp4a}-route $135$; the sensitivity is now stated with
both values and engine-checked.  The application composites quoted from
the companion (\S\ref{sec:impl}, \S\ref{sec:l3}) are updated to its
Table~11 at the conservative reconstruction cost ($1.9\times$ LOBPCG,
$1.6\times$ blocked QR, $1.2\times$ multifrontal on Rubin), with the
codesign-target values kept alongside, and a stale table pointer is
corrected.  Smaller repairs: the B200 anchor efficiency is quoted
against the current $281$-TFLOPS INT8 ceiling; the $1.4$-TFLOPS B300
\fp{64} figure is labelled by its NVL72 derivation rather than as a
datasheet value; two appendix passages (Garner cost ``negligible'';
the ${\approx}r$ reconstruction hosting stated as available) are brought
into line with the body's charged and hedged treatment; the coverage
claim's two strongest phrasings are hedged to the coverage-framework
reading the paper elsewhere insists on; the Part-3 platform sentence
no longer implies a committed second team; release wording for the
artifact and Zenodo DOI is made accurate; and revision references to
the companions name their current identities (Ozaki~2.5 Draft~27,
Part~2 Revision~10).  No numerical result of this paper changes.

\section{Background and Related Work}
\label{sec:background}

We summarise the two Ozaki schemes at the level of detail required for the
performance model.  Readers familiar with~\cite{ozaki2012,ozaki2025_scheme2}
can skim this section.

\subsection{Error-Free Transformations and the Ozaki Scheme}

Let $a,b\in\mathbb{F}$ be working-precision floats.  Veltkamp/Dekker splitting
expresses $a=a_h+a_\ell$ with $a_h$ holding the leading bits and $a_\ell$ the
trailing bits, all in $\mathbb{F}$; the product $ab$ then equals
$a_h b_h + a_h b_\ell + a_\ell b_h + a_\ell b_\ell$, computed exactly when each
factor fits in $\mathbb{F}$.  The Ozaki scheme~\cite{ozaki2012} generalises this
idea to dense matrices, replacing two-word splitting by $s$-way slicing along
the inner-product direction.

\subsection{Ozaki Scheme I: Mantissa Slicing}
\label{sec:ozaki1}

Given $A\in\mathbb{F}^{m\times k}$ and $B\in\mathbb{F}^{k\times n}$, Ozaki~I
forms decompositions
\begin{equation}
A=\sum_{p=1}^{S_A}A^{(p)},\qquad B=\sum_{q=1}^{S_B}B^{(q)},
\label{eq:ozaki1-decomp}
\end{equation}
in which each $A^{(p)}$ and $B^{(q)}$ has a bounded mantissa width compatible
with the target tensor format.  The reconstruction is
\begin{equation}
C \;\approx\;\sum_{p=1}^{S_A}\sum_{q=1}^{S_B} A^{(p)}B^{(q)},
\label{eq:ozaki1-recon}
\end{equation}
summed in working precision, at dominant cost $S_A\!\cdot\!S_B$
low-precision GEMMs.

\paragraph{Substrate-specific scaling.}  The three candidate tensor-core
substrates---\fp{16}, \inteight, and \fp{8}---differ in how
\eqref{eq:ozaki1-decomp}--\eqref{eq:ozaki1-recon} is actually computed.

\begin{itemize}[leftmargin=2em,itemsep=2pt]
\item \emph{\fp{16} tensor cores.}  Each $A^{(p)}$ and $B^{(q)}$ is stored
   directly as an \fp{16} matrix, and the tensor-core MMA accumulates
   $A^{(p)}B^{(q)}$ in \fp{32}.  No additional integer scaling is required.
\item \emph{\inteight tensor cores.}  Each slice carries a \emph{floating-point}
   mantissa range, but the \inteight engine consumes signed integers.  The
   decomposition therefore associates each slice with a per-slice
   power-of-two exponent: $A^{(p)} = 2^{e_p}\!\cdot\!\widetilde{A}^{(p)}$
   and $B^{(q)} = 2^{f_q}\!\cdot\!\widetilde{B}^{(q)}$, where
   $\widetilde{A}^{(p)},\widetilde{B}^{(q)}\!\in\!\mathbb{Z}^{8}$ are the
   signed-integer mantissa slices fed to the \inteight engine, and the
   reconstruction accumulates $2^{e_p+f_q}\!\cdot\!(\widetilde{A}^{(p)}
   \widetilde{B}^{(q)})$ in \fp{64}~\cite{mukunoki2020,ootomo2024}.  The
   integer products themselves are computed exactly in \intthirtytwo.
\item \emph{\fp{8} tensor cores.}  \fp{8} (E4M3) consumes only a 3+1-bit
   mantissa per element, so the slicing of an \fp{64} mantissa requires
   more slices than the \inteight case; in addition, the \fp{8} engine
   accumulates in \fp{32} rather than \intthirtytwo, and a separate
   quantisation correction is required to handle the implicit
   normalisation of the \fp{8} format.  The corresponding scheme is
   described in detail in \cite{mukunoki2025_fp8} and is the basis of the
   FP8 variant of Ozaki~II that we discuss in
   \S\ref{sec:ozaki2-fp8}.
\end{itemize}

\paragraph{Slice counts: accumulator-bound analysis.}  The required slice
count is governed jointly by the input mantissa width and by the
accumulator precision.  For \fp{64}-accurate output with summation length
$k$, the requirement that the accumulator hold the sum of $k$ products of
$b$-bit slices exactly is
\begin{equation}
2b + \lceil \log_2 k \rceil \;\leq\; w_{\text{acc}},
\label{eq:slice-bits}
\end{equation}
where $w_{\text{acc}}$ is the number of significand bits available in the
accumulator (24 for \fp{32}, 31 for \intthirtytwo signed).  Inverting
\eqref{eq:slice-bits} gives the accumulator-limited payload $b^\star\!=\!
\lfloor(w_{\text{acc}}
\!-\!\lceil\log_2 k\rceil)/2\rfloor$ bits per slice; the usable payload
is additionally capped by the input format's significand budget,
$b = \min(b^\star, b_{\text{input}})$, and the slice count needed to
cover the 53-bit \fp{64} mantissa is then $S = \lceil 53/b \rceil$.
Equation~\eqref{eq:slice-bits} is a \emph{sufficient} condition for
exact slice-product accumulation under stated worst-case assumptions
(sign-aligned slices, exact slice products, no cancellation model); it
is \emph{not} a necessary minimum for delivering \fp{64}-grade results
on given data.  Three quantities must accordingly be kept distinct: the
worst-case-sufficient count of \eqref{eq:slice-bits}; a proven
error-bound count for a named Ozaki-I variant and rounding rule; and
empirical counts that meet a specified test tolerance on specified
inputs.  The cited empirical values below are of the third kind and are
frequently \emph{smaller} than the worst-case-sufficient value
(\fp{16}: $3$--$4$ against $7$--$11$; \fp{8}: $\geq\!11$ against
$14$), because production data rarely realises the worst-case
alignment; they are not evidence against the constraint, and the
constraint is not a claim about their necessity.

Table~\ref{tab:slicecount} reports the resulting sufficient
worst-case slice counts at representative inner lengths $k$.  (Values in earlier versions
of this paper did not follow \eqref{eq:slice-bits} as printed and are
corrected here.)  Three observations follow.

\begin{table}[h]
\caption{Slice count \emph{sufficient} for exact slice-product
accumulation in Ozaki~I under the worst-case assumptions of the text,
derived from
\eqref{eq:slice-bits} with $b = \min(b^\star, b_{\text{input}})$.  The
input mantissa width is the engine's payload-bit budget (\fp{16}: 11 bits;
\inteight: 7 signed; \fp{8}~E4M3: 4 bits with implicit normalisation).
When $b^\star$ exceeds $b_{\text{input}}$ the input format is the
binding width---the slice count is then $k$-independent at
$\lceil 53/b_{\text{input}}\rceil$ (it does \emph{not} collapse to one
slice).
Empirical values from~\cite{mukunoki2020,mukunoki2025_fp8} are noted in
the rightmost column.}
\label{tab:slicecount}
\centering
\footnotesize
\setlength{\tabcolsep}{6pt}
\begin{tabular}{lccccc}
\toprule
\textbf{Substrate} & $w_{\text{acc}}$
                   & $k\!=\!256$ & $k\!=\!1024$ & $k\!=\!4096$ & $k\!=\!16384$
                   \\
\midrule
\fp{16} / \fp{32}-accum  & 24 &  $7$ & $8$ & $9$ & $11$ \\
\inteight / \intthirtytwo-accum & 31 & $8$ & $8$ & $8$ & $8$ \\
\fp{8}    / \fp{32}-accum & 24 & $14$ & $14$ & $14$ & $14$ \\
\bottomrule
\end{tabular}
\\[2pt]
{\scriptsize Reported empirical values: \fp{16}: $S\!\approx\!3$--$4$ at
moderate $k$~\cite{mukunoki2020}; \inteight: $S\!\approx\!7$--$10$ at
$k\!\leq\!16384$~\cite{mukunoki2020,ootomo2024}; \fp{8}: $S\!\geq\!11$ in
the configurations of~\cite{mukunoki2025_fp8}.  These meet stated test
tolerances on their benchmark data with fewer slices than the
worst-case-sufficient count---a different contract, not a violation of
the constraint (see text).}
\end{table}

\begin{enumerate}[leftmargin=2em,itemsep=2pt]
\item For \fp{16} tensor cores with \fp{32} accumulation, the bound is
   tight: at $k\!=\!4096$, only about 6 bits per slice can be safely
   accumulated, even though \fp{16} can carry 11.  In other words, at
   large $k$, the \fp{16}/\fp{32} substrate is \emph{accumulator-bound}
   rather than input-bound, and the slice count grows accordingly.
\item For \inteight cores with \intthirtytwo accumulation, the bound is
   loose: at $k\!=\!4096$, the safe payload is 9.5 bits, which exceeds the
   7-bit \inteight input. Consequently, the slice count is governed by
   the \emph{input} mantissa, not by the accumulator, and grows only
   logarithmically with $k$.  At large $k$, the \inteight substrate
   therefore requires comparable or fewer slices than \fp{16}, contrary
   to a naive ordering by input precision.
\item For \fp{8} tensor cores, the small input mantissa (effectively 4
   bits) combined with the \fp{32} accumulator constraint produces the
   highest slice count of the three; Mukunoki's reported
   $S\!\geq\!11$ for FP8~\cite{mukunoki2025_fp8} is consistent with
   Table~\ref{tab:slicecount}.
\end{enumerate}

\paragraph{Quadratic scaling.}  The total number of low-precision GEMMs
scales as $\Theta(S^2)$, which is the dominant disadvantage of Ozaki~I for
bandwidth-limited problems: even with fused decomposition, the temporal
volume of arithmetic grows quadratically in the slice count.  This is
the principal motivation for the linearly-scaling Ozaki~II scheme
described next.

\subsection{Ozaki Scheme II: Modular Arithmetic via the CRT}
\label{sec:ozaki2}

The Chinese Remainder Theorem-based Ozaki~II scheme, proposed by Ozaki,
Uchino and Imamura in 2025~\cite{ozaki2025_scheme2}, replaces the slice
decomposition by a residue decomposition over a set of pairwise-coprime moduli.
The algorithm consists of three phases.

\paragraph{Phase~1: Integer scaling.}  Diagonal scaling matrices
$D\!\in\!\mathbb{F}^{m\times m}$ and $E\!\in\!\mathbb{F}^{n\times n}$, chosen
with power-of-two diagonals to be exactly invertible in \fp{64}, are used to
form
\begin{equation}
\tilde A = \lfloor DA\rceil\in\mathbb{Z}^{m\times k},\qquad
\tilde B = \lfloor BE\rceil\in\mathbb{Z}^{k\times n},
\label{eq:scale}
\end{equation}
where the quantisation $\lfloor\,\cdot\,\rceil$ denotes either
round-to-nearest integer (the rule of the published error
analysis~\cite{ozaki_error_analysis_2026}) or truncation toward zero
(the shipping-library convention, adopted by the companion
paper's audited accounting with the centred tie rule stated
there~\cite{matsuoka2026ozaki25}).  The two rules differ by at most one
unit in the last integer place; every deconstruction count in this
paper is rule-independent, and wherever an error bound is invoked we
cite the theorem matching the rule in force---round-to-nearest
for~\cite{ozaki_error_analysis_2026}, with the truncation variant's
bound carried as a stated companion obligation.  The integer
matrices $\tilde A,\tilde B$ have magnitudes bounded by the chosen scaling
factor, typically $2^{p-1}-1$ for a representation width of $p$ bits.
Table~\ref{tab:contract} fixes the variant vocabulary used
throughout.

\begin{table}[h]
\caption{The numerical-contract variants.  The digit-centring tie
rule is a property of the residue decomposition, not of the
floating-point quantiser; the two are specified independently.}
\label{tab:contract}
\centering
\scriptsize
\setlength{\tabcolsep}{4pt}
\begin{tabular}{p{2.9cm}p{3.4cm}p{4.3cm}p{2.2cm}}
\toprule
\textbf{variant} & \textbf{\fp{64}$\to$digit rule} & \textbf{available guarantee} & \textbf{status} \\
\midrule
Ozaki-II-RN (reference) & round to nearest & the componentwise theorem of~\cite{ozaki_error_analysis_2026}, under its hypotheses & theoretical \\
Ozaki-II-TZ (implementation) & truncation toward zero, centred tie rule per the companion & no automatic inheritance from RN; new bound or measured error distribution required & proposed / unvalidated \\
Codesigned A/E/D & per-set rule~\cite{matsuoka2026ozaki25} & range/exactness script-checked exhaustively for all A/E/D planes and the six square S moduli; the six nonsquare S-tail layouts lie outside the \emph{canonical} E4M3 envelope $m\le321$ at every base, but are placed in E4M3 by the carry-corrected redundant split (envelope $m\le577$), checked over all $m^2$ ordered centred pairs per modulus (\S\ref{sec:frozen-kernel}); full accuracy an explicit obligation & discharged (layouts) / obligation (accuracy) \\
Native fallback & hardware/library \fp{64} & library/IEEE contract as applicable & available \\
\bottomrule
\end{tabular}
\end{table}

(Notationally, $D$ and $E$ carry the entire scaling: where earlier
versions wrote separate factors $\sigma_A,\sigma_B$, these are the
power-of-two magnitudes of $D = \sigma_A D_0$ and $E = \sigma_B E_0$
and are removed in the reconstruction by applying $D^{-1}, E^{-1}$
exactly---no additional hidden divisor exists.)

\paragraph{Phase~2: Modular GEMMs.}  Choose pairwise-coprime moduli
$m_1<m_2<\cdots<m_r$ large enough that
\begin{equation}
M = \prod_{i=1}^r m_i \;>\;
2\cdot \max_{ij}\bigl|(\tilde A\tilde B)_{ij}\bigr|,
\label{eq:M-bound}
\end{equation}
which guarantees that the integer product can be uniquely reconstructed from
its residues.  The bound is on the \emph{magnitude}: equivalently, writing
$\mu \ge \max_{ij}\lvert(\tilde A\tilde B)_{ij}\rvert$ for any bound on the
absolute value of the integer product entries, the requirement is
$M > 2\mu$, so that the symmetric residue range $[-M/2, M/2)$ contains
every entry with its sign.  For each $i\in\{1,\dots,r\}$ compute
\begin{equation}
C^{(i)} \;=\; \bigl(\tilde A \bmod m_i\bigr)\;\bigl(\tilde B \bmod m_i\bigr)
\bmod m_i.
\label{eq:modgemm}
\end{equation}
On \inteight tensor cores each $(\tilde A\bmod m_i)\cdot(\tilde B\bmod m_i)$ is
a standard \inteight GEMM accumulated in \textsc{int32}.  When using \fp{8} tensor
cores the modular reduction must be carried out in scaled floating point
following the technique of~\cite{uchino2026_fp8}.

\paragraph{Phase~3: CRT reconstruction (Garner's algorithm).}  We apply
Garner's algorithm~\cite{garner1959,knuth_taocp2} to recover the integer
product element-wise.  Writing $C_{ij}=v_1+v_2 m_1+v_3 m_1m_2+\cdots$ in mixed
radix, the digits $v_k$ are computed iteratively as
\begin{equation}
\begin{aligned}
v_1 &= C^{(1)}_{ij},\\
v_k &= \biggl(C^{(k)}_{ij}-\sum_{j=1}^{k-1}v_j\prod_{\ell=1}^{j-1}m_\ell\biggr)
       \cdot\biggl(\prod_{\ell=1}^{k-1}m_\ell\biggr)^{-1} \pmod{m_k},
\quad k\ge 2.
\end{aligned}
\label{eq:garner}
\end{equation}
The modular inverses $\bigl(\prod_\ell m_\ell\bigr)^{-1}\pmod{m_k}$ are
precomputed once.  The recovered integer is finally rescaled back to floating
point via $C \approx D^{-1}\tilde C E^{-1}$, which removes the entire
scaling exactly (the scalar magnitudes $\sigma_A,\sigma_B$ live inside
$D,E$; \S\ref{sec:ozaki2}, Phase~1).

\paragraph{Linear scaling.}  The cost is $r$ low-precision GEMMs plus the
$O(r^2)$ element-wise reconstruction.  Critically, $r$ scales \emph{linearly}
in the required exponent range for FP64-equivalent dynamic range; published
parameter sets use $r\!\in\![13,16]$ on \inteight cores~\cite{ozaki2025_scheme2}
and $r\!\in\![8,12]$ on \fp{8} cores~\cite{uchino2026_fp8}, with the exact
value driven by the desired error guarantee.

\subsection{The \fp{8} Variant of Ozaki~II}
\label{sec:ozaki2-fp8}

Uchino, Ozaki and Imamura observed in early 2026 that the original Ozaki~II
algorithm \emph{cannot be directly adapted to \fp{8} matrix-multiply-accumulate
units}, because modular reduction is fundamentally an integer
operation~\cite{uchino2026_fp8}.  They introduced a quantisation trick that
emulates modular arithmetic over \fp{8} by exploiting the exact-integer
behaviour of the format---noting carefully that E4M3's maximum finite
magnitude of $448$ and its \emph{consecutive exact-integer range} are
different properties: with a $3{+}1$-bit significand, E4M3 represents
consecutive integers exactly only up to $\pm16$, with spacing growing at
higher exponents, which is precisely why each modular value must be
carried across multiple \fp{8} planes with a correction scheme rather
than as a single \fp{8} number.  This
adaptation is the reason that Ozaki~II remains viable on Blackwell~Ultra and
on the upcoming NVIDIA Rubin GPU, both of which significantly downgrade \inteight
in favour of \fp{8}/\fp{4}~\cite{mukunoki2025_fp8,uchino2026_fp8}.

This paper assumes the existence of this \fp{8} adaptation but does not
re-derive it.  Our TME model is parameterised so that the choice of underlying
tensor format simply rescales the compute multiplier $\alpha$ and the
reconstruction latency $\gamma$, without altering the model's structure.

\paragraph{Moduli count and the $(3r+1)$ FP8 cost
structure.}  The performance projections in later sections adopt
$r\!=\!12$ moduli for the \fp{8} Ozaki~II variant, the value
recommended by Mukunoki~\cite{mukunoki2025_fp8} (who
report $r\!\in\![11,14]$ for \fp{64}-equivalent accuracy on the \fp{8}
(E4M3) substrate), together with the full $(3r{+}1)$ \fp{8} op count.
The latter, clarified by Imamura (private communication
and~\cite{matsuoka2026fft}), reflects that each \fp{64}-equivalent
residue product on the FP8 substrate expands into $(3r{+}1)$ FP8 MMAs
rather than $r$, because the Karatsuba structure used internally to
emulate signed \textsc{int}9 on FP8 introduces a factor-of-three
multiplier on the residue planes (the trailing $+1$ being the
max-magnitude estimation pass).  For the INT8 substrate the count is
$(r{+}1)$---one MMA per modulus plus the max pass, since a centred
residue fits an \inteight{} operand directly and no Karatsuba split is
needed---so a modulus is indeed ${\sim}3\times$ cheaper there.
\emph{That per-modulus saving does not transfer to the ceiling at equal
$r$}, and revisions of this paper before the July 2026 correction cycle were wrong to quote
$\alpha_{\textsc{int8}}=r{+}1=13$ alongside $\alpha=3r{+}1=37$ as if the
two shared an $r$.  An \inteight{} operand holds a centred residue only
for $m_i\!\le\!256$, and fourteen moduli $\le\!256$ cannot reach the CRT
range that the $r{=}12$ \fp{8} set supplies: even ignoring coprimality,
the fourteen largest admissible integers give
$\sum_{j=243}^{256}\log_2 j = 111.5$ bits against that set's $111.8$.
The smallest \inteight{}-native system is therefore the all-byte
candidate~A of the companion~\cite{matsuoka2026ozaki25}---%
$r_{\textsc{i}}\!=\!15$ moduli $\le\!256$, $117.8$ bits---which is also
where the published \inteight{} parameter sets sit
($r\!\in\![13,16]$~\cite{ozaki2025_scheme2}).  At matched accuracy the
comparison is thus $\alpha=37$ against
$\alpha_{\textsc{int8}}=r_{\textsc{i}}{+}1=16$, a factor $2.31$ and not
$2.85$, and Table~\ref{tab:int8vsfp8} is charged that way throughout.
Thus the \fp{8} compute multiplier
is $\alpha=3r{+}1=37$ at $r\!=\!12$, and \emph{all} dense
compute-bound ceilings in \S\ref{sec:arch}--\S\ref{sec:model} use this
value (an earlier draft used the moduli-only count $\alpha=r$, which
overstated the \fp{8} compute ceiling by $\approx\!3\times$; we thank
the NVIDIA libraries/DevTech team for flagging this).  Crucially, the
correction is confined to the compute-bound regime: it scales the dense
ceiling by $r/(3r{+}1)$ but leaves the parity-row structure of
Table~\ref{tab:speedups} and the H100-relative scaling of
Table~\ref{tab:h100baseline} unchanged, because those rows are governed
by HBM-bandwidth ratios, not by the compute-ceiling constant.  This is
the analytic content of the dense-vs-memory-bound distinction we make
explicit in Case~C of \S\ref{sec:model}.

\paragraph{The scheme in one place.}  Algorithm~\ref{alg:ozaki2-phases}
collects the three phases into the object the rest of this paper prices,
and writing it out earns three things the prose above leaves implicit.
\emph{First}, the two halves of the numerical contract that
Table~\ref{tab:contract} insists are specified independently are
literally two different lines: the quantiser is line~\ref{lin:quant} and
the centred residue layout is line~\ref{lin:dec}, and neither reads the
other.  (The table's fourth variant, the native fallback, is the
\textbf{if} on the admissibility test, which is why it is a contract
variant at all and not an implementation detail.)  \emph{Second}, the
substrate is confined to Phase~2: lines~\ref{lin:dec}
and~\ref{lin:mma} are the only two steps that differ between
\inteight{} and \fp{8}, and they differ in one respect only---how a
centred residue is laid out, and therefore how many products a modulus
costs.  That is the whole of $\alpha$, which is why the substrate
reaches the performance model through that constant and nothing else,
and why line~\ref{lin:dec} is the single place where the numerical
contract and the substrate meet.  \emph{Third}, the
phases map one-to-one onto the terms of that model,
Eq.~\eqref{eq:psvc}---Phase~1 is charged per \emph{input} element and
supplies the deconstruction numerator of the two host terms, Phase~2 is
the \emph{arithmetic} MMA roof $P_{\fp{8}}/\alpha$ ($473$~TFLOPS on Rubin
for the published set) together with the operand traffic, and
Phase~3 is charged per \emph{output} and enters those same two host
terms as the $N_{\text{rec}}(r)/2k$ addend that Eq.~\eqref{eq:psvc}
superscripts by host---$N_\gamma$ on the SIMT path, which is the hosting
this paper adopts and prices in \S\ref{sec:impl}---a tax proportional to
$1/k$ rather than a multiplier.  That input/output asymmetry is the whole of the duality
remarked on in \S\ref{sec:model}; it is also why what the model needs
here is a phase-level reading rather than a kernel listing, and the
fused kernels of \S\ref{sec:impl} are that same algorithm with the
phase boundaries dissolved into registers.

\begin{algorithm}[!t]
\caption{Ozaki~II dense GEMM at phase level (\S\ref{sec:ozaki2}).  The
substrate is confined to lines~\ref{lin:dec} and~\ref{lin:mma}, which
fix the residue layout and hence $\alpha$; the two independently
specified halves of the numerical contract are lines~\ref{lin:quant}
and~\ref{lin:dec}.  Phase~2 is charged $\alpha$
passes over the $mnk$ multiply volume; Phase~3 is charged $N_\gamma$
narrow operations per \emph{output}, hence a $1/k$ tax.  Fusing the
phases so the planes never reach HBM is \S\ref{sec:impl}.}
\label{alg:ozaki2-phases}
\begin{algorithmic}[1]
\State \textbf{input:} $A\in\mathbb{F}^{m\times k}$,
   $B\in\mathbb{F}^{k\times n}$ in \fp{64}; modulus count $r$;
   substrate $\in\{\inteight,\fp{8}\}$; contract variant
   \Comment{Table~\ref{tab:contract}}
\State \textbf{output:} $C \approx AB$ in \fp{64}
\Statex \emph{Phase~1 --- integer scaling}, Eq.~\eqref{eq:scale}
\State scan the exponent spans of $A,B$; choose power-of-two diagonals
   $D,E$ \Comment{exactly invertible; amortised}
\State \label{lin:quant} $\tilde A \gets \lfloor DA\rceil$,\quad
   $\tilde B \gets \lfloor BE\rceil$
   \Comment{RN or TZ: the contract's quantiser}
\State $\mu \gets k\,\max_{ij}|\tilde A_{ij}|\cdot\max_{ij}|\tilde B_{ij}|$
   \Comment{any valid bound serves; the scan supplies this one}
\State choose coprime $m_1<\cdots<m_r$ with $M=\prod_i m_i>2\mu$
   \Comment{Eq.~\eqref{eq:M-bound}: $[-M/2,M/2)$ holds every entry
   \emph{with its sign}}
\If{no admissible set of size $r$ exists}
   \State raise $r$, or fall back to native \fp{64} on the offending rows
      \Comment{ESC/ADP~\cite{schwarz2025}; proposed for Ozaki~II, not
      established}
\EndIf
\Statex \emph{Phase~2 --- modular GEMMs}, Eq.~\eqref{eq:modgemm}
\For{$i=1,\dots,r$}
   \State \label{lin:dec} $A^{(i)} \gets \tilde A \bmod m_i$,\quad
      $B^{(i)} \gets \tilde B \bmod m_i$, centred
      \Comment{per \emph{input}; the substrate fixes the layout---one
      \inteight{} operand iff $m_i\!\le\!256$, else \fp{8} digit planes,
      \S\ref{sec:frozen-kernel}}
\EndFor
\For{$i=1,\dots,r$}
   \State \label{lin:mma} $C^{(i)} \gets
      \bigl(A^{(i)}B^{(i)}\bigr)\bmod m_i$
      \Comment{\inteight: $1$ MMA/modulus, exact \intthirtytwo{},
      $\alpha=r{+}1$---but only at the larger $r$ a byte-modulus set
      needs ($r\!\ge\!15$, Table~\ref{tab:int8vsfp8}).
      \fp{8}: $3$ plane products, $\alpha=3r{+}1$}
\EndFor
\Statex \emph{Phase~3 --- CRT reconstruction}, Eq.~\eqref{eq:garner}
\For{each output entry $(i,j)$}
   \State $v_1,\dots,v_r \gets$ mixed-radix digits of $C_{ij}$ by Garner
      \Comment{$N_\gamma = r(r{-}1)/2+2r = 90$ at $r{=}12$}
   \State $\tilde C_{ij} \gets v_1 + v_2m_1 + v_3m_1m_2 + \cdots$
\EndFor
\State $C \gets D^{-1}\tilde C E^{-1}$
   \Comment{$D,E$ powers of two: the scaling is removed exactly}
\end{algorithmic}
\end{algorithm}

\subsection{The frozen plane kernel}
\label{sec:frozen-kernel}

Reviewers of the previous revision observed, correctly, that this paper
quoted three different per-modulus plane counts in three different
places---$p_{\text{store}}$ in the resource ledger, $3r$ in the roof
denominator, and $N_{\text{acc}}$ in the TMEM budget---without ever
writing the identity that relates them.  We therefore fix one kernel
here, with every operand and coefficient printed, and every later
section refers to it rather than restating it.  All claims in this
subsection are verified exhaustively---over \emph{every} centred
residue pair of \emph{every} modulus in the S, E and A sets---by
\texttt{verify/plane\_algebra.py} and \texttt{verify/nacc\_table.py} of
the artifact.

\paragraph{Three distinct counts.}  Let $r$ be the modulus count.  We
distinguish, once and for all:
\begin{itemize}
\item $p_{\text{store}}$---\emph{stored planes per scalar}: how many
  \fp{8}/\inteight{} bytes each input element occupies in the residue
  workspace.  This drives operand bandwidth and SMEM capacity.
\item $p_{\text{mma}}=3r$---\emph{plane-product MMAs issued}: the
  instruction count in the inner loop.  With the max-magnitude pass this
  is the roof denominator $\alpha=3r{+}1$.
\item $N_{\text{acc}}$---\emph{accumulator tiles that must persist
  across a $k$-slab}.  This drives TMEM occupancy, and it is
  \emph{neither} of the other two.
\end{itemize}
For the published S set ($r{=}12$: six squares $1089,1024,961,841,625,529$
and the nonsquare tail $511,509,503,499,491,487$),
$p_{\text{store}}=6(2)+6(3)=30$ while $p_{\text{mma}}=36$.  The gap is
not an inconsistency; it is the square-modulus shortcut, which we now
write out.

\paragraph{Square moduli: two planes, three products.}  For $m=s^2$ write
the centred residue in balanced base~$s$, $d = d_1 s + d_0$.  Then
\begin{equation}
  de \;=\; d_0e_0 \;+\; s\,(d_1e_0 + d_0e_1) \;+\; s^2\,d_1e_1
      \;\equiv\; d_0e_0 + s\,(d_1e_0 + d_0e_1) \pmod{s^2},
\label{eq:sq-shortcut}
\end{equation}
because the $d_1e_1$ term is annihilated by the modulus.  \emph{Two}
stored planes therefore feed \emph{three} plane products---the two
cross terms $d_1e_0$ and $d_0e_1$, not a diagonal pair.  This is the
sole source of the $p_{\text{store}}\!<\!p_{\text{mma}}$ gap, and it is
why every plane count in this paper must name which of the three
quantities it means.

\paragraph{Nonsquare moduli: three planes, three products.}  For
nonsquare $m$ no term is annihilated, so a base-$2^b$ split
$d=d_12^b+d_0$ needs all four products unless Karatsuba is used.  With
the third, sum plane $d_0{+}d_1$,
\begin{equation}
  de \;=\; (1-2^b)\,d_0e_0 \;+\; 2^b (d_0{+}d_1)(e_0{+}e_1)
        \;+\; (2^{2b}-2^b)\, d_1e_1 ,
\label{eq:kar}
\end{equation}
three products from three stored planes.  Note that the coefficients
are $(1-2^b)$, $2^b$ and $2^b(2^b-1)$---\emph{not} powers of two, a
fact that governs the accumulator count below.

\paragraph{The e4m3 layout envelope.}  \textsc{e4m3} represents
\emph{consecutive} integers exactly only on $[-16,16]$
(\S\ref{sec:ozaki2-fp8}), but it represents a great many larger
ones---write $\mathbb{E}$ for the exactly representable integers,
\ie those of the form $M\!\cdot\!2^{e}$ with $|M|\le15$ and magnitude
at most $448$, so that $18, 20, 22, 24, 26, \dots$ are exact while
$17, 19, 21, 23$ are not.  Both facts bear on the layout, and earlier
revisions of this paper used only the first.

Suppose the layout is \emph{canonical}: $d_0$ is the balanced base-$B$
digit of $d$ and $d_1=(d-d_0)/B$.  Writing $A=\max|d_0|$ and
$A'=\max|d_1|$, injectivity on the centred range requires
$(2A{+}1)(2A'{+}1)\ge m$, and sum-plane exactness requires
$\max_d|d_0{+}d_1|\in\mathbb{E}$.  Earlier revisions replaced that
second condition by $A+A'\le16$ and maximised to obtain $A=A'=8$ and
$m\le289$.  That substitution is too strong twice over.  The extremes
of $d_0$ and $d_1$ are not attained at the same residue, so
$\max_d|d_0{+}d_1|$ can be strictly smaller than $A+A'$; and the
admissible target is $\mathbb{E}$, not the interval $[-16,16]$.
Searching bases exhaustively (\texttt{verify/stail\_layout.py}) gives
the true envelope:
\begin{equation}
\begin{aligned}
  m &\le 321 &&\text{(nonsquare, canonical, at $B{=}17$)},\\
  m &\le 577 &&\text{(nonsquare, redundant, at $B{=}18$)},\\
  m &\le 33^2 = 1089 &&\text{(square)}.
\end{aligned}
\label{eq:envelope}
\end{equation}
At $m{=}321$, $B{=}17$ the digits reach $A{=}8$ and $A'{=}9$, so
$A{+}A'=17$ while $\max_d|d_0{+}d_1|$ is only $16$, and injectivity
holds with $17\cdot19 = 323 \ge 321$.  Even at the base $16$ the old
derivation assumed, the correct canonical bound is $303$, not $289$.

Three consequences follow, and all three are load-bearing for this
paper.  \emph{(i)}~The square bound is exactly tight at $s{=}33$---now
under membership of $\mathbb{E}$ rather than the narrower consecutive
criterion, which does not move it---and this is why $33^2$ tops the
published set.  \emph{(ii)}~The byte moduli of routes~E and~A (all
$\le 251$) sit inside the canonical bound with headroom; the artifact
confirms exact layouts at $b{=}4$ for $m\ge223$ and $b{=}3$ for
$m\le211$, with zero identity mismatches over all $m^2$ centred pairs.
\emph{(iii)}~The published S \emph{nonsquare} tail---all six moduli
$\ge 487$---lies outside the \emph{canonical} bound, and does so at
every base.  Earlier revisions of this paper concluded from this that
the six layouts were not merely open but infeasible in \textsc{e4m3}.
That inference was wrong: what fails is the canonical restriction, not
\textsc{e4m3}.  The companion's carry-corrected rule---take the
balanced base-$16$ digits and, if $d_0{+}d_1\notin\mathbb{E}$, shift
one unit of base weight,
$(d_0,d_1)\mapsto(d_0-16\varsigma,\ d_1+\varsigma)$ with
$\varsigma=\operatorname{sign}(d_0)$---preserves $d=d_0+16d_1$
identically and places all three planes in $\mathbb{E}$ for every
nonsquare $m\le513$, with $|d_0|\le14$, $|d_1|\le16$,
$|d_0{+}d_1|\le22$~\cite{matsuoka2026ozaki25}.  All six tail moduli
are therefore exact in \textsc{e4m3}, verified over all $m^2$ ordered
centred pairs per modulus with zero mismatches, at a cost of one
predicated add-pair on ${\approx}5\%$ of residues.  Nothing downstream
moves: the unmerged plane products stay at $22^2 = 484$, far inside
\eqref{eq:accum-exact}, while the merged nonsquare accumulator
$d_0e_0-2^bd_1e_1$ reaches $4095$ on the tail and so is exact at
$K_{\text{slab}}{=}4096$ by a single term---a margin too thin to build
on, which is why we keep the tail at $N_{\text{acc}}{=}3$.

What does \emph{not} change is the routing.  Every tensor route in our
menu still sends its \emph{reduction} GEMMs through \inteight{}, for a
constraint that has nothing to do with the layout envelope: the
reduction constants reach $826$ and the operand bytes span $[0,255]$,
neither of which is \textsc{e4m3}-representable
(\S\ref{sec:cqladder}).  What changes is the narrower claim about the
Ozaki \emph{product} passes.  Sets E and A are faster, and cheaper in
accumulators; they are not, as we previously wrote, the only layouts
that make an \fp{8} realisation exact at all.

\paragraph{Accumulators: $N_{\text{acc}}=2$ per modulus.}  A plane
product may share an accumulator with another only if their epilogue
coefficients differ by a signed power of two, since only then can the
difference be absorbed as an exponent-only edit of a stored plane---exact,
and free, because it happens once at conversion time rather than in the
$k$-loop.  Applying this to the two identities:
\begin{itemize}
\item Square, \eqref{eq:sq-shortcut}: the coefficients are $\{1,s,s\}$.
  The two cross terms already \emph{share} the coefficient $s$, so they
  accumulate together; $d_0e_0$ cannot join them because $s$ is not a
  power of two.  Hence $N_{\text{acc}}=2$.
\item Nonsquare, \eqref{eq:kar}: the ratio of the $d_1e_1$ and $d_0e_0$
  coefficients is $(2^{2b}-2^b)/(1-2^b) = -2^b$ exactly.  Storing the
  $d_1$ plane pre-scaled by $-2^b$ therefore merges those two products
  into one accumulator carrying the common weight $(1-2^b)$.  The
  pre-scaled plane is itself E4M3-exact---its per-modulus extremes take
  the four values $128/112/104/96$ across the byte sets, all on the exact
  grid---so the merge costs nothing.  The sum-plane coefficient $2^b$ is not a
  power-of-two multiple of $(1-2^b)$ (the ratio is $-16/15$ at
  $b{=}4$), so it cannot merge further.  Hence $N_{\text{acc}}=2$, and
  $2$ is a floor, not merely an achieved value.
\end{itemize}
So $N_{\text{acc}}=2r$ on the \fp{8} substrate: $26$ for~E and $30$
for~A.  S is the exception, at $6(2)+6(3)=30$, and the reason is
accumulator margin rather than representability.  Its six 9-bit
nonsquare tail moduli do admit the merge on \textsc{e4m3}: the
pre-scaled plane is exactly $-256=-2^{8}$ on all six, on the exact grid,
and the merged epilogue reproduces $d\,e$ over every ordered centred
pair.  But the merged term then reaches $4095$, so the schedule is exact
at $K_{\text{slab}}{=}4096$ by a single term, and we decline a margin of
one, keeping those six at $N_{\text{acc}}=3$
(\texttt{verify/stail\_merge.py}).  On \inteight{} the merge would be
unavailable there in any case---the pre-scaled plane leaves
$[-128,127]$---but that is a property of \inteight{}, which is not the
product substrate on any route in the menu.

\paragraph{Accumulation is exact, and the obligation discharges.}
Previous revisions deferred the exactness of in-loop accumulation to
Part~3 against a loosely stated ``\intthirtytwo/$2^{19}$'' bound.  It
does not need deferring.  \fp{32} accumulates integers exactly to
$2^{24}$, so the requirement is simply
\begin{equation}
  K_{\text{slab}} \cdot \max|\text{term}| \;\le\; 2^{24},
\label{eq:accum-exact}
\end{equation}
and the four accumulator kinds have $\max|\text{term}|$ equal to $256$
($d_0e_0$, square), $512$ ($d_1e_0{+}d_0e_1$), $225$
($(d_0{+}d_1)(e_0{+}e_1)$) and $1088$ ($d_0e_0 - 2^b d_1e_1$, the
merged nonsquare accumulator and the binding case).  At
$K_{\text{slab}}=4096$ the largest accumulator reaches $4\,456\,448 =
0.27\cdot2^{24}$, a $3.76\times$ margin; exactness would survive
$K_{\text{slab}}$ up to $15\,420$.  This is now a checked inequality,
not an obligation.

\subsection{Error Analysis}

Both Ozaki~I and Ozaki~II provide \emph{provable} error bounds for the
emulated product, in contrast to ad-hoc mixed-precision schemes.  For Ozaki~II
with $r$ moduli chosen by the criterion~\eqref{eq:M-bound}, the worst-case
componentwise relative error is bounded by the working-precision unit
round-off $u_{\textsc{fp64}}\approx 2^{-53}$ \emph{plus} the floating-to-integer
rounding introduced in~\eqref{eq:scale}~\cite{ozaki_error_analysis_2026}.  In
practice, for inputs with bounded condition number, the observed relative
error is within $2$--$10\,u_{\textsc{fp64}}$~\cite{ozaki2025_scheme2};
the ADP work of Schwarz
\emph{et al.}~\cite{schwarz2025} reports less than 10\% overhead for
componentwise-accurate DGEMM emulation on Blackwell in the Ozaki-I
setting (its extension to Ozaki~II is proposed, not yet demonstrated;
see the contract below).

We refer the reader to~\cite{ozaki_error_analysis_2026} for the full
componentwise analysis; the practical implication for this paper is that the
\emph{kernels we discuss inherit the same error bound as the underlying
emulated GEMM}, modulo the trivial summation rounding incurred at the
reconstruction step.

\paragraph{The numerical contract of this paper.}  Because ``exact'' and
``\fp{64}-accurate'' are different claims at different layers, we fix
the vocabulary once and use it throughout.  \emph{(i)~Exact core:} the
modular products and the multi-limb CRT/Garner reconstruction are exact
integer arithmetic within their stated ranges---algebraic statements,
independent of inputs.  \emph{(ii)~Grade~A (componentwise):} the
returned \fp{64} entries satisfy the componentwise bound
of~\cite{ozaki_error_analysis_2026} under the scaling assumptions of
Appendix~\ref{app:conditioning} \emph{and under that theorem's
round-to-nearest quantiser} (the truncation variant and the
codesigned sets carry the bound as a stated obligation,
Table~\ref{tab:contract}); this is the accuracy grade meant
whenever this paper says ``\fp{64}-accurate'' about a matrix product.
\emph{(iii)~Grade~B (compensated/normwise):} paths such as
\fp{32}+Kahan reductions carry a compensated error bound at the working
precision---strictly weaker than Grade~A---and are always labelled as
such; the tensor-resident Kulisch route of \S\ref{sec:cqladder} is
the exact upgrade.  \emph{(iv)~IEEE behaviour:} propagation of
NaN/Inf, subnormal handling, signed zero, and rounding-mode semantics
are \emph{not} claimed equivalent to native \fp{64} and are routed to
detection-plus-fallback (\S\ref{sec:falsification});
\emph{(v)~reproducibility} is claimed only where the exact
(Kulisch-class) reduction is used; and \emph{(vi)~fallback}: inputs
whose exponent span exceeds what the chosen $r$ certifies fall back to
the surviving native pipe, with frequency and cost an explicit
measurement target.  $r{=}12$ is a tested working point for the stated
distributions, $r{=}15$ the conservative guaranteed-range setting; no
claim in this paper should be read above its grade.

\paragraph{Dynamic range and the choice of $r$ are not free, but they
are bounded and automatable.}  A fair objection to any fixed-slice
emulation is that integer/fixed-point residue arithmetic carries no
exponent, so the safe number of moduli $r$ depends on the dynamic range
of the specific operands and cannot be assumed constant: a single
optimistic $r$ may under-serve ill-conditioned inputs, and determining
the right $r$ per operation is itself work that must be charged.  We
make two points.  First, the value used throughout this
paper, $r=12$, is the \fp{64}-equivalent setting recommended by
Mukunoki~\cite{mukunoki2025_fp8} ($r\!\in\![11,14]$) rather
than a throughput-maximising guess; the projections should be read as a
\emph{function} of $r$ (each unit of $r$ scaling the dense compute
ceiling by $\approx r/(3r{+}1)$), with the larger end of the range the
conservative choice for guaranteed accuracy.  Second, the per-input
selection of $r$ need not be hand-tuned or expensive: the
Exponent-Span-Capacity (ESC) estimator and Automatic Dynamic Precision
(ADP) framework of Schwarz \emph{et al.}~\cite{schwarz2025} choose the
slice count per GEMM at runtime, fully GPU-resident, with automatic
fallback to native \fp{64}, and report less than $10\%$ overhead even
under worst-case configurations, validated against adversarial
BLAS-grading tests---demonstrated for Ozaki-I; extending ESC/ADP to
Ozaki~II's modulus count and error model is a proposed part of the
validation programme, not established evidence.  The TME model accommodates this directly by letting
$\alpha=(3r{+}1)$ track an ESC-chosen $r$ rather than a fixed constant
(\S\ref{sec:futurework} and the per-row treatment of
Appendix~\ref{app:fusion-details}).  The residual question---how often
ADP must fall back on real scientific inputs, and whether that fallback
materially erodes the memory-roof advantage---is genuinely empirical and
is one of the measurements we prioritise in \S\ref{sec:futurework};
notably, on B300 the fallback target is the small \emph{surviving}
native \fp{64} pipe, not a zero-\fp{64} part, so an occasional fallback
degrades gracefully rather than catastrophically.

\subsection{Other FP64 Emulation Approaches}

Several alternative paths exist for emulating high-precision arithmetic on
low-precision tensor units.  Markidis \emph{et al.}~\cite{markidis2018}
explored \fp{16} tensor cores for \fp{32}-equivalent accuracy, the natural
ancestor of all modern emulation work.  Iterative-refinement solvers
\cite{haidar2018,anzt2018} use low-precision GEMMs as inner kernels and
correct residuals in high precision, achieving high accuracy on solvers but
not on bare GEMM-like primitives.  More recently, MixPert and similar
adaptive-precision frameworks compose low-precision tensor operations with
correction terms.

Ozaki-style emulation is distinctive in that it provides
\textbf{provable componentwise} error guarantees on the bare matrix product
itself, without requiring an outer iteration or a problem-specific tuning
loop.  This property is essential for drop-in replacement of cuBLAS DGEMM
in scientific codes, which is exactly the use case NVIDIA targeted with the
October~2025 cuBLAS integration~\cite{nvidia_cublas_emulation}.

\subsection{Tensor Cores for Memory-Bound Kernels: Prior Art}

There is a parallel literature, largely independent of Ozaki, on
\emph{making memory-bound kernels use tensor cores at all}---irrespective of
precision considerations.  TCStencil~\cite{liu2022_tcstencil},
SPTCStencil~\cite{gu2025_sptcstencil} and SparStencil~\cite{wang2025_sparstencil}
reformulate stencil sweeps as block-sparse GEMMs amenable to dense or
sparse tensor cores.  These works treat the precision question orthogonally
(typically retaining \fp{32} or \fp{16}).  Our contribution sits at the
\emph{intersection} of the two threads: we want both (i) the tensor-core
mapping that the stencil community has developed, and (ii) the \fp{64}-accurate
emulation that the Ozaki community has developed.  The TME model is the
analytic tool that lets us reason about both simultaneously.

Finally, the present work sits within a broader, ongoing reassessment of
the relationship between AI-oriented hardware and scientific computing, of
which recent strategic surveys of how HPC must adapt to AI-era
silicon~\cite{dongarra2026ride} are representative.  That literature
documents the shift---native \fp{64} ceasing to be the assumed foundation
of HPC---and its consequences for the field; our aim is complementary and
more specific, namely to supply the primitive-level theory that shows the
shift is not merely survivable but sound, with the FP8 matrix op as the
foundation onto which the established kernel taxonomy maps.

\section{Architectural Analysis: The FP64 Cliff}
\label{sec:arch}

We now establish the architectural baselines used throughout the rest of the
paper.  All numbers in Table~\ref{tab:specs} are taken from the official
NVIDIA datasheets and corroborated by independent
microbenchmarks~\cite{nvidia_blackwell_ultra,nvidia_b200_datasheet,jarmusch2025,dickenmann2026}.
We report \emph{dense} throughput per single GPU unless explicitly marked as
sparse, because real HPC workloads do not generally exhibit the 2:4 structured
sparsity required to reach the sparse rates.

\begin{table}[h]
\caption{Per-GPU architectural parameters used in the TME model.  Tensor
throughput numbers are dense unless noted.  Vector \fp{64} refers to
non-tensor SIMT pipes.  Sources: NVIDIA Blackwell~Ultra
datasheet~\cite{nvidia_blackwell_ultra}, HGX~B200
datasheet~\cite{nvidia_b200_datasheet},
microbenchmark study~\cite{jarmusch2025}, NVIDIA Rubin specifications
as published~\cite{nvidia_rubin_blog,theregister_rubin,lockwood_rubin},
and---for the Rubin \fp{8}/\fp{16} rates---the NVIDIA DGX Rubin NVL8
specification of June~30, 2026~\cite{nvidia_dgx_rubin_nvl8}.}
\label{tab:specs}
\centering
\footnotesize
\setlength{\tabcolsep}{4pt}
\begin{tabular}{l|cccc}
\toprule
\textbf{Metric} & \textbf{H100} & \textbf{B200} & \textbf{B300}\textsuperscript{$\star$} & \textbf{R200} \\
                & (Hopper)      & (Blackwell)   & (Blackwell Ultra) & (Rubin) \\
\midrule
FP64 Vector (TFLOPS)             & 34   & 40       & $\sim$1.3 & $\sim$33 \\
FP64 Tensor, native (TFLOPS)     & 67   & 40       & $\sim$1.2 & ---\textsuperscript{*} \\
FP64 Matrix, emulated (TFLOPS)   & ---  & ---      & via Ozaki~II & $\sim$200 \\
\midrule
FP16/BF16 Tensor (TFLOPS, dense) & 989  & 2{,}250  & 2{,}500   & 4{,}000 \\
TF32 Tensor (TFLOPS, dense)      & 494  & 1{,}100  & 1{,}250   & 2{,}000 \\
FP8 Tensor (TFLOPS, dense)       & 1{,}979 & 4{,}500 & 5{,}000 & \textbf{17{,}500}\textsuperscript{$\ddagger$} \\
FP6 Tensor (TFLOPS, dense)       & ---  & ---      & ---       & 17{,}500 \\
INT8 Tensor (TOPS, dense)        & 1{,}979 & 4{,}500 & $\sim$165 & $\sim$250 \\
FP4 / NVFP4 (TFLOPS, dense)      & ---  & 7{,}000  & 10{,}000  & 35{,}000 \\
\midrule
HBM Bandwidth (TB/s)             & 3.35 & 7.7--8.0 & 8.0       & \textbf{22} \\
HBM Capacity (GB)                & 80   & 180--192 & 279--288  & 288 \\
\midrule
\textbf{FP8 : native FP64 ratio} & \textbf{30:1} & \textbf{113:1} & \textbf{3800:1} & \textbf{530:1\textsuperscript{$\dagger$}} \\
\textbf{Memory ridge \flops{}/B (native)} & 10.1 & 5.0 & 0.16 & 1.5 \\
\bottomrule
\end{tabular}
\\[2pt]
{\scriptsize \textsuperscript{*}~Rubin specifications list FP64 matrix
performance under an explicit ``Emulated DGEMM'' column rather than as a
native tensor-core rate~\cite{nvidia_rubin_blog,lockwood_rubin}.
\textsuperscript{$\dagger$}~FP8 to native vector FP64; against the
published Emulated-DGEMM figure ($200$~TFLOPS) the ratio is
$\sim\!88{:}1$, and against the model ceiling
$P_{\fp{8}}/(3r{+}1)\approx473$~TFLOPS it is $\sim\!37{:}1$ ($r{=}12$).
\textsuperscript{$\ddagger$}~NVIDIA DGX Rubin NVL8 specification, June~30,
2026 (140~PFLOPS dense FP8/FP6 per 8~GPUs)~\cite{nvidia_dgx_rubin_nvl8};
supersedes the $\sim\!4{,}000$ used in versions of this paper through
July~3, 2026 (see \S\ref{sec:cqladder}).
\textsuperscript{$\star$}~The B300 column carries pre-freeze
illustrative constants; from \S\ref{sec:cqladder} onward the analysis
uses the companion's \emph{frozen} GB300 (NVL72-derived per-GPU) values
---$5.0$~PF \fp{8}, $166.7$~\tops{} \inteight{}, $1.4$~TF \fp{64},
$8$~TB/s---which differ from the HGX~B300 SKU; deltas $\le 7\%$ (see
the platform note in Table~\ref{tab:glossary}).}
\end{table}

\paragraph{Reading the table.}  Four columns---H100, B200, B300, and
R200---trace a clear architectural trajectory.  Native \fp{64} compute is
no longer growing with each generation; on B300 it has effectively been
removed, and on Rubin NVIDIA has chosen to expose ``Emulated DGEMM'' as
an explicit, first-class column in the official
specifications~\cite{nvidia_rubin_blog}.  At the same time, the
low-precision matrix pipes (\fp{8}, NVFP4) and the HBM bandwidth have
each grown by factors of $2$--$3$ per generation, with Rubin's $22$~TB/s
HBM4 marking a $2.75\times$ jump over B300.  Any performance model for
scientific computing on these architectures must therefore treat
emulation \emph{not as an optimisation} but as the \emph{native execution
model} for \fp{64} matrix operations.

\paragraph{Why Ozaki~II runs through \fp{8} on these architectures.}
The \inteight rate on Blackwell Ultra and Rubin has not scaled with
\fp{8} or NVFP4---in fact it has \emph{decreased} relative to Hopper, as
silicon area has been redirected toward low-precision floating-point
matrix units.  The natural path to high emulation performance on these
architectures therefore runs through
\fp{8}~\cite{uchino2026_fp8,mukunoki2025_fp8}.  On \fp{8} the B300
delivers $5$~\pflops{} dense and $10$~\pflops{} sparse; with
$r\!=\!12$ moduli and the full $(3r{+}1)$ \fp{8} cost
($\alpha=37$), the
\emph{effective upper bound} on emulated \fp{64} throughput is
$5{,}000/37 \approx 135$~TFLOPS dense.  On Rubin, the official
$17.5$~\pflops{} dense \fp{8}~\cite{nvidia_dgx_rubin_nvl8}
with the same parameters yields $\approx 473$~TFLOPS; NVIDIA's
published $\sim 200$~TFLOPS ``Emulated DGEMM'' figure sits at
$\approx\!42\%$ of this bound---a margin compatible with a delivered,
accuracy-guaranteed library figure~\cite{schwarz2025,ozaki_error_analysis_2026},
but also, as \S\ref{sec:rubinuplift} quantifies, exactly the deconstruction-limited branch
the deconstruction term alone produces at moderate problem sizes
($\bar{n}\!\approx\!500$--$900$)---a distinction a single size
sweep decides
(\S\ref{sec:rubinuplift}; versions of this paper through July~3, 2026
used a superseded $4$-\pflops{} figure under which the published number
appeared to exceed the bound).

\paragraph{Where the native memory ridge sits.}  Combining the native
\fp{64} compute roof with HBM bandwidth gives ridge points of
$10.1$~\flops{}/B on H100, $5.0$~\flops{}/B on B200, $0.16$~\flops{}/B on
B300, and $1.5$~\flops{}/B on Rubin.  These four numbers tell three
distinct stories.  On H100 and B200, the ridge sits above the
operational intensity of every standard memory-bound primitive (SpMV
$\sim 0.2$, stencils $\sim 0.5$, batched GEMV $\sim 0.5$--$2$), so those
kernels are bandwidth-bound natively and the architecture is balanced
for the workload class.  On B300, the ridge has collapsed to
$0.16$~\flops{}/B, below the operational intensity of \emph{every}
standard scientific kernel; the native pipe is the bottleneck even for
SpMV.  On Rubin, the picture is genuinely mixed: the bandwidth jump to
$22$~TB/s combined with the modest $33$~TFLOPS vector \fp{64} pushes
the ridge to $1.5$~\flops{}/B, which puts SpMV ($\sim 0.2$), stencils
($\sim 0.5$), and small-batch GEMV back into the memory-bound regime
where native \fp{64} vector is once again usable, while batched GEMV at
$B{=}8$ ($\mathrm{OI}\sim 2$) and dense GEMM
sit at or above the ridge and benefit from the Emulated DGEMM path.
In other words, Rubin partially restores the native-\fp{64}-viable
regime for the lowest-OI kernels, and Ozaki~II picks up the rest.

\section{The Tensor--Memory Equilibrium Model}
\label{sec:model}

We now develop the analytic performance model used in the rest of the paper.
Let a kernel perform $W$ \fp{64}-equivalent floating-point operations on
$Q$ bytes of memory traffic, so that its operational intensity is $\mathrm{OI}=W/Q$.
We adopt OI in preference to ``operational intensity'' to avoid notational
collision with AI throughout this paper.

\subsection{Native FP64}

The Roofline model~\cite{williams2009_roofline} predicts the native execution
time as
\begin{equation}
T_{\text{nat}} \;=\; \max\!\left(\frac{W}{P_{\textsc{fp64}}},\frac{Q}{B_\text{mem}}\right)
              \;+\; L_{\text{mem}},
\label{eq:tnat}
\end{equation}
where $P_{\textsc{fp64}}$ is the peak \fp{64} throughput, $B_\text{mem}$ the
HBM bandwidth and $L_\text{mem}$ the cold-start latency (assumed amortised
for the workload sizes considered).  The kernel is memory-bound when
$\mathrm{OI}<P_{\textsc{fp64}}/B_\text{mem}$, compute-bound otherwise.

\subsection{Emulated Execution}

Replacing native \fp{64} by Ozaki~II introduces four overhead parameters.
(Earlier versions of this paper carried only the first three; the fourth,
the input-side deconstruction cost, was identified as a first-order
omission in the technical review of Bayraktar, Gunnels, and
Caday~\cite{bayraktar2026tme} and is incorporated here---see
\S\ref{sec:whyupdate}.)

\begin{definition}[Emulation parameters]
For a fixed Ozaki~II configuration on a given hardware target,
\begin{description}[leftmargin=2em,topsep=2pt,itemsep=2pt]
\item[$\alpha$] $\;=\;$ number of low-precision tensor-core MMAs required per
   \fp{64} fused multiply--add.  For the \fp{8} Ozaki~II variant this is
   $3r{+}1$ with $r$ the number of moduli (three Karatsuba \fp{8} MMAs
   per modulus plus the max-magnitude pass; \S\ref{sec:model}
   Crossover Analysis), modulo constant overhead from index arithmetic.
\item[$\beta$] $\;\ge 1\;$ is the \emph{bandwidth multiplier}: the ratio of
   bytes moved by the emulated kernel to the bytes that the equivalent native
   \fp{64} kernel would move: with $Q_0$ the native reference bytes and
   $Q_{\text{emu}}$ the complete emulated bytes, $\beta = Q_{\text{emu}}/Q_0$,
   from the complete byte ledger.  $\beta=1$ for fully fused on-chip decomposition
   (residues live on chip and never touch global memory); in the
   unfused-materialised regime $\beta \approx (8+2p)/8$ per streamed
   operand ($p$ the stored plane bytes per scalar, $30$--$44$ for the
   candidate sets---write plus read), scheduling-dependent; $r$ is kept
   strictly as the modulus count.
\item[$\gamma$] $\;\ge 0\;$ is the per-output reconstruction latency, in
   seconds per output element, capturing the cost of Garner's algorithm.
\item[$c_q$] $\;\ge 0\;$ is the per-input \emph{deconstruction} cost: the
   number of SIMT integer instructions required to convert one streamed
   operand element into its residue representation for \emph{one} modulus
   (scaling, rounding, modular reduction, and---on the \fp{8}
   substrate---the Karatsuba plane conversions), so that the total
   deconstruction work is $c_q\, r$ instructions per converted element.
   $c_q$ and the sustained integer-issue rate $P_{\text{int}}$ that
   executes it are architecture and implementation parameters of the
   model, not constants.  For dense GEMM the conversion \emph{count}
   is written explicitly as
   $N_{\text{conv}} = \lambda_A\,mk + \lambda_B\,kn$, with
   $\lambda_A, \lambda_B \ge 1$ the schedule-dependent conversion
   multiplicities (\S\ref{sec:hwproposal}); the streamed-kernel form
   $\varphi\,n_{\text{in}}$ with $\varphi \in [0,1]$ is the special
   case in which conversions track streamed elements.  The conversion
   ledger and the HBM byte ledger are kept separate and are related
   only under a stated schedule.
\end{description}
\end{definition}

The emulated time's \emph{service-tail endpoint}---a $\max$ over the mutually
overlapped services followed by an unoverlapped reconstruction tail, \emph{not}
a full serialisation---is then
\begin{equation}
T_{\text{serial}} \;=\;
   \max\!\left(\frac{\alpha\, W_{\text{mma}}}{P_{\text{low}}},\;
               \frac{\beta\, Q_0}{B_\text{mem}},\;
               \frac{c_q\, r\, \varphi\, n_{\text{in}}}{P_{\text{int}}}\right)
   \;+\; \gamma\, n_\text{out},
\label{eq:temu}
\end{equation}
where $P_{\text{low}}$ is the relevant low-precision tensor throughput
($P_{\textsc{fp8}}$, $P_{\textsc{int8}}$, etc.), $n_\text{out}$ is the
number of output elements, $n_{\text{in}}$ is the number of streamed
input elements ($n_{\text{in}} \approx Q/8$ for \fp{64}-streamed
operands), and $\varphi \in [0,1]$ is the fraction of streamed elements
that must actually be converted (elements reused from an on-chip residue
copy, or constant across the kernel as in the stencil's coefficient
table, are converted once and re-used, so $\varphi<1$ where such reuse
exists).  In $T_{\text{serial}}$ the per-output reconstruction is charged
fully serially after the $\max$.  The companion's per-record engine and
every table in this paper instead use the overlapped \emph{host-service}
rate, in which each per-output reconstruction demand is added to whichever
host pipe (\textsc{int8} or SIMT) carries it and overlapped inside a
single $\min$ over service rates, with an explicit serial term only for
non-pipelinable dependencies.  Because that engine has accreted terms as
successive obligations were discharged, we state it once here in full;
\emph{these seven terms are the entire model}, and every projected rate in
either paper is one evaluation of Eq.~\eqref{eq:psvc}:
\begin{equation}
\begin{aligned}
P_{\text{svc}}(m,n,k) \;=\; \min\Big(
&\underbrace{\tfrac{P_{\fp{8}}}{\alpha}}_{\text{(1) MMA roof}},\;
\underbrace{\tfrac{P_{\text{int}}}{\frac{\mu\lambda_{\text{eff}}}{\bar n}
   + \frac{N^{I}_{\text{rec}}(r)}{2k}}}_{\text{(2) \inteight{} host}},\;
\underbrace{\tfrac{P_{\textsc{simt}}}{\frac{c_q^{\text{res}} r
   \lambda_{\text{eff}}}{\bar n}
   + \frac{N^{S}_{\text{rec}}(r)}{2k}}}_{\text{(3) SIMT host}},\\[1.2ex]
&\underbrace{\tfrac{B_{\text{hbm}} W}{Q_0 + Q_{\text{re}} + Q_{\text{red}}}}_{\text{(4) HBM}},\;
\underbrace{\tfrac{B_{\textsc{l2}}}{\phi\,p_{\text{store}}/T}}_{\text{(5) plane feed, \textsc{l2}}},\;
\underbrace{\tfrac{B_{\text{hbm}}}{p_{\text{store}}/T}}_{\text{(6) plane feed, HBM}},\;
\underbrace{\tfrac{B_{\textsc{l2}} W}{Q_{\text{red}}}}_{\text{(7) slab reduction}}
\Big).
\end{aligned}
\label{eq:psvc}
\end{equation}
Here $\mu$ is the route's deconstruction MACs per input element, $c_q^{\text{res}}$
its SIMT residual, $\lambda_{\text{eff}}=(m\lambda_A{+}n\lambda_B)/(m{+}n)$ the
size-weighted re-split multiplicity of \S\ref{sec:frozen-kernel},
$N^{I}_{\text{rec}}$/$N^{S}_{\text{rec}}$ the reconstruction op counts on the two
hosts, $Q_0$ the \fp{64} operand and output traffic, $Q_{\text{re}}$ the
\fp{64} re-reads incurred by on-the-fly re-splitting, $Q_{\text{red}}$ the
$k$-slab residue-reduction traffic, $T$ the tile edge, and
$\phi\in\{\tfrac12,1\}$ selects the hybrid (small operand resident) or fully
materialised variant.  Terms (5)--(7) are alternatives selected by the
schedule, not simultaneous; the engine takes the best-scoring schedule and
then the $\min$ within it.  A candidate eighth term---the SMEM operand
bandwidth needed to feed $p_{\text{store}}$ planes per scalar---is
\emph{provably} never binding: at a route's own roof its demand-to-supply
ratio is exactly $p_{\text{store}}/\alpha$, independent of platform and tile
size ($0.81/0.83/0.96$ for S/E/A), so it is checked in
\S\ref{sec:frozen-kernel} and then omitted.  These are model
\emph{endpoints}, not rigorous bounds on measured time: with perfect service
overlap the time is $T_{\text{svc}}$; the genuine no-overlap serialisation is
the \emph{sum} of the service times (which upper-bounds $T_{\text{serial}}$);
and imperfect overlap moves $T_{\text{meas}}$ between $T_{\text{svc}}$ and---absent
the assumed service-independence---potentially past $T_{\text{serial}}$.  We
report $T_{\text{svc}}$ and treat the gap as a sensitivity.  We write
$T_{\text{emu}}$ for the emulated time generically when the two endpoints
coincide, as they do in the memory-bound limit where the
$\gamma\,n_{\text{out}}$ term vanishes and both reduce to
$Q/B_{\text{mem}}\,(1+o(1))$.  Two structural remarks, both due
to~\cite{bayraktar2026tme}.  \emph{First, the duality:} reconstruction is
charged per \emph{output} ($\gamma\,n_{\text{out}}$) and deconstruction
per \emph{input} ($c_q r\,n_{\text{in}}/P_{\text{int}}$); for memory-bound
kernels $n_{\text{in}} \gg n_{\text{out}}$, which is precisely why the
term is first-order and why its earlier omission mattered.  \emph{Second,
tile granularity:} the MMA term must charge the \emph{executed} work, not
the useful work.  The \fp{8} MMA atom has a minimum accumulator width of
$n{=}8$ columns, so a kernel with $B<8$ right-hand sides executes
\begin{equation}
W_{\text{mma}} \;=\; \max\!\left(1,\,\tfrac{8}{B}\right) W,
\label{eq:wmma}
\end{equation}
paying for the padded lanes whether or not they carry data; at $B{=}1$
this raises the \fp{8} keep-up requirement eightfold (still second to the
deconstruction term at reference constants; Appendix~\ref{app:sparse}).

\paragraph{The conversion-bound regime and $\mathrm{OI}^{*}$.}
The deconstruction term creates a third binding regime alongside
compute-bound and memory-bound: a kernel is \emph{conversion-bound} when
$c_q r \varphi\, n_{\text{in}}/P_{\text{int}}$ dominates
Eq.~\eqref{eq:temu}.  Against native \fp{64}, emulation can win only if
converting an element costs less time than simply doing its \fp{64} work,
which for \fp{64}-streamed operands ($n_{\text{in}}=Q/8$) reduces
to~\cite{bayraktar2026tme}
\begin{equation}
\mathrm{OI} \;>\; \varphi\,\mathrm{OI}^{*}, \qquad
\mathrm{OI}^{*} \;=\; \frac{c_q\, r\, P_{\fp{64}}}{8\, P_{\text{int}}},
\label{eq:oistar}
\end{equation}
and any kernel below the threshold is capped at
$S \le \varphi^{-1}\,\mathrm{OI}/\mathrm{OI}^{*}$ times native
performance regardless of $\beta$, occupancy, or tensor throughput.
Similarly, the emulated kernel reaches the memory \emph{roof} only if the
conversion keeps pace with the stream, i.e.\ only if
$c_q \le 8P_{\text{int}}/(r\varphi B_{\text{mem}})$; the shortfall factor
$c_q r \varphi B_{\text{mem}}/(8P_{\text{int}})$ is independent of the number of
right-hand sides, because conversion is charged per streamed element.
\textbf{A note on the division of labour in this paper.}  Everything
up to this point in the subsection is \emph{architecture-independent
theory}: the four-term model, the duality, the tile-granularity charge,
and the thresholds of Eq.~\eqref{eq:oistar} hold on any machine
executing Ozaki~II, with $c_q$, $P_{\text{int}}$, and $\varphi$ as
symbolic parameters.  What follows---the reference constants
below, and the sensitivity analysis, implementation strategies, and
architectural observations of \S\ref{sec:rubinuplift}---is the
\emph{instantiation on current GPUs}, deliberately quarantined so that
the model survives hardware revisions unchanged while the constants
page turns with each generation (as this version's Rubin correction
illustrates).  \textbf{Reference constants.}  The review of~\cite{bayraktar2026tme}
counted, on the shipping cuBLAS INT8-emulation path,
$c_q = 16$ instructions per modulus per element (13 counted from the
shipping SASS plus 3 for the \fp{8} Karatsuba plane converts), with
$P_{\text{int}} \approx 75$~\tops{} on B300; the memo's
non-confidential stage structure and normalisation are summarised in
Appendix~\ref{app:memo}.  One provenance caution attaches: the released
cuBLAS GEMM emulation is publicly described as Ozaki-I
slicing~\cite{emugemm2026}, while NVIDIA's cuEST emulation stack now
exposes \emph{both} schemes (slice-count and modulus-count controls,
selected by compute capability)~\cite{nvidia_cuest}; the memo's counted
stage structure is CRT-flavoured, and the confirmation request to its
authors (Appendix~\ref{app:memo}) asks which realisation the counted
SASS belongs to.  Statements that read shipping \emph{anchors} through
CRT constants are conditional on that attribution.  At this paper's working
point ($r{=}12$, $P_{\fp{64}}{=}1.3$~TFLOPS) these give
$\mathrm{OI}^{*} \approx 0.42$~\flops{}/B and a roof-shortfall factor of
$\approx 2.6$; at the fully conservative $r{=}15$ (the source's guaranteed-range
setting~\cite{uchino2026_fp8}) and the NVL72-derived $P_{\fp{64}}{=}1.4$~TFLOPS,
$\mathrm{OI}^{*} \approx 0.56$ and the shortfall is $\approx 3.2$.
\textbf{A reconciliation note on $P_{\text{int}}$.}  The spectral
companion~\cite{matsuoka2026fft} adopts a different normalisation for
the same pipe: $41.7$~T~inst/s, derived as the published B300 \fp{32}
TFLOP/s figure divided by two on the assumption that \fp{32} and
integer issue share lanes on Blackwell (an FMA counts as two flops on
one issue slot).  The $75$~T~inst/s used here is the memo's own stated
normalisation, adopted as the reference constant because it is the one
the memo's instruction counts were made against; the two differ by
$1.8\times$, and \emph{which is physically right is precisely the
integer-pipe census that Part~3 is scoped to measure}.  We therefore
carry the spec-derived value as an explicit sensitivity: at
$P_{\text{int}} = 41.7$~T~inst/s and this paper's working point,
$\mathrm{OI}^{*} \approx 0.75$~\flops{}/B ($\approx 0.81$ at the
NVL72-derived $P_{\fp{64}} = 1.4$~TFLOPS, the companion's own figure),
which widens the conversion-bound band---the SpMV and single-RHS GEMV
verdicts of Table~\ref{tab:speedups} become \emph{more} conservative,
not less---and likewise tightens every $P_{\text{int}}$-bounded quantity
in the companion's dense analysis (the SIMT keep-up budget of
\S\ref{sec:cqladder} falls from $6.25$ to ${\approx}3.5$ instructions
per modulus on GB300 and from $2.3$ to ${\approx}1.3$ on Rubin, and the
SIMT-residual knees move outward accordingly).  The direction is
uniform: the $41.7$ case makes the software-only picture harder and the
co-design case stronger, never the reverse.  The tensor-side quantities
($473$~TFLOPS; the companion's $0.497$ ratio of the deconstruction-$\lambda$
term to the roof) do not depend on $P_{\text{int}}$, but the dense
\emph{floor} does, because it is a minimum over terms: the companion's
engine re-run at $41.7$~T~inst/s finds the SIMT-residual term binding on
Rubin, with the large-square floor at ${\approx}148$~TFLOPS ($0.31$ of
$473$) in place of the $0.50$ quoted at $75$, and GB300 at route~E's own
$125$~TFLOPS roof in place of the \texttt{dp4a}-route $135$; route~E
leads on both parts under either normalisation (companion,
Appendix~B reconciliation note).  Until the census reports, the
$75$-T~inst/s verdicts are the ones to quote and the $41.7$ case is the
stated downside.  We
emphasise the model reading: $c_q{=}16$ is a \emph{memo-counted property
of a library path never engineered to minimise deconstruction}, not a
floor;
roof recovery requires $c_q \lesssim 5$--$6$ at $r{=}12$ (the
published set's refined two-limb residual $5.8$ sits at this
threshold; the all-byte codesign's $5.3$ runs at its own $r'{=}15$
and delivers $0.95$ of the \emph{memory} roof~\cite{matsuoka2026ozaki25}) and single-RHS GEMV
native-parity $c_q \lesssim 7$--$10$ (depending on $r$ and
$P_{\fp{64}}$), and~\cite{bayraktar2026tme} itself states the
two-sided decision criterion in exactly this form.  Determining the
engineered floor of $c_q$---and whether the integer-issue budget
$P_{\text{int}}$ can be supplemented by split issue across pipes or by
routing the (bitwise-linear) reduction onto the tensor cores
themselves---is the opening experiment of the Part~3 measurement
programme (\S\ref{sec:futurework}).  Throughout this paper we therefore
state each claim twice: at the fused, conversion-ideal bound, and at the
reference constants above.

\paragraph{Evidence labels.}  Because this paper mixes vendor
specifications, a source-counted memo, script-reproduced algebra, and
model projections, every quantitative claim carries (explicitly or by
context) one of six evidence labels, shared with the companion
paper~\cite{matsuoka2026ozaki25}: \emph{published specification} (a
vendor peak or ``up to'' value, not a measured kernel rate);
\emph{measured} (timed on named hardware and software---in these two
papers, only the companion's call-shape traces); \emph{source-counted}
(derived from an instruction trace with stated provenance, e.g.\ the
memo-counted $c_q{=}16$); \emph{script-checked} (algebra or arithmetic
reproduced by a script shipped in the companion's artifact, e.g.\ its
modulus-set obligations and supply bounds); \emph{ledger-projected} (an
instruction-count projection for an uncompiled route, e.g.\ the
$6.3/5.8/5.3/4.3$ residuals); and \emph{modeled projection} (the
output of a stated equation under stated assumptions).  No
ledger-projected or modeled value in this paper is a delivered rate,
and none is labelled ``verified'' or ``measured.''

\paragraph{The formal statement of the thesis.}  The machinery above
lets the paper's central claim (\S\ref{sec:formalclaim}) be stated as a
conditional theorem whose conditions are exactly the seams between
theory and implementation.

\begin{proposition}[Ideal-overlap roof-attainment criterion for the
atomic primitive, under the TME model]
\label{prop:atomic}
Let a kernel perform $W$ \fp{64}-equivalent operations over $Q$ bytes of
HBM traffic with $n_{\text{out}}$ outputs, realised through Ozaki~II on
an \fp{8} matrix engine under the model of
Eqs.~\eqref{eq:temu}--\eqref{eq:oistar}, \emph{with the model's ideal
overlap of the compute, memory, and conversion streams taken as an
assumption} (the $\max$ in Eq.~\eqref{eq:temu} is a service-time lower
bound, not a delivered-time identity), and suppose
\begin{itemize}[leftmargin=2.5em,itemsep=1pt,topsep=2pt]
\item[\textbf{(C1)}] \emph{fused traffic:} $\beta = 1$;
\item[\textbf{(C2)}] \emph{deconstruction keep-up:}
   $c_q\, r\, \varphi\, n_{\text{in}} / P_{\text{int}} \le
   Q/B_{\text{mem}}$ (equivalently $\chi \le 1$);
\item[\textbf{(C3)}] \emph{tensor keep-up:}
   $\alpha W_{\text{mma}} / P_{\text{low}} \le Q/B_{\text{mem}}$ in the
   memory-bound regime;
\item[\textbf{(C4)}] \emph{reconstruction amortisation:}
   $\gamma\, n_{\text{out}} \ll Q/B_{\text{mem}}$.
\end{itemize}
Then $T_{\text{emu}} = Q/B_{\text{mem}} + \gamma n_{\text{out}} =
(1 + o(1))\,Q/B_{\text{mem}}$, the $o(1)$ taken along any problem
family with $n_{\text{out}}/Q \to 0$ at fixed architecture (e.g.\
growing inner dimension $k$): the
emulated kernel runs at the memory roof up to the vanishing
reconstruction term, i.e.\ asymptotically at the time an ideal machine
with the same bandwidth and \emph{unbounded native \fp{64} compute}
would achieve---no performance is sacrificed by the reduction to the
\fp{8} primitive.  When (C3) fails because
$\mathrm{OI} > P_{\text{low}}/(\alpha B_{\text{mem}})$, the kernel is
compute-bound and attains the emulation ceiling
$P_{\text{low}}/(\alpha \cdot W_{\text{mma}}/W)$---which equals
$P_{\text{low}}/\alpha$ only when the tile-granularity inflation of
Eq.~\eqref{eq:wmma} is absent ($W_{\text{mma}} = W$)---the
corresponding no-sacrifice statement for that regime.
\end{proposition}

The proposition itself is immediate from Eq.~\eqref{eq:temu}; it is a
\emph{conditional ideal-overlap bound}, not a proof that (C1)--(C4)
are jointly attainable on any given silicon---overlap attainability,
issue coexistence, and the storage-mode coupling of
\S\ref{sec:hwproposal} are exactly what the measurement programme
tests.  Its
content lies in two places.  First, the analytic sections of this
paper (\S\ref{sec:impl}--\S\ref{sec:l3} and
Appendices~\ref{app:fusion-details}--\ref{app:sparse}) verify, class
by class across the surveyed taxonomy, that (C1)--(C4) are
\emph{satisfiable in the model} with stated margins---this is what
establishes low-precision MM as a viable atomic primitive at the level
of theory, and it is the claim on which this paper stakes its title.
Second, the conditions are precisely where \emph{implementation
artifacts} live: delivered $\beta$ versus (C1), engineered $c_q$ and
effective $P_{\text{int}}$ versus (C2), delivered tile efficiency
versus (C3), reconstruction throughput versus (C4).  Whether a given
piece of silicon satisfies them today is an empirical question about
constants, not about the theorem---quantified for 2026 silicon in
\S\ref{sec:cqladder} and measured systematically in Part~3.

\subsection{Crossover Analysis}

Let $\rho \;=\; P_{\text{low}}/P_{\textsc{fp64}}$ be the precision ratio
(\eg{} $\rho\approx 3800$ for \fp{8} on B300).  The compute multiplier
$\alpha$ is the number of low-precision MMAs per high-precision op.  On
the \fp{8} substrate this is $\alpha=(3r{+}1)$: each of the $r$ residue
planes expands into three \fp{8} MMAs (the Karatsuba structure used to
emulate signed \textsc{int}9), plus one MMA for the max-magnitude
estimation pass.  At the recommended $r\!=\!12$ this gives
$\alpha=37$.  (On the INT8 substrate a centred residue fits the format
directly whenever $m_i\!\le\!256$, so the Ozaki~II cost is
$\alpha_{\textsc{int8}}=r_{\textsc{i}}{+}1$---one MMA per modulus plus
the max pass, ${\sim}3\times$ cheaper \emph{per modulus}, but charged at
the $r_{\textsc{i}}\!=\!15$ that a byte-modulus set needs for matched
accuracy, hence $16$ and not $13$; see \S\ref{sec:int8vsfp8}.)

\paragraph{Case~A: Native compute-bound, emulation memory-bound.}
This is the regime that the B300 collapses into for nearly every standard
kernel.  Equations \eqref{eq:tnat} and \eqref{eq:temu} give
\begin{equation}
T_{\text{nat}} = \frac{W}{P_{\textsc{fp64}}},\qquad
T_{\text{emu}} = \frac{\beta Q_0}{B_\text{mem}}+\gamma n_\text{out},
\end{equation}
and the emulation is profitable when
\begin{equation}
\frac{W}{P_{\textsc{fp64}}} \;>\; \frac{\beta Q_0}{B_\text{mem}} + \gamma n_\text{out}.
\label{eq:case-a-condition}
\end{equation}
Ignoring $\gamma$ momentarily, and assuming $\beta=1$ (the fused case),
the speedup of emulation over native is
\begin{equation}
\frac{T_\text{nat}}{T_\text{emu}}
\;=\;\frac{W/P_{\textsc{fp64}}}{Q/B_\text{mem}}
\;=\;\frac{I\cdot B_\text{mem}}{P_{\textsc{fp64}}},
\label{eq:case-a-speedup}
\end{equation}
that is, the speedup equals the kernel's distance into the compute-bound
region of the native roofline.  On the B300 with $I=0.5$ (a 7-point stencil),
$B_\text{mem}=8$~TB/s, $P_{\textsc{fp64}}=1.3$~TFLOPS, this gives a
speed-up of $0.5\cdot 8/1.3 \approx 3.1\times$.  This bound holds only
where the deconstruction condition of Eq.~\eqref{eq:oistar} is also met:
at the reference constants the same stencil ($\varphi\!\approx\!0.5$,
because each grid element's residues are reused across neighbouring
outputs and the coefficient table is converted once) is
conversion-capped at $\varphi^{-1}\mathrm{OI}/\mathrm{OI}^{*} \approx
2.4\times$, and kernels below $\varphi\,\mathrm{OI}^{*}$ see no gain at
all (\S\ref{sec:eval}).

\paragraph{Case~B: Both regimes memory-bound (small $I$).}
When the native kernel is already memory-bound, emulation cannot improve
time-to-solution beyond $T_\text{nat}$ in the best case ($\beta=1$).  More
precisely, $T_{\text{emu}}/T_{\text{nat}}\to\beta$, \emph{provided} the
deconstruction keep-up condition
$c_q \le 8P_{\text{int}}/(r\varphi B_{\text{mem}})$ also holds; otherwise the
ratio degrades further by the conversion shortfall factor of
Eq.~\eqref{eq:oistar} and emulation is strictly slower than native even
at $\beta=1$.  Hence fused decomposition
($\beta=1$) yields parity, and unfused decomposition is strictly worse.  This
gives the important practical guidance: \textbf{for genuinely memory-bound
kernels in the small-$I$ regime, on-chip tile fusion \emph{and} a
low-$c_q$ deconstruction path are not just optimisations; they are jointly a
correctness requirement for emulation to be viable at all}.

\paragraph{Case~C: Both regimes compute-bound (dense GEMM).}
Here the speedup is $\rho/\alpha$, and the bandwidth roof is irrelevant.
On the B300 with \fp{8} cores at $r\!=\!12$ ($\alpha=3r{+}1=37$), this gives a
throughput ceiling of $5{,}000/37 \approx 135$~TFLOPS, which is
$\approx 104\times$ over
B300 native \fp{64} ($1.3$~TFLOPS).  This is the regime in which
Ozaki~II for dense GEMM has been most extensively
studied~\cite{uchino2026_fp8,mukunoki2025_fp8,schwarz2025}.  Note that
the $(3r{+}1)$ \fp{8} cost structure bites \emph{only} in this
compute-bound case, where the kernel runs at the emulation ceiling and
the throughput scales as $1/\alpha$.  In Cases~A and~B the kernel is
bandwidth-limited and never reaches the compute ceiling, so the same
factor merely lowers the \emph{emulation ridge}---the operational
intensity below which a kernel stays memory-bound---from
$P_{\fp{8}}/(rB_\text{mem})$ to $P_{\fp{8}}/((3r{+}1)B_\text{mem})$;
on B300 this is a shift from $\approx 52$ to $\approx 17$~\flops{}/B,
still far above the intensity of every memory-bound kernel surveyed
(\S\ref{sec:eval}).

\subsection{The TME Picture}

Figure~\ref{fig:crossover} draws the TME roofline for both B300 and Rubin.
The native \fp{64} roof on B300 is flat at $1.3$~TFLOPS, and on Rubin at
$\sim 33$~TFLOPS for vector \fp{64}.  Ozaki~II with fused decomposition
follows the memory roof up to its own compute ceiling: $\approx 135$~TFLOPS
on B300 (dense \fp{8} at $5$~PFLOPS, $r{=}12$, $\alpha=3r{+}1=37$) and
$\approx 473$~TFLOPS
on Rubin (dense \fp{8} at the official $17.5$~PFLOPS with the same
parameters; \S\ref{sec:cqladder}).  Generationally, \emph{both} axes
now scale: the \fp{8} ceiling grows $3.5\times$ over B300 and the memory
roof $2.75\times$ (8~$\to$~22~TB/s), so the \fp{8} compute available per
byte streamed actually \emph{rises} from $\approx\!625$ to
$\approx\!795$ \fp{8}-ops/B---the emulation margin widens with the
generation rather than merely tracking bandwidth.  (Earlier versions of
this paper, using the superseded $4$-PFLOPS \fp{8} figure, described the
Ozaki~II ceiling as essentially flat between B300 and Rubin; the
official rate retires that reading.)  NVIDIA's published Emulated-DGEMM
figure for Rubin ($\sim 200$~TFLOPS) now sits \emph{inside} the model
bound---at $\approx\!42\%$ of the $r{=}12$ ceiling ($\approx\!53\%$ of
the conservative $r{=}15$ ceiling of $\approx\!380$~TFLOPS)---which is
exactly the relationship one expects between a delivered, guaranteed
library figure (with margin for componentwise-error
guarantees~\cite{schwarz2025,ozaki_error_analysis_2026} and delivered
efficiency) and an analytic peak; the earlier puzzle of the published
figure \emph{exceeding} the model bound is resolved by the corrected
\fp{8} rate.

Two observations follow directly from the figure and are the central
quantitative message of this paper.

\paragraph{Memory-bound regime: parity with the memory roof.}  In the
bandwidth-bound region (left of the ridge point) the Ozaki~II/\fp{8}
curve overlaps the memory roof exactly.  This is precisely what
scientific computing requires: \emph{whatever bandwidth the architecture
provides, the emulated path can consume}.  On B300 this lifts memory-bound
kernels from the collapsed $1.3$~TFLOPS native floor up to the $8 I$~TFLOPS
memory roof---a gain that ranges from $1.2\times$ at SpMV-grade
intensities ($I\!\approx\!0.2$) to $\approx\!12\times$ at batched-GEMV
intensities ($I\!\approx\!2$ at $B{=}8$).  These are the fused,
conversion-ideal bounds; the region below
$\varphi\,\mathrm{OI}^{*}$ (marked in Figure~\ref{fig:crossover}) is
conversion-bound at the reference constants and there the surviving
native pipe is the correct path until the engineered $c_q$ is
demonstrated (\S\ref{sec:eval}).  On Rubin, the same emulated curve
follows a $22 I$~TFLOPS memory roof, $2.75\times$ higher in absolute
throughput than B300 across the entire memory-bound region.  This
bandwidth-driven gain is the dominant Rubin advantage and is felt by
\emph{every} HPC kernel with $I \lesssim P_{\fp{8}}/((3r{+}1)\cdot
B_{\text{mem}}) \approx 17$~\flops{}/B (the corrected emulation ridge on
B300 at $r{=}12$; an earlier draft placed this at $\approx 52$ using
$\alpha=r$).

\paragraph{Compute-bound regime: surpassing B200 by a wide margin.}
In the compute-bound region (right of the ridge point) the Ozaki~II/\fp{8}
curve saturates at $\sim 135$~TFLOPS on B300 and $\sim 473$~TFLOPS on
Rubin (dense \fp{8}$/(3r{+}1)$ at $r{=}12$)---the \emph{theoretical} roof; a
large dense \emph{square} output delivers half the Rubin figure (the
${\approx}235$-TFLOPS deconstruction-$\lambda$ floor, \S\ref{sec:rubinuplift})
until the co-design of \S\ref{sec:hwproposal} lifts it, while tall/skinny and
small-batch shapes are projected to reach the roof on shipping hardware.  For reference, B200's native \fp{64} ceiling sits at
$40$~TFLOPS---so the emulated path on B300 exceeds B200's \emph{native}
roof by $\approx 3.4\times$ and on Rubin by $\approx 11.8\times$.  The
comparison basis matters and is stated here once for the paper: against
B200's own \emph{best available} path---the shipping INT8-substrate
emulated DGEMM at $\approx\!150$~TFLOPS delivered
(\cite{nvidia_cublas_emulation}; consistent with the
$281$-TFLOPS INT8 ceiling of Table~\ref{tab:int8vsfp8} at the measured
$\approx\!53\%$ efficiency)---B300's \fp{8}-only ceiling of $135$ is
\emph{not} a generational gain: B300 cannot run the INT8 path its
predecessor used ($165$ vs $4{,}500$~TOPS), and its \fp{8} route only
restores, not exceeds, the prior best.  This is itself a sharp
illustration of the substrate-transition cost; the unambiguous
generational gain arrives with Rubin's $473$.  In other
words, the TME roofline shows that emulation does not merely
\emph{compensate} for the \fp{64} collapse; it carries the entire
operational-intensity spectrum above the best native
\fp{64} performance offered on the prior generation, even after the
full $(3r{+}1)$ \fp{8} cost is charged.  This is the
quantitative argument for treating Ozaki~II as the new baseline rather
than as a fallback.

\begin{figure}[htbp]
\centering
\begin{tikzpicture}
\begin{axis}[
    width=14cm, height=8cm,
    xmode=log, ymode=log,
    xlabel={Operational intensity $\mathrm{OI}$ (FLOPs / Byte)},
    ylabel={Modeled upper-bound throughput (TFLOPS)},
    xmin=0.05, xmax=200,
    ymin=0.2, ymax=5000,
    grid=both,
    legend style={
        at={(0.02,0.98)},
        anchor=north west,
        font=\scriptsize,
        cells={anchor=west},
        draw=black!50,
        fill=white,
        fill opacity=0.85,
        text opacity=1.0
    },
    legend cell align={left}
]
\addplot[color=violet, very thick, dotted, domain=0.05:200] {22*x};
\addlegendentry{Rubin mem.\ roof (22 TB/s)}
\addplot[color=black, very thick, domain=0.05:200] {8*x};
\addlegendentry{B300 mem.\ roof (8 TB/s)}
\addplot[color=orange, dashed, very thick] coordinates {(0.05,0.4) (5,40) (200,40)};
\addlegendentry{B200 native FP64 roofline $\min(B\cdot\mathrm{OI},40)$}
\addplot[color=violet!70!black, dashed, very thick] coordinates {(0.05,33) (200,33)};
\addlegendentry{Rubin native FP64 vec.\ (33)}
\addplot[color=red, very thick] coordinates {(0.05,1.3) (200,1.3)};
\addlegendentry{B300 native FP64 (1.3)}
\addplot[color=blue!70!black, very thick] coordinates {(0.05,0.4) (16.9,135) (200,135)};
\addlegendentry{B300 Ozaki~II/FP8 ($r{=}12$, $\alpha{=}37$)}
\addplot[color=teal, very thick] coordinates {(0.05,1.1) (21.5,473) (200,473)};
\addlegendentry{Rubin Ozaki~II/FP8 ($r{=}12$, $\alpha{=}37$)}
\addplot[color=teal!55!black, densely dashed, very thick] coordinates {(50,235) (200,235)};
\addlegendentry{Rubin large-square deliv.\ floor (235)}
\node[teal!55!black, anchor=south, font=\scriptsize] at (axis cs:110,238) {deconstruction-$\lambda$ floor};
\addplot[mark=none, gray, dashed, forget plot] coordinates {(0.2,0.2) (0.2,5000)};
\node[black, rotate=90, anchor=south] at (axis cs:0.2,0.6) {\scriptsize SpMV};
\addplot[mark=none, gray, dashed, forget plot] coordinates {(0.5,0.2) (0.5,5000)};
\node[black, rotate=90, anchor=south] at (axis cs:0.5,0.6) {\scriptsize Stencil};
\addplot[mark=none, gray, dashed, forget plot] coordinates {(2,0.2) (2,5000)};
\node[black, rotate=90, anchor=south] at (axis cs:2,0.6) {\scriptsize bGEMV};
\addplot[mark=none, red!70!black, densely dotted, thick, forget plot] coordinates {(0.42,0.2) (0.42,5000)};
\node[red!70!black, rotate=90, anchor=south] at (axis cs:0.42,30) {\scriptsize $\mathrm{OI}^{*}$ (ref.\ $c_q$)};
\addplot[mark=none, gray, dashed, forget plot] coordinates {(50,0.2) (50,5000)};
\node[black, rotate=90, anchor=south] at (axis cs:50,0.6) {\scriptsize GEMM};
\end{axis}
\end{tikzpicture}
\caption{TME roofline projection for B300 and Rubin.  Each platform has
its own memory roof (black for B300, dotted violet for Rubin) and its
own native \fp{64} ceiling (red for B300, dashed violet for Rubin
vector; B200 in dashed orange for reference).  The Ozaki~II/\fp{8}
curves (blue for B300, teal for Rubin) trace the memory roof in the
bandwidth-bound regime and saturate at the emulation compute ceiling in
the compute-bound regime.  The Ozaki~II/FP8 ceiling is $\approx
135$~TFLOPS on B300 and $\approx 473$~TFLOPS on Rubin (each = $P_{\fp{8}}/(3r{+}1)$
with $r=12$, $\alpha=37$; Rubin at the official 17.5-PFLOPS dense \fp{8}
rate of June~30, 2026~\cite{nvidia_dgx_rubin_nvl8}); NVIDIA's published
``Emulated DGEMM'' figure
for Rubin, $\sim 200$~TFLOPS~\cite{nvidia_rubin_blog,lockwood_rubin},
sits at $\approx\!42\%$ of this bound---compatible with a delivered
library figure, but also exactly the branch the deconstruction term
alone produces at moderate problem sizes
(Eq.~\eqref{eq:pdense}, \S\ref{sec:rubinuplift}).  The teal $473$ line is
the \emph{theoretical} roof; for a large dense \emph{square} output the
delivered rate plateaus at half of it (the dashed teal
${\approx}235$-TFLOPS deconstruction-$\lambda$ floor in the dense-GEMM
region), which the hardware co-design of \S\ref{sec:hwproposal} (Option~C)
lifts back to the roof---tall/skinny and small-batch GEMMs are
projected to reach the roof on shipping hardware.
The dotted red vertical marks the conversion-bound threshold
$\mathrm{OI}^{*}\!\approx\!0.42$~\flops{}/B at the reference
deconstruction constants ($c_q{=}16$, $P_{\text{int}}{=}75$~\tops,
$r{=}12$; Eq.~\eqref{eq:oistar}): left of it the emulated curves are
\emph{not attained} at those constants---the binding resource is the
integer pipe---and the surviving native \fp{64} is the correct path
pending an engineered $c_q$.
The B200 reference is drawn as its full roofline
$\min(B\cdot\mathrm{OI}, 40)$ at its own $8$~TB/s: in the
bandwidth-bound region B300's emulated curve essentially
\emph{coincides} with B200's roofline (equal bandwidth), Rubin's
advantage there is its bandwidth ratio, and neither is an emulation
effect; the emulation advantage proper appears in
the compute-bound region.  All curves are modeled upper bounds, not
measurements.}
\label{fig:crossover}
\end{figure}

\section{Implementation Strategies for Memory-Bound Kernels}
\label{sec:impl}

The crossover analysis of \S\ref{sec:model} establishes that the
\emph{kernel design discipline}---and specifically the value of $\beta$---is
what determines whether Ozaki~II is profitable on memory-bound workloads;
in the vocabulary of Proposition~\ref{prop:atomic}, this section is the
class-by-class verification that condition (C1) is satisfiable in the
model for the streaming kernels.
We now describe how to drive $\beta\to 1$ for the three canonical
bandwidth-limited primitives.

\subsection{On-Chip Tile Fusion: the Common Pattern}

The unifying technique is to perform the decomposition (\eqref{eq:scale} and
the residue reduction $\bmod\,m_i$) \emph{after} the working-precision
operand has arrived in registers, and to perform the reconstruction
(\eqref{eq:garner}) on the accumulator before the result is stored.  This
keeps the residue planes off the HBM bus.  In CUDA terms:
\begin{enumerate}[leftmargin=2em,itemsep=2pt]
\item Each thread block reads a tile of $A$ (and possibly $B$) in
   \fp{64} from HBM into shared memory.
\item Within the tile, each warp computes the $r$ residue planes in
   registers, using a small precomputed table of moduli stored in
   \texttt{constant} memory.
\item The warp issues $r$ low-precision \texttt{wmma::mma\_sync} (or
   equivalent \texttt{tcgen05}) instructions, each operating on the $i$th
   residue plane.
\item After the tile loop, the warp reconstructs the integer accumulator
   via Garner's algorithm \eqref{eq:garner}, rescales by $D^{-1}E^{-1}$,
   and stores one \fp{64} result.
\end{enumerate}

\paragraph{Two residency regimes, stated explicitly.}  The
register-resident language of the streamed-kernel algorithms below
(Algorithms~\ref{alg:fused-gemv}--\ref{alg:fused-spmv}) is deliberate and is
\emph{not} in tension with the dense-GEMM re-hosting of
Appendix~\ref{app:beta-blackwell}.  For the streamed kernels---batched
GEMV, stencil via im2col, blocked-ELLPACK SpMV---each thread carries
only a handful of residue \emph{fragments} (bytes per thread, not a
tile), so register residency is the intended and feasible regime, and
$\beta\!\to\!1$ follows directly.  The register-file-pressure failure
that Appendix~\ref{app:beta-blackwell} analyses is specifically the
\emph{dense-GEMM tile} case, where a full $128\times256$ operand tile
$\times\,r$ planes would spill; there the planes are re-hosted to
SMEM/TMEM and the register file holds only addresses and control.
Table~\ref{tab:kernelresid} makes the residence, lifetime, conversion
multiplicity, capacity limit, and evidence status explicit per kernel.

\begin{table}[h]
\caption{Per-kernel residence of the residue state (design values,
unmeasured unless noted).  ``fragments'' = a few bytes per thread;
``planes'' = a full per-modulus tile.  $\lambda$ is the conversion
multiplicity; the capacity limit is the resource that bounds the
regime.}
\label{tab:kernelresid}
\centering
\scriptsize
\setlength{\tabcolsep}{3.5pt}
\begin{tabular}{p{2.5cm}p{2.7cm}p{2.2cm}p{1.0cm}p{2.4cm}p{1.7cm}}
\toprule
kernel & operand/plane residence & accumulator & $\lambda$ & capacity limit & evidence \\
\midrule
batched GEMV ($B{=}8$) & fragments in registers & registers$\to$\fp{64} out & 1 & regs/thread ($\approx$fragments) & ledger \\
stencil (im2col) & fragments in registers, halo in SMEM & registers & 1 & SMEM halo band & ledger \\
SpMV (blocked-ELL) & value fragments in registers; $x$-residues L2-band & registers & 1 & L2 $x$-band, shared & ledger \\
dense-GEMM tile & planes in SMEM/TMEM; regs hold addresses/control & TMEM ($r$ sets) & $\lambda_A,\lambda_B$ & SMEM $\approx\!150$--$165$~KB & App.~\ref{app:beta-blackwell} \\
\bottomrule
\end{tabular}
\end{table}

Because the residue planes never reach HBM, the bandwidth multiplier
$\beta$ equals~1 (modulo a negligible constant for the moduli table and the
precomputed Garner inverses).  The compute multiplier
$\alpha$ equals $3r{+}1$ on the \fp{8} substrate (the three Karatsuba
\fp{8} MMAs per modulus plus the max-magnitude pass; a single MMA per
modulus suffices only on an INT8 substrate with byte moduli, where
$\alpha_{\textsc{int8}}=r_{\textsc{i}}{+}1=16$ at the
$r_{\textsc{i}}\!=\!15$ that matched accuracy then costs,
\S\ref{sec:background}).  The reconstruction latency
$\gamma$ is $O(r^2)$ per output element, but since each output element is
the result of an inner-product reduction of length~$k$, the per-FMA overhead
is $O(r^2/k)$.  The asymptotic
alone is not a rate argument, and the ``negligible at $k\gtrsim 10^3$''
aside of earlier revisions was too optimistic for the naive hosting, so
we replace it with a route-specific rate bound.  Garner at
$r{=}12$ costs $N_\gamma \approx r(r{-}1)/2 + 2r \approx 90$ narrow
multiply-adds per output; hosted on the SIMT integer pipe and charged
against $P_{\text{int}} = 75$~\tops{}, its service time is a
\emph{fraction} $45\,P_{\text{emu}}/(k\,P_{\text{int}})$ of the emulated
GEMM's---which at $k = 1000$ is $8.1\%$ on B300 and $28.4\%$ on Rubin
(the faster the emulated roof, the larger the tax), and falls below
$1\%$ only at $k \gtrsim 8.1{\times}10^3$ (B300) /
$2.84{\times}10^4$ (Rubin).  SIMT-hosted Garner is therefore a genuine
small-$k$ wall, exactly as the review notes, and reconstruction is
\emph{retained} for stream-rate outputs ($k \lesssim r^2$, e.g.\
single-RHS short-row SpMV, the corner Appendix~\ref{app:sparse}
brackets).  We therefore \emph{adopt} the SIMT--Garner cost
($N_\gamma\approx90$ narrow multiply--adds per output; below $1\%$
only at $k\gtrsim 8.1{\times}10^3$ on B300 / $2.84{\times}10^4$ on
Rubin) as the \emph{conservative reconstruction cost ceiling}---the most
expensive implementable reconstruction, hence a \emph{performance floor}
\emph{within the model}---a lower bound on the modeled rate, not a measured
one, since every other term in the composite remains a projection---and all companion
trace composites are reported at it.  The $\approx\!r$-op \textsc{int8}
hosting is a \emph{codesign target}, not a delivered count: an exact
reconstruction of the ${\approx}112$-bit CRT state requires either a
multi-limb integer recombination ($\approx\!rL$ byte-MACs with
$L\approx14$; by op-count $rL{\approx}168$ exceeds Garner's $\sim\!90$ narrow
ops, so the comparison is of host-pipe \emph{times}, $rL/P_{\text{byteMAC}}$ vs
$N_\gamma/P_{\text{narrow}}$, which the companion engine finds comparable to
or below Garner---no decisive gain) or a floating-point fractional-CRT
($\approx\!r$ \fp{32} MACs) whose \fp{64}-exactness is a Part-3
validation obligation; until one of these is demonstrated, the
$k\gtrsim 977$ (B300) / $2270$ (Rubin) thresholds are an explicit
sensitivity/target rather than an earned result.  The companion's per-record engine carries both
hostings and applies the tax to the emulated envelope and the shipping
baseline alike; at the conservative reconstruction cost floor the companion composites are
unchanged from the $\approx\!r$ target for the compute-advantaged
records ($k \gg$ threshold) but lower for the small-$k$, memory-bound
fronts---some shipping-relative ratios move by up to ${\approx}20\%$ (LOBPCG
$b{=}64$: Rubin $2.43\!\to\!1.94$, GB300 $1.60\!\to\!1.67$; UMFPACK GB300
$1.00\!\to\!1.19$; dense-QR Rubin $2.06\!\to\!1.60$; the companion's
Table~11 against its codesign-target table), while the
vs-native multipliers of the small-$k$ multifrontal and dense-LU
records on Rubin fall---which the codesign target would restore,
pending Part-3 validation.

\paragraph{$\beta=1$ is an upper bound, not a delivered number.}
We stress that $\beta=1$ is an \emph{idealisation} that the kernel
engineering must earn, not a property the model is entitled to assume.
Two costs that the abstract accounting above suppresses are real and
must ultimately be charged.  First, holding $r$ residue planes plus the
integer CRT-reconstruction operands in registers competes for the same
register file that bounds occupancy; at realistic GEMM tile sizes a
naive implementation can spill (raising the effective $\beta$ above~1)
or lose occupancy, so the achievable $\beta$ is an \emph{empirical}
quantity governed by the tile-size/register-budget tradeoff rather than
a constant---the model should be read as charging $\beta\!\ge\!1$ with
equality only in the fully fused limit, not as assuming unbounded
registers.  Second, the per-operand preprocessing---the
floating-to-integer scaling of~\eqref{eq:scale}, the per-modulus
reduction, and the max-magnitude estimation pass (the trailing $+1$ in
$3r{+}1$)---is genuine data-touching work.  As of this version that work
is no longer an informal caveat: it is the deconstruction term
$c_q r\,\varphi\,n_{\text{in}}/P_{\text{int}}$ of Eq.~\eqref{eq:temu},
charged explicitly on the integer pipe rather than folded into $\beta$,
and at the reference constants it---not the register file---is the
binding constraint for the lowest-intensity kernels
(\S\ref{sec:whyupdate}).  The honest statement of our claim is therefore
that the emulated kernel is memory-bound \emph{provided the
deconstruction keep-up condition holds}, and measuring the engineered
$c_q$ on real kernels is the first item of the programme of
\S\ref{sec:futurework}.  None of this
is hypothetical: the reference library
GEMMul8~\cite{gemmul8_github} already realises the
quantise~$\rightarrow$~modular-GEMM~$\rightarrow$~CRT pipeline with
both INT8 and \fp{8} backends and reports strong throughput on
GH200/Blackwell, so the open question is not \emph{whether} on-chip
fusion is achievable but \emph{how much} of the $\beta=1$ ideal survives
at production tile sizes---precisely the quantity we intend to measure
(\S\ref{sec:futurework}); Appendix~\ref{app:beta-blackwell} quantifies the
per-kernel $\beta$ budgets this question must meet and the Blackwell-class
dataflow (shared-memory operands, tensor-memory accumulators, asynchronous
copies) that the model projects meets them.  In the framing of \S\ref{sec:dwarfs}, $\beta$
is the canonical second-stage variable: should a kernel prove to sit at
$\beta>1$ for some structured class (large stencils, say), the design
response is itself a codesign question---add architectural registers and
compiler fusion to drive $\beta$ back toward~1, retain a small native
\fp{64} unit for that class, or emulate it at a higher-precision substrate
such as \fp{32}---and it is one the bound lets us pose precisely rather
than in the abstract.

\paragraph{The unifying goal: extending the roofline ceiling.}
It is worth stating the strategy of this entire section in one
sentence, because it frames everything that follows.  On a
\fp{64}-collapsed part the native compute ceiling sits at
$\sim\!1.3$~TFLOPS, \emph{below} the memory roof across essentially the
whole operational-intensity spectrum, so kernels that are physically
memory-bound are forced compute-bound by the silicon.  Ozaki~II does not
make these kernels run faster than bandwidth allows; rather, it
\emph{lifts the achievable compute ceiling}---from the collapsed
$1.3$~TFLOPS native floor up to the emulated $P_{\fp{8}}/(3r{+}1)\approx
135$~TFLOPS ceiling on B300---so that the memory roof, not the
\fp{64} pipe, once again becomes the binding constraint.  Equivalently,
emulation \emph{extends the compute-bound region of the roofline to the
right}, pushing the emulated ridge out to $\approx\!17$~\flops{}/B
(\S\ref{sec:model}) and so \emph{recovering the memory-bound character}
of every kernel whose intensity falls below it.  The four primitives
treated next (batched GEMV, stencil, SpMV)---and, in the companion
paper, the 3-D FFT~\cite{matsuoka2026fft}---are exactly the cases for
which this recovery is the goal; how far the ceiling can be extended in
practice, against the register-pressure and latency limits noted above,
is the empirical question driving the measurement program of
\S\ref{sec:futurework}.

\subsection{Strategy 1: Batched GEMV}
\label{sec:gemvstrategy}

Standard GEMV ($y=Ax$) cannot use tensor cores at all because the operands
have a one-dimensional axis.  Batched GEMV, $Y=AX$ with
$X\in\mathbb{F}^{n\times B}$ for batch size~$B$, exposes a small $B$ that
maps onto the 16- or 32-wide tensor-core $n$-dimension.  Algorithm
\ref{alg:fused-gemv} sketches the fused Ozaki-II GEMV kernel.

\paragraph{The abstract per-modulus product.}  To keep the pseudocode
substrate-honest without repeating the plane schedule in every
listing, all three algorithms call one abstract routine,
\[
  \texttt{fp8\_modular\_product}(a^{(i)}, x^{(i)}, m_i),
\]
whose \fp{8} realisation is: for a \emph{square} modulus, two stored
operand planes and the corresponding plane-product MMAs; for a
\emph{nonsquare} modulus, three stored planes ($d_0$, $d_1$,
$d_0{+}d_1$) and \emph{three} plane-product MMAs with the Karatsuba
correction, all with \fp{32} accumulation (exact in the small-integer
range), followed by the deferred exact integer reduction of the
accumulator to the residue interval in the epilogue.  Summed over the
modulus set this is the $3r{+}1$ \fp{8}-MMA count of \S\ref{sec:background}
(the $+1$ being the amortised max-magnitude pass); on the \inteight{}
substrate the same routine collapses to a single MMA per modulus with
native \intthirtytwo{} accumulation, but only for $m_i\!\le\!256$, so
$\alpha_{\textsc{int8}} = r_{\textsc{i}}{+}1 = 16$ at the byte-modulus
set's $r_{\textsc{i}}\!=\!15$ (\S\ref{sec:background}).
Every term of $\alpha$ therefore maps to a line of this routine, and
the single \texttt{mma\_sync} calls in the listings below are
invocations of it, not single hardware MMAs.

\begin{algorithm}[t]
\caption{Fused Ozaki-II batched GEMV (pseudocode).}
\label{alg:fused-gemv}
\begin{algorithmic}[1]
\State \textbf{Input:} matrix $A\in\mathbb{F}^{M\times N}$ (\fp{64}), batch
$X\in\mathbb{F}^{N\times B}$ (\fp{64}), moduli table $\{m_i\}_{i=1}^r$,
scaling diagonals $D,E$.
\State \textbf{Output:} $Y=AX\in\mathbb{F}^{M\times B}$.
\State Allocate registers $\{a^{(i)}_\text{frag}\}_{i=1}^r$, $\{x^{(i)}_\text{frag}\}_{i=1}^r$, $\{C^{(i)}_\text{acc}\}_{i=1}^r$.
\State Zero all accumulators $C^{(i)}_\text{acc}$.
\For{tile $k = 0,\dots,N/T_k-1$}
    \State Cooperatively load $A_\text{tile}$ and $X_\text{tile}$ in \fp{64} into shared memory.
    \For{$i = 1,\dots,r$}
        \State $a^{(i)}_\text{frag} \gets \mathrm{residue}\bigl(\lfloor D\,A_\text{tile}\rceil,\;m_i\bigr)$
        \Comment{in registers}
        \State $x^{(i)}_\text{frag} \gets \mathrm{residue}\bigl(\lfloor X_\text{tile}\,E\rceil,\;m_i\bigr)$
    \EndFor
    \For{$i = 1,\dots,r$}
        \State $C^{(i)}_\text{acc} \gets \texttt{fp8\_modular\_product}(a^{(i)}_\text{frag},x^{(i)}_\text{frag},m_i)$
        \Comment{2/3 plane MMAs, \fp{32} accum.\ (see text); 1 MMA on \inteight}
    \EndFor
\EndFor
\State $\tilde C \gets \text{GarnerReconstruct}(C^{(1)},\dots,C^{(r)};\{m_i\})$
\State $Y \gets D^{-1}\tilde C E^{-1}$ \Comment{rescale to \fp{64}}
\State Store $Y$.
\end{algorithmic}
\end{algorithm}

\paragraph{Key analysis.}  The matrix $A$ is read once per output row (cached
behaviour assumed), and the input batch $X$ is read $\lceil M/T_m\rceil$
times.  For \fp{64}-streamed $A$ (8~B per element, the dominant traffic
at large $M,N$) the operational intensity is $2B$ flops per element over
8~B, i.e.\ $\approx B/4$~\flops{}/Byte---$2$~\flops{}/B at
$B{=}8$.  (Earlier versions used $B/2$; the value is corrected here
following~\cite{bayraktar2026tme}.)  With $\beta=1$ and the
deconstruction condition met, the emulated kernel hits
the memory roof at $16$~TFLOPS on the B300, versus $1.3$~TFLOPS native
\fp{64}---a $\approx 12\times$ speedup.  At the reference deconstruction
constants the kernel is instead conversion-capped at
$\mathrm{OI}/\mathrm{OI}^{*} \approx 4.8\times$ native
(Eq.~\eqref{eq:oistar}); note that increasing $B$ raises the cap
proportionally (more useful flops per converted element) but does
\emph{not} close the gap to the memory roof, which is
$B$-independent.  The fused bound is realised only when the
register pressure of $r$ residue planes is tolerable; in practice we expect
$B$ to be limited to about $4$--$8$ before register spilling forces
$\beta>1$.

\subsection{Strategy 2: Stencil via im2col}
\label{sec:stencil}

A 7-point stencil computes, at every grid point $(i,j,k)$ of a
3-D array $u$,
\begin{equation}
v_{ijk} \;=\; c_0 u_{ijk} + c_1\!\left(u_{i\pm 1,j,k} + u_{i,j\pm 1,k}
                                       + u_{i,j,k\pm 1}\right),
\label{eq:stencil7pt}
\end{equation}
or, more generally, a fixed-coefficient linear combination of a small
neighbourhood.  At fp64 the operational intensity is
$\mathrm{OI}\!\approx\!0.5$~\flops{}/Byte (7 multiply-adds per output
across $\sim\!28$ bytes of HBM traffic when neighbour reads are reused
via the L1/shared-memory cache).

\paragraph{Mapping to GEMM via im2col.}  The standard tensor-core
mapping~\cite{liu2022_tcstencil} flattens the stencil into a GEMM by
treating the coefficient vector $c\!\in\!\mathbb{R}^{7}$ as a row matrix
and the neighbourhood values as columns of a $7\!\times\!N_\text{tile}$
matrix $U_\text{im2col}$ for a tile of $N_\text{tile}$ output points:
\begin{equation}
v_\text{tile} \;=\; c \cdot U_\text{im2col}\,,\qquad
U_\text{im2col}\!\in\!\mathbb{R}^{7\times N_\text{tile}}.
\label{eq:stencil-gemm}
\end{equation}
This produces a tall-and-skinny matrix product whose inner dimension is
the stencil width.  The coefficient vector $c$ is constant across the
entire kernel and can therefore be pre-decomposed once into its $r$
residue planes
$\{c^{(i)} = \mathrm{residue}(\lfloor sc\rceil,\,m_i)\}_{i=1}^r$ and
held in constant memory; the grid values must be residue-decomposed on
each tile.

\paragraph{Fused Ozaki-II 7-point stencil.}  Algorithm~\ref{alg:fused-stencil}
gives the fused kernel.  The key structural property is that the
$r\!\times\!7$ pre-decomposed coefficients fit in constant memory once
for the entire kernel run; on each tile the kernel reads
$N_\text{tile}\!+\!2$ planes from HBM (for the
three-axis halo), residue-decomposes the values into
residue planes in registers, invokes the $r$ modular products (2/3
plane MMAs each on \fp{8}; \S\ref{sec:gemvstrategy}) to compute
$c^{(i)} \cdot U^{(i)}_\text{im2col}$, and reconstructs
one fp64 output per grid point via Garner.

\begin{algorithm}[t]
\caption{Fused Ozaki-II 7-point stencil (pseudocode).
The pre-decomposed coefficient table $\{c^{(i)}\}$ lives in constant memory;
all residue planes are register-resident, giving $\beta\!\to\!1$.}
\label{alg:fused-stencil}
\begin{algorithmic}[1]
\State \textbf{Input:} grid $u\!\in\!\mathbb{F}^{N_x\times N_y\times N_z}$ (\fp{64}),
       coefficient vector $c\!\in\!\mathbb{F}^7$ (\fp{64}), moduli table
       $\{m_i\}_{i=1}^r$, scaling factor $s$.
\State \textbf{Output:} stencil image $v$.
\State \textbf{Precompute (once):} $c^{(i)}\!\gets\!\mathrm{residue}(\lfloor sc\rceil,\,m_i)$ for $i\!=\!1,\dots,r$;
       store $\{c^{(i)}\}$ in constant memory.
\For{tile $t$ partitioning the grid}
   \State Cooperatively load $u_\text{halo}$ for tile $t$ (including
          $\pm 1$ neighbourhood) into shared memory.
   \For{each output point $p$ in the tile (one warp per warp-tile)}
      \State Assemble neighbourhood vector
             $U_p\!=\![u_p,u_{p\pm \hat x},u_{p\pm \hat y},u_{p\pm \hat z}]\!\in\!\mathbb{F}^7$
             from shared memory.
      \For{$i = 1,\dots,r$}
         \State $U_p^{(i)}\!\gets\!\mathrm{residue}(\lfloor s U_p\rceil,\,m_i)$
                \Comment{in registers}
         \State $C_p^{(i)}\!\gets\!\texttt{fp8\_modular\_product}(c^{(i)},U_p^{(i)},m_i)$
                \Comment{$1\!\times\!7\!\times\!N_\text{tile}$ shape; 2/3 plane MMAs per modulus}
      \EndFor
      \State $v_p\!\gets\!\mathrm{GarnerReconstruct}(C_p^{(1)},\dots,C_p^{(r)})\,/\,s^2$
   \EndFor
\EndFor
\State Store $v$.
\end{algorithmic}
\end{algorithm}

\paragraph{Key analysis.}  Each grid point is read once (cached, with
the $\pm\,1$ neighbours hit in shared memory) and written once.
One ledger, three labelled points, used consistently hereafter: the
\emph{compulsory} minimum is $16$~B per output (one \fp{64} read +
one \fp{64} write; $\mathrm{OI} = 14/16 \approx 0.88$); the
\emph{tuned} estimate of Appendix~\ref{app:fusion-details} charges
imperfect halo reuse for $\approx\!24$~B ($\mathrm{OI} \approx
0.58$); and the \emph{reference} ledger used in every claims table
additionally charges a store-miss line fill on the output stream (a
stated assumption about the cache transaction, not a measurement) and
coefficient/boundary traffic for $\approx\!28$~B per output
($\mathrm{OI} \approx 0.5$)---the conservative value behind the
$4$-TFLOPS roof quoted below.  Compute per output is the $r$ modular
products---$(3r{+}1)$ plane MMAs on \fp{8}
(\S\ref{sec:gemvstrategy})---over the 7-element inner product, a
total of $\approx\!(3r{+}1)\!\cdot\!7$ \fp{8}
multiply-adds per output, which on B300's $5$~PFLOPS \fp{8} tensor
ceiling is $\sim\!5\cdot10^{-14}$~s per output and therefore
comfortably below the
${\approx}3.5$~ps/output ($28$~B $/$ $8$~TB/s) that the
reference-ledger memory roof
affords---a ${\sim}70\times$ margin (an earlier version misprinted
the stream time in nanoseconds).  Provided the register
pressure of $r$ residue planes is tolerable, the coefficient
constant-memory traffic does not saturate the constant cache, and the
deconstruction keep-up condition (C2) holds, the
$\beta\!=\!1$ discipline is realised and the kernel runs at the
memory roof of $\approx\!4$~TFLOPS on B300, against $\approx\!1.3$~TFLOPS
under native \fp{64}.  This is the $3.1\times$ fused-bound speedup reported in
Table~\ref{tab:speedups}; at the reference deconstruction constants the
conversion cap is $2.4\times$ ($\varphi\approx0.5$).

\paragraph{Sparsity exploitation (future work).}  The im2col matrix is
structurally sparse: at most 7 non-zeros per row for a 7-point stencil.
This invites the use of NVIDIA's 2:4 sparse tensor cores; SPTCStencil
and SparStencil exploit this directly for
\fp{32}/\fp{16}~\cite{gu2025_sptcstencil,wang2025_sparstencil}.
Combining the structured-sparsity tensor cores with the Ozaki-II
residue decomposition is a natural next step but is outside the scope
of this paper; we leave it open for follow-up (\S\ref{sec:futurework}).

\subsection{Strategy 3: SpMV via Blocked-Ellpack}
\label{sec:spmv}

SpMV computes $y\!=\!Ax$ for sparse $A\!\in\!\mathbb{F}^{M\times N}$ and
dense $x\!\in\!\mathbb{F}^N$.  It is the most challenging of the three
strategies because the underlying access pattern is irregular and the
operational intensity is the lowest ($\mathrm{OI}\!\sim\!0.2$~\flops{}/Byte
for typical PDE-discretisation sparsity).

\paragraph{Mapping to tensor cores via Blocked-Ellpack.}  The
\textbf{Blocked-ELL} format groups non-zeros into row-major blocks of a
uniform block-column width $\mathit{bw}$ (typically 16 or 32, chosen
to match the tensor-core $k$-dimension).  Sparse rows that have fewer
than $\mathit{bw}$ non-zeros are padded with structural zeros; row
permutations may be applied to balance block density.  Each block of
$\mathit{bw}$ columns becomes a small dense GEMM
$y_\text{tile}\!=\!A_\text{block} x_\text{gather}$, where
$x_\text{gather}\!\in\!\mathbb{F}^{\mathit{bw}}$ is a gather of the
relevant entries of $x$.  Padding wastes compute, but the wasted compute
is \fp{8}/\inteight compute, which is roughly four orders of magnitude
cheaper than wasted \fp{64} compute on B300.  Even a 90\%-wasteful
padding scheme remains profitable provided $\beta$ stays close to~1.

\paragraph{Fused Ozaki-II Blocked-ELL SpMV.}  Algorithm~\ref{alg:fused-spmv}
gives the fused kernel.  Each block row issues a small GEMV that maps
onto the tensor-core $\mathit{bw}\!\times\!1$ MMA shape, with $r$
residue planes computed and reconstructed in registers.  The dense
input vector $x$ is shared across all block rows and remains in shared
memory; the sparse pattern dictates the gather indices.

\begin{algorithm}[t]
\caption{Fused Ozaki-II Blocked-Ellpack SpMV (pseudocode).
Each block row issues one $\mathit{bw}\!\times\!1$
\texttt{fp8\_modular\_product} per modulus (2/3 plane MMAs each on
\fp{8}); $\beta\!\to\!1$ provided shared-memory tiling keeps
the gathered $x_\text{gather}$ register-resident.}
\label{alg:fused-spmv}
\begin{algorithmic}[1]
\State \textbf{Input:} sparse matrix $A$ in Blocked-ELL layout with
       block-column width $\mathit{bw}$ (data array $A_\text{val}$,
       column-index array $A_\text{col}$),
       dense vector $x\!\in\!\mathbb{F}^N$,
       moduli table $\{m_i\}_{i=1}^r$, scaling factors $D,E$.
\State \textbf{Output:} $y\!=\!Ax$.
\State Allocate registers
       $\{C_\text{row}^{(i)}\}_{i=1}^r$, zero each.
\For{block row $b$ of $A$}
   \State Load $A_\text{val}[b,\cdot]\!\in\!\mathbb{F}^{\mathit{bw}}$
          and indices $A_\text{col}[b,\cdot]$ for this block row.
   \State Gather $x_\text{gather}\!\in\!\mathbb{F}^{\mathit{bw}}$
          using $A_\text{col}[b,\cdot]$ (shared-memory access if
          locality permits).
   \For{$i = 1,\dots,r$}
      \State $A^{(i)}_\text{frag}\!\gets\!\mathrm{residue}\bigl(\lfloor D\,A_\text{val}[b,\cdot]\rceil,\,m_i\bigr)$
      \State $x^{(i)}_\text{frag}\!\gets\!\mathrm{residue}\bigl(\lfloor x_\text{gather}\,E\rceil,\,m_i\bigr)$
      \State $C_\text{row}^{(i)}\!\gets\!\texttt{fp8\_modular\_product}(A^{(i)}_\text{frag},x^{(i)}_\text{frag},m_i)$
   \EndFor
   \State $\tilde{y}_b\!\gets\!\mathrm{GarnerReconstruct}(C_\text{row}^{(1)},\dots,C_\text{row}^{(r)})$
   \State $y_b\!\gets\!D^{-1}\tilde{y}_b E^{-1}$
\EndFor
\State Store $y$.
\end{algorithmic}
\end{algorithm}

\paragraph{Key analysis.}  Bandwidth is dominated by the streaming
read of $A_\text{val}$ plus the gather of $x$; with a moderate
shared-memory tile of $x$ the gather hits cache most of the time, so
the effective intensity matches the structural value of
$\sim\!0.2$~\flops{}/Byte.  At $\beta\!=\!1$ the emulated kernel
saturates HBM at $\approx 1.6$~TFLOPS on B300 and $\approx 4.4$~TFLOPS
on Rubin, against $\approx 1.3$~TFLOPS native B300 and $\approx 4.4$~TFLOPS
native Rubin (the latter is at parity because Rubin's higher HBM ridge
already accommodates the SpMV intensity).  The B300 speedup is
modest ($\sim 1.2\times$) precisely because the kernel is so deeply
bandwidth-bound that the native pipe is already a tolerable match;
the point of including it is not the speedup magnitude but the
\emph{equality}: emulation gives up nothing.  That statement must now be
qualified by the deconstruction condition: SpMV sits at
$\mathrm{OI}\!\approx\!0.2 < \mathrm{OI}^{*}\!\approx\!0.42$, i.e.\
squarely in the conversion-bound region at the reference constants
($\varphi\!=\!1$: every element of $A$ is used once), where the emulated
path is capped at $\approx 0.48\times$ native and the surviving native
pipe is the correct route.  Recovering the fused $1.2\times$ bound
requires the engineered $c_q \lesssim 5$--$6$ that Part~3 measures;
until then the SpMV row of the claims tables reads ``native fallback''
at reference constants (\S\ref{sec:eval}).

\paragraph{Caveat: padding inflation and non-ELL-amenable structure.}  The Blocked-ELL mapping inflates
$Q$ by the inverse of the row density.  This is the closest we come in
this paper to a regime where $\beta\!>\!1$ is fundamentally
unavoidable: $\beta$ is bounded below by the padding ratio.  For
matrices with strongly heterogeneous row lengths, hybrid CSR-ELL or HYB
formats are required to keep $\beta$ acceptable; otherwise the model
predicts a regression rather than parity.  See
Appendix~\ref{app:fusion-details} for the precise mapping of
padding ratio to $\beta$.  More broadly, the Blocked-ELL analysis is a
\emph{best case} for tensor-core SpMV, and several important sparse
structures do not fit it: Monte-Carlo / particle codes with
irregular, dynamically generated connectivity; matrices whose
nonzero distribution resists ELL blocking even after reordering; and
dual-mesh or unstructured-stencil discretisations whose access
pattern is gather-dominated rather than block-dense.  For these the
padding ratio---and hence $\beta$---can be large enough to erase the
emulation benefit, and the right comparison may instead be against
B300's surviving native \fp{64} on the (already memory-bound) kernel.
We expect that many element distributions can be clustered into
ELL-amenable blocks by row/column reordering---or composed through a
hybrid Ellpack/CSR or blocked formulation so that the dominant work is
again a matrix multiplication, building on the substantial body of
high-performance sparse matrix--matrix (SpGEMM) work that casts irregular
sparse computation into dense-block products~\cite{matsuoka2018blocked}---but
this is a hypothesis to be tested on real matrices,
not an assumption of the model.  In the terms of \S\ref{sec:dwarfs}, the
Blocked-Ellpack analysis establishes the \emph{upper bound} for the
canonical sparse dwarf; whether a given irregular matrix attains it is the
second-stage format question, and the genuinely irregular cases above
belong in the empirical campaign of \S\ref{sec:futurework} rather than
in the analytical projection.

\section{Performance Projections}
\label{sec:eval}

We instantiate the TME model on the architectures of Table~\ref{tab:specs}
and compute the projected speedup of Ozaki~II/\fp{8} over native \fp{64}
for the four canonical workloads.  We use the standard abbreviation
\emph{OI} for operational intensity, measured in FLOPs per byte of HBM
traffic; values are taken from the Williams Roofline
literature~\cite{williams2009_roofline} and the workload-specific
analyses in \S\ref{sec:impl}.  All projections assume the conditions of
Proposition~\ref{prop:atomic} at the bound---(C1) $\beta=1$ (fused
decomposition) and (C4) $\gamma$ amortised---with $r=12$ moduli for the \fp{8} variant at the full
$(3r{+}1)=37$ \fp{8} cost
\cite{uchino2026_fp8,mukunoki2025_fp8}---the value recommended by
Mukunoki~\cite{mukunoki2025_fp8} for \fp{64}-equivalent accuracy---and $\gamma$ amortised to zero for
inner-product lengths $k\!\gtrsim\!100$.  Each B300
projection is stated twice: at the fused, conversion-ideal bound, and as
the conversion-bound cap $\varphi^{-1}\mathrm{OI}/\mathrm{OI}^{*}$ at the
shipping reference constants of \S\ref{sec:model} ($c_q{=}16$,
memo-counted, Appendix~\ref{app:memo};
$P_{\text{int}}{=}75$~\tops{}, $\mathrm{OI}^{*}{\approx}0.42$~\flops{}/B);
rows whose cap falls below $1\times$ are re-assigned to the surviving
native \fp{64} pipe.  Because the shipping count is an engineering
artifact and not a floor, \S\ref{sec:cqladder} first fixes the
\emph{realistic} constants---the corrected platform rates, the keep-up
budgets they imply, and the companion paper's script-checked,
ledger-projected $c_q$
ladder~\cite{matsuoka2026ozaki25}---and only then do we project; the
staged assembly of \S\ref{sec:step7} reads every cap against what
software demonstrably recovers.  The batched-GEMV operational
intensities are the corrected $B/4$ values (\S\ref{sec:impl}).

\subsection{The Realistic Constants: the Rubin Correction, the Keep-Up
Budgets, and the Audited $c_q$ Ladder}
\label{sec:cqladder}

\paragraph{The corrected Rubin rates.}  Versions of this paper through
July~3, 2026 (arXiv v1--v3) used $\sim\!4$~PFLOPS for Rubin's
dense \fp{8} tensor rate, traced to early third-party specification
compilations~\cite{lockwood_rubin} that, in retrospect, placed Rubin's
FP16/BF16 rate in the \fp{8} slot.  NVIDIA's official product
specification for the DGX Rubin NVL8---published June~30, 2026, for the
platform announced at CES~2026---lists \emph{140~PFLOPS of dense
FP8/FP6 training performance across eight Rubin GPUs}, i.e.\
\textbf{17.5~PFLOPS of dense \fp{8} per GPU}~\cite{nvidia_dgx_rubin_nvl8},
$4.4\times$ the figure used previously (NVIDIA marks the values
preliminary and subject to change).  Because the publication postdates
v1--v2 of this paper and was missed by days in v3, we record it here
explicitly as a mandated correction, propagated through
Tables~\ref{tab:specs}, \ref{tab:speedups}, \ref{tab:h100baseline}
and~\ref{tab:int8vsfp8} and Figure~\ref{fig:crossover}.  Its
consequences are uniformly favourable to the thesis.  \emph{(i)}~The
Rubin Ozaki~II ceiling rises to $\approx\!473$~TFLOPS at $r{=}12$
($\approx\!380$ at the conservative $r{=}15$), and the emulated ridge
moves out to $17{,}500/(37\cdot22)\approx21.5$~\flops{}/B---every
surveyed kernel is deeper inside the memory-bound region on Rubin than
on B300.  \emph{(ii)}~The published $\sim\!200$-TFLOPS Emulated-DGEMM
figure, which under the old rate \emph{exceeded} the model bound, now
sits at $\approx\!42\%$ ($r{=}12$) to $\approx\!53\%$ ($r{=}15$) of the
analytic ceiling---a margin compatible with a delivered,
accuracy-guaranteed library number, so the model and NVIDIA's official
figure are now mutually consistent, itself a nontrivial cross-check of
the $(3r{+}1)$ cost structure (consistency is not yet causality;
\S\ref{sec:rubinuplift} sharpens it into a falsifiable size-sweep
hypothesis).  \emph{(iii)}~The \fp{8} compute available per byte of
HBM stream \emph{rises} generationally, from $\approx\!625$
\fp{8}-ops/B on B300 to $\approx\!795$ on Rubin---emulation margin
grows with the generation rather than merely riding bandwidth.
\emph{(iv)}~All tensor-side keep-up margins widen: streaming \fp{64}
operands at 22~TB/s (2.75~Telem/s) demands $\approx\!204$~TFLOPS of
\fp{8} work at $\alpha=37$, i.e.\ $1.2\%$ of peak---$\approx\!9\%$
even at the $B{=}1$ tile-granularity worst case of
Eq.~\eqref{eq:wmma}.

\paragraph{The two keep-up budgets.}  A kernel that saturates the HBM
stream consumes \fp{64} elements at rate $B_{\text{mem}}/8$; dividing
each pipe's throughput by that rate gives the \emph{per-element
budget} each pipe can spend without falling behind the stream:

\begin{itemize}[leftmargin=2em,itemsep=2pt]
\item \emph{Tensor budget} $= 8P_{\fp{8}}/B_{\text{mem}}$:
   $\approx\!5{,}000$ \fp{8}-ops per element on B300, and
   $\approx\!6{,}400$ on Rubin.  It \emph{grows} with the generation
   (consequence~(iii) above).
\item \emph{SIMT budget} $= 8P_{\text{int}}/B_{\text{mem}}$:
   $75$ instructions per element on B300 ($6.25$ per modulus at
   $r{=}12$), but only $\approx\!27$ on Rubin ($2.3$ per modulus;
   $1.8$ at $r{=}15$)---because bandwidth grew $2.75\times$ while the
   SIMT integer rate presumably did not.  Even a $2\times$
   integer-pipe scaling reaches only $4.5$ per modulus.  It
   \emph{shrinks} with the generation.
\end{itemize}

Against these budgets stands the shipping deconstruction cost:
$c_q = 16$ instructions per modulus, memo-counted on the shipping
cuBLAS path by the NVIDIA review~\cite{bayraktar2026tme}---an
instruction-count estimate, never a throughput measurement; the
non-confidential stage structure behind it is summarised in
Appendix~\ref{app:memo}.  The comparison is stark: \emph{on
Rubin-class bandwidth, SIMT-only deconstruction at anything near the
shipping count cannot keep up with the stream under any plausible
engineering, while the tensor pipe idles at $\ge\!97\%$}.  The fourth
term becomes more binding with the generation even as \fp{8} headroom
widens; the uplift simultaneously \emph{funds} a migration of the
work and \emph{mandates} it.  Everything that follows turns on one
question: \emph{how much of $c_q$ can leave the SIMT pipe?}

\begin{figure}[htbp]
\centering
\includegraphics[width=\linewidth]{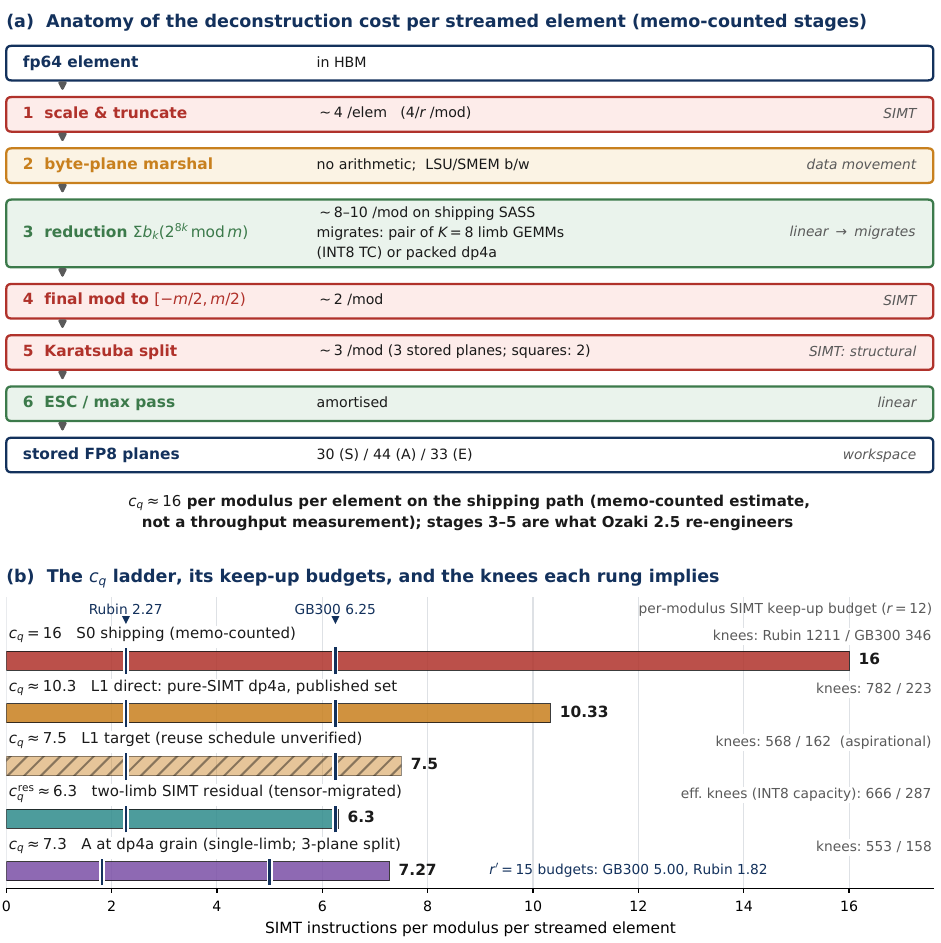}
\caption{Anatomy of the deconstruction cost, after the companion
paper's script-checked accounting~\cite{matsuoka2026ozaki25}.
\textbf{(a)}~The six stages that turn one streamed \fp{64} element into
its $3r$ \fp{8} operand planes, colour-coded by disposition: the
elementwise nonlinearities (round-to-integer, final modular reduction,
Karatsuba plane split) stay on the SIMT pipe or fixed function (red);
the \emph{linear} stages---the per-modulus byte-plane reduction, the
bulk of the memo-counted cost, and the ESC/max pass---migrate to the tensor
pipes or \texttt{DP4A} (green); byte-plane marshaling is data movement
on the LSU/SMEM path, an arithmetic-free cost that FLOPs cannot buy
(amber).  \textbf{(b)}~Five rungs of the audited software ladder, each
drawn against the per-modulus SIMT keep-up budget
$8P_{\text{int}}/(r B_{\text{mem}})$ that applies to \emph{that} rung
(paired ticks; \textsf{GB300} labels the frozen SKU behind this paper's
B300 columns).  The budget moves with $r$: the four $r{=}12$ rungs
(the memo-counted $16$; the direct \texttt{dp4a} route
$\approx\!10.3$ and its unverified $\approx\!7.5$ reuse-schedule
target; the two-limb residual $6.3$) are charged against $6.25$ per
modulus on B300 and $2.27$ on Rubin, while the all-byte codesign at
\texttt{dp4a} grain ($\approx\!7.3$) runs at $r'{=}15$ and is charged
against $5.00$ and $1.82$.  Every bar ends to the right of both of its
own ticks; the narrowest miss drawn is $6.3$ against $6.25$.  The
refined ($5.8$), lazy ($4.3$) and one-pass all-byte ($5.3$) floors are
results of the accounting rather than rungs of this ladder and are not
drawn here; the text compares them with these same budgets, and its
conclusion---software fits B300's budget, nothing realisable fits
Rubin's---is the gap the hardware options of \S\ref{sec:hwproposal}
close.}
\label{fig:anatomy}
\end{figure}

\paragraph{The audited answer: what migrates, and what cannot.}
Since the first version of this paper, that question has been worked
out in full in a standalone companion paper---\emph{Ozaki~2.5},
submitted to arXiv~\cite{matsuoka2026ozaki25}---whose script-checked
accounting (ledger-projected for the uncompiled routes)
this section imports rather than re-derives;
Figure~\ref{fig:anatomy} summarises it.  Of the six stages that turn
one streamed \fp{64} element into its $3r$ \fp{8} operand planes,
three are elementwise nonlinearities that a linear engine cannot
host---scale-and-truncate ($\sim\!4$ per element, amortised $4/r$ per
modulus; truncation is the shipping convention), the final modular
reduction ($\sim\!2$ per modulus), and the Karatsuba plane split
($\sim\!3$ per modulus)---and the split is moreover \emph{structural}
for the \fp{8} substrate: the companion's exhaustive supply bounds
prove that no pairwise-coprime system of \fp{8}-significand-sized
moduli spans the required CRT range (sub-64 systems cap at $89.9$ of
the needed $111.8$ bits), so some piece handling survives every
reformulation.  Byte-plane marshaling is data movement on the
LSU/SMEM path---a bandwidth cost that \fp{8} FLOPs cannot buy
(Appendix~\ref{app:beta-blackwell}).  The bulk of the counted cost,
the per-modulus byte-plane reduction ($\sim\!8$--$10$ of the $13$
counted instructions), is \emph{linear over the byte planes},
$x \bmod m = \sum_k b_k\,(2^{8k} \bmod m) \bmod m$, and
migrates---to packed \texttt{dp4a} dot products on the SIMT pipe, or
to the tensor side---subject to one exactness subtlety established in
the companion: the reduction constants reach $826$ for the real
$9$--$11$-bit moduli, so an exact one-pass byte GEMM is
impossible---the exact realisation is a \emph{pair} of $K{=}8$ limb
GEMMs on the \inteight{} tensor pipes (operand bytes are not
\textsc{e4m3}-representable, so the \fp{8} pipes cannot host them),
recombined at one extra SIMT instruction per modulus.  (The sixth
stage, the amortised ESC/max pass, is a linear tile reduction and
migrates trivially.)

The audited ladder of engineered $c_q$ that results---each rung a
shipping-silicon software route, and each a Part~3
kernel~\cite{matsuoka2026ozaki25}---is: $16$, the memo-counted
shipping path (S0); $\approx\!10.3$, the direct \texttt{dp4a} route
(L1; two limbs, published set; an earlier draft's $\approx\!7.5$
survives only as an explicitly unverified reuse-schedule target);
$6.3$ uniform / $5.8$ refined, the SIMT residual after the two-limb
tensor migration with the final reduction inline (L2); $\approx\!5.3$,
the all-byte \emph{codesigned} modulus set, which legalises a one-pass
reduction at an $\alpha'{=}46$ pass-count surcharge that is free
precisely where kernels are memory-bound; $\approx\!4.3$, L2 with the
\emph{lazy-reduction} $\alpha$-trade (L3), deferring the final
reduction to the epilogue at $\alpha'\!\approx\!6r{+}1{=}73$
(likewise benign here: even at $\alpha'{=}73$ the Rubin ceiling is
$\approx\!240$~TFLOPS, above the published $200$, and the ridge
$\approx\!11$~\flops{}/B); and $2$, the speculative piece-free floor.
Summing the non-migratable stages gives the two software floors
\begin{equation*}
c_q^{\text{res}} \;\approx\; \frac{4}{r} + 1 + 2 + 3 \;\approx\; 6.3,
\qquad
c_q^{\text{res,lazy}} \;\approx\; \frac{4}{r} + 1 + 3 \;\approx\; 4.3
\ \text{instructions per modulus at } r{=}12,
\end{equation*}
where the $+1$ is the recombination that two-limb exactness forces
(the companion's refined per-modulus accounting gives $5.8$ for the
published set; an earlier draft of this paper printed $5.3/3.3$ from
the invalid one-pass assumption---the all-byte codesign is what
legitimately restores a $5.3$-class \emph{per-modulus} residual,
though at its own $r'{=}15$ its per-element total, $\approx\!79$
instructions, slightly exceeds the published set's $70$--$76$, so its
streamed dividend is realised through the \texttt{dp4a} and lazy
variants rather than the inline tensor-migrated residual).
Comparing floors with
budgets yields the pivotal result (Figure~\ref{fig:anatomy}(b) draws
the ladder rungs against the budget each of them is actually charged
at; the refined, lazy and one-pass floors named next are not rungs of
that ladder and are compared with the same budgets here rather than on
the panel): \emph{the software floors sit at or
inside B300's $6.25$-per-modulus budget (the refined $5.8$ inside,
the uniform $6.3$ marginally outside; $4.3$ inside with
margin; the all-byte $5.3$, read against its own $r'{=}15$ budget of
$5.0$, delivers $0.95$ of the B300 \emph{memory} roof), but no realisable software
floor fits Rubin's $2.3$}
($6.3$ delivers $0.36$ of the Rubin \emph{memory} roof, $5.8$ delivers $0.39$,
$4.3$ with its $\alpha'$ charge $0.53$; only the speculative
piece-free $2$ would fit).  Software alone can recover the roof on
B300; on Rubin-class bandwidth something must change in the
hardware---which is what \S\ref{sec:hwproposal} proposes.

\paragraph{The division of labour, and the Kulisch carriers.}
The ladder induces a coherent software end-state worth naming---an
\emph{FP8-resident Ozaki~II pipeline}: the \fp{8} tensor cores carry
the heavy work (the Ozaki product passes); the residual \inteight{}
tensor pipes carry the two-limb reduction GEMMs and the
reconstruction epilogues; and the SIMT pipe touches each element only
for the audited residual, with \texttt{dp4a} as the tensor-free
fallback (\S\ref{sec:int8vsfp8} develops the substrate reading).  The
reconstruction side deserves its own note, because it generalises
well beyond Garner (L4).  An \fp{8} MMA with \fp{32} accumulation is
an exact small-integer engine (products of $\le\!4$-bit-significand
operands accumulate exactly for up to $\sim\!2^{16}$ terms per
window), so a Kulisch-style $S$-slice fixed-point accumulation costs
$\approx\!2S$ \fp{8}-MACs per term---under $0.4\%$ of Rubin's peak at
$S{\approx}11$; the \inteight{}-tensor carrier is weaker in rate but
\emph{exact by construction} (native INT32 accumulate, no
exactness-window bookkeeping), and reconstruction's demand is small
enough that even the deprioritised \inteight{} pipes suffice
($\approx\!13\%$ of B300's $166$~\tops{} residual rate,
$\approx\!24\%$ of Rubin's $250$).  The route---the concrete
resolution of the Part-2 suggestion, first developed for the
companion FFT analysis~\cite{matsuoka2026fft}---upgrades BLAS-1 dot
products and norms from \fp{32}+Kahan (compensated) to \emph{exact};
yields bitwise-reproducible global reductions (directly addressing
the reproducibility corner case of \S\ref{sec:falsification}); hosts
the $\gamma$ reconstruction epilogues of all Ozaki~II kernels
(relaxing the $k \lesssim r$ corner of Appendix~\ref{app:sparse})
and the long-horizon drift accumulations of \S\ref{sec:l3}(f); and
relieves the fourth term systemically, since every reconstruction
moved off the SIMT pipe returns budget to the one cost that cannot
leave it.  All ladder figures remain first-order and unmeasured---
they charge neither lane-marshaling nor carry resolution---and the
knees, dispatch windows, and measured call-shape evidence behind
them are the companion's~\cite{matsuoka2026ozaki25}.

\subsection{Projections Across the Kernel Spectrum}
\label{sec:spectrum}

With the constants fixed, Table~\ref{tab:speedups} assembles the
projection across the four canonical workloads.

\begin{table}[h]
\caption{Projected speedups of Ozaki~II/\fp{8} over native \fp{64}.  All
projections use the TME model of \S\ref{sec:model}, the architectural
data of Table~\ref{tab:specs}, and assume on-chip tile fusion
($\beta=1$).  Memory-bound speedups are upper-bounded by the memory roof
$I\cdot B_{\text{mem}}$ divided by the native \fp{64} compute roof;
compute-bound speedups are upper-bounded by $P_{\fp{8}}/((3r{+}1)\cdot
P_{\fp{64}})$ at $r=12$ ($\alpha=37$).  The ``B300 conv.-cap'' column
gives the conversion-bound cap $\varphi^{-1}\mathrm{OI}/\mathrm{OI}^{*}$
at the shipping reference constants
(Eq.~\eqref{eq:oistar})---the delivered ceiling of the
\emph{shipping} path; the audited software ladder of
\S\ref{sec:cqladder} supplies the engineered $c_q$ that lifts these
caps, pending Part-3 measurement (Table~\ref{tab:combined}, S1).
Entries below
$1\times$ are re-assigned the verdict ``native'' (the surviving \fp{64} pipe is the
correct path).  For Rubin we report two columns: one against the native
\fp{64} vector roof ($33$~TFLOPS), the other against NVIDIA's
conservative published Emulated-DGEMM figure of $\sim 200$~TFLOPS.
\textsuperscript{$\star$}~B300 columns use the frozen GB300
(NVL72-derived per-GPU) constants of Table~\ref{tab:specs}, distinct
from the HGX~B300 SKU.}
\label{tab:speedups}
\centering
\scriptsize
\setlength{\tabcolsep}{2.5pt}
\begin{tabular}{lcccccccc}
\toprule
\textbf{Workload} & \textbf{OI} & $\varphi$ & \textbf{H100} & \textbf{B200} & \textbf{B300}\textsuperscript{$\star$}
& \textbf{B300}\textsuperscript{$\star$} & \textbf{R200} & \textbf{R200} \\
                  & \textbf{(FL/B)} & &        &               & \textbf{($\beta{=}1$, ideal)}
& \textbf{conv.-cap} & \textbf{vs.\ nat.} & \textbf{vs.\ Emul.} \\
\midrule
Dense GEMM (comp.-bound) & $\ge 50$ & $\to 0$ & $\sim 0.8\times$ & $\sim 3.0\times$ & $\sim 104\times$ & not binding & $\sim 14\times^{\dagger}$ & $\sim 2.4\times$ \\
Batched GEMV ($B{=}8$)     & $\sim 2$  & $\sim 1$ & $\sim 1.0\times$ & $\sim 1.0\times$ & $\sim 12\times$  & $\sim 4.8\times$ & $\sim 1.3\times$ & mem.~bound \\
Batched GEMV ($B{=}2$)     & $\sim 0.5$& $\sim 1$ & $\sim 1.0\times$ & $\sim 1.0\times$ & $\sim 3.1\times$ & $\sim 1.2\times$ & $\sim 1.0\times$ & mem.~bound \\
7-point Stencil            & $\sim 0.5$& $\sim 0.5$ & $\sim 1.0\times$ & $\sim 1.0\times$ & $\sim 3.1\times$ & $\sim 2.4\times$ & $\sim 1.0\times$ & mem.~bound \\
SpMV (CSR/ELL hybrid)      & $\sim 0.2$& $1$ & $\sim 1.0\times$ & $\sim 1.0\times$ & $\sim 1.2\times$ & nat.\ ($0.48\times$) & $\sim 1.0\times$ & mem.~bound \\
\bottomrule
\end{tabular}
\\[2pt]
{\scriptsize ``$\sim 1.0\times$'' indicates parity: emulation neither
hurts nor helps because the kernel is already at the memory roof under
native execution.  $\varphi$ is the converted fraction of streamed
elements (Eq.~\eqref{eq:temu}); the stencil's $\varphi\!\approx\!0.5$
reflects residue reuse across neighbouring outputs and the
once-converted coefficient table.  ``native ($0.48\times$)'' means
emulation is conversion-bound \emph{below} native at reference constants
and the kernel routes to the surviving \fp{64} pipe.  On Rubin, native
vector FP64 has been partially
restored ($33$ vs.\ B300's $1.3$~TFLOPS), so memory-bound kernels with
$I<33/22 = 1.5$~FLOPs/B remain at parity even before considering
the Emulated-DGEMM path.\\[2pt]
$^{\dagger}$The Rubin dense-GEMM ${\sim}14\times$ is the \emph{theoretical}
roof ($473/33$). A real large dense \emph{square} output is re-split once per
thread-block cluster and delivers ${\approx}\tfrac12$ of it---the
${\approx}235$-TFLOPS deconstruction-$\lambda$ floor, ${\approx}7\times$ vs.\
native (Table~\ref{tab:densedgemm})---until the co-design of
\S\ref{sec:hwproposal} (Option~C) retires the deconstruction term;
tall/skinny and small-batch GEMMs are unaffected and are projected to reach the roof on shipping hardware.}
\end{table}

\paragraph{What the projections show.}  Table~\ref{tab:speedups} carries
two simultaneous messages.  On B300, emulation delivers substantial
speedups across the operational-intensity spectrum at the fused
bound---from $\sim 1.2\times$ on bandwidth-saturating SpMV to
$\sim 104\times$ on dense GEMM---while the conversion-cap column shows
what survives at the reference deconstruction constants: the
compute-bound and mid-intensity kernels retain material gains
($104\times$ for dense GEMM above the size crossover later quantified
in \S\ref{sec:hwproposal}, $4.8\times$, $2.4\times$, with dual-RHS
GEMV holding a
conversion-capped $\sim\!1.2\times$), the lowest-intensity tail (SpMV
and single-RHS GEMV) is conversion-bound and correctly routes to
the surviving native pipe, and the distance between the two columns is
governed by the single engineered constant $c_q$.  Read against the
audited ladder of \S\ref{sec:cqladder}, the cap column is a statement
about the shipping path, not a fate: at the two-limb residual
($6.3$ uniform, $5.8$ refined) B300's caps lift essentially to the
memory roof---software alone restores the ideal column (stage S1 of
Table~\ref{tab:combined})---while on Rubin even the full software
ladder holds only $0.53$--$0.56$ of the \emph{memory} roof for fully
streamed kernels ($0.53$ at the refined two-limb residual, $0.56$ under
the all-byte lazy split; Table~\ref{tab:combined}), the gap the hardware path of \S\ref{sec:hwproposal}
closes.  A unified Ozaki~II
library can serve the whole spectrum, with the runtime dispatching on
$\mathrm{OI} \gtrless \varphi\,\mathrm{OI}^{*}$ exactly as it already
dispatches on the ridge point.  On
Rubin, NVIDIA has already incorporated emulated DGEMM as the official
path, exposing it as $\sim 200$~TFLOPS in published specifications; at
the official 17.5-PFLOPS dense \fp{8} rate the model ceiling
dense-\fp{8}$/(3r{+}1)$ at $r{=}12$
sits at $\sim 473$~TFLOPS (\S\ref{sec:arch}, \S\ref{sec:cqladder})---the
\emph{theoretical} \emph{arithmetic} roof---a different quantity from the
\emph{memory} roof the streamed-kernel fractions above are taken against.
For a large dense \emph{square} output, however, the
deliverable on today's Rubin is half of that $473$, the ${\approx}235$-TFLOPS
deconstruction-$\lambda$ floor (\S\ref{sec:hwproposal},
Table~\ref{tab:densedgemm}), which the hardware co-design would lift back toward
the roof; the published $\sim\!200$ thus sits just under that delivered
large-square floor---a conservative delivered figure---rather than far below the
raw roof.  On H100 and B200 the
projections show \emph{parity or modest gain}: emulation neither
materially hurts nor helps because
native \fp{64} is healthy.  This parity property is what makes a unified
library implementation feasible---the same Ozaki~II code path is safe to
ship on all four architectures, with the runtime selecting emulation only
where it pays.

\paragraph{The positive reading of the memory-bound parity rows.}
Several rows in Table~\ref{tab:speedups} show $1.0\times$ on H100, B200,
and the Rubin-vs-native column.  This is exactly the behaviour
scientific computing wants: in the memory-bound regime, emulation is
\emph{indistinguishable in performance from native \fp{64} on a
\fp{64}-healthy chip}.  Combined with the compute-bound speedups (e.g.\
$\sim 3\times$ on B200, $\sim 104\times$ on B300, $\sim 14\times$ on
Rubin---the theoretical roof, ${\approx}7\times$ for a real large \emph{square}
output at the deconstruction-$\lambda$ floor until the co-design of
\S\ref{sec:hwproposal} lifts it), the message is unambiguous: at the fused bound, switching the entire numerical
stack to Ozaki~II costs essentially nothing on memory-bound workloads
and gains order-of-magnitude throughput on the compute-bound rest---with
conversion-bound kernels dispatched to the surviving native pipe until the
engineered $c_q$ closes condition (C2).

\paragraph{Where emulation does not currently help.}  In two scenarios,
emulation still gives no benefit.  First, on a \fp{64}-healthy GPU
(H100, B200, Rubin-vector-only), memory-bound kernels are already at
the memory roof, so there is no headroom.  Second, kernels whose
dominant cost is host--device data movement, or whose effective
operational intensity is bounded by ELL-padding inflation pushing
$\beta$ above unity, will see no improvement.  Bringing those into the
emulation envelope is a separate problem that we revisit in
\S\ref{sec:futurework}.

\subsection{Generational Performance: H100 as the Baseline}
\label{sec:h100baseline}

Table~\ref{tab:speedups} reports speedups within each GPU---i.e., how
much Ozaki~II/\fp{8} accelerates each workload relative to that GPU's
own native \fp{64} performance.  This is the right view for evaluating
emulation \emph{on a single chip}, but it is not the right view for
evaluating whether \fp{64}-emulated execution \emph{regresses} or
\emph{progresses} relative to the prior-generation HPC baseline.  The
appropriate baseline for that question is the H100, the last
data-centre NVIDIA GPU whose architecture was balanced for HPC rather
than for AI inference.  Table~\ref{tab:h100baseline} therefore reports
absolute modeled upper-bound \fp{64}-equivalent throughput for the
same five
workloads, with H100 native \fp{64} explicitly set as the unit.

\begin{table}[h]
\caption{Modeled upper-bound \fp{64}-equivalent throughput per workload, in
TFLOPS, and relative to H100 native \fp{64} (last column block, in
parentheses).  Native throughput uses the \fp{64} tensor path for dense
GEMM and the \fp{64} vector path for the memory-bound primitives.
Ozaki~II throughput uses \fp{8} tensor cores with $r=12$ moduli, the
$(3r{+}1)=37$ \fp{8} cost structure, and
$\beta=1$; for Rubin, the dense-GEMM Ozaki entry is the model-derived
upper bound ($P_{\fp{8}}/(3r{+}1) = 473$~TFLOPS at the official
17.5-PFLOPS dense \fp{8} rate), to be compared with NVIDIA's
conservative published Emulated-DGEMM figure of $\sim 200$~TFLOPS.
Ozaki~II rows are the fused, conversion-ideal bounds; at the reference
deconstruction constants the memory-bound B300 Ozaki entries are
conversion-capped instead (SpMV $0.63$, bGEMV
$B{=}2$ $1.6$, stencil $3.1$, bGEMV $B{=}8$ $6.3$~TFLOPS; the roof
shortfall is OI-independent)---see the
generational-claim discussion in the text.}
\label{tab:h100baseline}
\centering
\footnotesize
\setlength{\tabcolsep}{3pt}
\begin{tabular}{lc|cccc|cccc}
\toprule
& & \multicolumn{4}{c|}{\textbf{Absolute throughput (TFLOPS)}}
  & \multicolumn{4}{c}{\textbf{Relative to H100 native}} \\
\textbf{Workload} & \textbf{Path}
  & H100 & B200 & B300 & R200
  & H100 & B200 & B300 & R200 \\
\midrule
\multirow{2}{*}{Dense GEMM (OI$\ge$50)}
  & Native    & 67   & 40   & 1.2  & 33    & $1.00\times$ & $0.60\times$ & $0.02\times$ & $0.49\times$ \\
  & Ozaki~II  & 53   & 122  & 135  & 473$^{*}$ & $0.80\times$ & $1.82\times$ & $2.02\times$ & $7.06\times$ \\
\midrule
\multirow{2}{*}{bGEMV $B{=}8$ (OI$\approx$2)}
  & Native    & 6.7  & 16   & 1.3  & 33    & $1.00\times$ & $2.39\times$ & $0.19\times$ & $4.93\times$ \\
  & Ozaki~II  & 6.7  & 16   & 16   & 44    & $1.00\times$ & $2.39\times$ & $2.39\times$ & $6.57\times$ \\
\midrule
\multirow{2}{*}{bGEMV $B{=}2$ (OI$\approx$0.5)}
  & Native    & 1.68 & 4.0  & 1.3  & 11    & $1.00\times$ & $2.39\times$ & $0.78\times$ & $6.57\times$ \\
  & Ozaki~II  & 1.68 & 4.0  & 4.0  & 11    & $1.00\times$ & $2.39\times$ & $2.39\times$ & $6.57\times$ \\
\midrule
\multirow{2}{*}{7-pt Stencil (OI$\approx$0.5)}
  & Native    & 1.68 & 4.0  & 1.3  & 11    & $1.00\times$ & $2.39\times$ & $0.78\times$ & $6.57\times$ \\
  & Ozaki~II  & 1.68 & 4.0  & 4.0  & 11    & $1.00\times$ & $2.39\times$ & $2.39\times$ & $6.57\times$ \\
\midrule
\multirow{2}{*}{SpMV (OI$\approx$0.2)}
  & Native    & 0.67 & 1.6  & 1.3  & 4.4   & $1.00\times$ & $2.39\times$ & $1.94\times$ & $6.57\times$ \\
  & Ozaki~II  & 0.67 & 1.6  & 1.6  & 4.4   & $1.00\times$ & $2.39\times$ & $2.39\times$ & $6.57\times$ \\
\bottomrule
\end{tabular}
\\[2pt]
{\scriptsize $^{*}$~For Rubin dense GEMM, NVIDIA's published Emulated
DGEMM specification of $\sim 200$~TFLOPS corresponds to $2.99\times$ on
the H100-relative scale and sits at $\approx\!42\%$ of the
dense-\fp{8}$/(3r{+}1)$
model bound shown here ($473$~TFLOPS, $7.06\times$ at $r{=}12$, at the
official 17.5-PFLOPS dense \fp{8} rate of June~30,
2026~\cite{nvidia_dgx_rubin_nvl8})---the expected margin for a
delivered, accuracy-guaranteed library figure (\S\ref{sec:cqladder}).
This $473$ is the \emph{theoretical} arithmetic roof.  On the current Rubin
architecture a real large dense \emph{square} GEMM is re-split once per
thread-block cluster and so delivers ${\approx}\tfrac12$ of it---the
${\approx}235$-TFLOPS deconstruction-$\lambda$ floor (${\approx}3.5\times$
H100; Table~\ref{tab:densedgemm})---until the hardware co-design of
\S\ref{sec:hwproposal} (Option~C) retires the deconstruction term, so the
delivered dense-\emph{square} entry is ${\approx}235$, not $473$, on today's
silicon; tall/skinny and small-batch GEMMs are unaffected and are
projected to reach the roof on shipping hardware.}
\end{table}

\paragraph{The key generational claim.}  Three patterns in
Table~\ref{tab:h100baseline} support a single thesis: \emph{Ozaki~II
does not regress performance against the H100 baseline; on the contrary,
it restores or improves the prior-generation scaling on every
workload.}

\begin{sloppypar}
\begin{enumerate}[leftmargin=2em,itemsep=2pt]
\item \textbf{Native B300 regresses catastrophically} for every
   compute-sensitive workload: $0.02\times$~H100 on dense GEMM,
   $0.19\times$ on batched~GEMV ($B{=}8$), and $0.78\times$ on
   batched~GEMV ($B{=}2$).  Only SpMV ends up faster than the H100:
   at $I\!=\!0.2$ the memory roof is $1.6$~TFLOPS, so B300's native
   pipe is compute-bound at its $1.3$~TFLOPS---short of the roof, but
   still $1.94\times$ the H100's $0.67$.  This is the
   regression that NVIDIA's silicon-area redirection imposes on every
   non-emulated HPC code running on B300.
\item \textbf{B300 with Ozaki~II returns to B200's scaling at the fused
   bound}: every memory-bound row of the Ozaki rows on B300 matches
   B200's row exactly ($2.39\times$~H100, the HBM-bandwidth ratio
   $8/3.35$).  Dense GEMM under emulation reaches $2.02\times$~H100,
   exceeding B200's native $0.60\times$ and emulated $1.82\times$.  In
   other words, the silicon area
   redirected from \fp{64} units to \fp{8} units is fully recovered
   through Ozaki~II---there is no net regression versus B200.
\item \textbf{Rubin with Ozaki~II scales as the HBM4 jump alone would
   predict}: the $6.57\times$~H100 figure for every memory-bound row
   is exactly the bandwidth ratio $22/3.35$.  The emulation path simply
   passes the bandwidth advantage through to the application without
   bottlenecking it on the compressed \fp{64} pipe.
\end{enumerate}
\end{sloppypar}

\paragraph{The generational claim at reference deconstruction
constants.}  The three patterns above are stated at the fused,
conversion-ideal bound, and honesty requires the second column.  At the
reference constants ($c_q{=}16$), B300's conversion-limited emulated
throughput is $\mathrm{OI}\cdot 8P_{\text{int}}/(\varphi c_q r) \approx
3.1\,\mathrm{OI}/\varphi$~TFLOPS, against H100's native
$3.35\,\mathrm{OI}$~TFLOPS---so the \emph{hybrid} best path on B300
(native below $\varphi\,\mathrm{OI}^{*}$, emulated above) delivers
between $0.93\times$ and $2.02\times$ H100 across every surveyed
workload: parity within $7\%$ in the worst case (the conversion-bound
$\varphi{=}1$ rows), and full gains on stencil ($1.86\times$), SpMV via
native ($1.94\times$), and dense GEMM ($2.02\times$).  The claim ``no
regression versus H100'' therefore survives the deconstruction term in
the form \emph{parity within $7\%$, everywhere, with no engineered
$c_q$ at all}; restoring the full $2.39\times$ bandwidth-ratio scaling
of the fused bound requires the roof-recovery threshold
$c_q \lesssim 5$--$6$---tighter than the ${\sim}7$ single-RHS-parity
decision criterion of~\cite{bayraktar2026tme}---and both
thresholds are the first measurements of Part~3.  The audited ladder
of \S\ref{sec:cqladder} already meets the roof-recovery threshold on
paper: the refined two-limb residual ($5.8$ at $r{=}12$) sits inside
B300's budget (the all-byte codesign, at its own $r'{=}15$, delivers
$0.95$ of the \emph{memory} roof), so Part~3 measures
an audited design rather than searching for one.

Put differently: Ozaki~II is not just a compensation mechanism for
NVIDIA's \fp{64} regression; it is the mechanism that \emph{converts}
the silicon-area savings into bandwidth-scaling and into AI-grade
tensor throughput, both of which the application then sees as faster
\fp{64}.  The H100 baseline view shown in
Table~\ref{tab:h100baseline} is, in the author's opinion, the
appropriate framing for procurement, application-porting, and
benchmarking decisions: the question is not ``does emulation match
B300 native \fp{64}?''---of course it does, by orders of magnitude---but
``does the post-\fp{64} stack continue the H100$\to$B200$\to$B300$\to$Rubin
generational improvement that HPC procurement has historically
relied on?'' Table~\ref{tab:h100baseline} answers \emph{yes}.

The table covers the four primitives---dense GEMM, batched GEMV,
stencils, SpMV---for which Ozaki~II is the appropriate emulation path.
The corresponding analysis for the fifth canonical primitive,
three-dimensional FFT, appears in a companion
paper~\cite{matsuoka2026fft}, which establishes that a full-\fp{64}
$1024^3$ FFT needs no \fp{64} arithmetic anywhere in its kernel, and
that its binding resource on B300 is a per-output integer-SIMT
epilogue: software alone projects $63$--$87$~ms against the $12.9$~ms
memory roof---$1.3$--$1.9\times$ the collapsed native path at peak
issue---and the roof is reached under that paper's minimal co-design
ladder, whose load-path deconstruction datapath is the Option-C
converter of \S\ref{sec:hwproposal}.  Together
with this paper's results, the integrated kernel coverage is the
subject of \S\ref{sec:l3}.

\paragraph{A note on FP8:FP64 ratios.}  The ratios across the four GPUs
are wildly different (30:1 on H100, 113:1 on B200, 3800:1 on B300,
$\sim$530:1 against native vector on Rubin at the official \fp{8}
rate---or $\sim$88:1 against the published Emulated-DGEMM figure and
$\sim$37:1 against the model ceiling).  A simple compute-balance
analysis~\cite{williams2009_roofline}---defining balance as
\emph{equal service times} for the two pipes, so that the GEMM
fraction $f_G$ emulated at $P_{\fp{8}}/\alpha$ and the residual
$(1\!-\!f_G)$ run natively at $P_{\fp{64}}$ finish together, which
gives the balanced ratio $P_{\fp{8}}/P_{\fp{64}} = \alpha\,f_G/(1\!-\!f_G)$
---shows that B300's 3800:1 ratio is balanced for
$f_G\!\geq\!0.990$, i.e.\ essentially pure-AI workloads, while
classical HPC mixes ($f_G\!\in\![0.6,0.9]$) call for ratios of
$\sim\!56$:1 to $\sim\!333$:1 at $\alpha{=}37$.  This is the headline tension that the
thesis of \S\ref{sec:l3} resolves: the
post-\fp{64} stack collapses the $(1\!-\!f_G)$ residue onto pipes that
\emph{are} appropriately provisioned (Ozaki~II for GEMM-like work,
Kulisch INT32 for FFT-class reductions, FP32+Kahan for BLAS-1),
making the native FP8:FP64 ratio operationally irrelevant for HPC.
Of the current generation Rubin's effective $\sim$37:1--88:1 (against
the model ceiling and the published Emulated-DGEMM figure respectively)
is the closest to a balanced HPC design point;
B300's $3800$:1 is, in isolation, the furthest---but in combination
with the software stack of this paper, it ceases to matter.

\subsection{INT8 and FP8: the Substrate Division of Labour}
\label{sec:int8vsfp8}

A recurring question is which low-precision tensor format should
underlie Ozaki~II.  The original
formulation~\cite{ozaki2025_scheme2,uchino2025_ijhpca} used
\inteight{}: integer modular arithmetic is the natural CRT fit, and
on \inteight-rich parts it remains the cheaper carrier---one MMA per
modulus against \fp{8}'s three, which after paying for the larger
byte-modulus set ($\alpha_{\textsc{int8}}=16$ against $\alpha=37$;
\S\ref{sec:background}) is still $2.31\times$---which is why the shipping cuBLAS emulation takes the
INT8 tensor path on Blackwell~\cite{nvidia_cublas_emulation}, as the
measured B200 anchors confirm.  But the AI workloads that drive
silicon economics have decisively favoured narrow
\emph{floating-point}: \fp{8}'s per-element exponent absorbs the
outlier-heavy activation and gradient distributions that fixed-point
scaling clips~\cite{nvidia_fp8_training_blog,micikevicius2022_fp8},
a preference that carried \fp{8} into mainstream LLM
training~\cite{deepseekv3} and now four-bit NVFP4
pretraining~\cite{nvidia_nvfp4_2025}---so B300 and Rubin cut the
\inteight{} tensor rate to a residual while scaling \fp{8} to
$5$--$17.5$~PFLOPS.  Table~\ref{tab:int8vsfp8} shows the consequence
for the emulation ceilings: on the \inteight-rich H100 and B200 the
INT8 substrate leads; on B300 and Rubin \fp{8} is the only strong
carrier.

\begin{table}[h]
\caption{Emulated dense \fp{64}-equivalent throughput under the two
candidate Ozaki~II substrates, each charged at \emph{its own} matched-accuracy
cost: \fp{8} at $\alpha=3r{+}1=37$ ($r{=}12$, the published nonsquare set,
$111.8$ bits), INT8 at
$\alpha_{\textsc{int8}}=r_{\textsc{i}}{+}1=16$
($r_{\textsc{i}}{=}15$, the all-byte set, $117.8$ bits---one MMA per
modulus plus the max pass, but no byte-modulus system of fewer than
fifteen moduli reaches $111.8$ bits, \S\ref{sec:background}).  Revisions
before the July 2026 correction cycle charged INT8 at $r{+}1=13$, i.e.\ at the \fp{8} set's $r$ on
moduli an \inteight{} operand cannot hold; the INT8 column and the
advantage column are $\sim\!23\%$ lower and higher respectively as a
result.  (Earlier
versions also mixed Ozaki-I slice notation and understated B200's INT8 rate
by $\sim\!29\times$---corrected from the HGX
specification.)  ``FP8 advantage'' is the ratio of the \fp{8} ceiling to
the INT8 ceiling.  On INT8-rich parts (H100, B200) the INT8 substrate
leads; on the \inteight-cut parts (B300 Ultra, Rubin) \fp{8} is the
only strong carrier---Ozaki~II rides whichever low-precision engine the
architecture favours.  The tabulated rates are \emph{theoretical} arithmetic
ceilings $P/\alpha$ with each column's own $\alpha$: on Rubin a real large dense \emph{square} GEMM delivers
${\approx}\tfrac12$ the \fp{8} ceiling---the ${\approx}235$-TFLOPS
deconstruction-$\lambda$ floor (Table~\ref{tab:densedgemm})---until the
co-design of \S\ref{sec:hwproposal} (Option~C) retires the deconstruction term,
tall/skinny and small-batch shapes reaching the ceiling today; GB300, whose
$135$-TFLOPS ceiling already sits at its floor, is roof-bound.
$^{a}$Measured B200 anchors~\cite{uchino2026_fp8}: $\approx$137--138
TFLOPS delivered on the INT8 path ($\approx\!49\%$ of the analytic
Scheme-II ceiling shown; these anchors are research Scheme-II runs,
distinct from the shipping cuBLAS path whose algorithm attribution is
discussed in \S\ref{sec:model}) and $\approx$61--65 on the \fp{8}
path ($\approx\!53\%$)---the
same delivered-to-ceiling band as NVIDIA's published Emulated-DGEMM
figures (\S\ref{sec:cqladder}), and consistent with the shipping
cuBLAS emulation using the INT8 tensor path on
Blackwell~\cite{nvidia_cublas_emulation}.}
\label{tab:int8vsfp8}
\centering
\footnotesize
\begin{tabular}{lccccc}
\toprule
\textbf{GPU} & $P_{\inteight}$ & $P_{\fp{8}}$
             & Ozaki/\inteight  & Ozaki/\fp{8}
             & FP8 \\
             & (TOPS)          & (TFLOPS)
             & ceiling (TFLOPS) & ceiling (TFLOPS)
             & advantage \\
\midrule
H100  & 1{,}979 & 1{,}979 & 123.7 & 53.5 & $0.43\times$  \\
B200  & 4{,}500  & 4{,}500 & 281$^{a}$ & 121.6$^{a}$ & $0.43\times$  \\
B300  & $\sim$165  & 5{,}000 & 10.3 & 135.1 & $13.1\times$  \\
R200  & $\sim$250  & 17{,}500 & 15.6 & 473.0 & $30.3\times$  \\
\bottomrule
\end{tabular}
\end{table}

The either/or framing that this table invites is, however, now
obsolete.  The companion paper's audited
accounting~\cite{matsuoka2026ozaki25} turns the substrate question
into a \emph{division of labour}, and thereby settles it.  In the
engineered pipeline of \S\ref{sec:cqladder} the two formats hold
fixed, complementary roles: the \fp{8} tensor cores carry the heavy
work---the $(3r{+}1)$ Ozaki product passes---at the rates the AI
market already pays for, while the \emph{residual} \inteight{} tensor
capability hosts precisely the work the \fp{8} pipes cannot: the
exact two-limb byte-plane reduction GEMMs (the reduction constants
reach $826$, and operand bytes are not \textsc{e4m3}-representable;
\S\ref{sec:cqladder}) and the exact-INT32 reconstruction epilogues,
with \texttt{dp4a}-class SIMT dot products as the tensor-free
fallback.  The integer demand is small---${\approx}13\%$ of B300's
$166$-\tops{} residual rate and ${\approx}24\%$ of Rubin's $250$ for
the reconstruction epilogues---but it is a real, \emph{binding}
capacity: the companion's capacity-aware analysis shows the residual
\inteight{} rate setting the effective dense crossovers
($n^{*}\!\approx\!290/670$; \S\ref{sec:hwproposal}), and the two
formats' concurrency on shared tensor hardware is itself unproved,
so the dense tensor-migrated routes carry an
independent-versus-serialised service bracket
(\S\ref{sec:hwproposal}).  Withering to a
minor residual and withering to zero are therefore architecturally
very different outcomes for the post-\fp{64} stack: the design
recommendation is not that narrow integer arithmetic is dispensable,
but that it be \emph{retained at minor scale} while \fp{8} carries
the multiplication (\S\ref{sec:falsification} states this as an
explicit boundary of the thesis).

The longstanding fear was that the AI-driven preference for \fp{8}
over \inteight{} and \fp{64} would \emph{compromise HPC}, forcing a
bifurcation of the product line into AI-optimised and
\fp{64}-preserving SKUs.  The central architectural message of this
paper is that the fear is misplaced: Ozaki~II is
substrate-agnostic---the reference library
GEMMul8~\cite{gemmul8_github} exposes both backends, and the original
INT8 formulation~\cite{ozaki2025_scheme2} and the
Uchino--Ozaki--Imamura \fp{8} adaptation~\cite{uchino2026_fp8} are
two instantiations of one precision-independent idea---so emulated
\fp{64} simply rides whichever low-precision engine the architecture
favours, and that engine is now \fp{8}.  Building on the substrate
the AI market funds recovers \fp{64}-equivalent throughput across
the surveyed matrix-dominated kernel classes; what the Ozaki-2.5
accounting adds is the precise, sized role of the integer residual.
The substrate question is thus no longer \emph{which format wins}
but \emph{how much of the other to keep}---and the answer, a few
tens of TOPS plus \texttt{dp4a}, is the cheapest line item in the
hardware discussion of \S\ref{sec:hwproposal}.

\subsection{The Deconstruction Path: Sensitivity, the Hardware
Proposal, and the Combined Projection}
\label{sec:rubinuplift}

This subsection is deliberately the paper's one
architec\-ture-specific deep dive: the implemen\-tation-side companion
to the symbolic fourth term of \S\ref{sec:model}.  In the vocabulary of
Proposition~\ref{prop:atomic}: the theory establishes the bound and the
satisfiability of conditions (C1)--(C4) in the model; \emph{this
subsection takes the realistic constants of \S\ref{sec:cqladder} as
given and studies what they make of condition (C2) on 2026
silicon}---how sensitive the delivered roofs are to the engineered
constant and the vendor dial, what hardware would close the Rubin
gap that no realisable software floor can, and what the combined
picture projects across the kernel spectrum.  The theory is invariant;
everything here is dated to the silicon it describes and is
falsifiable cell by cell in Part~3.  We proceed in four movements: a
one-parameter sensitivity analysis that compresses the whole question
into a single dimensionless group; the hardware proposal---three
concrete options, sized and compared; the dense linear-algebra
dividend of the deconstruction path, with the published Rubin
Emulated-DGEMM figure as a testable case and the sparse variations of
the Ozaki-2.5 discipline; and the combined projection, mapped claim
by claim to Part~3 measurements.

\subsubsection{Sensitivity: One Dimensionless Group}
\label{sec:step5}

The budgets and ladder of \S\ref{sec:cqladder} involved two
quantities the vendor controls
($P_{\text{int}}$, and implicitly the issue rates behind
\texttt{DP4A}) and one the software controls ($c_q$).  Nothing in the
\emph{algorithm} fixes either: the algorithm's irreducible demand is
$O(r)$ narrow bit-operations per element, whose silicon cost is
minuscule; the binding arises because GPUs provision SIMT integer
issue for control and addressing, not for data-plane transformation of
every streamed byte---a dial the vendor has moved nearly every
generation (Turing added a dedicated concurrent INT32 pipe on
instruction-mix grounds; Ampere re-merged it; Blackwell rebalanced
again).  The conversion-bound regime is an artifact of that dial's
current position, and the correct analytical response is sensitivity
analysis.  All delivered quantities depend on $(c_q, P_{\text{int}})$
only through the dimensionless shortfall group
\begin{equation}
\chi \;=\; \frac{c_q\, r\, \varphi\, B_{\text{mem}}}{8\,P^{\text{eff}}_{\text{int}}},
\qquad \text{roof fraction} = \min(1,\,\chi^{-1}),\qquad
\mathrm{OI}^{*} = \chi \cdot \frac{P_{\fp{64}}}{\varphi\,B_{\text{mem}}},
\label{eq:chi}
\end{equation}
so a $2\times$ on the effective integer budget is worth exactly a
$2\times$ on the engineered $c_q$, and the levers compound.
$P^{\text{eff}}_{\text{int}}$ aggregates every way the budget can
grow: added narrow-integer lanes, verified split issue across the
\fp{32} lanes (Appendix~\ref{app:beta-blackwell}), and the tensor
offload of the ladder, which removes work from the denominator's pipe.
Table~\ref{tab:pintsens} evaluates Eq.~\eqref{eq:chi} across the
scenario grid; the $c_q$ rows are exactly the rungs of the audited
ladder of \S\ref{sec:cqladder}.

\begin{table}[h]
\caption{Sensitivity of the fourth term to the effective integer-issue
budget: fraction of the memory roof delivered by a fully streamed
($\varphi{=}1$) emulated kernel, $\min(1,\chi^{-1})$, at $r{=}12$.
Rows are engineered-$c_q$ scenarios (per modulus per element,
ledger-projected counts of the companion
paper~\cite{matsuoka2026ozaki25}): $16$ = memo-counted shipping path
(S0); $10.3$ = direct \texttt{dp4a} route (ladder L1, two limbs);
$6.3$ = post-tensor-migration two-limb residual (L2, inline final
reduction; $5.8$ refined); $5.3$ = the all-byte codesign's one-pass
residual, the one row evaluated at its own $r'{=}15$
($\approx\!79$ instructions per element; its $\alpha'{=}46$ tensor
charge is benign for
memory-bound kernels); $4.3$ = with the lazy-reduction trade also
applied (L3, charging $\alpha'{\approx}73$---likewise benign here);
$2$ = residual with piece handling also removed (speculative
piece-free formats).  Columns scale $P^{\text{eff}}_{\text{int}}$
from the $75$-\tops{} baseline.  The roof fraction is independent
of the kernel's OI---one table covers the entire streamed class---and
on Rubin-class bandwidth no plausible $c_q$ reaches the roof
without either the tensor migration (row~3) or a budget multiple.}
\label{tab:pintsens}
\centering
\footnotesize
\setlength{\tabcolsep}{5pt}
\begin{tabular}{lcccccc}
\toprule
& \multicolumn{3}{c}{\textbf{B300} (8 TB/s)} & \multicolumn{3}{c}{\textbf{Rubin} (22 TB/s)} \\
\cmidrule(lr){2-4}\cmidrule(lr){5-7}
\textbf{$c_q$ scenario} & $1\times$ & $2\times$ & $3\times$ & $1\times$ & $2\times$ & $3\times$ \\
\midrule
16 (memo-counted)        & 0.39 & 0.78 & \textbf{1.0} & 0.14 & 0.28 & 0.43 \\
10.3 (direct \texttt{dp4a}, L1) & 0.61 & \textbf{1.0} & \textbf{1.0} & 0.22 & 0.44 & 0.66 \\
6.3 (two-limb tensor-migr., L2; 5.8 refined) & 0.99 & \textbf{1.0} & \textbf{1.0} & 0.36 & 0.72 & \textbf{1.0} \\
5.3 (all-byte codesign at its own $r'{=}15$; $\alpha'{=}46$) & 0.95 & \textbf{1.0} & \textbf{1.0} & 0.34 & 0.69 & \textbf{1.0} \\
4.3 (L2 + lazy red.\ L3, $\alpha'{\approx}73$) & \textbf{1.0} & \textbf{1.0} & \textbf{1.0} & 0.53 & \textbf{1.0} & \textbf{1.0} \\
2 (piece-free)           & \textbf{1.0} & \textbf{1.0} & \textbf{1.0} & \textbf{1.0} & \textbf{1.0} & \textbf{1.0} \\
\bottomrule
\end{tabular}
\end{table}

The vs-native threshold improves twice as fast as the table suggests,
because doubling $P^{\text{eff}}_{\text{int}}$ also halves
$\mathrm{OI}^{*}$: on B300 at $2\times$, even the \emph{memo-counted}
$c_q{=}16$ brings SpMV to $0.96\times$ native (from $0.48\times$), and
$3\times$ takes it past parity.  The table's message for the hardware
proposal: on
B300 the tensor-migrated rung essentially reaches the roof ($0.99$ at
the uniform two-limb count, $1.0$ refined; $0.95$ for the all-byte
codesign at its own $r'{=}15$) and a $2\times$ budget certainly
suffices; on Rubin the
roof requires \emph{both} the migration and a budget multiple---or
hardware that removes the term at the source.

\subsubsection{The Hardware Proposal: Three Concrete Options}
\label{sec:hwproposal}

A word on stance. This paper is deliberately hardware-\emph{independent}:
its contribution is the Tensor--Memory Equilibrium model, and Ozaki~II runs
on whatever low-precision engine the architecture provides. But the very act
of dividing delivered throughput into the model's terms, and then locating
\emph{which} term is the binding one where the theory falls short of the
roof, is what identifies the co-design points \emph{precisely}: the fourth
(deconstruction) term is the sole obstruction for the conversion-bound
sparse kernels, so $c_q$---and nothing else---is the coordinate a hardware
change must move here. A model that did not separate the terms could name no
such target. In that spirit the proposal below is offered not as a
requirement but as the precise, model-localised ask that follows from the
theory; it is what a hardware partner would act on, and it is stated so that
NVIDIA can.

\begin{figure}[htbp]
\centering
\includegraphics[width=\linewidth]{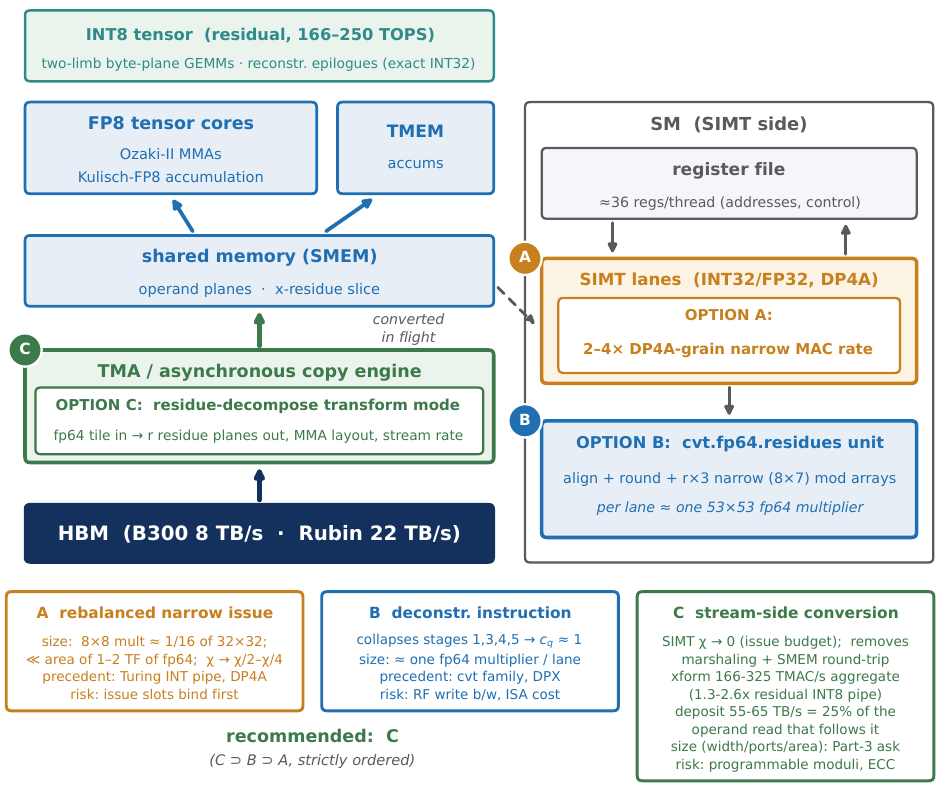}
\caption{Where the three hardware options of this proposal sit (schematic
Blackwell/Rubin-class SM and memory path).  \textbf{Option~A} scales
the narrow (\texttt{DP4A}-grain) MAC rate inside the existing SIMT
lanes; \textbf{Option~B} adds a \texttt{cvt.fp64.residues} datapath
beside the register file, collapsing conversion to one issue at a
per-lane cost of roughly one \fp{64} multiplier; \textbf{Option~C}
(recommended) adds a residue-decompose transform mode to the
TMA/asynchronous copy engine, so \fp{64} tiles are converted \emph{in
flight} between HBM and shared memory---residue planes arrive in MMA
operand layout at stream rate, uniquely eliminating the marshaling and
SMEM round-trip costs along with the arithmetic ($\chi\to0$ in the
SIMT issue budget).  Option~C zeroes the compute-pipe term, \emph{not}
conversion time: the C card states the resulting vendor requirement
rather than a delivered device count---the transform must sustain
$166$--$325$~TMAC/s aggregate ($1.3$--$2.6\times$ the platform's
entire residual \inteight{} tensor throughput) and deposit
$55$--$65$~TB/s of plane writes, the latter being exactly $T/R=64/256$
of the operand read the MMA already performs.  Both are evaluated per
route \emph{at its own} roof; datapath width, ports, area and power
are a Part-3 obligation.  The
left column shows the tensor-side consumers of the converted planes
(Ozaki~II MMAs, Kulisch-\fp{8}
accumulation) and the residual INT8 tensor pipe carrying the exact
reconstruction epilogues.  Annotation cards give each option's size,
$\chi$ effect, precedent, and principal risk; options are strictly
ordered $C \supset B \supset A$.  \emph{Correction to earlier drafts:}
the C card previously read ``size: ${\approx}300$ narrow MACs\,/\,engine''
and ``deposit b/w $(p/8){\times}$stream provisioned''; the first is a
per-engine width that is not derivable without a clock and a
copy-engine count, and the second priced the deposit off a stream rate
rather than off each route's own roof.  Generated by
\texttt{render\_hw\_options.py}, which imports the numbers from
\texttt{params.py} at render time so the artwork cannot drift from the
requirement table of the companion~\cite{matsuoka2026ozaki25}.}
\label{fig:hwoptions}
\end{figure}

We now state the deconstruction-path proposal concretely, as three
options of increasing vendor commitment.  Each is specified by what it
is, where it sits, its first-order size, its effect on $\chi$, its
precedent in the vendor's own design vocabulary, and its principal
risk.  Table~\ref{tab:hwoptions} summarises,
Figure~\ref{fig:hwoptions} places the three options at their physical
locations on the SM and memory path, and prose follows.

All three target one resource---$c_q$, the deconstruction cost per
element---and that is the deliberate scope of \emph{this} paper's
hardware proposal, because $c_q$ is the lever the \emph{sparse and
conversion-bound} kernels depend on: SpMV, single-RHS GEMV, and stencils
meet $c_q$ directly (below $\mathrm{OI}^{*}$, \S\ref{sec:model}), with no
re-split floor to escape, so only a lower $c_q$---in software through
persistent (Mode~P) deconstruction, or in hardware through Options
A--C---moves their verdict.  The \emph{dense}-DGEMM ceiling is bounded by
a different resource, the on-the-fly re-split floor $R\,P_{\text{int}}/(c_q r)$
set by a per-element deconstruction load $c_q r/R$ at cluster reach $R$,
whose escape adds two levers this proposal does not cover---cluster reach
(CTAs per cluster, TMEM tile) and L2 service bandwidth---developed as a
companion hardware co-design section in
Ozaki~2.5~\cite{matsuoka2026ozaki25}. The two sections are cross-referenced
and target complementary silicon: $c_q$ here (sparse-critical, dense-shared),
reach and L2 there (dense-specific). $c_q$ appears in both because it lifts
the dense floor at every reach as well.

\begin{table}[h]
\caption{The three hardware options for the deconstruction path,
summarised.  Area figures for A and B are first-order gate-count
comparisons, not layout estimates; all presume retention of the minor
narrow-integer capabilities (residual INT8 tensor, \texttt{DP4A}) per
\S\ref{sec:cqladder}.  $\chi$ effects at reference constants, $r{=}12$.
For C we deliberately give no device count: the aggregate transform
rate is derivable from the route roofs and the reach, a per-engine
width is not (it needs a clock and a copy-engine count, neither
published).  C's two quantitative entries are engine-derived at each
route's own roof, per the companion's requirement table and
\texttt{verify/optionc\_req.py}~\cite{matsuoka2026ozaki25}.}
\label{tab:hwoptions}
\centering
\scriptsize
\setlength{\tabcolsep}{3pt}
\begin{tabular}{p{2.6cm}p{3.3cm}p{2.6cm}p{2.6cm}p{2.9cm}}
\toprule
\textbf{option} & \textbf{what / where} & \textbf{first-order size} & \textbf{effect on $\chi$} & \textbf{precedent / risk} \\
\midrule
\textbf{A. Rebalanced narrow-integer issue} & $2$--$4\times$ \texttt{DP4A}-class dot-product rate per SM (INT8-grain MAC lanes) & $8{\times}8$ mult.\ $\approx 1/16$ of $32{\times}32$; total $\ll$ area of 1--2 TFLOPS \fp{64} & $\chi/2$--$\chi/4$; B300 roof at memo-counted $c_q$; Rubin roof with L2 & Turing INT pipe; \texttt{DP4A} since Pascal.  Risk: issue slots bind before ALUs \\
\textbf{B. Deconstruction instruction} (\texttt{cvt.fp64. residues}) & SM instruction: one \fp{64} register in, $r$ packed residues (or $3r$ \fp{8} pieces) out; align+round+$r$ parallel narrow mod arrays & per lane $\approx$ one \fp{64} multiplier ($r{=}15{\times}3$ pieces ${\times}\,8{\times}7$ array $\approx$ a $53{\times}53$ array) & $c_q \to \sim\!1$ (issue-limited); $\chi \to \chi/16$-class & \texttt{cvt} family, Hopper DPX.  Risk: register-file write bandwidth for $r$ planes; ISA cost \\
\textbf{C. Stream-side conversion} (TMA residue-decompose mode) & Transform mode in the async-copy engine: \fp{64} tile in HBM $\to$ residue planes in SMEM, MMA-ready layout, at stream rate & \emph{aggregate rate stated, width not}: $166$--$325$~TMAC/s of narrow ($8{\times}7$) MACs, i.e.\ $1.3$--$2.6\times$ the residual \inteight{} tensor pipe.  Per-engine width, ports, area, power: Part-3 & SIMT $\chi \to 0$; removes stage-2 marshaling and the SMEM round-trip; plane deposit $55$--$65$~TB/s, which is $T/R{=}\tfrac14$ of the operand read the MMA already performs & TMA swizzle transforms; Blackwell decompression engine (decompress-on-copy $\Rightarrow$ deconstruct-on-copy); DPX.  Risk: moduli set must be descriptor-programmable; verification/ECC \\
\bottomrule
\end{tabular}
\end{table}

\emph{Option A---rebalanced narrow-integer issue}---is the incremental
insurance: multiply the \texttt{DP4A}-class narrow dot-product rate per
SM by $2$--$4\times$, in exactly the $8$-bit grain the anatomy of
\S\ref{sec:cqladder} showed the workload needs.  Because multiplier area scales roughly quadratically
in width, this is an order of magnitude cheaper than the equivalent
INT32-lane scaling, and it composes with every software rung.  Its
risk is microarchitectural: if instruction \emph{issue} slots, not ALU
count, are the true limiter on the unified Blackwell lanes, added MACs
buy nothing---precisely what the Part~3 pipe census measures first.

\emph{Option B---a deconstruction instruction}---is the ISA-clean
general solution: a \texttt{cvt}-class instruction taking one \fp{64}
operand (or a short vector) plus a moduli descriptor, and emitting the
$r$ residues---or directly the $3r$ \fp{8} Karatsuba pieces---into
registers or shared memory.  The datapath is an exponent align, a
round, and $r$ parallel mod-by-invariant arrays in $8{\times}7$-bit
logic; a gate-count comparison makes its scale vivid: at $r{=}15$ the
whole per-lane array ($15 \times 3$ pieces $\times$ $8{\times}7$
partial products $\approx 2{,}500$ units) is \emph{about one
$53{\times}53$ \fp{64} multiplier} ($\approx 2{,}800$ units).  That
is, full-width deconstruction hardware costs roughly what a
$1$--$2$-TFLOPS \fp{64} vector pipe costs---which B300 already
carries.  It collapses the scale/truncate, reduction, final-mod, and
split stages of the anatomy (Fig.~\ref{fig:anatomy}) into one issue
($c_q \to \sim\!1$), leaving only stage-2 marshaling in software.  Its
risks are the register-file write bandwidth of $r$ output planes and
the usual cost of an ISA addition; both are Part~3 study items.

\begin{figure}[htbp]
\centering
\includegraphics[width=\linewidth]{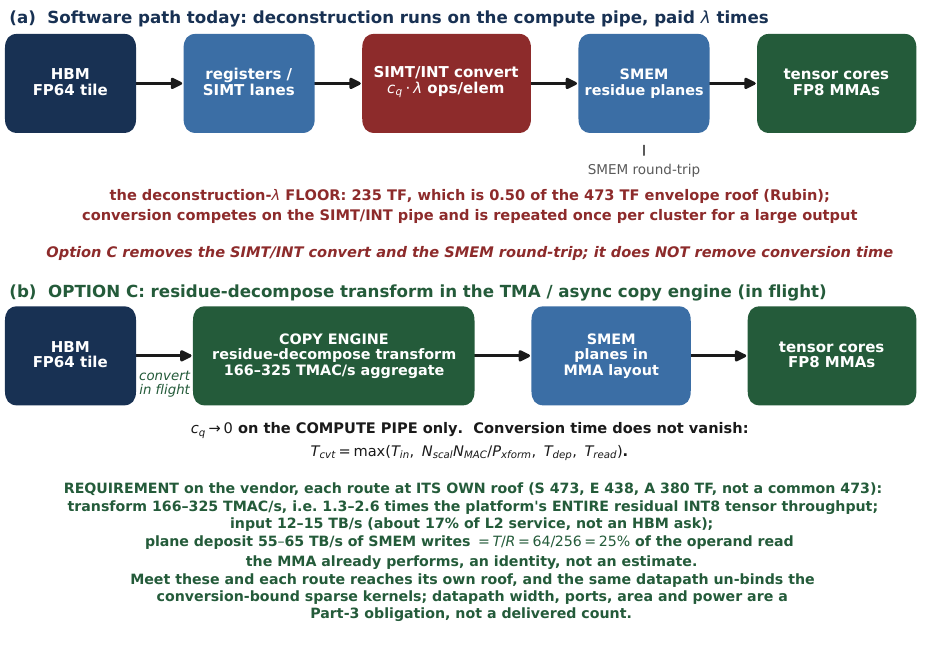}
\caption{Option~C, the central hardware co-design target, illustrated.
\emph{(a)}~Today the residue conversion runs on the SIMT/integer pipe,
competes with the \fp{8} MMA stream, and round-trips through SMEM---the
deconstruction cost the fourth TME term charges. \emph{(b)}~Option~C performs
the residue-decompose transform in the async copy engine, converting each
\fp{64} tile in flight into planes in MMA operand layout, so $c_q$ leaves the
compute budget ($c_q{\to}0$, hence $\chi{\to}0$); it retires both the conversion arithmetic and
the SMEM round-trip that software cannot remove. For the companion Ozaki~2.5
this is what lifts the large-DGEMM deconstruction-$\lambda$ floor to the roof;
for the conversion-bound sparse kernels here it is what moves them back above
$\mathrm{OI}^{*}$---one datapath closing both regimes.}
\label{fig:optionc}
\end{figure}

\emph{Option C---stream-side conversion on the copy path}---is the
complete solution, and our recommendation for the Part~3 codesign
study (illustrated in Fig.~\ref{fig:optionc}).  Extend the asynchronous
bulk-copy engine (TMA) with a
\emph{residue-decompose} transform mode: an \fp{64} tile enters from
HBM, and residue planes---placed in MMA operand layout by the
transform---land in shared memory at stream rate.  Uniquely among the
options, C also eliminates \emph{stage-2 marshaling} and the SMEM round
trip (the two costs software can never fully remove,
Appendix~\ref{app:beta-blackwell}), because the transform happens in
flight and deposits operands where the MMA reads them.  It retires the
\emph{SIMT} term identically ($\chi \to 0$ in the issue budget).

But a finite converter is not free, and it is worth being precise about
which part of ``not free'' we can actually derive.  Option~C zeroes the
compute-pipe term $c_q$; it does \emph{not} zero conversion \emph{time},
which becomes $T_{\text{cvt}}=\max(T_{\text{input}},\,
N_{\text{scalar}}N_{\text{MAC}}/P_{\text{transform}},\,
T_{\text{deposit}},\,T_{\text{tensor-read}})$.  The companion states the
resulting bill of materials as a requirement table, evaluated per Rubin
GPU at \emph{each route's own} roof under the $4{\times}4$ cluster of
$64^2$ tiles (reach $R{=}256$), with script
\texttt{verify/optionc\_req.py}~\cite{matsuoka2026ozaki25}; three of its
rows matter here.  \emph{The deposit is small, and exactly so}: the SMEM
plane-write rate is $p_{\text{store}}/R$ bytes per useful \fp{64} FLOP
against an operand read of $p_{\text{store}}/T$ that the MMA already
performs, so the new write traffic is exactly $T/R = 64/256 = \tfrac14$
of traffic that already happens---an identity, not an estimate, and
$55$--$65$~TB/s in absolute terms.  \emph{The input side is an L2 ask,
not an HBM ask}: $8/R$ bytes per FLOP is $12$--$15$~TB/s, ${\approx}17\%$
of L2 service, because the panels are L2-resident under the fused
schedule.  \emph{The transform datapath is the large ask}: at each
route's own roof it must sustain $166$--$325$~TMAC/s, which is
$1.3$--$2.6\times$ the platform's \emph{entire} residual \inteight{}
tensor throughput.  Earlier drafts of both papers described this block as
``${\approx}300$ narrow MACs per copy engine''; that is a per-engine
\emph{width}, and it is not derivable without a clock and a copy-engine
count, neither of which is a published quantity.  We therefore state the
aggregate rate, which is derivable, and leave width, ports, area and
power to the vendor---a Part-3 obligation, not a delivered count.  What
the software does buy is that the datapath is \emph{narrow}: because the
codesigned modulus sets reduce conversion to byte extraction plus a
handful of mod-by-invariant multiply--truncates, it is a fixed-function
residue block rather than a general converter, and the moduli, the
two-limb reduction and the exact reconstruction all stay in software.
The honest form of the claim is therefore: Option~C removes the SIMT
issue demand, the marshaling, and the round trip, and each route then
reaches \emph{its own} roof \emph{provided} the transform rate and the
deposit path above are provisioned---a stated area/bandwidth line item of
the Part-3 study rather than an identity.  With that provision the claims
tables collapse to their fused columns on any
bandwidth.  And it is squarely inside the vendor's established
memory-pipeline vocabulary---TMA already applies layout/swizzle
transforms on copy, Blackwell's decompression engine already runs a
\emph{decompress-on-copy} datapath of far greater complexity, and
Hopper's DPX shows the willingness to add domain instructions.  The
design risks---a fixed versus descriptor-programmable moduli set, and
verification/ECC of an arithmetic transform in the memory path---are
exactly the questions the Part~3 area/energy study
is scoped to answer.

The options are strictly ordered: C subsumes B's arithmetic and adds
the marshaling; B subsumes A's throughput and adds the fusion of
stages.  The recommendation is therefore staged, mirroring the
sensitivity table: A is worth taking on any rebalancing opportunity
(near-free insurance); B suffices for B300-class bandwidth
indefinitely; C is what Rubin-class and later bandwidth requires for
roof-parity streamed emulation, and is the piece of silicon this paper
formally proposes.  None of the three resembles the large
native-\fp{64} ALU provisioning of the holy-grail position: the
\emph{most ambitious} option is a fixed-function residue block on the
copy path, whose aggregate rate we state ($166$--$325$~TMAC/s) and whose
width we do not.

\paragraph{The dense-LA dividend: peak attainment at all sizes.}
The three options are motivated by the streamed, memory-bound kernels,
where the fourth term binds first; but they pay an immediate dividend
in \emph{compute-bound dense linear algebra} as well.  Deconstruction
is asymptotically amortised in dense Ozaki~II GEMM---$O(n^2)$ operand
elements against $O(n^3)$ MMA work---but the amortisation has a
crossover, and it is further out than the asymptotic intuition
suggests.  For a $m{\times}k$ by $k{\times}n$ product with each
operand element deconstructed once and its residue planes reused
thereafter (so conversion is charged per unique operand element, not
per streamed occurrence), setting the conversion time
$c_q r\,(mk{+}kn)/P_{\text{int}}$ against the tensor time
$\alpha\cdot 2mkn/P_{\fp{8}}$ shows that the fourth term stops binding
only when the \emph{harmonic mean} of the two outer dimensions clears
a critical size:
\begin{equation}
\frac{2mn}{m+n} \;\ge\; n^{*} \;=\;
\frac{c_q\, r\, P_{\fp{8}}}{\alpha\, P_{\text{int}}},
\label{eq:densecrossover}
\end{equation}
independent of the inner dimension $k$.  At the memo-counted
$c_q = 16$ (with $r = 12$, $\alpha = 37$,
$P_{\text{int}} \approx 75$~\tops{}) this gives
$n^{*}\approx 350$ on GB300\footnote{The companion
paper~\cite{matsuoka2026ozaki25} freezes the Blackwell-Ultra
datapoint to GB300 dense per-GPU rates ($5$~PFLOPS \fp{8},
$166$~\tops{} \inteight{}, ${\approx}1.4$~TFLOPS \fp{64}); the
``B300'' columns elsewhere in this paper predate that freeze and are
generation-illustrative.} ($P_{\fp{8}}{=}5$~PFLOPS) and
$n^{*} \approx 1{,}200$ on Rubin at the official 17.5-PFLOPS
rate---the uplift again raising the bar, exactly as with the keep-up
budgets of \S\ref{sec:cqladder}: below
these sizes even nominally compute-bound emulated GEMM is
conversion-bound and cannot reach the $P_{\fp{8}}/(3r{+}1)$ ceiling.
That operating region is not exotic.  It is precisely the tall-skinny
panel updates of blocked LU/QR factorisations (panel widths
$64$--$256$ put the harmonic mean at $\lesssim\!512$ regardless of how
tall the panel is), the block operations of communication-avoiding
Krylov and LOBPCG solvers (\S\ref{sec:l3},
Appendix~\ref{app:l3-exemplars}), and batched small GEMMs in tensor
contractions.  The corrected software ladder moves the crossover to
$n^{*}\!\approx\!290/670$ (GB300/Rubin: the two-limb route at its
\inteight{}-capacity-aware effective knee; the companion's
shape-switched codesign dispatch attains each set's roof from
${\approx}150$--$620$ depending on platform and
band~\cite{matsuoka2026ozaki25}; the L3 lazy-reduction rung is
excluded here, since its $\alpha'\!\approx\!73$ would halve the
very ceiling being sought); Option~A divides these by its
$2$--$4\times$ rebalance; Option~B ($c_q\!\to\!\sim\!1$) brings them
to ${\approx}22/76$; and Option~C removes the modeled external
deconstruction bottleneck (launch overhead, delivered tile
efficiency, and reconstruction remain), so
the Ozaki-II ceiling is projected at \emph{every} size the tensor
pipeline itself can sustain---with the converted planes additionally
landing in MMA-ready layout, freeing the SIMT lanes for the address
generation and scheduling work that dense pipelines still need.  The
same hardware that rescues SpMV from the conversion-bound regime is
thus also what lets dense Ozaki~II behave as a drop-in \texttt{DGEMM}
at peak from the first panel, not only at asymptotic sizes; the
$n^{*}$ crossover of Eq.~\eqref{eq:densecrossover} is directly
measurable and joins the Part~3 validation map on both counts.

\paragraph{Where the planes live: storage modes, and what each
charges.}  The convert-once accounting above must state where the
stored residue planes reside, because conversion multiplicity and
plane traffic are coupled.  The companion paper makes this an explicit
dispatch variable~\cite{matsuoka2026ozaki25}, now with the corrected
accounting (an earlier draft charged only the plane \emph{read}; the
write is not optional): transient materialisation through an HBM
workspace (mode M) mandates $Q_{\text{plane}} = 2p\,n_{\text{uniq}}$
bytes ($p = 30$--$44$ stored plane bytes per scalar), giving the
necessary plane-only bound $P \le B_{\text{mem}}\,\bar{n}/(2p)$, whose
obstruction ceases only above $\bar{n}_{\text{mat}} =
2p\cdot(\text{that set's own roof})/B_{\text{mem}} \approx 1290$--$1522$
on Rubin and $1014$--$1196$ on GB300.  That plane-only figure is a
\emph{necessary} obstruction only; roof \emph{attainment} is the
stronger \emph{additive} condition $P \le B_{\text{mem}}/((8{+}c)/\bar
n + 4/k)$, $c \in \{2p, p\}$ (both operands materialised, or one under
the fused hybrid), whose crossovers are $\bar n_{\text{add}} \approx
1462$--$1695$ (Rubin, full) and $817$--$917$ (fused, one operand).  The
one-operand fused-hybrid crossover above uses the \emph{square}-case
plane term $p/\bar n$; the general shape-dependent denominator is
$\tfrac{8}{\bar n}+\tfrac{4}{k}+\tfrac{p}{\ell}$ with $\ell=\max(m,n)$,
and the companion's per-record engine charges the exact
$2pk\min(m,n)$ plane traffic.

Auditing that engine against this bound is what exposed a defect in
it, and we record the correction rather than the earlier, false
reassurance.  Previous revisions of this paper stated that the
companion's engine had always capped each materialised mode by the
additive traffic $Q_0+Q_{\text{plane}}$, and that no trace record took
an HBM-materialised branch.  Neither held: the HBM branch charged a
plane-\emph{feed} cap but never the $c=2p$ write-plus-read itself, and
the two terms cross at $\bar n = 128$, so the omission was invisible
on square shapes and material on tall-and-skinny ones---$276$ CCSD
records at $\bar n \approx 42$ took the branch, the worst credited
$38.2$~TF against an additive bound of $12.3$~TF.  The companion's
L2-resident branch carried the same omission in milder form, charging
a streaming feed smaller than one pass over the workspace whenever
$\bar n < 64$.  With both terms charged, all $276$ records fall to the
on-the-fly schedule and the audit claim becomes true---and is now
enforced by a test in the companion artifact rather than asserted in
prose.  One composite moves: CCSD on Rubin, ${\approx}1.73\times$ to
${\approx}1.69\times$ over shipping; the companion's Table~11 is
otherwise unchanged.  The output-$C$ traffic in $Q_0$ is
charged as a single write under the overwrite convention
$\beta_{\text{BLAS}}{=}0$ that the interposer records assume; a nonzero
$\beta_{\text{BLAS}}$ would add an $8mn$-byte read of $C$ (a triangular
read for \textsc{syrk}), which the current trace format does not record.
Below those sizes, three schedules
carry the work, each with a stated eligibility predicate.  \emph{Split-k fused}:
when the output fits a single cluster
($T_mT_n \le C$, i.e.\ $\max(m,n) \lesssim 256$ square), every element is
converted once and the cluster's CTAs share each $k$-slab's planes
through distributed shared memory---no L2 or HBM plane storage (Gram-type
blocks, small fronts).  \emph{L2-resident materialisation}: when the plane
workspace fits L2 ($p\,k(m+n) \le C_{\text{L2}}$; at an assumed
$C_{\text{L2}} = 126$~MB, square problems to
$k \approx 1449/1196/1381$ for S/A/E), planes never touch HBM.
\emph{Fused-hybrid}: the streamed operand converted
$\lambda = \lceil \min(m,n)/256 \rceil$ times (the $256$-element cluster
reach), the smaller operand's
planes materialised (L2 if they fit, else HBM at $2p\,kn$)---a
\emph{tall/skinny} schedule whose HBM surcharge
$\delta = 2pkn/Q_0$ is $p/12 \approx 2.5$--$3.75$ for square shapes
and small only where $m/n \gtrsim 36$--$55$ (at
$\delta \le 0.2$, $k \gg n$) or the planes are L2-resident; the
earlier ``$\le$10--20\%'' figure was a tall/skinny value and is
withdrawn for general shapes.  Composed, and evaluated over a \emph{selected boundary-oriented suite}
($\pm 1$ probes around the $\lambda$ transitions $\min(m,n){=}256j$ and the
constructed L2-capacity boundaries, clamped to the declared range, coverage
ratio taken within one reconstruction world), the switched dispatch covers
the enumerated shapes at a cluster-aligned minimum of $\mathbf{0.50}$ of the
top ($473$-TFLOPS) roof on
Rubin (at large squares such as $m{=}n{=}666$) and $1.00$ on GB300, whose
$135$-TFLOP roof sits at or below its own re-split floor.  The operative
bound on the multi-cluster shapes is \emph{not} L2 plane-feeding but the
on-the-fly re-split (deconstruction-$\lambda$) load: because the planes are
generated on-chip and never materialised, the minimum is a compute bound,
\emph{insensitive} to the L2 service rate---replacing the intermediate
``$0.79$, L2-bandwidth-conditional'' figure, which had priced a
materialised schedule the design does not use (and undercounted the
re-split through a $1024$- rather than $256$-element cluster reach).  Two
conditions on the floor, both Part-3 measurables: it is a large-$k$ value
(the Garner residual drags it to ${\approx}0.35$ at $k{=}256$, recovering
by $k{\gtrsim}1024$) and it assumes the $16$-CTA cluster opt-in (an
$8$-CTA-portable $4{\times}2$ cluster lowers the reach to
$R{=}170.67$ and the floor to ${\approx}157$~TFLOPS).  The
single-cluster fused regime ($T_mT_n\le C$, $\lesssim256^2$ square) keeps
$\lambda{=}1$ and follows the $\bar n$-crossover.\footnote{The
\emph{strict} enumerated minimum under rigid integer-$\lceil\cdot\rceil$
re-splitting is $0.36$ (Rubin) at $(513,513)$, a $+1$ ragged boundary where
a one-element sliver is charged a full extra re-split; a ragged-edge-aware
schedule is \emph{modeled} to hold $0.54$ there---a Part-3 validation
target, since we enumerate it but do not construct it as a minimum.  We
therefore headline the cluster-aligned $0.50$ plateau and report $0.36$ as
the only rigid-schedule worst case the enumeration establishes.}  This
replaces the earlier, too-strong ``no matrix-size regime is left
uncovered'' and its blanket $0.83$.  The
$\bar{n}{=}512$ switched cell of Table~\ref{tab:densedgemm} remains
schedule-conditional, and the per-record engine and enumerator that
apply these predicates to every measured trace record (and recompute
the companion's application composites accordingly) are supplied with this
submission as part of the companion's reproduction artifact (arXiv
ancillary files); an archival Zenodo DOI and commit hash will be added
in the first arXiv revision.

\paragraph{A delivered-throughput estimate for emulated DGEMM, and
the published Rubin figure.}
The crossover admits a throughput estimate---a \emph{conditional
upper envelope}, neglecting plane traffic outside its mode's validity
(previous paragraph), launch overhead, delivered tile efficiency, and
tensor-service contention---and the estimate
makes a sharp, testable statement about NVIDIA's own published number.
With conversion overlapped behind the MMA stream (the best software
can do), the envelope rate of an emulated DGEMM with outer-dimension
harmonic mean $\bar{n} = 2mn/(m{+}n)$ is
\begin{equation}
P_{\text{DGEMM}}(\bar{n}) \;=\;
\min\!\left(\ \frac{P_{\fp{8}}}{3r{+}1},\ \
\frac{\bar{n}\,P_{\text{int}}}{c_q\, r}\ \right);
\label{eq:pdense}
\end{equation}
with conversion fully serialised (no overlap) the delivered rate is
instead the ceiling divided by $(1 + n^{*}/\bar{n})$, and real kernels
land between the two brackets.  Table~\ref{tab:densedgemm} evaluates
the estimate on Rubin, and two readings follow.

\begin{table}[h]
\caption{Convert-once \emph{upper envelope} for emulated DGEMM on Rubin from
Eq.~\eqref{eq:pdense} (overlapped bracket; TFLOPS; $r{=}12$,
$P_{\fp{8}}{=}17.5$~PFLOPS, $P_{\text{int}}{=}75$~\tops{}, ceiling
$473$).  $\bar{n}$ is the harmonic mean of the two outer dimensions.
\emph{Read the software rows (S0/L1/L2/switched) as an upper bound, not a
delivered rate for large square:} the bold $473$ at large $\bar n$ is the
convert-once envelope, reached only within a single cluster's reach
($\bar n{\le}256$, i.e.\ a ${\le}256^2$ output) or, to within a mild clip, for
tall/skinny shapes ($\lambda_{\mathrm{eff}}$ bounded at $O(1)$: $1.25$--$1.5$
under square cluster dispatch, $1.06$--$1.25$ shape-matched); a large
\emph{square} output is re-split
on the fly and \emph{delivered at the deconstruction-$\lambda$ floor}---block~(b),
${\approx}235$~TFLOPS best-route ($0.50$ of the $473$-TFLOPS arithmetic roof;
per-\emph{rung} S0/L2 floors are $99/182$).  The table is split into two
blocks precisely because $P_{\text{dec-only}}$ and $P_{\text{svc}}$ are
different functions of different arguments; no row mixes them.  Option~B and Option~C are the
co-design that removes the floor, so their $473$ \emph{is} delivered at every
size; bold marks where a rate meets the $473$ ceiling.  Rows use the
companion's
ledger-projected conservative counts~\cite{matsuoka2026ozaki25};
$^{\ast}$the switched row dispatches among the codesigned modulus
sets and rides set-specific roofs ($438$ at $\bar{n}{=}512$ is the
hybrid's roof, not the published set's $473$).  For calibration, the
serialised bracket at $c_q{=}16$ reads $141/217/297/365$~TFLOPS at
$\bar{n}{=}512/1024/2048/4096$.  All rows are ideal-overlap
envelopes assuming independent
FP8/\inteight{} tensor service and an eligible storage schedule
(see text); under \inteight-tensor serialisation the dispatch holds
the \texttt{dp4a} family ($352$ at $\bar{n}{=}512$, $176$ at
$256$), under additional SIMT--\fp{8} serialisation the endpoints
fall to ${\approx}183/120$ against a similarly serialised shipping
baseline of ${\approx}141/83$ (consistent-worlds floor
${\approx}1.3$--$1.5\times$), and the $\bar{n}{=}512$ switched cell is
schedule-conditional.  All entries are model projections;
the size sweep that tests them is a Part~3 measurement.}
\label{tab:densedgemm}
\centering
\small
\setlength{\tabcolsep}{5pt}
\begin{tabular}{lc|ccccc}
\toprule
 & {\footnotesize modeled} & \multicolumn{4}{c}{convert-once upper envelope$^{\dagger}$} & ceiling from \\
\cmidrule(lr){2-2}\cmidrule(lr){3-6}
rung & 256 & 512 & 1024 & 2048 & 4096 & $\bar{n}\ \ge$ \\
\midrule
\multicolumn{7}{l}{\emph{(a)} $P_{\text{dec-only}}(\bar n)$: deconstruction-limited
  envelope; $k{\to}\infty$, \textbf{no} reconstruction charge$^{\dagger}$}\\
S0: memo-counted $c_q{=}16$ & 100 & 200 & 400 & \textbf{473} & \textbf{473} & 1211 \\
L1: direct \texttt{dp4a}, $c_q{\approx}10.3$ & 155 & 310 & \textbf{473} & \textbf{473} & \textbf{473} & 782 \\
L2: two-limb tensor-migr.\ (capacity-aware) & 182 & 364 & \textbf{473} & \textbf{473} & \textbf{473} & 666 \\
Switched codesign dispatch (A/E/S)$^{\ast}$ & 243 & 438 & \textbf{473} & \textbf{473} & \textbf{473} & --- \\
\midrule
\multicolumn{7}{l}{\emph{(b)} $P_{\text{svc}}(\bar n,\bar n,k)$: reconstruction-aware
  service rate at \emph{finite} $k$, large \emph{square} output$^{\ddagger}$}\\
\quad deep-$k$ slab, $k{=}4096$ & 235 & 235 & 235 & 235 & 235 & --- \\
\quad cubic, $k{=}\bar n$ & 168 & 202 & 234 & 235 & 235 & --- \\
\midrule
Option B: $c_q{\to}\sim\!1$ (co-design) & \textbf{473} & \textbf{473} & \textbf{473} & \textbf{473} & \textbf{473} & 76 \\
Option C: stream-side conv.\ (co-design) & \multicolumn{5}{c}{\textbf{473} at every size} & --- \\
\bottomrule
\end{tabular}
\\[2pt]
{\footnotesize \textbf{Blocks (a) and (b) are two different functions and must
not be read across.}  $^{\dagger}$Block~(a) is the convert-once envelope
$P_{\text{dec-only}}(\bar n)$: it charges deconstruction only, takes
$k{\to}\infty$ so the Garner/CRT reconstruction amortises to zero, and is
therefore an \emph{upper bound}, not a rate any finite problem delivers.  Its
entries are attained as written only up to a single cluster's reach
($\bar n{\le}256$) or for tall/skinny shapes.  In particular the $243$ at
$\bar n{=}256$ is an envelope value (route~A at its own roof), \emph{not} a
delivered rate---the corresponding delivered numbers are the $235$/$168$ of
block~(b), and an earlier version of this table carried $243$ in the delivered
row, which conflated the two functions.
\ $^{\ddagger}$Block~(b) is $P_{\text{svc}}(m,n,k)$ at $m{=}n{=}\bar n$: the
rate a large \emph{square} output actually delivers under on-the-fly
re-splitting ($\lambda_{\mathrm{eff}}{=}$edge$/256$) \emph{with} the finite-$k$
reconstruction charge included---the deconstruction-$\lambda$ floor,
${\approx}235$~TFLOPS best-route ($k{=}4096$; per-\emph{rung} S0/L2 floors
$99/182$, i.e.\ $0.50$ of the $473$-TFLOPS arithmetic roof at best route).
The two $k$ rows bracket the shape dependence: at $k{=}\bar n$ the
reconstruction is amortised over only $\bar n$ steps, which costs $29\%$ at
$\bar n{=}256$ and vanishes by $\bar n{=}2048$.  Removed by Option~B/C.}
\end{table}

\emph{First, the software mitigation ladder transfers to dense GEMM
essentially for free.}  The migrated stage-3 reduction, executed as
the exact two-limb pair of L2, costs $16r \approx 192$ useful
\inteight{}-MACs per operand element against the $37\bar{n}$ \fp{8}
ops per element that the main MMA stream performs---a tensor tax of
$192/(37\bar{n})$, about $2\%$ at $\bar{n}{=}256$ and falling as
$1/\bar{n}$---and runs on the residual \inteight{} tensor pipe
(\S\ref{sec:int8vsfp8}).  Whether that work contends with the \fp{8}
stream is an open microarchitectural question: advertised
per-precision rates do not establish independent execution resources,
so the tensor-migrated values below are stated at the
independent-service endpoint, with the fully serialised endpoint
$P_{\text{ser}} = (1/P_{F} + 1/P_{I})^{-1}$ as the other bracket
(${\approx}206$--$228$~TFLOPS at $\bar{n}{=}512$ on Rubin).  The
dispatch also holds the pure-\texttt{dp4a} routes, which are
\emph{INT8-tensor-contention-free} (they use no tensor \inteight{}
capacity), though not unconditionally contention-immune: they still
assume SIMT--\fp{8} overlap, parameterised
$\rho_{\text{simt,fp8}} \in [0,1]$, whose fully serialised endpoint
is $(1/352 + 1/380)^{-1} \approx 183$~TFLOPS for the all-byte route
and ${\approx}187$ for L1 at $\bar{n}{=}512$.  The like-for-like
comparison must then serialise the shipping baseline too---S0's own
conversion runs on the SIMT pipe under the same overlap assumption,
so a consistently serialised S0 delivers
$(1/200 + 1/473)^{-1} \approx 141$~TFLOPS at $\bar{n}{=}512$
(${\approx}83$ at $256$)---an \emph{equal-overlap-degradation
sensitivity} (not an unconditional floor: it conditions on both sides
sharing one overlap coefficient) of ${\approx}1.30\times$ ($512$) to
$1.46\times$ ($256$).  Its three ingredients stay separate: the route
endpoints (ideal $352/310$, serialised $183/187$), the announced
$200$ of unknown size/algorithm provenance, and the matched-degradation
ratio just computed; comparing serialised routes against the announced
$200$ mixes worlds and is not claimed.
Structurally, SIMT--tensor overlap is the weaker of the two
concurrency assumptions to doubt (a tensor MMA occupies an issue slot
once per many tensor-busy cycles), but it is an argument, not a
measurement; the paired-rate experiments join the Part-3 plan.  On
GB300 the \texttt{dp4a} route reaches
the full $135$-TFLOPS ceiling from $\bar{n}\!\approx\!223$, so the
GB300 conclusions are INT8-tensor-contention-free with the same
SIMT-overlap caveat.  The residual \inteight{} rate
is in any case a real capacity, and it is what
sets the effective knees above~\cite{matsuoka2026ozaki25}.  The
stage-2 marshaling runs at element throughput
$P_{\fp{8}}/(37\bar{n})$, roughly a third of the full-stream rate at
$\bar{n}\!\approx\!500$, comfortably inside the SMEM budget that
Appendix~\ref{app:beta-blackwell} shows only the fully streamed case
exceeding.  The software rungs therefore apply to dense MM unchanged,
and in the conversion-bound region modeled throughput scales as
$1/c_q$: \emph{software alone is worth ${\approx}2.4\times$
uniformly across the width-bounded band under the switched codesign
dispatch, and a conditional $1.8$--$2.2\times$ at the announced
operating region}---worth ${\approx}1.76\times$ under
INT8-tensor serialisation alone (the \texttt{dp4a} bracket
$[352, 438]$ at $\bar{n}{=}512$, $[176, 243]$ at $256$), with a
consistent-worlds floor of ${\approx}1.30$--$1.46\times$ if even
SIMT--\fp{8} overlap is denied to both sides.

\emph{Second, the published figure sits exactly where the fourth term
predicts a deconstruction-limited branch.}  Reading Eq.~\eqref{eq:pdense} at the reference
$c_q{=}16$: the delivered rate is $200$~TFLOPS at $\bar{n}{=}512$
overlapped, and $200$~TFLOPS at $\bar{n}\!\approx\!900$
serialised---NVIDIA's published $\sim\!200$-TFLOPS Emulated-DGEMM
figure~\cite{nvidia_rubin_blog} is reproduced by the deconstruction
term \emph{alone} for benchmark sizes anywhere in the
$\bar{n}\!\approx\!500$--$900$ window, with the \fp{8} tensor
pipes idle $58\%$ of the time.  We state this as a hypothesis, not a
finding.  The same figure would be explained by $\bar{n}$-independent
artifacts (delivered tile efficiency, sustained-versus-boost clocks)
if it was benchmarked at large $\bar{n}$, and the B200 anchor's
similar delivered-to-ceiling ratio (${\approx}49$--$53\%$ against its
$281$-TFLOPS INT8 ceiling, \S\ref{sec:model}), on a path whose own knee sits much
nearer, cautions against over-attribution.  But the two explanations
carry orthogonal signatures: the fourth term predicts a delivered rate
\emph{linear in $\bar{n}$} up to a knee at $n^{*}$ that moves with
$c_q$, while efficiency artifacts are flat in $\bar{n}$.  A single
DGEMM size sweep on Rubin silicon therefore decides the question,
fits $c_q$ from the knee position, and prices the remedy in the same
experiment---the cheapest, highest-leverage entry on the Part~3
validation map.  If the knee is confirmed, the consequence is
immediate and, to our knowledge, unremarked: the switched codesign
dispatch, with conversion overlapped, lifts official Emulated DGEMM
from $200$ to $438$~TFLOPS at $\bar{n}{=}512$ (a conditional
$2.2\times$, software only, no hardware change; the two-limb route
alone reaches $364$, and the dispatch delivers $243$ at
$\bar{n}{=}256$, $2.4\times$ over S0), and Options~B/C hold the full
ceiling down to $\bar{n}\!\approx\!76$ and to every size
respectively---making the deconstruction path, not the tensor cores,
the pivotal lever on official dense emulated-\fp{64} performance as
well as on the streamed kernels.  The full development---exact
two-limb kernels and their proof obligations, the codesigned modulus
sets, the capacity-aware dispatch on GB300 and Rubin, and the
falsifiable validation campaign---is the subject of the companion
paper~\cite{matsuoka2026ozaki25}, which also grounds the shape
claims empirically: measured call-shape traces of six instrumented
configurations show LOBPCG at block width $64$ placing $100\%$ of
its GEMM work at $\bar{n} \le 256$, netlib blocked QR placing half
its BLAS-3 work at $\bar{n} \approx 64$, and multifrontal sparse-LU
fronts sitting at median $\bar{n} \approx 670$---squarely on the
hypothesised branch.

\paragraph{Sparse variations of the Ozaki-2.5 discipline.}
Three of the companion's findings port directly to the sparse and
streamed workloads that are this paper's subject, and one variation
is specific to them.  \emph{(i)~Operand-stationary sparse
operators.}  For iterative solvers with a fixed sparse $A$ (CG,
GMRES, block-Krylov, LOBPCG, iterative refinement), the convert-once
discipline (O1 of~\cite{matsuoka2026ozaki25}) can be applied to the
\emph{value array}: deconstruct $A$'s nonzeros once at setup into
stored residue planes.  Stated honestly, this is a
\emph{bandwidth-for-instructions trade, not a free amortisation}: the
planes are re-read every iteration, changing the dominant value+index
stream from ${\approx}12$ to ${\approx}34$~B/nnz for the published
set (${\times}2.83$), which caps delivered throughput near $0.35$ of
the raw-value memory roof.  On GB300, where the fused on-the-fly
residual ($5.8$--$6.3$ per modulus) fits the SIMT budget, fused
conversion is strictly better and operand-stationary planes should
\emph{not} be used; on Rubin the two are comparable ($0.35$ against
the fused $0.36$--$0.39$ of that same memory roof), and the stationary variant earns
its keep only when the plane array is L2-resident (small operators)
or when SIMT issue is contended by other kernel work.  The streamed
vector data---fresh $x$ and $y$ panels each iteration---is charged
identically in both variants, at the corrected residuals
($6.3/5.8$ inline; $5.3$ under the all-byte codesign, whose
$\alpha'$ surcharge is free in the memory-bound regime; $4.3$
lazy).  \emph{(ii)~The dense fraction of sparse solvers.}
Supernodal and frontal update GEMMs, block-Krylov tall-skinny
products, and factorisation panels are precisely the
width-bounded class the companion's switched dispatch recovers, and
its traces show these are the shapes real sparse codes issue: the
schedule-filtered, memory-roof-capped projections there (each record
assigned an eligible storage schedule and capped by its own memory
roof, with the shipping baseline capped identically) give
GEMM-portion composites of
${\approx}1.9\times$ (LOBPCG, $b{=}64$), ${\approx}1.6\times$
(blocked QR panels), and ${\approx}1.2\times$ (multifrontal LU) on
Rubin versus the shipping path at the conservative reconstruction cost
(the companion's Table~11; ${\approx}2.4\times$, $2.1\times$ and
$1.2\times$ at its ${\approx}r$ codesign target)---the multifrontal
figure down from an earlier unfiltered $2.1\times$ because its many
small fronts are bandwidth-bound, where emulated and shipping paths
coincide---gains
a sparse direct or eigensolver
library collects by dispatching on $\bar{n}$ at the BLAS boundary,
with no change to the solver.  \emph{(iii)~SpMV/SpMM proper.}  For
sparse-times-dense-block products the dense operand's planes convert
once per block application and the stationary values once per setup,
so the binding conversion is again the streamed-vector share---the
regime Eq.~\eqref{eq:chi} already covers; the Ozaki-2.5 reduction
machinery (two-limb \inteight{} GEMMs or \texttt{dp4a}) simply
replaces the scalar stage-3 sequence inside the fused kernels of
\S\ref{sec:impl} at the audited counts.  \emph{(iv)~The
sparse-specific codesign dividend.}  Because streamed kernels are
memory-bound, the all-byte set's $24\%$ pass-count surcharge costs
nothing, and in its \texttt{dp4a} realisation ($c_q \approx 7.3$ per
modulus, no unverified terms: $109$ instructions per element at
$r'{=}15$ against the published set's $124$) and its lazy variant
($49$ against $52$) it lowers the one constant that binds---so for
sparse kernels dispatched to those realisations the codesigned set is
the better default.  The inline tensor-migrated residual, by
contrast, still favours the published set ($70$--$76$ against $79$
per element), so the choice is realisation-dependent rather than
uniform; confirming this ranking on a fused SpMV/SpMM
kernel joins the Part-3 validation map.

\subsubsection{The Combined Projection, and the Part-3 Validation Map}
\label{sec:step7}

Finally we assemble the audited ladder (\S\ref{sec:cqladder}), the
sensitivity (\S\ref{sec:step5}), and
the hardware proposal (\S\ref{sec:hwproposal}) into a projected
delivered-performance picture
for the whole surveyed spectrum.  Table~\ref{tab:combined} traces four
cumulative stages: S0, today's memo-counted constants; S1, the
software-only combined pipeline, defined to include the full ladder
L2{+}L3{+}L4---exact two-limb tensor-side deconstruction, lazy
reduction with its $\alpha'\!\approx\!73$ tensor charge (benign for
memory-bound kernels: tensor keep-up remains a few percent of peak),
and Kulisch-carrier reconstruction---for a SIMT residual
$c_q^{\text{res}}\!\approx\!4.3$ at the companion's audited
counts~\cite{matsuoka2026ozaki25} (without L3 the residual is
$6.3/5.8$, which sits at B300's budget but delivers only
$0.36$--$0.39$ of the Rubin \emph{memory} roof; the all-byte codesign restores
$5.3$ inline and ${\approx}3.3$ with lazy reduction \emph{per
modulus}, at its own $r'{=}15$---per-element $49$ against the
published set's lazy $52$, a ${\approx}6\%$ edge---its larger
$\alpha$ surcharge again benign here); S2, S1 plus a $2\times$
effective integer budget (Option~A-class); S3, the deconstruction
datapath (Option~B/C).

\begin{table}[h]
\caption{Projected fraction of the memory roof delivered across the
kernel spectrum under the combined pipeline, by stage (B300~/~Rubin
emulated path; $r{=}12$).  S0 = memo-counted constants today; S1 =
software-only combined pipeline (the streamed-row range spans the
published set at the audited lazy residual $4.3$, $0.53$ of that memory
roof, to
the all-byte codesign's lazy $3.3$ at its own $r'{=}15$, $0.56$ of the
same~\cite{matsuoka2026ozaki25}); S2 = S1 with
$2\times$ effective integer budget (Option~A); S3 = the
deconstruction datapath, shown as one column because Options B and C
share the modeled roof, though their conditions differ: S3B (the
\texttt{cvt} instruction) retains ${\approx}1$ conversion issue per
value, S3C (stream-side conversion) assumes the provisioned
converter of \S\ref{sec:hwproposal}.  On Rubin, native \fp{64} (33~TFLOPS) already holds the
roof for $\mathrm{OI}<1.5$, so the emulated fractions there matter for
the \fp{64}-less design question rather than for Rubin itself.  BLAS-1
under the Kulisch route is \emph{exact and bitwise-reproducible} at
the roof, versus compensated at S0.  The FFT row reports the
companion's final analysis~\cite{matsuoka2026fft}: its binding term is
a per-output integer epilogue outside this table's streamed-kernel
model, no software stage reaches its roof, and the roof is reached by
the companion's own H1--H3 co-design ladder, whose H2 load-path
datapath is this table's S3C converter.
All S1--S3 entries are unmeasured model projections---the Part~3
scenario ladder maps to these columns stage by stage.  For the
streamed kernels of this table both tensor demands---the Ozaki \fp{8}
MMAs and the \inteight{} reduction GEMMs---are a few percent of their
pipes' peaks, so the FP8/INT8 concurrency bracket of
\S\ref{sec:hwproposal} is immaterial here; it binds only the dense
routes.}
\label{tab:combined}
\centering
\footnotesize
\setlength{\tabcolsep}{3pt}
\begin{tabular}{lcccccc}
\toprule
\textbf{kernel class} & OI & $\varphi$ & \textbf{S0} & \textbf{S1} & \textbf{S2} & \textbf{S3} \\
\midrule
SpMV / GEMV (any $B$), streamed & 0.2--2 & 1 & 0.39 / 0.14 & 1.0 / 0.53--0.56 & 1.0 / 1.0 & 1.0 / 1.0 \\
7-pt stencil (residue reuse)    & 0.5 & $\sim$0.5 & 0.78 / 0.28 & 1.0 / 1.0 & 1.0 / 1.0 & 1.0 / 1.0 \\
BLAS-1 (\texttt{ddot}, \texttt{dnrm2}) & 0.125--0.25 & 1 & 1.0$^{\dagger}$ / 1.0$^{\dagger}$ & 1.0$^{\ast}$ / 1.0$^{\ast}$ & 1.0$^{\ast}$ / 1.0$^{\ast}$ & 1.0$^{\ast}$ / 1.0$^{\ast}$ \\
3-D FFT$^{\P}$                  & --- & --- & 0.15--0.20 / --- & 0.17--0.22 / --- & \multicolumn{2}{c}{roof via H1--H3} \\
Dense GEMM                      & $\ge$50 & $\to$0 & \multicolumn{4}{c}{ceiling-bound: 135 / 473 TFLOPS$^{\S}$} \\
\bottomrule
\end{tabular}
\\[2pt]
{\scriptsize $^{\dagger}$~at the roof via \fp{32}+Kahan
(compensated).  $^{\P}$~the spectral companion's numbers; S2/S3 reach its roof only
via its own co-design ladder, $0.86$--$1.0$ of roof
(\cite{matsuoka2026fft}; its H2 datapath is this table's S3C
converter), not an S-stage of this paper.  $^{\ast}$~at the roof via the Kulisch carriers, \emph{exact}
and bitwise-reproducible.  $^{\S}$~for outer dimensions above the
crossover of Eq.~\eqref{eq:densecrossover} at that stage's $c_q$
(for dense GEMM the L3 lazy-$\alpha$ trade is excluded, so S1 uses
the L2 two-limb residual, $6.3/5.8$; \S\ref{sec:hwproposal}).  This is
the convert-once envelope: a large \emph{square} output at S0--S2 is re-split
on the fly and clipped to the deconstruction-$\lambda$ floor (Rubin
${\approx}235$~TFLOPS, $0.50$ of the $473$-TFLOPS arithmetic roof;
Table~\ref{tab:densedgemm}); only
tall/skinny shapes reach the ceiling above the crossover, and only
Option~C (S3) attains it at every size.}
\end{table}

In absolute terms the S1 column reads, on B300: SpMV $1.6$~TFLOPS
($1.23\times$ native), stencil $4.0$ ($3.1\times$), batched GEMV
$B{=}8$ $16$ ($12.3\times$), BLAS-1 $2.0$ exact, dense GEMM $135$,
FFT $58$--$78$~ms against its $12.9$-ms memory roof for $1024^3$ (the
companion's software-plus-H0 endpoint, $1.3$--$1.9\times$ B300's
collapsed native path at peak issue)---i.e.\ \emph{the
software-only combined pipeline restores every fused-bound claim of
Table~\ref{tab:speedups} on B300 except the FFT's}, collapsing the
conversion-cap column
back into the ideal column while upgrading the reduction kernels from
compensated to exact; the FFT's remaining $4.5$--$6.1\times$ is the
companion's integer-epilogue wall, removed only by its co-design
ladder.  On Rubin the same software stage delivers
$0.53$--$0.56$ of the \emph{memory} roof for fully streamed kernels (published set
to all-byte codesign at its own $r'{=}15$; moot for Rubin itself,
whose native pipe holds that region, but decisive for the
\fp{64}-less design question), and Option~A---or the Option~B/C
datapath---completes the spectrum at $22$-TB/s bandwidth with
$2.75\times$ B300's absolute throughput everywhere.  The projection to
hold in mind is the top-right corner: at S2/S3 on Rubin-class parts,
\emph{every} surveyed kernel class runs at its memory roof at full
\fp{64} accuracy---with two bounded exceptions: large dense square
DGEMM, capped at a cluster-aligned deconstruction-$\lambda$ floor at
${\approx}0.50$ of the $473$-TFLOPS \emph{arithmetic} roof (a compute
limit, and note the change of denominator: this fraction is not of the
memory roof the rest of this sentence is about; L2-insensitive;
Fig.~\ref{fig:dgemm}); and the 3-D FFT, which additionally needs the
companion's H1 integer-tensor rung before the S3C datapath
pays~\cite{matsuoka2026fft}---with exact
reproducible reductions, on a chip whose
only \fp{64}-specific silicon is a small vector pipe and (at S3) a
fixed-function residue block on the copy path.  That corner is the
quantitative form of this paper's title, and every cell on the path to
it is a falsifiable Part~3 measurement: S0 fixes the constants ($c_q$
derby including the \texttt{DP4A} variant; pipe census for
$P^{\text{eff}}_{\text{int}}$ and issue limits), S1 builds and
measures the ladder, S2 is decided by the census, and S3 is priced by
the measured S1--S2 gap in the Part-3 area/energy study.

Whether the system-level tradeoffs ultimately favour a single
converged \fp{8}-centric architecture over separate \fp{64}-preserving
SKUs---accounting for delivered emulation efficiency, the
register-pressure and preprocessing costs of \S\ref{sec:impl}, occupancy
and latency, and the silicon-area and code-porting economics---is exactly
the question the validation programme of \S\ref{sec:futurework} is meant
to settle empirically.  The present paper makes the analytic case that the
convergence is achievable; building and measuring it is the next step.

\section{Composite Kernels Reduce to the Primitive (Layer L3)}
\label{sec:l3}

The previous section established that the basic numerical kernels---the
dwarfs of layer L2---reduce to the FP8 primitive and run at the memory
roof on B300 at the fused bound, with the conversion-bound tail routed
to the surviving native pipe at reference constants.  The thesis of \S\ref{sec:dwarfs}, however, is stronger:
it asserts that the \emph{composite} kernels of layer L3, which form the
inner loops of real applications, inherit the same property by
composition.  This section substantiates L3 directly.  Because every L3
kernel is a composition of L2 dwarfs, and each dwarf reduces to the FP8
matrix op, an L3 kernel can fail to reach the memory roof only if its
particular composition introduces a \emph{new} compute-bound bottleneck
that is not itself one of the recovered dwarfs.  We therefore audit the
composite and solver kernels that dominate production scientific codes
and ask, for each, whether any such residual bottleneck survives.

\paragraph{This is a coverage argument organised by the Berkeley dwarf taxonomy.}  We stress at the
outset that the kernels examined below are not an illustrative selection
but a \emph{systematic enumeration} anchored to the Berkeley dwarf
taxonomy~\cite{asanovic2006berkeley,colella2004dwarfs}.  The original
dwarfs partition scientific computing into a small, heuristic spanning
set (with explicitly enumerated exclusions: native-\fp{64} fallback,
\fp{32}/Kulisch reductions, and the unvalidated codesign routes) of
computation/communication classes; an argument that ranges over all of
them, plus the composite solver kernels that combine them in production
codes, is therefore an argument about the \emph{surveyed classes} of scientific
computing to first order, not about a fortunate subset.  Table~\ref{tab:coverage}
makes the mapping explicit: each dwarf class is named, paired with the
concrete kernel(s) through which we examine it, the L2 primitive its work
reduces to, and the verdict.  The seven composite-kernel categories
(a)--(g) that follow then cover the ways these dwarfs are combined into
the inner loops of real applications---iterative solvers, dense and sparse
direct factorisations, lattice field theory, and stochastic methods---so
that the audit is organised, by the Berkeley dwarf taxonomy, over both
the basic classes and their dominant compositions.  Two worked exemplars, tracing a composite kernel
all the way down the L3$\rightarrow$L0 chain, are given in
Appendix~\ref{app:l3-exemplars}.

\begin{table}[t]
\centering
\small
\caption{Coverage of the Berkeley dwarf classes and their compositions.
Each computational dwarf is examined through a concrete kernel whose work
reduces to an L2 primitive (dense GEMM, SpMV/SpMM, FFT, or stencil) and
thence, via Ozaki~II or the Kulisch FFT route, to the L0 FP8 matrix op.
The right column gives the recovery verdict on B300 at full \fp{64}
accuracy \emph{at the fused, conversion-ideal bound}; at the reference
deconstruction constants, sub-$\mathrm{OI}^{*}$ kernels (notably
single-RHS SpMV) route to the surviving native pipe instead
(\S\ref{sec:eval}), and panel-type GEMMs below the dense crossover of
Eq.~\eqref{eq:densecrossover} await an engineered $c_q$ or the
hardware options of \S\ref{sec:hwproposal}.  ``\fp{32}+K'' denotes the FP32-with-Kahan path for BLAS-1
reductions.  The enumeration is organised by the dwarf taxonomy as a
coverage framework (with the exclusions noted in the text), not a
closed partition.}
\label{tab:coverage}
\setlength{\tabcolsep}{4pt}
\footnotesize   
\begin{tabular}{@{}llll@{}}
\toprule
\textbf{Dwarf class} & \textbf{Examined via} & \textbf{Reduces to} & \textbf{B300 verdict} \\
\midrule
Dense linear algebra   & GEMM, LU/\texttt{trsm} panel   & dense GEMM (L2)        & emul.\ ceiling, $104\times$ (\S\ref{sec:eval}) \\
Sparse linear algebra  & SpMV, SpMM, sparse-direct      & SpMV/SpMM (L2)         & memory roof (App.~\ref{app:sparse}) \\
Spectral methods       & 3-D FFT                        & FFT (L2)~\cite{matsuoka2026fft} & epilogue wall; roof w/ ladder \\
Structured grids       & stencil sweeps                 & stencil/im2col (L2)    & memory roof (\S\ref{sec:eval}) \\
Unstructured grids     & FEM SpMV, dual-mesh            & SpMV/SpMM (L2)         & memory roof / corner (\S\ref{sec:falsification}) \\
Dense reductions (BLAS-1) & \texttt{ddot}, \texttt{dnrm2} & \fp{32}+K            & not binding ($<\!5\%$) \\
Structured stencil (QCD) & Dirac inverter               & stencil/GEMM (L2)      & memory roof (\fp{32} inner) \\
Monte Carlo / particle & FCI-QMC, beam dynamics         & --- (algorithmic)      & corner (\S\ref{sec:falsification}) \\
\bottomrule
\end{tabular}
\end{table}

Tables~\ref{tab:speedups} and~\ref{tab:h100baseline} establish that
the four basic kernels of L2---dense GEMM,
batched GEMV, structured stencils, and SpMV---are either restored to
their memory-roof performance on B300 through Ozaki~II, were
already memory-bound on a healthy native pipe (in which case
emulation costs nothing), or---for the conversion-bound tail at
reference deconstruction constants---run within a bounded distance of
the roof on the surviving native pipe, with full recovery contingent on
the engineered $c_q$ (\S\ref{sec:eval}).  The companion analysis on spectral
workloads~\cite{matsuoka2026fft} treats the fifth case, the
three-dimensional FFT, via an exact Kulisch fixed-point Phase~B
route~\cite{kulisch1977,kulisch_miranker} that runs the per-output
forward-CRT reduction on B300's surviving integer SIMT pipe; its
final per-instruction-class floors put that integer-SIMT epilogue at
$25$--$35\,B_{\text{mem}}$, which B300's $5.2\,B_{\text{mem}}$ pipe
misses by $4.9$--$6.7\times$, while the tensor-class floors (FP8-TC
$444\,B_{\text{mem}}$, FP16-TC $96\,B_{\text{mem}}$) are both met
(full derivation and the roof-recovering co-design ladder
in~\cite{matsuoka2026fft}; summary in \S\ref{sec:h100baseline}).
This leaves the natural question: \emph{are there any other kernels
that would still make a typical scientific simulation compute-bound on
B300?}  We enumerate the candidates by category.

\paragraph{(a) Reductions and dot products.}  BLAS-1
inner-product kernels (\texttt{ddot}, \texttt{dnrm2}, CG residuals,
Krylov orthogonalisation) have per-kernel operational intensities of
$1/8$ (\texttt{ddot}: $2$ flops per $16$ streamed bytes), $1/4$
(\texttt{dnrm2}: one stream), and $1/12$ (\texttt{daxpy}-class
updates with the write), and typically
constitute $<\!5\%$ of CG/GMRES/BiCGStab time-to-solution; the
dominant cost is the SpMV plus preconditioner, both addressed by
Ozaki~II.  When present, the dot products themselves can be executed
in \fp{32} with Kahan compensation~\cite{kahan1965}---a Grade-B
compensated contract, whose adequacy is a solver-level property
(convergence rate, attainable residual, iteration count) to be
confirmed per solver in the validation suite rather than asserted;
explicitly re-orthogonalising methods tolerate it most readily, and
the Kulisch route of \S\ref{sec:cqladder} upgrades the reductions to
exact where it does not.  B300's \fp{32} vector pipe at
$\sim\!60$--$80$~TFLOPS is well above the BLAS-1 memory-roof
requirement of at most $0.25 \cdot 8 = 2$~TFLOPS ($1$~TFLOPS for
\texttt{ddot}).  Not binding.

\paragraph{(b) FFT.}  Handled in detail by the companion
paper~\cite{matsuoka2026fft} and summarised above: B300 fails the
native FP64 floor by ${\sim}9\times$, and the emulated route---a
full-\fp{64} $1024^3$ FFT with no \fp{64} arithmetic anywhere in its
kernel---projects $63$--$87$~ms in software against the $12.9$~ms
memory roof at full \fp{64} mantissa width: $1.3$--$1.9\times$ the
$116$~ms collapsed native path at peak integer issue (a wash at
realised issue), and $4.9$--$6.7\times$ off the roof, walled by a
per-output integer-SIMT epilogue that no evaluated substrate
reassignment removes.  The wall's floor is closed-form and names its
own escape: the companion's co-design ladder---minor rungs (restore
B200's INT8 tensor core with positional accumulation; add the load-path
deconstruction datapath of \S\ref{sec:hwproposal}'s Option~C and
multiword-carry ISA idioms) to $16.0$--$23.5$~ms, and one moderate rung
(modular reduction at the MMA output) to $12.9$--$15.0$~ms---the roof
at the optimistic end of the epilogue interval.  The projection has not been
measured, and building the kernel and its epilogue instruction ledger
is a principal item in \S\ref{sec:futurework}.

\paragraph{(c) Triangular solve and LU panel factorisation.}  The
inner kernels of dense direct solvers (\texttt{dtrsm}, \texttt{dgetrf}
panel) are sequential along one dimension, with operational intensity
$\sim\!2$--$5$ for typical panel widths.  On B300 the panel falls
into the compute-bound regime in native \fp{64}, but the surrounding
\texttt{dgemm} update is $O(n^3)$ and dominates by a factor
proportional to the matrix size; Ozaki~II handles the outer update at
full speed, leaving the panel as a sub-percent residual.  This is
the \emph{HPL-pattern} that NVIDIA's cuBLAS Ozaki
integration~\cite{nvidia_cublas_emulation} and the 2.3$\times$ B200
HPL speedup~\cite{nvidia_hpc_benchmarks} already demonstrate
empirically.  Not binding at production scale.

\paragraph{(d) Sparse direct solvers.}  MUMPS-class~\cite{mumps}
multifrontal factorisations decompose into many small dense frontal
matrices plus large dense GEMM-like extend-add updates.  The GEMM
portion is recovered by Ozaki~II; the frontal-matrix portion has the
same panel-residual structure as~(c) and the same conclusion.
Provided the maximum frontal matrix is $\geq\!64\!\times\!64$,
tensor-tile granularity on B300 is not binding at the fused bound; at
reference constants the small-frontal regime sits below the dense
crossover of Eq.~\eqref{eq:densecrossover} and is exactly the
batched-small-GEMM case that an engineered $c_q$ or the hardware
options of \S\ref{sec:hwproposal} unlock.  Very ill-conditioned,
deeply nested supernodal
structures may remain problematic and require empirical investigation.

\paragraph{(e) Lattice QCD Dirac inversion.}  The Dirac
operator~\cite{lqcd_review} is a structured stencil with $4{\times}4$
or $3{\times}3$ complex-matrix-valued links, $\mathrm{OI}\!\approx\!1$,
GEMM-decomposable.  Production LQCD codes already use mixed precision
(FP32 inner, FP64 polish) with iterative refinement~\cite{lqcd_mixed_precision};
an FP8-emulated \fp{32}-accuracy Ozaki~II mode (a reduced-$r$
configuration, distinct both from native \fp{32} and from full
\fp{64} emulation, with the accuracy contract stated per grade)
addresses the inner solve at memory-roof
speed, and the \fp{64} polish is BLAS-2 and falls under~(a).

\paragraph{(f) Small-inner-kernel codes (latency-bound).}  A small but
real class of codes---accelerator beam dynamics over $10^6$ turns,
geodynamo simulations, long-time-integration storm-resolving
climate---show drift that mixed precision cannot fully tame and whose
dominant inner kernel may be neither FFT nor GEMM.  These have
historically been read as a B300 counter-example, but on closer
inspection the issue is algorithmic and GPU-agnostic: when the
per-time-step inner kernel is a few hundred sites, the wall-clock
cost is dominated by kernel-launch latency on \emph{any} accelerator
(including H100 native), not by FP64 throughput.  Ozaki~II is not
appropriate here because the kernel is too small for the crossover to
bite, but neither is native FP64 silicon: the binding constraint is
per-step latency, which a future generation would address through
persistent kernels or graph-style execution, not through additional
FP64 silicon.  This residue~\cite{mfp64_whitepaper} is an open problem
for the field, but it is not evidence that B300 is the wrong target
for HPC; the same problem exists on H100, B200, and Rubin.

\paragraph{(g) Stochastic methods (FCI-QMC, AFQMC).}  Population
dynamics with extreme dynamic range that resist low-precision
reformulation~\cite{fci_qmc}.  Active research area; appropriate
mitigation is algorithmic rather than architectural.

\paragraph{Verdict.}  Of the seven composite-kernel categories, (a),
(c), (d), and (e) admit a bandwidth-bound execution path on B300 at
full \fp{64} accuracy via the combination of Ozaki~II (this paper) and
FP32+Kahan compensation; (b) admits a software path---the
Ozaki--Bailey + Kulisch Phase~B FFT~\cite{matsuoka2026fft}---that
beats B300's collapsed native pipe by $1.3$--$1.9\times$ at peak
issue but reaches its memory roof only under the companion's
co-design ladder (minor rungs plus one moderate); (f) is latency-bound and GPU-agnostic (the
binding constraint is kernel-launch latency, not \fp{64} throughput, and
it is present identically on native-\fp{64} parts); (g) is an
algorithmic-research question whose mitigation is likewise architectural-
neutral.  Taken with the dwarf-class coverage of Table~\ref{tab:coverage},
this completes the \emph{surveyed} coverage over the basic
computational classes and their dominant compositions.  It is a
coverage framework with stated exclusions, not a closure theorem:
categories (f) and (g)---and the Monte-Carlo/particle row of
Table~\ref{tab:coverage}, which maps to no matrix primitive---are
carried as explicit, algorithmically-mitigated exclusions, and the
claim is exactly this: \emph{every surveyed dwarf class,
and every composite kernel surveyed, admits a bandwidth-bound execution
path on B300 at Grade-A accuracy for the matrix work (Grade-B compensated
for the BLAS-1 reductions; \S\ref{sec:background})}---subject, for the
sub-$\mathrm{OI}^{*}$ tail, to the deconstruction condition of
\S\ref{sec:model} (with the surviving native pipe as the interim
route), and, for the FFT, to the companion's co-design ladder (minor
rungs plus one moderate; its software path stopping at $1.3$--$1.9\times$ the collapsed
native pipe, $4.9$--$6.7\times$ off the roof)---with the only
residue---(f) and
(g)---being algorithmic rather than a deficiency of \fp{64} silicon and
therefore not curable by adding \fp{64} units.  Because the dwarf taxonomy
is, to first order, the accepted coverage map of scientific computing,
this is the strongest form the claim can take short of measured
demonstration: not that a sample of kernels works, but that every
category the taxonomy names has been priced---a coverage statement, as
\S\ref{sec:l3} is careful to say, not a closure theorem.  Native FP64 silicon is, on this
evidence, not the holy grail it has historically been taken to
be---abundant \fp{8} tensor throughput, combined with the right
algorithmic scaffolding, is sufficient for the HPC workload
spectrum, modulo the engineering effort required to build the Ozaki~II
and Kulisch kernels and, for the FFT, the companion's co-design
ladder.  That effort, as argued in \S\ref{sec:futurework}, is a
well-scoped, tractable implementation programme.  The architectural recommendation that emerges
is the layered codesign rule of~\cite{matsuoka2026fft}: any
post-Rubin GPU intended to serve spectral scientific workloads should
provide either $\eta_{\fp{64}\text{-vec}} \geq 1.56\,B_{\text{mem}}$
(the safe native target, met by Rubin within $4\%$), or, as an
engineered fallback, the route-complete emulated set: FP8-TC
$444\,B_{\text{mem}}$ and FP16-TC $96\,B_{\text{mem}}$ (both met on
B300) \emph{together with} the integer-epilogue floor of
$25$--$35\,B_{\text{mem}}$---which no shipping part meets in
software, and whose recommended form is therefore not a wider scalar
pipe but the companion's co-design ladder (H1--H3), whose load-path
rung is this paper's Option~C.
The tensor floors are comfortably met on current datacentre GPUs---a
direct consequence of NVIDIA's deliberate scale-up of FP8 silicon for
AI workloads---so the binding constraint is the epilogue floor, not
the tensor path.  Cutting both the FP64 and INT32 pipes simultaneously
would close both escape routes; further cuts to FP8 would invalidate
the FP8-is-enough thesis altogether.  To these boundaries this version
adds a third, load-bearing one from \S\ref{sec:int8vsfp8}:
\emph{retain the minor narrow-integer capabilities}---the residual
INT8 tensor rate and the \texttt{DP4A}-class SIMT dot
products---because retaining them asks for no new silicon (they
already ship on every current part) and they carry
the deconstruction and reconstruction side-work of the entire emulated
stack (13--24\% utilisation suffices at current rates), and their
complete removal would push that work onto pipes $16\times$ heavier
than the job requires.  This is the design boundary that future
post-\fp{64} generations should respect.

\section{Discussion}
\label{sec:discussion}

\subsection{The Native-FP64 Alternative}

Not every vendor has taken the emulation path; the principal alternative
is to keep investing in native \fp{64} silicon, a position articulated by
AMD around the MI430X generation~\cite{hpcwire_amd}.  This is a coherent
strategy, and a portfolio in which some hardware provides native \fp{64}
while other hardware leans on the FP8 primitive will be the practical
near-term picture.  The audit of \S\ref{sec:l3}, however, narrows the
technical case for native \fp{64} as a \emph{requirement}.  The classic
argument was that real codes contain non-GEMM hot spots---FFT above all,
plus sort, scan, and atomics---where emulation does not apply; of these
the FFT is now fully analysed by the spectral
companion~\cite{matsuoka2026fft}---no \fp{64} arithmetic anywhere in
its kernel, at worst parity with B300's collapsed native pipe in
software, at the memory roof under minor hardware plus one moderate
addition---and the
integer-dominated kernels need no \fp{64} silicon in the first
place.  What remains for the native-\fp{64} case is procurement
convenience---no FP8 kernels to build---and a long tail of codes not yet
engineered for the primitive, both of which weaken as the FP8 kernel
ecosystem matures (\S\ref{sec:futurework}).

\paragraph{The deconstruction term as a codesign equation.}  The fourth
TME term reframes what the low-intensity tail actually asks of future
silicon, and the answer is stated concretely in
\S\ref{sec:hwproposal}: three
options of increasing commitment---(A)~rebalanced narrow-integer
issue at the \texttt{DP4A} grain, (B)~a
\texttt{cvt.fp64.residues}-class deconstruction instruction whose
per-lane array costs roughly one \fp{64} multiplier, and (C)~a
residue-decompose transform mode on the asynchronous copy path (the
recommended Part~3 codesign target, SIMT $\chi\to0$ provided the
transform sustains $166$--$325$~TMAC/s and the deposit path
$55$--$65$~TB/s)---each sized, compared, and mapped to a measurement in
Table~\ref{tab:hwoptions}, alongside the always-available fallback of
a small surviving native \fp{64} pipe, which is what B300's
$1.3$~TFLOPS already is.  Three strategic points bear repeating at
the discussion level.  First, the sharpened form of this paper's
thesis is not ``FP8 alone'' but \emph{``FP8 plus a narrow
deconstruction path is all you need''}, with the retained slivers of
narrow integer arithmetic carrying the side-work.  Second, none of
the options is specified here as a large native-\fp{64} ALU array.
For Option~C the model derives an aggregate $166$--$325$~TMAC/s
transform requirement; per-engine width, copy-engine count, ports,
area, and power remain Part-3 obligations.  Third, the choice is
generation-dependent in
one direction only: bandwidth is growing faster than SIMT integer
issue, so the per-element budget shrinks ($6.25$ instructions per
modulus on B300, $2.3$ on Rubin-class bandwidth;
\S\ref{sec:cqladder}), and what is optional insurance today graduates to
effectively mandatory for roof-parity streamed emulation on
Rubin-class and later parts.

\subsection{Implications for Science on AI-Class Hardware}

That major scientific-computing programmes have begun to identify Ozaki
emulation as a path to \fp{64} on AI-centric
accelerators~\cite{hpcwire_genesis} indicates that the technique is moving
from research curiosity toward strategic infrastructure.  In the author's
view this is a healthy development: it lets science continue to exploit
the cost and energy efficiency of AI-class hardware without abandoning the
precision discipline that distinguishes simulation from inference.  It
also places a burden on the numerical-software community to verify that
the error bounds hold on \emph{real} application inputs, not just on the
well-conditioned synthetic matrices typical of benchmark suites.  The
deeper trade is that emulation exchanges silicon area---the missing
\fp{64} units---for software complexity, the kernel work of building the
FP8 primitive path.  That complexity is, however, of a highly structured
and repetitive kind, and as \S\ref{sec:futurework} argues it is
increasingly amenable to AI-assisted implementation, which tilts the
silicon-versus-software trade steadily in favour of relying on the
post-\fp{64} stack rather than insisting on native \fp{64} hardware.

\subsection{From Formalisation to Measurement}
\label{sec:futurework}

This paper is, by design, the formalisation half of a two-part
programme: it establishes the theoretical framework and its upper
bound---that the canonical kernel taxonomy reduces to the FP8
primitive without performance sacrifice at the bound
(Proposition~\ref{prop:atomic}).  The complementary half is
empirical---to \emph{build} the
L1--L2 kernels on production silicon and measure where the delivered
constants place each kernel relative to the bound.  We sketch that follow-on programme
here at the level of what must be measured, deferring its execution to
subsequent work.

\paragraph{Validating the reductions.}  As of this version the single
most consequential quantity to measure is the \emph{engineered
deconstruction cost} $c_q$: the reference value ($16$ instructions per
modulus per element) is a property of a shipping library path never
optimised for this purpose, the parity and roof thresholds sit at
$c_q \lesssim 7$--$10$ and $\lesssim 5$--$6$ respectively
(\S\ref{sec:model}), and~\cite{bayraktar2026tme} states the two-sided
decision criterion in exactly these terms.  The candidate routes down are now enumerated with audited counts in
the companion paper~\cite{matsuoka2026ozaki25}---the exact two-limb
tensor reduction ($c_q^{\text{res}} \approx 6.3/5.8$), the direct
\texttt{dp4a} route ($\approx\!10.3$, with an unverified $7.5$ reuse-schedule
target), and the codesigned modulus sets (all-byte $5.3$, its
$\alpha'$ surcharge free for the memory-bound class)---joined by
ESC-adaptive $r$ (well-conditioned inputs need fewer than the
conservative $r{=}15$ moduli, shrinking $c_q r$ and the
packed-residue storage alternative simultaneously); each is a
concrete kernel to build and count, as is the
\emph{pipe census} that determines whether the integer-issue budget
$P_{\text{int}}$ can be supplemented by split issue
(Appendix~\ref{app:beta-blackwell} assumes it can; the assumption is
unverified).  Alongside $c_q$, the central quantity remains the
\emph{delivered} bandwidth multiplier $\beta$: how much of the
fully fused $\beta=1$ ideal survives once the $r$ residue planes and the
integer reconstruction operands compete for the register file at
production tile sizes (\S\ref{sec:impl}).  Closely related are the
delivered compute multiplier $\alpha$ (how much of the $3r{+}1$ FP8 work
is genuinely hidden behind the memory stream), the delivered MMA tile
efficiency at small right-hand-side counts ($W_{\text{mma}}$ at
$B \in \{1,2,4,8\}$), the converted fraction $\varphi$ on real composed
solvers, and, where adaptive
precision is used, the frequency with which an ESC/ADP
scheme~\cite{schwarz2025} must raise the slice count or fall back---a
fallback that, on a part retaining a small native \fp{64} pipe, degrades
gracefully.  Each of these is a measurement against a reference
implementation such as the open-source GEMMul8
library~\cite{gemmul8_github} or the vendor cuBLAS Ozaki
path~\cite{nvidia_cublas_emulation}, on the basic kernels of L2 and then
on representative L3 compositions and full L4 applications running at
\fp{64} accuracy.  The encouraging structural fact is that this kernel
work is repetitive and algorithmically shallow---the translation of a
known pattern ($r$-fold residue decomposition, tensor-core MMA, Garner
reconstruction) onto each kernel's tile shape and register
budget---so the validation programme is well-scoped rather than
open-ended.

\paragraph{Part 3.}  The architecture-specific half of this programme is
already underway and will not wait for a full campaign to conclude: a
\textbf{Part~3} companion paper---its scope organised by the
technical review that isolated the deconstruction term and the
tile-granularity correction~\cite{bayraktar2026tme}---is in preparation
for rapid publication, with substantially expanded experiments and
adversarial scrutiny.  Its scope is exactly the material
this paper defers: measured $c_q$ across candidate conversion-kernel
designs (with SASS/\texttt{ncu}-counted instruction budgets against the
pre-agreed thresholds), the integer-pipe census that fixes today's true
$P^{\text{eff}}_{\text{int}}$, delivered
$W_{\text{mma}}(B)$ including DMMA-versus-DFMA comparisons on
\fp{64}-strong parts, delivered-$\beta$ decompositions for the fused
kernels, $\varphi$ on composed solvers, an HPCG end-to-end study, and
the area/energy evaluation of the three hardware options of
\S\ref{sec:hwproposal} (Table~\ref{tab:hwoptions}).  The experiments are organised as the
S0--S3 scenario ladder of Table~\ref{tab:combined}: first measure the
S0 constants, then build and measure the S1 software pipeline
(byte-plane deconstruction, Kulisch-\fp{8} reconstruction), then
determine by census whether S2's effective-budget doubling already
exists in the silicon, leaving S3 as the codesign recommendation whose
value the measured gap prices.  The experimental platform is GB200-class
hardware, beginning with RIKEN's GB200~NVL4 system (\emph{Rikyu}),
with results recorded in a per-run schema designed so that any
independently built model, should a partner offer one, can be compared
cell by cell.  This paper
deliberately retains the architecture-independent formulation and hands
the constants to Part~3.

\paragraph{Extending the framework.}  Several refinements of the TME model
itself remain open.  First, $\beta$ is treated here as a discipline
question (fused or not), but real kernels live on a continuum: a fuller
model would parameterise $\beta(r,T_k,T_m,T_n,\text{regs})$ explicitly and
predict the optimal tile shape jointly with the moduli count, dropping out
of careful instrumentation of the reference
implementations~\cite{nvidia_cublas_emulation,gemmul8_github}.  Second,
the marriage of Ozaki~II with structured 2:4 sparse tensor cores is
non-trivial: applying the sparsity mask at the residue-plane level may
invalidate the modular reduction of~\eqref{eq:modgemm} because the mask is
data-dependent, and a clean construction is, to our knowledge, open.
Third, the non-GEMM, non-FFT hot spots---integer-dominated sort and scan,
atomic-heavy graph traversals, and stochastic methods with extreme dynamic
range (\S\ref{sec:l3})---fall outside the present kernel set and warrant
their own $(\alpha,\beta,\gamma,c_q)$ treatment.  Finally, the speedups
projected here are largest on \fp{64}-starved parts; quantifying how the
benefit narrows as the silicon mix evolves is itself a question the
framework should answer for future generations.

\section{Falsifiability and the Limits of the Claim}
\label{sec:falsification}

A claim as strong as ``the FP8 matrix op is the dominant matrix-work
substrate for the surveyed classes'' earns
its keep only if it can be wrong, and only if the conditions under which
it would be wrong are stated plainly.  We do so here, and we address in
one place the principal objections the thesis invites, rather than
threading defensiveness through the analysis.  The unifying observation is
that almost every objection reduces to a statement about the bandwidth
multiplier $\beta$, about the deconstruction constant $c_q$, or about a
small, identifiable set of corner cases---and
in all three forms it is a second-stage question of architecture and
engineering layered on top of the bound, not a flaw in the underlying
numerics.  The machinery is symmetric: the same fit of architectural
parameters that would expose a failure also certifies a success, so each
objection below resolves into a two-way test---decidable analytically
where the margins are wide, by measurement where they are not.  The
deconstruction case is also the demonstration that this
section is not rhetorical: the review of~\cite{bayraktar2026tme} used
exactly this machinery to expose a first-order omission in the published
model, and this version incorporates the result
(\S\ref{sec:whyupdate}).

\paragraph{What would falsify the thesis.}  The claim is operational and
admits four clean failure modes, each of which the TME model is built to
expose.  \emph{(i) A missed comparator.}  Failure means missing the
kernel's \emph{stated comparator}, not merely being compute-bound:
for the streamed and memory-bound kernels the comparator is the
memory roof (a kernel driven compute-bound on the FP8 ceiling
\emph{below} that roof fails); for dense GEMM---intentionally
compute-bound---the comparator is the $P_{\fp{8}}/(3r{+}1)$ ceiling
above the size crossover, and running at that ceiling is success, not
failure.  The model checks the streamed class per kernel
and finds substantial margin (for the most stringent sparse case, the
ridge sits roughly two orders of magnitude above the kernel's intensity;
Appendix~\ref{app:sparse}).  \emph{(ii) A composition that breaks
$\beta$.}  If composing the emulated building blocks forces $\beta$
persistently above the memory-bound threshold with no architectural
escape, an L3 or L4 kernel could fail even though each constituent dwarf
succeeds.  \emph{(iii) A missing primitive.}  If any kernel demands an
arithmetic primitive that is neither an FP8 matrix op nor the bounded,
peripheral INT32 reduction of Garner reconstruction, the ``sole
primitive'' claim is false as stated.  \emph{(iv) An irreducible
deconstruction cost.}  If the engineered $c_q$ cannot be brought below
the parity threshold $c_q^{*} = 8P_{\text{int}}\,\mathrm{OI}/(r
P_{\fp{64}})$ for the sub-$\mathrm{OI}^{*}$ kernels---i.e.\ if the
reference $c_q{=}16$ proves to be a floor rather than an artifact of an
unoptimised code path---then the FP8-only claim fails for the
low-intensity tail on parts without hardware deconstruction support, and
those kernels permanently require one of the three hardware options of
\S\ref{sec:hwproposal} (rebalanced narrow-integer issue, a
deconstruction instruction, or stream-side conversion---or the
fallback of a small surviving native pipe).  We adopt the two-sided
decision criterion
of~\cite{bayraktar2026tme} (Appendix~\ref{app:memo}) verbatim: a
conversion kernel whose counted
cost is below ${\sim}7$ instructions per modulus per element refutes
their bound; a best-effort kernel at or above it confirms the
conversion-bound verdicts of Table~\ref{tab:speedups} as
delivered---and we commit to publishing whichever outcome the
measurement returns (\S\ref{sec:futurework}).  None of the kernels
surveyed in
\S\ref{sec:impl} or \S\ref{sec:l3} exhibits failure modes (i)--(iii),
and mode (iv) is bounded as described; that this
survives the model's scrutiny is the theoretical content of the paper.
Three of the four failure modes are conditions of
Proposition~\ref{prop:atomic} read in the negative---(i) is (C3), (ii)
is (C1), (iv) is (C2), with (C4) folded into (i)/(ii) through
$\gamma$---while (iii) would contradict the primitive definition of
\S\ref{sec:dwarfs} itself.  The mapping cuts both ways: any outcome,
positive or negative, lands on a named condition, so a confirmation
certifies exactly which mechanisms carry the thesis and a failure
identifies exactly which term is responsible; the follow-on
implementation work (\S\ref{sec:futurework}) is what
converts ``survives the model'' into ``demonstrated on hardware.''

\paragraph{The composition overhead, made precise.}  The live form of
failure mode~(ii) is the cost of decomposing, reconstructing, and fusing
the emulated blocks, captured entirely by $\beta$.  Here $\beta=1$ is the
fully fused, on-chip-resident ideal, and the genuine open question is how
far above~1 a real kernel lands at production tile sizes.  The question
can now be posed with per-kernel numbers: since delivered memory-bound
performance is $\mathrm{OI}\cdot B/\beta$, each claim of this paper
implies a $\beta$ \emph{budget}, and Appendix~\ref{app:beta-blackwell}
computes them---the binding case is SpMV, which must hold
$\beta\!\le\!1.23$ to keep beating B300's own native \fp{64} (and
$\le\!1.10$ for the strict memory-roof claim), while stencils, batched
GEMV, and H100-parity carry $2.4$--$12\times$ of slack.  ($\beta$
feasibility is necessary but not sufficient: the deconstruction keep-up
condition of \S\ref{sec:model} is an independent constraint on the same
kernels, and at reference constants it, not $\beta$, is what binds the
SpMV and single-RHS GEMV rows.)  The same appendix
shows that a register-file-centric kernel design genuinely cannot meet the
SpMV budget---confirming the sharpest form of the reviewer criticism---and
that the Blackwell-class dataflow (shared-memory-sourced operands,
accumulators in dedicated tensor memory, asynchronous bulk copies, and the
two-dimensional instance parallelism of real workloads) meets every budget
except the strict roof claim on adversarially padded sparse matrices, a
format rather than a register property.  This remains a codesign question
with a known design space---more on-chip state and better fusion drive
$\beta$ toward~1; a small surviving native \fp{64} unit absorbs kernels
that resist fusion; emulation at a higher-precision substrate such as
\fp{32} trades a little of the $3{:}1$ Karatsuba penalty for a shorter
residue chain.  The point is not that $\beta=1$ is guaranteed, but that
the question is now quantified and bounded, and the delivered-$\beta$
decomposition is exactly what the validation programme measures.

\paragraph{Sparse coverage.}  The most pointed objection is that the Ozaki
decomposition is naturally a dense technique, and that mapping it onto
irregular sparse structure is harder than a roofline argument conveys.
The objection conflates two questions that should be separated.  The
first---whether, once a sparse kernel is in tensor-core-amenable form, its
emulation stays beneath the memory roof---is the $\beta$ question, and the
Blocked-Ellpack treatment of \S\ref{sec:impl} answers it at the bound
level, establishing the upper bound for the canonical sparse dwarf;
Appendix~\ref{app:sparse} shows in detail that for sparse kernels the
critical efficiency lever is \emph{not} a large dense block but the
reduction length and the moduli count, so that the bandwidth multiplier
remains near unity even for one-nonzero-per-row patterns, with the FP8
unit needing only a single-digit percentage of its peak to stay hidden
behind HBM.  The second question---whether a \emph{given} irregular matrix
can be put into that form without inflating data movement---is an
implementation matter about formats, not numerics, and a concrete route
exists: hybrid Ellpack/CSR or blocked formulations, drawing on established
high-performance SpGEMM techniques for casting irregular sparse work into
dense-block products~\cite{matsuoka2018blocked}, compose the dominant work
as matrix multiplication, after which the question again reduces to
$\beta$.  Whether it holds for the most adversarial sparsity patterns is
precisely the kind of claim to be settled by measurement on real matrices;
we hold the bound and assign the
attainment to the implementation programme.

\paragraph{Application semantics: the identifiable corner cases.}  A bound
on emulated-matrix throughput and accuracy is not the same as a guarantee
of full IEEE-\fp{64} \emph{application} behaviour, and the distinction
should be stated rather than blurred.  Even granting \fp{64}-level
accuracy on the matrix product, that does not by itself deliver correct
\texttt{NaN}/\texttt{Inf} and exception semantics, bitwise reproducibility
across runs and decompositions, stable convergence of preconditioned
iterative solvers, or robustness on severely ill-conditioned
inputs---properties production software depends on that cannot be inferred
from a roofline model.  These are corner cases, not refutations, and they
form a small and \emph{identifiable} set.  Some are already addressed: an
ESC/ADP scheme~\cite{schwarz2025} supplies a per-operation accuracy
guarantee with automatic native-\fp{64} fallback for inputs whose dynamic
range exceeds what a fixed slice count can certify, and the exact Kulisch
fixed-point reconstruction of the companion analysis~\cite{matsuoka2026fft}
gives a reproducible, exactly-rounded accumulation for the spectral case.
Others---exception semantics across a full application, convergence of
emulated-kernel solvers, and the genuinely ill-conditioned regimes of
Appendix~\ref{app:conditioning}---remain open, but they are cheaply
detectable (for instance via the exponent-span check ESC already performs)
and routable to a safe path.  Bracketing them deliberately is what allows
the central claim to be stated cleanly; characterising their frequency and
cost on real workloads is part of the validation programme.  We expect
some of these bounds to be shown only partially attainable in practice,
and we regard mapping that boundary as the natural continuation of this
work rather than a concession against it.

\section{Conclusion}
\label{sec:conclusion}

The post-\fp{64} era is here.  On B300 the native \fp{64} pipe has fallen
to the point that nearly every standard scientific kernel becomes
compute-bound on physical silicon, while the memory bandwidth those
kernels were designed around remains available but uncollectable; on
Rubin, NVIDIA has listed ``Emulated DGEMM'' as a first-class column in the
official specifications~\cite{nvidia_rubin_blog,nvidia_hgx},
positioning emulation as the primary high-throughput \fp{64} matrix
path---a column whose published value, if the size-sweep test of
\S\ref{sec:rubinuplift} confirms its fourth-term origin, the
software deconstruction routes of the companion Ozaki~2.5
paper~\cite{matsuoka2026ozaki25} would conditionally roughly double
at the announced operating region ($1.8$--$2.2\times$), with the
codesign dividend accruing to this paper's sparse and streamed
kernels as well (\S\ref{sec:hwproposal}).  This
paper has argued that the right response is not to mourn native \fp{64}
but to recognise what its disappearance reveals: that the FP8 tensor-core
matrix-multiply is the candidate dominant matrix-work substrate for the
surveyed classes of double-precision science, under the stated conditions.

The argument is structural.  Organised as the hierarchy of
\S\ref{sec:dwarfs}, the canonical numerical kernels---the Berkeley
dwarfs---are realised as compositions of the Ozaki~II emulated product
(L1), which is itself a composition of FP8 matrix operations (L0); the
composite solver kernels of real application inner loops (L3) and the
applications above them (L4) inherit the same reduction by composition.
Because the dwarf taxonomy is a heuristic spanning set for the surveyed
matrix-dominated kernel classes, exhibiting
the reduction for every surveyed dwarf class---which the kernel-coverage audit of
\S\ref{sec:l3} does, together with the companion spectral
analysis~\cite{matsuoka2026fft} and FP32+Kahan compensation for
reductions---establishes the claim for the surveyed classes rather than
for a sample.  The Tensor--Memory Equilibrium model is the instrument that
makes the claim testable in both directions: it could have shown a dwarf
driven compute-bound
on the FP8 ceiling, or a composition forced past the memory-bound
threshold, and it does not---while the same machinery has
already functioned once as designed, absorbing the deconstruction term
that the NVIDIA review exposed~\cite{bayraktar2026tme} and emerging with
the claims bounded rather than broken (\S\ref{sec:whyupdate}).  In the
bandwidth-bound regime the emulated
path matches the memory roof exactly at the fused bound; in the compute-bound regime it rises
above the previous generation's native ceiling, exceeding B200's
$40$~TFLOPS \emph{native} dense roof by $\approx 3.4\times$ on B300 and $\approx
11.8\times$ on Rubin (at the official 17.5-PFLOPS dense \fp{8} rate
published June~30, 2026; against B200's own best \emph{emulated} path
the honest comparison is stated in \S\ref{sec:model}: B300 restores
rather than exceeds it, and the unambiguous generational gain arrives
with Rubin; \S\ref{sec:cqladder}).  Re-baselined against the last \fp{64}-balanced
data-centre GPU, every workload on every GPU studied---via its best
native/emulated path---matches or
exceeds H100 at the fused bound, and stays within $7\%$ of H100 scaling
even at the currently memo-counted deconstruction constants via the hybrid
native/emulated path---in contrast to native B300's regression of up to
$50\times$: no penalty where the application is memory-bound, substantial
gain where it is compute-bound, and no regression relative to the
architecture HPC was last comfortable with.

The concrete B300 projections, all at \fp{64}-equivalent accuracy and
bounded by the memory roof, are---at the fused, conversion-ideal
bound---$\sim\!1.2\times$ for SpMV, $\sim\!3\times$
for stencils, $\sim\!12\times$ for batched GEMV at a reasonable batch
size ($B{=}8$, at the corrected $B/4$ intensity),
and $\sim\!104\times$ for dense GEMM at $r{=}12$ with the full $(3r{+}1)$
\fp{8} cost; at the reference deconstruction constants the delivered
column reads $\sim\!2.4\times$ for stencils, $\sim\!4.8\times$ for
batched GEMV, $\sim\!104\times$ for dense GEMM above the size
crossover of \S\ref{sec:rubinuplift}, with SpMV and
single-RHS GEMV conversion-bound and routed to the surviving native
pipe pending the engineered $c_q$ that Part~3 measures.  What enables
the gains is the quiet asymmetry on which the
whole thesis rests: the very FP8 tensor cores whose scale-up for AI caused
the \fp{64} collapse supply throughput so far in excess of HBM bandwidth
that emulating \fp{64} matrix work behind the memory wall costs essentially
nothing---once, and only once, each operand has been deconstructed into
its residues, which is why the engineered cost of that single step is
now the pivotal open constant.  The shift the industry feared turns out
to be the foundation the field can build on.

\paragraph{What this paper establishes, and what comes next.}  We close
by restating the contribution in the structural terms of
\S\ref{sec:dwarfs}, with the program/result discipline of
\S\ref{sec:formalclaim} kept explicit.  The research \emph{program}
named by the title is deliberately strong---FP8 tensor-core matrix
multiplication as the dominant computational primitive of
double-precision scientific computing---and the \emph{present result} is
its boxed, conditional form: for the surveyed matrix-dominated kernel
classes, and under the stated storage, overlap, and numerical
conditions, the FP8 op is the candidate dominant multiplication
substrate, with fixed-width integer side-work, \fp{32}/Kulisch
reductions, movement/control, and a native-\fp{64} fallback remaining
explicit auxiliary mechanisms.  The Berkeley dwarfs catalogued the
canonical kernel classes precisely so that the field could be reasoned
about at the right level of abstraction; this paper advances the
question they left open by exhibiting the single hardware primitive onto
which all of them map at \fp{64}-grade accuracy under those conditions,
with \fp{64} demoted from a hardware requirement to a derived guarantee
produced by composition over FP8.  The
contribution is the \emph{formal framework} that establishes this: the
four-term TME model with its parameters held symbolic, and
Proposition~\ref{prop:atomic}, which states the thesis as a conditional
theorem---at the model's upper bound, under conditions (C1)--(C4), the
reduction to the \fp{8} primitive sacrifices no performance---together
with the class-by-class verification that those conditions are
satisfiable across the taxonomy.  The model is a symmetric instrument:
with the key architectural parameters fitted in, it can \emph{confirm
or refute} the thesis---analytically for most classes, and, where
analysis alone cannot decide, by implementation measurement that
identifies the exact term, and hence the exact cause, on which the
verdict turns.  The division it enforces is the paper's
discipline: what the theory presents is the bound; what separates any
given silicon from the bound is an implementation artifact,
parameterised by the same model, quantified for today's parts in
\S\ref{sec:eval}, and measured---with the hardware options that
would retire the largest artifacts---in Part~3.  One consequence of
the fourth-term analysis deserves separate recognition: deconstruction
merits engineering attention for \emph{dense} linear algebra as much
as for the sparse and streamed kernels that first exposed it.  Below
the size crossover of Eq.~\eqref{eq:densecrossover}---which at
memo-counted constants already reaches $\bar{n}\!\approx\!1{,}200$ on
Rubin---dense Ozaki~II cannot attain its theoretical $(3r{+}1)$ peak,
and the same deconstruction optimisation that rescues SpMV is what
restores it: to moderate sizes by software alone, to every size by
Option~C.  Emulated \texttt{DGEMM} at the theoretical peak is thus
not a given but a deliverable of the deconstruction path.  What this paper does not do, and does
not claim to do, is measure the result on delivered silicon; that is the
role of the follow-on implementation work, which realises the kernels and
quantifies the efficiency with which each reduction is achieved.  The open
questions that remain---chiefly the engineered deconstruction cost
$c_q$ and the composition overhead captured by
$\beta$, together with the identifiable corner cases of coverage and
application semantics set out in \S\ref{sec:falsification}---are
second-stage matters of architecture and engineering, generated and
organised \emph{by} the structural claim rather than obstacles to it.
The measurement of those constants, and the evaluation of the hardware
remedies the deconstruction term motivates, are the subject of the
rapidly forthcoming Part~3.  Two placement
questions now stand explicitly ahead of the delivered benefit, and
both are measurable: \emph{where the residue planes live}
(fused, materialised, or persistent---each mode with its own
conversion multiplicity and traffic charge, \S\ref{sec:hwproposal}),
and \emph{how the FP8 and \inteight{} tensor workloads
contend} (the $\theta$-bracket, whose serialised endpoint the
INT8-tensor-contention-free \texttt{dp4a} routes floor at
${\approx}1.76\times$ the shipping path in the width-bounded band,
${\approx}1.3$--$1.5\times$ in the consistently serialised world).
Under ideal overlap the routes of this paper are upper envelopes;
under materialised or shared-pipe schedules they are lower; the
implementation campaign determines the achievable schedule, the
numerical contract, and the dispatch regions.  What stands, on
today's evidence, is a \emph{parameterised design theory and a
falsifiable research agenda} for FP8 as the candidate dominant matrix-work
substrate of the surveyed double-precision science---with its enumerated auxiliary
substrate priced, its failure modes stated, and its every projection
mapped to a measurement.  We
hold that a top-down, primitive-level treatment of this kind has been
missing from the reduced-precision and AI--HPC convergence debates, and we
offer it with conviction: \fp{8}---together with a narrow deconstruction
path and the retained slivers of narrow integer arithmetic that carry
the side-work---is not merely adequate for science but is
its sufficient foundation, and the theory established here---the bound,
the conditions, and the price list of everything between silicon and the
bound---is the case for building
on it.

\paragraph{Acknowledgements.}  The author thanks Katsuhisa Ozaki, Yuki
Uchino, Toshiyuki Imamura, and Daichi Mukunoki for the body of work on
which this analysis rests; any errors in interpretation are the author's.
The author is particularly grateful to Katsuhisa Ozaki for detailed
review comments on \S\ref{sec:ozaki1}, which led to the
substrate-specific treatment of integer scaling and to the
accumulator-bound slice-count analysis in Table~\ref{tab:slicecount}.
The author thanks Jack Dongarra for a detailed critical reading of the
posted preprint, Jens Domke for technical feedback on the architectural
and kernel analysis, and Matt Walters for discussions that helped sharpen
the framing of the thesis; together with the NVIDIA libraries and DevTech
teams, whose technical review is gratefully acknowledged, their questions
directly sharpened the scope and the second-stage research agenda set out
here.  The author is especially indebted to Harun Bayraktar, John
Gunnels, and Peter Caday, whose technical note~\cite{bayraktar2026tme}
identified the deconstruction term and the tile-granularity correction
incorporated in this version---a model case of adversarial review
improving a falsifiable claim---and to Dan Ernst and Matthew Martineau
for discussions that helped shape the experimental programme.  The author also thanks the broader RIKEN R-CCS team, the Institute
of Science Tokyo faculty, and the NVIDIA cuBLAS team for technical
discussions that shaped this paper.  This work was undertaken as part of
the FugakuNEXT project and related R-CCS initiatives on AI for Science.

\paragraph{Disclosure of AI-assisted writing.}  This manuscript was
prepared with substantial assistance from large language models.  The
present version was authored with Anthropic's Claude (Opus~4.8 and
Fable~5, with the final pre-submission pass of Revision~37 by
Fable~5.1), which contributed draft generation and revision, the
model-consistency and sensitivity analyses executed under the author's
direction, figure generation, and \LaTeX{} mechanics; OpenAI's GPT-5.6
provided independent critical reviews of the present version, several
of whose findings are incorporated here; earlier versions were
prepared with assistance from Claude (Opus~4.7) and Google's Gemini~3.
All scientific arguments,
architectural interpretations, performance projections, and conclusions
were directed, reviewed, and validated by the author, who takes full
ownership of and responsibility for the content of this paper, including
any errors of fact or judgment.

\bibliographystyle{plain}
\bibliography{references}

\appendix

\section{Garner's Algorithm: Detailed Derivation}
\label{app:garner}

Equation \eqref{eq:garner} is the iterative formulation of Garner's
algorithm.  We expand it here for the case $r=3$ to make the dependence
pattern explicit.

Let $C\in\mathbb{Z}$ with $0\le C<m_1m_2m_3$, and let $C^{(i)}=C\bmod m_i$.
We seek mixed-radix digits $v_1,v_2,v_3$ such that
\begin{equation}
C \;=\; v_1 + v_2 m_1 + v_3 m_1 m_2.
\label{eq:mixedradix}
\end{equation}
Reducing \eqref{eq:mixedradix} modulo $m_1$, $m_2$, $m_3$ in turn:
\begin{align}
v_1 &\equiv C^{(1)} \pmod{m_1},\\
v_2 &\equiv (C^{(2)}-v_1)\, m_1^{-1} \pmod{m_2},\\
v_3 &\equiv \bigl((C^{(3)}-v_1)\,m_1^{-1} - v_2\bigr)\, m_2^{-1} \pmod{m_3}.
\end{align}
The inverses $m_1^{-1}\!\pmod{m_2}$ and $m_2^{-1}\!\pmod{m_3}$ are
precomputed once and stored in constant memory.  The arithmetic in
\eqref{eq:garner} is done in 32-bit or 64-bit signed integers; modular
reductions are implemented either as branch-and-correct or as Barrett
reduction.  The total cost is $O(r^2)$ small-integer multiplications per
reconstructed output, which for $r=12$ is $\sim 100$ small
multiplications.  This is \emph{not} negligible on the SIMT pipe: charged
as the body does (\S\ref{sec:impl}), it is $8.1\%$ of the emulated GEMM's
work at $k=1000$ on B300 and $28.4\%$ on Rubin, falling below $1\%$ only
at $k \gtrsim 8.1{\times}10^{3}$ and $2.84{\times}10^{4}$ respectively.

\section{Pseudocode for the Fused Ozaki-II GEMV Kernel}
\label{app:code}

The pseudocode in Listing~\ref{lst:fused-gemv-corrected} makes the
structural points of the fused kernel explicit: $r$ \emph{distinct}
fragment objects (one per modulus), $r$ \emph{distinct} accumulators,
and a sized constant-memory moduli table.  The listing is illustrative
C++17/CUDA pseudocode; production code would use the
architecture-appropriate interfaces---\texttt{tcgen05}/CUTLASS
\texttt{cute} on Blackwell-class (SM100) parts, WGMMA/GMMA on
Hopper---rather
than the legacy \texttt{wmma} API.  For brevity the listing is written
with \texttt{int8\_t} fragment types as in the original Ozaki~II/INT8
formulation; the FP8 variant of Uchino et al.~\cite{uchino2026_fp8}
that we recommend for B300 and Rubin has the same structural skeleton
with \texttt{int8\_t} replaced by the appropriate FP8 fragment type
and \texttt{int32\_t} accumulators replaced by FP32 accumulators
with the quantisation correction described in
\cite{uchino2026_fp8,mukunoki2025_fp8}.

\begin{lstlisting}[caption={Illustrative pseudocode for the fused
Ozaki-II batched GEMV kernel, shown in its \emph{INT8-substrate}
realisation (one MMA per modulus, \textsc{int8} operands,
\textsc{int32} accumulation---type-correct as written); the \fp{8}
realisation replaces each \texttt{mma\_sync} with the 2/3-plane
\texttt{fp8\_modular\_product} of \S\ref{sec:gemvstrategy} (\fp{32} accumulation, exact
integer epilogue as in Listing~2).  Not
production-ready.},label={lst:fused-gemv-corrected}]
#include <mma.h>
using namespace nvcuda;

constexpr int R = 12;                  // number of moduli
__constant__ int32_t c_moduli[R];      // pairwise-coprime moduli

template <int BATCH>
__global__ void ozaki_gemv_kernel(const double* __restrict__ A,
                                  const double* __restrict__ X,
                                        double* __restrict__ Y,
                                  int M, int N) {
    // r distinct fragments and accumulators (one per modulus).
    wmma::fragment<wmma::matrix_a,      16, 16, 16,
                   int8_t, wmma::row_major> a_frag[R];
    wmma::fragment<wmma::matrix_b,      16, 16, 16,
                   int8_t, wmma::col_major> x_frag[R];
    wmma::fragment<wmma::accumulator,   16, 16, 16,
                   int32_t> c_acc[R];

    #pragma unroll
    for (int i = 0; i < R; ++i) wmma::fill_fragment(c_acc[i], 0);

    for (int k = 0; k < N; k += 16) {
        // Load FP64 tiles cooperatively into shared memory (omitted)
        double a_val = /* gather from shared */;
        double x_val = /* gather from shared */;

        // Register-level decomposition into r residues.
        int8_t a_res[R], x_res[R];
        #pragma unroll
        for (int i = 0; i < R; ++i) {
            a_res[i] = compute_residue(a_val, c_moduli[i]);
            x_res[i] = compute_residue(x_val, c_moduli[i]);
        }

        // Pack residues into fragments via warp shuffles (omitted)
        // and issue r tensor-core MMAs.
        #pragma unroll
        for (int i = 0; i < R; ++i) {
            pack_into_frag(a_frag[i], a_res[i]);
            pack_into_frag(x_frag[i], x_res[i]);
            wmma::mma_sync(c_acc[i], a_frag[i], x_frag[i], c_acc[i]);
        }
    }

    // CRT reconstruction (Garner's algorithm; see Appendix A).
    // Each accumulator is a full WMMA tile: stage tiles to correctly
    // sized, aligned buffers, then reconstruct per output element.
    __shared__ int32_t acc_tiles[R][TILE_M * TILE_N];  // one tile/residue
    #pragma unroll
    for (int i = 0; i < R; ++i)
        wmma::store_matrix_sync(acc_tiles[i], c_acc[i], TILE_N,
                                wmma::mem_row_major);
    for (int out = lane_id(); out < TILE_M * TILE_N; out += warpSize) {
        int32_t digits[R];
        #pragma unroll
        for (int i = 0; i < R; ++i) digits[i] = acc_tiles[i][out];
        // garner_reconstruct: narrow per-digit arithmetic, multi-limb
        // (e.g. 5xINT32) mixed-radix value, one rounded cvt to fp64.
        Y[global_index(out)] = garner_reconstruct(digits, c_moduli);
    }
}
\end{lstlisting}

\section{Numerical Conditioning and the Choice of Scaling}
\label{app:conditioning}
\label{app:scaling}

The integer scaling in \eqref{eq:scale} is the only place where the
emulated kernel can lose accuracy beyond the working-precision unit
round-off.  The standard practice~\cite{ozaki2025_scheme2,uchino2026_fp8}
is to choose the diagonal scaling matrices $D,E$ with power-of-two diagonal
entries (so that the rescaling $D^{-1}\tilde C E^{-1}$ is itself
error-free in \fp{64}), with each diagonal entry set so that the largest
absolute value in the corresponding row/column of $\tilde A$ (resp.\
$\tilde B$) reaches $2^{p-1}-1$ for an integer width $p$.  This maximises
the signal-to-noise ratio of the integer quantisation.

For inputs with rows of strongly heterogeneous magnitude, this row-wise
scaling can introduce noticeable error if a single row contains both very
large and very small entries.  Adaptive precision schemes such as
ADP~\cite{schwarz2025} adjust the number of moduli $r$ (or equivalently the
\emph{slice} count for Ozaki~I) per GEMM call, from the operands'
measured exponent spans (per-row and per-column scaling diagonals
absorb the intra-matrix variation), to recover the lost bits.  The TME
model can be extended to capture ADP-style adaptation by allowing
$\alpha$ to vary with the call (or, for blocked variants, the panel);
the projections in
Table~\ref{tab:speedups} use a single global $r=12$ for simplicity.

\section{On-Chip-Fusion Details for Stencil and SpMV Kernels}
\label{app:fusion-details}

This appendix gives the \emph{first-order} per-kernel breakdown of HBM
traffic for the fused kernels of \S\ref{sec:stencil}
and~\S\ref{sec:spmv}---the accounting that fixes the operational
intensities used in Table~\ref{tab:speedups}---together with the
padding-ratio analysis for SpMV.  It deliberately stops short of the
register-pressure question.  Whether the fused state can actually stay
resident, and hence whether the $\beta$ figures implied here are
delivered, is a property of the kernel \emph{dataflow}, not of the
traffic count; that question is answered quantitatively in
Appendix~\ref{app:beta-blackwell}, whose $\beta$ budgets and refined
per-kernel $\beta$ values are authoritative wherever the two appendices
could be read to differ.

\paragraph{Stencil $\beta$ accounting.}  For the 7-point fused stencil
kernel of Algorithm~\ref{alg:fused-stencil}, the per-output HBM traffic
breaks down as: $8$~B for the centre-point read, $\approx\!8$~B for the
amortised neighbour reads (each neighbour is reused by $\sim\!6$
neighbouring outputs in a shared-memory tile of $\sim\!32^3$ points),
and $8$~B for the write.  Total $\approx\!24$~B per output, against
$\sim\!7\!\times\!2 = 14$ fp64-equivalent ops per output, giving
$\mathrm{OI}\!\approx\!0.58$~\flops{}/Byte by this accounting---the
``tuned'' point of the three-point ledger of \S\ref{sec:stencil};
Table~\ref{tab:speedups} and the conversion-cap arithmetic of
\S\ref{sec:eval} use the conservative reference point
$\mathrm{OI}\!\approx\!0.5$ ($\sim\!28$~B of HBM traffic per output,
including the assumed store-miss line fill and coefficient traffic).

The register footprint per warp is $r$ residue accumulators---one per
\emph{modulus}, not one per plane, and this is a scalar path, where the
per-$k$ fold into a single residue is possible---plus the working tile,
a first-order count of $\approx\!4r$~B, i.e.\ $r$ INT32 registers, per
output beyond the working set.  On the tensor-core path the
corresponding count is not $r$ but $N_{\text{acc}}$
(\S\ref{sec:frozen-kernel}): the MMA accumulates \emph{inside} the
tensor core over $K_{\text{mma}}$, so the plane products cannot be
folded between $k$ steps and each carries its own accumulator.
Whether this state can remain resident without spilling is not decided
by the count but by the dataflow, and the two candidate designs differ
qualitatively: Appendix~\ref{app:beta-blackwell} shows that a
register-file-centric (Hopper-style) kernel has \emph{no} register
allocation inside the $\beta$ budgets of
Table~\ref{tab:betabudgets}---spills convert directly into HBM traffic
---whereas the Blackwell-native dataflow (shared-memory-sourced
operands, TMEM-resident accumulators) removes the register file from
the problem entirely and holds the fused stencil at
$\beta\!\approx\!1.04$.  The modeled $\beta$ for this kernel is
therefore governed by Appendix~\ref{app:beta-blackwell}; the present
appendix contributes only the traffic accounting above.

\paragraph{SpMV padding and $\beta$.}  For the Blocked-ELL Ozaki-II SpMV
kernel of Algorithm~\ref{alg:fused-spmv}, the bandwidth multiplier
inherits the padding ratio of the underlying Blocked-ELL format.
Let $\rho_\text{pad} \in [1,\infty)$ be the average ratio of
``stored'' to ``actual'' non-zeros, where structural zeros count as
stored.  The padding ratio enters the composed bandwidth multiplier of
Appendix~\ref{app:beta-blackwell} as its leading factor,
$\beta_{\text{pad}}\!=\!\rho_\text{pad}$, and bounds $\beta$ from below:
$\beta \ge \rho_\text{pad}$ regardless of how well the residue
generation, slicing, and spill terms of that appendix are engineered.
For a
3-D 7-point Laplacian discretisation on a regular grid at block width
$\mathit{bw}\!=\!8$, a 7-entry row occupies $8$ stored slots, so
$\rho_\text{pad} = 8/7 \approx 1.14$---already most of the strict
$10\%$ roof budget of Appendix~\ref{app:beta-blackwell} before residue
generation or spill, which is why width selection (e.g.\
$\mathit{bw}\!=\!7$ where the format allows it, giving
$\rho_\text{pad}\!=\!1$) or HYB fallback is part of the dispatch, not
an optimisation;
for a typical finite-element matrix with row densities in
$[6, 24]$ at $\mathit{bw}\!=\!32$ one obtains $\rho_\text{pad}\!\approx\!2$, so
the kernel's effective intensity halves and the speedup of
Table~\ref{tab:speedups} should be read against half the listed
intensity column.

For strongly heterogeneous row lengths
($\sigma(\text{nnz/row})/\mu(\text{nnz/row})\!>\!1$), hybrid CSR-ELL or
HYB formats are required: long rows fall back to CSR (no padding) and
short rows use Blocked-ELL.  The TME model accommodates this by
treating $\beta$ as a per-row property in the spirit of the ADP
extension in Appendix~\ref{app:scaling}.

\section{On-Chip Fusion: \texorpdfstring{$\beta$}{beta} Budgets, Blackwell-Native Dataflow}
\label{app:beta-blackwell}

A natural objection to the on-chip-fusion pattern of
\S\ref{sec:impl}---whose first-order traffic accounting appears in
Appendix~\ref{app:fusion-details}, and which earlier versions of this
paper asserted was register-feasible without analysis---raised in
technical review of the
posted preprint---is that it cannot be achieved in practice: the
decomposition precomputation is too expensive to hide behind the tensor-core
MMA; the CRT-reconstruction operands are too large to hold in registers or
shared memory at realistic tile sizes; and the TME model, as published,
effectively assumes an unbounded register file.  This appendix answers the
objection quantitatively.  The short version: the criticism is correct
\emph{for a Hopper-style, register-file-centric kernel design}, and it is
dissolved by the dataflow that Blackwell-class hardware itself
provides---while the analysis exposes the \emph{actual} binding resource,
which is not the register file.  (Since the July~3 version this appendix
must be read together with the fourth TME term of \S\ref{sec:model}: the
residue-\emph{generation} throughput analysed below is the
dataflow-level view of the deconstruction cost $c_q r/P_{\text{int}}$,
and the $\beta$ budgets here are necessary but not sufficient---a kernel
can meet every $\beta$ budget and still be conversion-bound at the
reference constants of~\cite{bayraktar2026tme}.)

\paragraph{First, the correct sensitivity question.}  The bandwidth
multiplier does not push a kernel over a roofline edge; delivered
memory-bound performance is $\mathrm{OI}\cdot B/\beta$, so $\beta>1$
slides performance \emph{down the bandwidth roof}.  The right question is
therefore the per-kernel $\beta$ \emph{budget} against each claim of this
paper.  Three thresholds matter: the memory-roof claim (allowing 10\%
slack, $\beta\le1.10$); the crossover beyond which emulation stops beating
B300's own collapsed native \fp{64}
($\beta \le \mathrm{OI}\cdot B/\min(\mathrm{OI}\cdot B,\,P_{\fp{64}})$);
and H100 parity, which for kernels memory-bound on both machines is
uniformly $\beta \le B_{\text{B300}}/B_{\text{H100}} = 2.39$.

\begin{table}[h]
\centering\small
\caption{Per-kernel $\beta$ budgets on B300.  The on-chip-fusion question
is binding for exactly one kernel class: SpMV.}
\label{tab:betabudgets}
\begin{tabular}{@{}lcccc@{}}
\toprule
\textbf{kernel} & \textbf{OI} & \textbf{roof claim} & \textbf{vs native B300} & \textbf{vs H100} \\
\midrule
SpMV (blocked-ELL)  & 0.2 & $\beta\le1.10$ & $\beta\le\mathbf{1.23}$ & $\beta\le2.39$ \\
Stencil (7-pt)      & 0.5 & $\beta\le1.10$ & $\beta\le3.08$          & $\beta\le2.39$ \\
Batched GEMV ($B{=}8$) & 2.0 & $\beta\le1.10$ & $\beta\le12$            & $\beta\le2.39$ \\
Dense GEMM          & $\gg$ridge & \multicolumn{3}{c}{compute branch; $\beta$ second-order} \\
\bottomrule
\end{tabular}
\end{table}

\paragraph{The register-file-centric design fails, as the objection says.}
A fused Ozaki-II SpMV that keeps all state in the register file needs
$\sim\!90$ registers per thread at $r{=}12$ (the residue accumulators
for the thread's output tile, double-buffered operand fragments, Garner
constants, gather addresses, pipeline control), limiting occupancy to
$\sim\!22$ warps/SM.
Forcing occupancy higher by under-allocating spills state to local memory,
which \emph{is} HBM traffic: with spilled values re-touched once per tile
($\sim\!32$ elements/thread) and 70\% L2 absorption, allocating 64
(resp.~48) registers yields $\beta \approx 1.24$ (resp.~$1.39$)---already
past the SpMV budget of Table~\ref{tab:betabudgets}.  \emph{On this design
there is no register sweet spot inside the budget}, and the reviewer
intuition is simply right.

\paragraph{The Blackwell-native dataflow removes the register file from
the problem.}  Blackwell's fifth-generation tensor core deliberately moved
matrix state out of the register file: MMA operands are sourced from
shared memory, accumulators reside in the dedicated 256\,KiB-per-CTA Tensor
Memory (TMEM), and a single thread issues the MMA.  The identical fused
kernel then needs only $\sim\!36$ registers per thread (gather addresses,
TMA descriptors, loop control, a compact Kulisch-style running partial):
the accumulator state that the objection identifies as the
blocker is precisely the state TMEM exists to hold---$N_{\text{acc}}$
tiles per output tile, not $r$ (\S\ref{sec:frozen-kernel}), so at the
$m16\,n8$ fp32 accumulator shape ($512$~B per tile) a set costs
$N_{\text{acc}}\!\cdot\!512$~B and $17$--$19$ concurrent sets fit, not
the ${\sim}40$ an $r$-deep set would suggest---and the staged A tile plus
residue-generation buffers and a persistent $x$-residue slice occupy
$\sim\!176$ of $228$\,KB of shared memory with double buffering.

\begin{figure}[htbp]
\centering
\includegraphics[width=\linewidth]{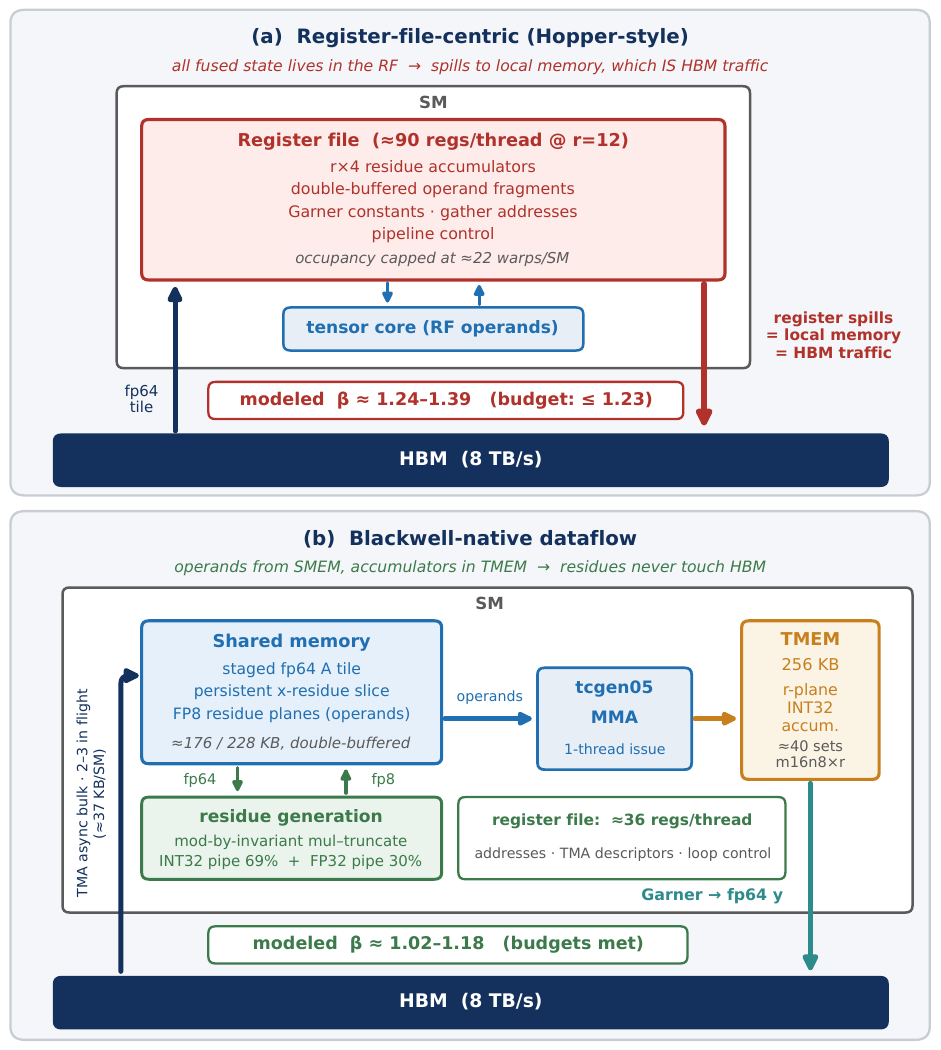}
\caption{How effective on-chip fusion is achieved.  \textbf{(a)}~The
register-file-centric (Hopper-style) design holds all fused state---$r$
residue accumulators, operand fragments, reconstruction constants---in
the register file; at $r{=}12$ this costs $\sim\!90$ registers per
thread, and any allocation inside the occupancy envelope spills to local
memory, which \emph{is} HBM traffic: modeled
$\beta\!\approx\!1.24$--$1.39$, outside the SpMV budget of
Table~\ref{tab:betabudgets}.  \textbf{(b)}~The Blackwell-native dataflow
removes the register file from the path: TMA streams the \fp{64} tile
into shared memory, residues are generated on the fly with the
mod-by-invariant multiply--truncate split across the INT32 ($\sim\!69\%$)
and FP32 ($\sim\!30\%$) pipes, MMA operands are sourced from shared
memory, and the $N_{\text{acc}}$ accumulator sets reside in TMEM.  The residue planes
never touch HBM; the only HBM write is the Garner-reconstructed \fp{64}
result, and the register file shrinks to $\sim\!36$ registers per thread
of addresses and control.  Modeled $\beta\!\approx\!1.02$--$1.18$
across the surveyed kernels, meeting the budgets.  \emph{Both $\beta$ bands
are modeled projections from the budget analysis of
Table~\ref{tab:betabudgets}, not measurements} (\S\ref{sec:model},
``Evidence labels'').  Generated by
\texttt{render\_fusion\_dataflow.py}, released with the paper.}
\label{fig:blackwell-dataflow}
\end{figure}

Figure~\ref{fig:blackwell-dataflow} contrasts the two dataflows and makes
the mechanism of \emph{effective} fusion visible: fusion is achieved not by
holding the residue planes in the register file---the design the objection
correctly rules out---but by never letting them exist anywhere the HBM bus
can see.  Operands stream in once by asynchronous bulk copy, residues are
generated on the fly and live only in shared memory as MMA operands and in
TMEM as accumulators, and the sole HBM write is the reconstructed \fp{64}
result.  Listing~\ref{lst:blackwell-native} gives the corresponding kernel
skeleton in illustrative CUDA-style pseudocode (production code would use
the \texttt{cute}/\texttt{tcgen05} interfaces, which are
Blackwell-class (SM100); cf.\ the Hopper-style WGMMA-era
skeleton of Listing~\ref{lst:fused-gemv-corrected}, whose fragment and
accumulator state is exactly what moves out of the register file
here).  The single-HBM-write property is a steady-state claim: it
presumes the $x$-residue column band stays L2-resident across the
CTAs that share it and excludes one-time setup traffic; the
residency and amortisation assumptions are those of the resource
ledger, Table~\ref{tab:resledger}.

\begin{lstlisting}[caption={Illustrative pseudocode for the
Blackwell-native fused Ozaki-II kernel skeleton (SpMM/stencil pattern).
Operands are staged in shared memory by TMA, residues are generated on
the INT32{+}FP32 pipes, plane-product MMAs source operands from shared
memory and accumulate in \fp{32} into Tensor Memory (TMEM)---exact in
the small-integer range---with the exact integer conversion, Karatsuba
recombination, and final modular reduction in the epilogue; the
register file holds only
addresses and control.  Not production-ready.},
label={lst:blackwell-native}]
__global__ void fused_ozakiII_blackwell_native(/*A_desc, x, y, moduli*/)
{
  // -- state placement: the whole point ------------------------------
  __shared__ TileBuf  smem_A[2];       // fp64 A tile (TMA dst, dbl-buf)
  __shared__ ResPlane smem_Ares[2][R]; // ResPlane = packed per-modulus
                                       //   plane set: 2 (square) or 3
                                       //   (nonsquare: d0, d1, d0+d1)
                                       //   FP8 byte planes per scalar
                                       //   (30 B/scalar over the set)
  __shared__ XSlice   smem_xres[R];    // persistent x-residue planes
                                       //   (L2-resident column band)
  // Accumulators: w(i) = 2 per modulus (Sec 2.3, eq. 8-9
  // [eq:sq-shortcut, eq:kar]), NOT 3 and NOT 1.  Square: {d0e0} and
  // {d1e0 + d0e1} (the two cross terms share the coefficient s).
  // Nonsquare: {(d0+d1)(e0+e1)} and
  // {d0e0 - 2^b d1e1} (merged, since the coefficient ratio is exactly
  // -2^b and the d1 plane is stored pre-scaled by -2^b at conversion).
  // EXCEPTION: the merge has TWO gates -- the pre-scaled plane must be
  // substrate-exact, and the merged term must leave accumulator margin
  // at K_slab.  On fp8 (the product substrate everywhere here) gate one
  // always passes: on S's six 9-bit nonsquare moduli the pre-scaled
  // plane is exactly -256 = -2^8, which is e4m3-exact.  Gate two is the
  // one that fails there: the merged term reaches 4095, exact at
  // K_slab = 4096 by ONE term, so we decline it.  Those keep w(i) = 3.
  //   N_acc = sum_i w(i) = 30 / 26 / 30 for S / E / A
  //           (S = 6*2 + 6*3, NOT 2r = 24;  E = 2r = 26;  A = 2r = 30).
  // Only one MODULUS BLOCK is live at a time.  Blocks partition MODULI,
  // not accumulators: a modulus contributes its whole w(i) to one block,
  // so the live count is the balanced-partition optimum (eq. 29
  // [eq:nblk]), NOT ceil(N_acc / NBLK) -- unattainable for odd r.
  // NBLK=2 -> 15/14/16 tiles = 240/224/256 KiB for S/E/A (A exactly at
  // budget).  NBLK=3 -> a uniform 10 tiles = 160 KiB, but each extra
  // block costs another 8/T B/FLOP of fp64 slab re-reads (eq. 30
  // [eq:blkcost]): 1/3 of L2 at NBLK=2, 2/3 at NBLK=3.  So 2, not 3.
  constexpr int NBLK = 2;              // modulus blocks per k-slab
  tmem_t acc[max_block_width(NBLK)];   // 15/14/16 tiles for S/E/A
  // acc_at(i) = prefix sum of w() over this block's earlier moduli.
  int32_t res[R];                      // one reduced residue per modulus
  // register file: gather addresses, TMA descriptors, loop control,
  // one compact running partial -- ~36 regs/thread in total.

  if (warp_is_producer()) {
    // One producer warp sustains the stream.  Little's law at 8 TB/s
    // and ~700 ns latency needs only ~37 KB in flight per SM,
    // i.e. 2-3 outstanding bulk transactions.
    for (int t = 0; t < num_tiles; ++t)
      tma::load_async(smem_A[t & 1], A_desc, t); // cp.async.bulk.tensor
  } else {
   // Modulus blocking: the planes of distinct moduli are DISJOINT, so
   // sweeping the k-slab once per block reads each plane byte exactly
   // once overall -- blocking costs no extra operand traffic, and it is
   // what keeps the 64x64 output tile (and hence the reach) affordable.
   for (int blk = 0; blk < NBLK; ++blk) {
    for (int t = 0; t < num_tiles; ++t) {
      tma::wait(t);                              // mbarrier arrive/wait
      // -- residue generation: one staging pass, on the fly ----------
      // fp64 SMEM tile -> regs -> FP8 planes in SMEM (this single pass
      // IS a SMEM round trip; what is infeasible is the multi-pass
      // staged variant, ~72 TB/s LSU vs ~34 available, beta ~ 2).
      // mod-by-invariant multiply-truncate, split ~69%/~30% across the
      // INT32 and FP32 pipes so generation keeps pace with the full
      // HBM stream (delta_gen ~ 0).
      double a = smem_A[t & 1].load(lane);
      #pragma unroll
      for (int i = blk; i < R; i += NBLK)   // this block's moduli only
        smem_Ares[t & 1][i].store(lane,
            residue_planes_fp8(a, m_inv_int32[i], m_inv_fp32[i]));
      barrier_sync();
      // -- MMA: SMEM-sourced operands, TMEM accumulators, issued by a
      // single elected thread (fifth-generation tensor-core model).
      // THREE products per modulus (p_mma = 3r over the set), paired as
      // CROSS terms -- a diagonal plane(p) x plane(p) pairing cannot
      // express d1*e0 or d0*e1 and would be silently wrong.
      if (elect_one_thread())
        #pragma unroll
        for (int i = blk; i < R; i += NBLK) {
          const int a2 = acc_at(i);              // += w(i) per modulus
          const auto A_ = smem_Ares[t & 1][i]; const auto X_ = smem_xres[i];
          if (is_square(i)) {               // eq. 8 [eq:sq-shortcut] {1,s,s}
            tcgen05::mma(acc[a2  ], A_.d0(), X_.d0());   // d0*e0
            tcgen05::mma(acc[a2+1], A_.d1(), X_.d0());   // d1*e0  }
            tcgen05::mma(acc[a2+1], A_.d0(), X_.d1());   // d0*e1  } share s
          } else if (merge_taken(i)) {           // eq. 9 [eq:kar], Karatsuba
            tcgen05::mma(acc[a2  ], A_.sum(), X_.sum()); // (d0+d1)(e0+e1)
            tcgen05::mma(acc[a2+1], A_.d0(),  X_.d0());  // d0*e0        }
            tcgen05::mma(acc[a2+1], A_.d1s(), X_.d1());  // -2^b*d1*e1   }
          } else {                  // S's 9-bit tail, merge declined: w=3
            tcgen05::mma(acc[a2  ], A_.sum(), X_.sum()); // (d0+d1)(e0+e1)
            tcgen05::mma(acc[a2+1], A_.d0(),  X_.d0());  // d0*e0
            tcgen05::mma(acc[a2+2], A_.d1(),  X_.d1());  // d1*e1, unmerged
          }                                      // d1s() is the plane stored
        }                                        // pre-scaled by -2^b: exact,
    }                                            // exponent-only, free.
    // -- epilogue, ONCE PER BLOCK: TMEM -> exact int -> mod ------------
    tcgen05::commit_wait();
    for (int i = blk; i < R; i += NBLK) {
      const int a2 = acc_at(i);   const int64_t B = (int64_t)1 << b_of(i);
      // Accumulators hold exact integers: K_slab * max|term| <= 2^24
      // (eq. 11 [eq:accum-exact]; worst case 1088*4096 = 0.27 * 2^24).
      // Apply the epilogue coefficients of eq. 8 [eq:sq-shortcut]
      // and eq. 9 [eq:kar], then reduce.
      const int64_t lo = exact_i64(tcgen05::ld(acc[a2  ], lane));
      const int64_t hi = exact_i64(tcgen05::ld(acc[a2+1], lane));
      const int64_t v = is_square(i)
        ? lo + (int64_t)s_of(i) * hi              // 1*d0e0 + s*(cross)
        : merge_taken(i)
          ? B * lo + (1 - B) * hi                 // 2^b*sum + (1-2^b)*merged
          : B * lo + (1 - B) * hi                 // 2^b*sum + (1-2^b)*d0e0
            + (B*B - B) * exact_i64(tcgen05::ld(acc[a2+2], lane));
      res[i] = final_mod(v, c_moduli[i]);         // to [-m/2, m/2)
    }
   }
    y[row_of(lane)] = garner_reconstruct(res, c_moduli) * Dinv * Einv;
  }
}
\end{lstlisting}

\begin{table}[htbp]
\caption{CTA/cluster resource ledger for the Blackwell-native fused
schedule (illustrative configurations backing the SMEM figures of
\S\ref{sec:hwproposal}; all entries are design values, unmeasured).
$\lambda_A,\lambda_B$ are conversion multiplicities;
lifetimes state where each operand's planes live and for how long.
The third column is the split-$k$ fused schedule for a \emph{single
cluster}: the $\lambda\!=\!1$, no-plane-storage branch applies only when
the output fits one cluster ($T_mT_n\le C$, with $C\!=\!8$ portable /
$16$ opt-in CTAs, i.e.\ $\max(m,n)\lesssim256$ square; $16$ tiles of a
$64\times64$ grid cover a $256^2$ output), whose CTAs share each
$k$-slab's once-converted planes through distributed shared memory (DSM),
with the inner dimension partitioned into $S_k$ $k$-slabs and per-slab
partials reduced in the residue domain through an L2 workspace.  Larger
outputs (e.g.\ $1024^2$) span multiple clusters and are re-split on the
fly---cluster-granular repeated conversion
($\lambda_A\!=\!\lceil n/256\rceil$, $\lambda_B\!=\!\lceil m/256\rceil$), or
else a materialised L2 plane workspace---so a $1024^2$ output is
\emph{not} a single $\lambda\!=\!1$ cluster tile.  The $k$-slab reduction
($S_k>1$) is charged under either tiling.  The engine gates the
no-plane-storage branch on $T_mT_n\le\texttt{CLUSTER\_CTAS}$ and tags
$S_k\!=\!1$ records \texttt{grid-fused-nosplit}.}
\label{tab:resledger}
\centering
\scriptsize
\setlength{\tabcolsep}{3.5pt}
\begin{tabular}{p{3.15cm}p{3.35cm}p{3.35cm}p{3.55cm}}
\toprule
 & \textbf{band, $n{=}256$} & \textbf{band, $n{=}512$} & \textbf{split-$k$ fused, single cluster ($\max(m,n)\!\lesssim\!256$)} \\
\midrule
CTA tile / $k$-slab & $128\times256$, $b_k{=}64$ & $128\times512$, $b_k{=}32$ & $64\times64$ output tile, $S_k$ $k$-slabs of $K_{\text{mma}}{=}64$ \\
warps / pipeline & 1 producer + 8 consumer; 2-stage TMA double buffer & same & same, $+$ residue-domain reduction epilogue \\
SMEM (raw tile + per-modulus planes + buffers) & $\approx\!165$~KB & $\approx\!150$~KB & $\approx\!140$~KB (small-side operand cached, not replicated) \\
TMEM & $N_{\text{acc}}^{\text{live}}$ accumulator tiles under modulus blocking (\S\ref{sec:frozen-kernel}, eq.~\eqref{eq:nblk}): at $n_{\text{blk}}{=}2$, $15/14/16$ tiles $=240/224/256$~KiB for S/E/A vs the $256$~KiB budget (A exactly at it) & same & same: $N_{\text{acc}}^{\text{live}}$ tiles, $4N_{\text{acc}}^{\text{live}}$ B/output ($60/56/64$ at $n_{\text{blk}}{=}2$)---\emph{not} $4r$, which the MMA's in-core accumulation over $K_{\text{mma}}$ forbids \\
registers / thread & $\approx\!36$ (addresses, control) & same & same \\
occupancy & 1 CTA/SM at this SMEM; latency hidden by TMA (Little's-law $\approx$37~KB in flight) & same & 1 CTA/SM; reduction tree in L2 \\
$\lambda_A$ (stream side) & 1 (CTA-owned; cluster to $n\!\approx\!256$) & 1 & 1 (each element in exactly one slab) \\
$\lambda_B$ / small-side planes & 1; $p\,k n \lesssim$ few MB, L2-resident, cluster/call lifetime & 1; same & none stored; reduction traffic $\approx\!4r/K_{\text{slab}}$ B/FLOP, L2-resident (worst traced $\approx\!2.44$ TB/s L2 reduction demand at CCSD, E route, within the assumed $4\times$-HBM L2 service model), $K_{\text{slab}}\!\ge\!4096$ \\
$x$-residues (streamed kernels) & L2-resident column band, shared across CTAs, call lifetime & same & n/a (dense) \\
\bottomrule
\end{tabular}
\end{table}

Per-CTA TMEM is $512$ columns $\times$ $128$ lanes $=256$~KiB, and the
accumulator footprint of a $64\times64$ output tile is
$64^2\cdot4\,N_{\text{acc}}^{\text{live}}$~B---the quantity the TMEM row
of Table~\ref{tab:resledger} charges.  Earlier revisions offered
two rival accountings here---one tile per modulus versus one tile per
plane product---and deferred the choice to Part~3.  Neither is correct,
and the choice does not need deferring: \S\ref{sec:frozen-kernel}
establishes $N_{\text{acc}}=2$ per modulus \emph{wherever the merging
pre-scale is representable}, which is every modulus of E and of A, and
the six square moduli of S---giving $N_{\text{acc}}=2r=26$ for~E and
$2r=30$ for~A.  S is the exception and must not be read as $2r=24$: its
six 9-bit nonsquare tail moduli stay at $N_{\text{acc}}=3$, so the S
total is $6(2)+6(3)=30$.  The reason is the accumulator margin
established there, not \textsc{e4m3} representability: the
pre-scaled plane reaches exactly $-256=-2^{8}$ on all six and is
\textsc{e4m3}-exact, so the merge is available and reproduces $d\,e$
over all $m^{2}$ ordered centred pairs with zero mismatches---but the
merged term reaches $4095$, which is exact at $K_{\text{slab}}{=}4096$
by a single term, and we decline a margin of one
(\texttt{verify/stail\_merge.py}).  The totals are
therefore $N_{\text{acc}}=30/26/30$ for S/E/A---numerically the same
$30$ for S and A, but reached by different arithmetic.  At a $64\times64$ tile that is
$480/416/480$~KiB, which exceeds the budget; at
$N_{\text{acc}}{=}r$ it would be $192/208/240$~KiB, which fits---but
$N_{\text{acc}}{=}r$ is not achievable, because $d_0e_0$ cannot share an
accumulator with the cross terms (the coefficient $s$ is not a power of
two) and the sum product cannot share with the merged pair (the
coefficient ratio is $-16/15$).

The resolution is neither a rival accounting nor a smaller tile but
\emph{modulus blocking}.  Sweep the $k$-slab once per block of the
modulus set, keeping only that block's accumulators resident and running
the epilogue per block.  The blocks partition \emph{moduli}, not
accumulators---a modulus contributes its whole width $w(i)\in\{2,3\}$ to
whichever block holds it---so the live count is the optimum of a
number-partitioning problem,
\begin{equation}
  N_{\text{acc}}^{\text{live}}(n_{\text{blk}}) \;=\;
  \min_{\text{partitions } \mathcal{B}} \;\max_{B\in\mathcal{B}}
  \;\sum_{i\in B} w(i)
  \;\;\ge\;\; \Big\lceil N_{\text{acc}}/n_{\text{blk}} \Big\rceil ,
\label{eq:nblk}
\end{equation}
and the bound on the right is \emph{not} generally attained.  It is worth
stating why, because an earlier version of this passage quoted the bound
as though it were the count.  When every $w(i)=2$ and $r$ is odd---E at
$r{=}13$, A at $r{=}15$---no two-way split of the moduli is even, so
$n_{\text{blk}}{=}2$ gives $2\lceil r/2\rceil = 14$ and $16$ tiles, not
the $13$ and $15$ the ceiling suggests.  S is the one route where the
bound happens to bind, and only because its mixed widths cooperate:
three square and three nonsquare moduli per block give $3(2)+3(3)=15$
(a squares-then-tail split would leave $18$ live).  The honest
$n_{\text{blk}}{=}2$ row is therefore $15/14/16$ tiles $=240/224/256$~KiB
for S/E/A---A exactly at the $256$~KiB budget, with \emph{zero}
headroom, which is not a configuration to design against.

Blocking is free in \emph{plane} traffic---the planes of distinct moduli
are disjoint, so each plane byte is still read exactly once across the
whole sweep---but it is not free, and an earlier version of this passage
was wrong to say so flatly.  An accumulator persists across an entire
$k$-slab, so the block loop must \emph{enclose} the $k$-loop; the slab is
therefore traversed $n_{\text{blk}}$ times, and each traversal re-reads
the \fp{64} source tile.  Conversion work does not multiply (each
traversal converts only its own moduli), and neither does plane traffic;
what multiplies is the \fp{64} operand read, at
\begin{equation}
  \Delta_{\text{blk}} \;=\; (n_{\text{blk}}-1)\,\frac{8}{T}
  \quad\text{bytes per useful \fp{64} FLOP,}
\label{eq:blkcost}
\end{equation}
served from \textsc{l2}, where a cluster's \fp{64} slab panels comfortably
sit.  At $T{=}64$ and the $235.3$-TFLOPS floor this is $29.4$~TB/s at
$n_{\text{blk}}{=}2$---a third of Rubin's assumed $88$~TB/s \textsc{l2}
service---and $58.8$~TB/s, two thirds, at $n_{\text{blk}}{=}3$.  So
$n_{\text{blk}}{=}2$ is the operating point, not $3$: the third block
would buy TMEM slack that is not needed on S or E and would spend a
further third of \textsc{l2} to get it.

That leaves A at exactly $256$~KiB, and we let it stand rather than
paper over it.  A zero-headroom fit is not a design point---it leaves
nothing for the double-buffered epilogue staging a real implementation
will want---and it is the \emph{first} of three independent reasons to
treat the all-byte route~A as an upper bound.  The second is the $0.96$
operand-bandwidth ratio derived below: A keeps four percent of margin at
its own roof, where S and E keep nineteen and seventeen.  The third is
that roof itself, $380$~TFLOPS against S's $473$---a fifth lower before
any of the rest is counted.  (Earlier revisions named a ``$2\%$
asymptotic \textsc{dsm} margin'' as the third reason.  It was neither
independent---it was the operand margin restated---nor correct, having
been computed at $p_{\text{store}}(\text{A}){=}45$ rather than the $44$
that one square in fifteen moduli actually buys.  The \textsc{dsm} share
is $(1{-}1/c)$ times the operand ratio, $0.72$ at $c{=}4$, and is
nowhere near binding.)  Together they are why A's $380$-TFLOPS ceiling
appears in this paper as an upper bound rather than a target.  S and E, the routes we
actually recommend, sit at $240$ and $224$~KiB with $16$ and $32$~KiB clear.
Equations~\eqref{eq:nblk} and~\eqref{eq:blkcost} are evaluated exactly,
for every $n_{\text{blk}}$ up to $6$, in \texttt{verify/blocking.py} of
the artifact.

This last point is what makes the difference load-bearing.  The obvious
alternative, halving the output tile to $64\times32$, also fits
($120/104/120$ tiles' worth, $240/208/240$~KiB), and a previous
revision offered it as the fallback while asserting that ``A returns to
$240$~KiB''---a statement that was arithmetically wrong under the
accounting it was attached to ($64\cdot32\cdot4\cdot45 = 360$~KiB).  It
is right under the frozen kernel, but it is still the wrong choice,
because the reach of a single cluster is set by the output-tile edge:
shrinking $64\times64$ to $64\times32$ drops the harmonic reach of a
$4\times4$ cluster from $256$ to $170.67$ and the Rubin floor from
$235.3$ to $156.9$~TFLOPS, a $33\%$ loss.  Modulus blocking keeps the
$64\times64$ tile and therefore keeps the floor.  The
reach-dependent floors reported here and in the companion are the
$64\times64$ numbers, and they are unchanged by the choice of
$n_{\text{blk}}$---blocking moves accumulator residency, not the
output-tile edge that sets reach.

One supply check remains, since blocking preserves operand traffic
rather than reducing it.  Operand demand at a $T\times T$ tile is
$p_{\text{store}}/T$ bytes per useful \fp{64} FLOP, while a machine that
reaches its own \fp{8} roof at that tile is by construction delivering
$P_{\fp{8}}/T$ bytes/s.  At a route's roof the ratio is therefore
$p_{\text{store}}/\alpha$---independent of platform and of tile
size---giving $30/37=0.81$, $33/40=0.83$ and $44/46=0.96$ for S, E
and~A.  All are below unity, so operand bandwidth is never the binding
term; but the margin for the all-byte route~A is only $4\%$ against
$19\%$ and $17\%$, which is one of the three independent reasons
(alongside its lower roof and the zero-headroom TMEM fit of
\S\ref{sec:frozen-kernel}) not to treat A's $380$~TFLOPS ceiling as
reachable.

\paragraph{The cluster map, concretely.}  The claim that a cluster converts
each element \emph{once} ($\lambda{=}1$) is a statement about a specific
producer/consumer assignment, so we state it rather than assert it.  Index the
$c{\times}c$ cluster by $(i,j)$, $0\le i,j<c$; CTA~$(i,j)$ owns the output tile
$C[iT{:}(i{+}1)T,\;jT{:}(j{+}1)T]$, so the cluster covers a $cT\times cT$
output ($256^2$ at $c{=}4$, $T{=}64$).  Partition the $k$-slab into $c$ chunks.
\emph{Producers:} CTA~$(i,j)$ converts $A[iT{:}(i{+}1)T,\,\text{chunk }j]$ and
$B[\text{chunk }i,\,jT{:}(j{+}1)T]$---and nothing else.  Every plane byte of
the slab is therefore produced exactly once, and the conversion load is exactly
balanced ($T\!\cdot\!K/c$ elements of each operand per CTA).
\emph{Consumers:} the $A$ row-panel~$i$ is read by the $c$ CTAs $(i,\cdot)$ and
the $B$ column-panel~$j$ by the $c$ CTAs $(\cdot,j)$.  Each CTA thus needs
$c{-}1$ remote panels it did not produce, which is exactly what distributed
shared memory is for: the producer publishes into its own
\texttt{shared::cta} window and the consumers pull with
\texttt{cp.async.bulk.shared::cluster.shared::cta.mbarrier::}\allowbreak\texttt{complete\_tx::bytes},
addressing the peer window through \texttt{mapa} and completing on a
\texttt{.cluster}-scope \texttt{mbarrier}; all three primitives are PTX~ISA~8.0
and require \texttt{sm\_90} or later~\cite{nvidia_ptx_isa}.  Operands from HBM
arrive once per cluster by TMA multicast, not once per CTA.

The byte charge follows from the counting, not from a simulation.  Of the
$p_{\text{store}}/T$ operand bytes per useful \fp{64} FLOP that each CTA
consumes, a fraction $1/c$ is locally produced and $(c{-}1)/c$ crosses the DSM
fabric, so
\begin{equation}
\frac{\text{DSM bytes}}{\text{useful \fp{64} FLOP}}
  \;=\; \Big(1-\tfrac1c\Big)\frac{p_{\text{store}}}{T},
\quad\text{and at a route's own roof}\quad
\frac{\text{demand}}{\text{supply}} \;=\; \Big(1-\tfrac1c\Big)\frac{p_{\text{store}}}{\alpha}.
\label{eq:dsm}
\end{equation}
At $c{=}4$ that is $0.61/0.62/0.72$ for S/E/A---three quarters of the operand
ratio computed above, and non-binding for the same reason.  In absolute terms
the fabric must carry $166/169/196$~TB/s at the S/E/A roofs against the
$P_{\fp{8}}/T = 273$~TB/s that a machine at its \fp{8} roof is by construction
moving through the operand path.  Note the direction of
Eq.~\eqref{eq:dsm}: the DSM share rises monotonically in $c$ toward
$p_{\text{store}}/\alpha$, so the $64$- and $128$-CTA rungs that lift the floor
to the roof (\S\ref{sec:hwproposal}) buy reach by spending fabric margin.  We
state the limit rather than overstate the risk: because $(1-1/c)<1$, the DSM
ratio is \emph{strictly below} the operand ratio at every $c$, so on this
accounting the fabric never becomes the binding term---but the margin falls
from $39\%$ at $c{=}4$ to the asymptote $1-p_{\text{store}}/\alpha$, i.e.\
$19\%/17\%/4\%$ for S/E/A.  That asymptote is not a fourth constraint: it is
the operand ratio of the preceding paragraph, approached from below as $c$
grows, and we count it once.  What Eq.~\eqref{eq:dsm} adds is the
\emph{direction}---A's fabric headroom is thinnest exactly at the large
cluster sizes the co-design proposal wants, so the two arguments against
quoting A's $380$-TFLOPS ceiling as reachable reinforce rather than merely
coexist.  The absolute
fabric rate is a Part-3 measurement; the ratio is not, since it is a counting
identity.

The split-$k$ reduction is L2-resident (no HBM traffic) but consumes
L2 service bandwidth, now charged explicitly in the companion engine
(\texttt{params.py}: \texttt{B\_L2}, \texttt{K\_SLAB}); the earlier
``$<\!1\%$ of $Q_0$'' entry was incorrect---as a fraction of $Q_0$ the
reduction is $\approx\!rn/(2K_{\text{slab}})$ (e.g.\ ${\sim}150\%$ for a
width-$1024$ square output)---and is removed, the correct per-FLOP
figure being $4r/K_{\text{slab}}$ (the earlier $8r/K_{\text{slab}}$
dropped the factor of two in the useful work $2mnk$).  The multiplier
$B_{\rm L2}=4B_{\rm HBM}$ is a modelling \emph{assumption}: the $126$-MB
L2 is a documented \emph{capacity}, not a documented $4\times$ sustained
reduction rate.  Under a $1/2/4\times$-HBM sensitivity the worst traced
reduction demand---$\approx\!2.44$~TB/s at CCSD $(441,441,4324)$ on the E
route---is $\approx\!11\%/6\%/3\%$ of the Rubin L2 service budget at
$1/2/4\times$ HBM ($22/44/88$~TB/s), i.e.\ within the assumed service model
even at the $1\times$ endpoint.

\paragraph{Plane-feeding vs.\ on-the-fly generation, and why materialisation loses.} The residue planes reach the tensor cores one of two ways, and the choice decides the bound. \emph{Materialise} them---store the $r$ planes and stream them back---or \emph{generate} them on the fly, splitting each block into planes in registers/TMEM immediately before its MMA (the schedule this appendix builds). Materialisation is \emph{feed-starved}: at the $64\times64$ TMEM-limited tile the reuse is $64$-fold, so streaming both operands costs $p/\mathrm{TILE}$ bytes per useful FLOP, and to hold set~S's roof ($473$~TF) that feed must sustain $p/\mathrm{TILE}\cdot473 = 222$~TB/s; the companion figures for E and A, $226$ and $262$~TB/s, are each taken at \emph{that set's own} roof ($438$ and $380$~TF), since the three sets do not share one. L2 at the assumed $4\times$HBM supplies $88$~TB/s and HBM $22$, so a materialised schedule caps at $B_{\mathrm{L2}}/(p/\mathrm{TILE}) = 188/171/128$~TFLOPS (L2-resident, S/E/A) or $47/43/32$ (HBM-resident)---both \emph{below} the on-the-fly rate. (Those are the \emph{feed} term alone; the companion's per-record engine also charges the one-off write of the planes into L2, which lowers the realised L2-resident cap slightly and tightens the inequality further.) So the design generates on the fly (as the pipe census of the next paragraph independently requires), and the planes never occupy L2 or HBM.

Two consequences follow. First, the emulated throughput is \emph{insensitive} to the $B_{\mathrm{L2}}$ service rate---a $4/2/1\times$ sweep moves the trace composites by $\le\!1$ native-$\times$---so the L2-bandwidth assumption a previous version made load-bearing no longer conditions the result. Second, on-the-fly generation costs \emph{re-splitting}: a $16$-CTA cluster covers a $256\times256$ output ($T_mT_n\le C$), inside which each block is split once and shared through DSM ($\lambda=1$); a larger output spans several clusters and---planes being on-chip and unshared across clusters---each block is re-split once per cluster it feeds, $\lambda_A=\lceil n/256\rceil$, $\lambda_B=\lceil m/256\rceil$, size-weighted $\lambda_{\mathrm{eff}}=(m\lambda_A+n\lambda_B)/(m{+}n)$. For a square output that is $E/256$; for tall/skinny it stays $O(1)$---bounded, not unity---and a cluster \emph{shaped to the output} (a rectangular $c_m{\times}c_n$, $c_mc_n\le C$) tightens it further by keeping every CTA productive: $\lambda_{\mathrm{eff}}{=}1.23$ at $m{=}4096,n{=}64$ under a fixed $4{\times}4$ grid versus $1.05$ for the matched $16{\times}1$, and $1.25/1.50$ at LOBPCG's $110{,}592\times64/128$ versus $1.06/1.25$ for the matched $16{\times}1$ and $8{\times}2$. The fixed $4{\times}4$ square cluster is thus a conservative upper bound and the shape-matched schedule is realisable \emph{and} at least as good; the tall/skinny entries of Table~\ref{tab:kernman} are the square-dispatch values.

The reach itself is an integer-geometry quantity: the continuous $T\sqrt C$ is exact only at square $C$, a $c_m{\times}c_n$ grid of aspect $\rho$ realising $2\sqrt\rho/(1{+}\rho)$ of it, so a $2{:}1$ grid loses $5.7\%$ (the shipping $C{=}16{=}4{\times}4$ reach $256$ and the $C{=}64{=}8{\times}8$ target $512$ are exact; the $C{=}128$ target must be $16{\times}8$ and reaches $683$, not $724$---still above the reach $666$ at which the floor meets the roof, so the companion's $128$-CTA $\to$ roof rung holds).

\begin{figure}[htbp]
\centering
\includegraphics[width=\linewidth]{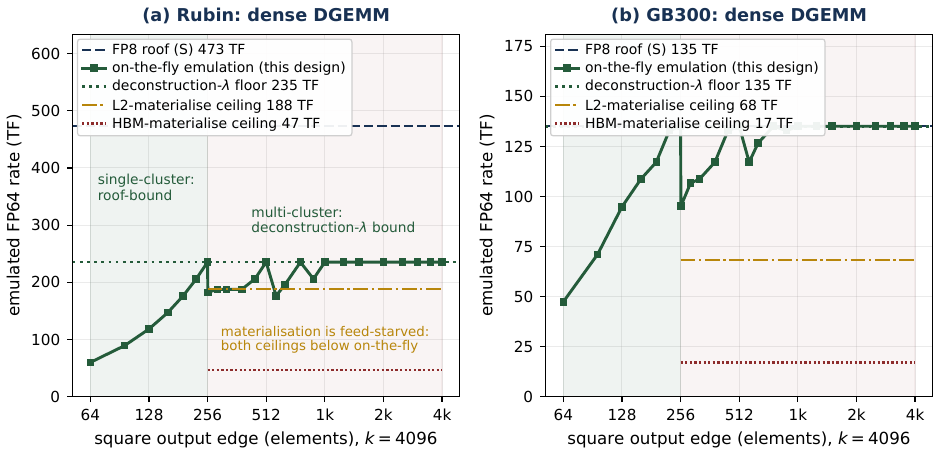}
\caption{Emulated dense DGEMM under on-the-fly residue generation. Up to one cluster's reach ($256^2$) the rate follows the single-conversion crossover; larger square outputs are pinned at the deconstruction-$\lambda$ floor (${\approx}235$~TFLOPS Rubin, $0.50$ of set~S's $473$-TFLOP arithmetic roof; GB300's floor coincides with its lower $135$-TFLOP roof, so GB300 is roof-bound throughout). Materialised alternatives (L2/HBM-resident, dash-dot/dotted) are feed-starved below the on-the-fly curve.}
\label{fig:dgemm}
\end{figure}

Figure~\ref{fig:dgemm} shows the consequence: outputs $\le\!256^2$ follow the $\bar n$-crossover of \S\ref{sec:model}, larger squares settle for cluster-aligned sizes at the deconstruction-$\lambda$ floor (${\approx}235$~TFLOPS on Rubin, $0.50$ of set~S's $473$-TF arithmetic roof; a sawtooth in $\lceil E/256\rceil$ whose aligned upper edge is the plateau and whose ragged worst case is the companion's disclosed ${\approx}0.36$ rigid-schedule minimum, modeled (not constructed) at ${\approx}0.54$ with panel predication as a Part-3 validation target; the floor binds on the INT8/TMAC deconstruction-MAC pipe, not the SIMT residual), and every materialised alternative sits below it. That floor is where NVIDIA's preliminary ${\sim}200$-TFLOPS ``Emulated DGEMM'' figure would place a large-square benchmark.

The crossover and the floor are thus one deconstruction curve, not two claims: the $\lambda{=}1$ crossover is the convert-once upper envelope, realised to the cluster reach, and the floor is what the achieved rate plateaus at beyond it. Crucially the floor equals $R\,P_{\text{int}}/(c_q r)$ at cluster reach $R$---the per-element deconstruction load $c_q r/R$ is what pins, and writing it without the factor $r$, as earlier drafts did, is dimensionally incomplete---so it is a \emph{co-design coordinate, not a wall}. The companion Ozaki~2.5~\cite{matsuoka2026ozaki25} quantifies its escape: a larger cluster or TMEM tile lifts it to the roof ($64$-CTA $\to0.92$, $128$-CTA $\to1.0$), a ${\approx}10\times$-HBM L2 reaches it through materialisation, and lowering $c_q$ (the deconstruction-path hardware of \S\ref{sec:hwproposal}) lifts it at every reach; each retained modulus set is rescued by a different knob. The sparse and streaming kernels meet the \emph{same} $c_q$ term from the other side---below the threshold intensity they are conversion-bound rather than re-split-bound---and their lever is \emph{persistent} (Mode~P) deconstruction: convert the stationary operand once and amortise its planes across the iteration (\S\ref{sec:hwproposal}).

\begin{table}[t]
\centering\small
\caption{How the deconstruction-$\lambda$ limit manifests per kernel (Rubin). Only large \emph{dense} square GEMMs with large $k$ sit at the floor; the traced workloads, governed by their real shapes, are largely immune. ``B/A'' is the composite ratio of the honest on-the-fly schedule (B) to the materialised one (A). The tall/skinny $\lambda_{\mathrm{eff}}$ entries assume the square $4{\times}4$ cluster dispatch the engine models; a shape-matched rectangular cluster lowers them to $1.06/1.25$, so that row is conservative.}
\label{tab:kernman}
\setlength{\tabcolsep}{4pt}
\footnotesize
\begin{tabular}{@{}llll@{}}
\toprule
kernel / regime & shape & mechanism & outcome \\
\midrule
Dense DGEMM $\le 256^2$ & 1 cluster & $\lambda{=}1$; SMEM-fed on-chip & crossover model \\
Dense DGEMM $\gtrsim 512^2$, lg.\ $k$ & multi-cluster & re-split $\lambda{=}E/256$; feed-starved & $\approx$235 TF ($0.50$ of $473$) \\
LOBPCG ($b{=}64/128$) & tall-skinny & $\lambda_A{=}1$; $\lambda_{\mathrm{eff}}{\approx}1.25/1.5$ (bounded) & mild ($0.89$--$0.94\times$) \\
Multifrontal LU & $\bar n$ med.\ 665 & memory-bound fronts, below floor & immune ($1.00\times$) \\
CCSD contractions & $\bar n$ 441, $k{\sim}1953$ & large-$k$, compute-bound ($53\%$ work) & \textbf{hit} ($0.82\times$) \\
Dense LU trailing & \emph{rank-}$64$ & small $k{=}64\Rightarrow$ memory-bound & immune ($1.00\times$) \\
Dense QR & $k\in\{32,3232\}$ & no large-square/large-$k$ product & near-immune ($0.95\times$) \\
Sparse SpMV/SpMM, stencil & streaming & converted once, consumed on-chip & exempt \\
\bottomrule
\end{tabular}
\end{table}

\paragraph{Where the binding resource actually moved.}  Working the
dataflow at the full 8\,TB/s stream exposes the genuine constraint, which
is residue-\emph{generation} throughput, not storage.  Three schedules must
be kept apart, because they are easily conflated.  \emph{(S1)~Multi-pass
materialised staging}---generate all planes to SMEM, then re-read them in
separate passes---requires $\sim\!72$\,TB/s of LSU shared-memory bandwidth
against $\sim\!34$ available and is \emph{infeasible} (effective
$\beta\!\approx\!2$); this is the variant we reject.  \emph{(S2)~Single-pass
on-the-fly staging} (Listing~\ref{lst:blackwell-native})---one generation pass
writes only the planes the MMA then sources from SMEM (the $B$/$x$ operand is
necessarily an SMEM descriptor under \texttt{tcgen05.mma})---is the schedule
this design \emph{uses}, and its single SMEM round-trip is not the rejected
$72$-\,TB/s path.  \emph{(S3)~Register-to-TMEM generation}---place the $A$
residues in TMEM instead of SMEM---makes the LSU traffic trivial but loads the
INT32 pipe at $\sim\!137\%$; splitting generation across the INT32 \emph{and}
FP32 pipes (mod-by-invariant multiply--truncate) lands at $\sim\!69\%/30\%$
utilisation---feasible at the full HBM rate, adding essentially no HBM traffic
($\delta_{\text{gen}}\!\approx\!0$ in the composition below).  The
tensor-operand feed itself (the MMA reading planes from SMEM/TMEM) travels the
specialised descriptor/collector path at rate $P_{\text{operand}}$, \emph{not}
the general LSU, so a literal ``$p/64$-per-CTA'' LSU read count over-charges
it; naming and measuring $P_{\text{operand}}$ is the honest replacement for
$\delta_{\text{gen}}\!\approx\!0$, and is a Part-3 measurable.  \textbf{Caveat
(unverified assumption).}  This split-pipe figure assumes the INT32 and
FP32 pipes issue independently at the stated rates; if the
Blackwell-class SM instead unifies integer and \fp{32} issue on shared
lanes, the available budget collapses to the single
$P_{\text{int}}\!\approx\!75$~\tops{} figure measured
by~\cite{bayraktar2026tme}, the generation margin shown here largely
disappears, and the per-element instruction counts of the fourth TME
term govern (\S\ref{sec:model}).  The pipe census that decides this is
an explicit early target of the Part~3 measurement programme, and until
it reports, the conversion-bound verdicts of Table~\ref{tab:speedups}
at reference constants are the ones to quote.  Note also that the
precomputation does not need to hide behind the MMA at all: for a
memory-bound kernel it needs to hide behind the \emph{memory stream}, and
the FP8-side keep-up requirement is small
(Appendix~\ref{app:sparse})---it is the \emph{integer-side} keep-up that
binds.

\paragraph{Occupancy and latency, dissolved by two-dimensional
parallelism.}  The implicit chain behind the objection---high register use
$\Rightarrow$ low occupancy $\Rightarrow$ exposed latency $\Rightarrow$
lost bandwidth---no longer holds on hardware with asynchronous bulk
copies.  By Little's law, saturating 8\,TB/s at $\sim\!700$\,ns requires
only $\sim\!37$\,KB in flight per SM, i.e.\ two to three outstanding TMA
transactions, sustainable by a single producer warp.  Latency hiding thus
ceases to be an occupancy property \emph{provided each SM always has
independent work}---which production workloads supply in two dimensions:
algorithmically (multiple right-hand sides from block-Krylov, $s$-step,
eigensolver blocks, and ensembles, so that the sparse kernel is in
practice SpMM), and by vector slicing ($x$ partitioned into L2-resident
column bands, independent row-band$\times$column-band sub-problems, with
partial-$y$ aggregation kept on chip)---the latter simultaneously
eliminating the gather-traffic component of $\beta$.  The same
decomposition applies to stencils (plane/tile decomposition, halos through
cluster shared memory) and trivially to batched GEMV.

\paragraph{Refined $\beta$ and the honest residual.}  Composing the terms
($\beta = \beta_{\text{pad}}(1+\delta_{\text{gen}}+\delta_{\text{slice}}
+\delta_{\text{spill}}+\delta_{\text{recon}})$, where each
$\delta$ is a \emph{fractional traffic increment} near zero, not a
standalone multiplier) for the Blackwell-native
design gives stencil $\approx\!1.04$ and batched GEMV $\approx\!1.02$
(all budgets met), and SpMV $\approx\!1.14$--$1.18$ at
$\beta_{\text{pad}}{=}1.10$: inside the native-crossover and H100 budgets,
but missing the strict 10\% roof claim---and the responsible term is the
Blocked-ELL \emph{padding} on adversarial sparsity patterns, a
format/dataset property (Appendix~\ref{app:fusion-details}), not a
register property; well-blocked matrices come in at $\approx\!1.06$ and
meet every budget.  This is exactly the corner case bracketed in
\S\ref{sec:falsification}.  All figures in this appendix are model
projections under stated architectural assumptions (per-SM resources of
the published Blackwell SM carried over to B300; INT32/FP32 pipe rates;
spill re-touch and L2-absorption parameters), and the delivered-$\beta$
decomposition---spill, gather, padding, generation---is precisely what the
validation programme of \S\ref{sec:futurework} measures.

\section{Worked Exemplars: Tracing Composite Kernels to the FP8 Primitive}
\label{app:l3-exemplars}

Section~\ref{sec:l3} argues that the composite (L3) kernels reduce to the
FP8 primitive by composition.  This appendix makes that reduction concrete
for two representative solvers, tracing each step of the inner loop down
the L3\,$\rightarrow$\,L2\,$\rightarrow$\,L1\,$\rightarrow$\,L0 chain and
identifying, at each step, the L2 dwarf it invokes and the operational
intensity that keeps it memory-bound.  The intent is to show that no step
introduces an arithmetic operation outside the FP8-matrix-op-plus-bounded-
INT32 set, and that no composition forces $\beta$ past the memory-bound
threshold.

\subsection{Exemplar 1: Preconditioned / block conjugate gradient}
\label{app:cg}

A single iteration of preconditioned CG (PCG) on a sparse system $Ax=b$
consists of one sparse matrix--vector product, one preconditioner
application, two inner products, and three \textsc{axpy} vector updates.
With a block right-hand side of width $s$ (block-CG, or $s$-step
communication-avoiding CG), the vector operands widen to $n\times s$ panels
and the products become matrix--matrix.  The reduction is as follows.

\begin{enumerate}[leftmargin=2.2em,itemsep=2pt]
\item \textbf{$q \gets A p$ (L3 step $\rightarrow$ L2 SpMM).}  With $s$
  right-hand sides this is sparse-matrix\,$\times$\,dense-panel
  (SpMM), the canonical sparse dwarf at $n\!=\!s$.  By
  Appendix~\ref{app:sparse} its operational intensity is
  $\sim\!0.2s$~\flops{}/B and it remains memory-bound for all practical
  $s$; the FP8 unit needs a single-digit percentage of peak to stay hidden
  behind HBM, and the $s\!\geq\!8$ block makes the tensor-core $n$-dimension
  fully used.  \emph{Reduces to:} Ozaki~II SpMM (L1) $\rightarrow$ FP8 MMA
  (L0).
\item \textbf{$z \gets M^{-1} r$ (preconditioner).}  For the common
  preconditioners the apply is itself one of the audited dwarfs: a
  sparse-triangular solve (block-Jacobi, ILU) reduces to small dense
  block solves of the form handled in \S\ref{sec:l3}(c); an algebraic-
  multigrid V-cycle is a sequence of SpMV/SpMM smoothing and
  restriction/prolongation operators, each an L2 sparse product; a
  polynomial (Chebyshev) preconditioner is a short series of SpMM
  applications.  \emph{Reduces to:} Ozaki~II SpMV/SpMM (L1) $\rightarrow$
  L0.
\item \textbf{$\rho \gets \langle r, z\rangle$ and $\langle p, q\rangle$
  (BLAS-1 reductions).}  These are the dense reductions of
  \S\ref{sec:l3}(a): operational intensity $1/8$ (\texttt{ddot}-class,
  two streams), $<\!5\%$ of
  iteration time, executed in \fp{32} with Kahan compensation on B300's
  healthy FP32 pipe.  This is Grade-B compensation, and standard PCG's
  short recurrence provides no re-orthogonalisation safety net, so the
  honest claim is solver-level: convergence rate, attainable residual,
  and iteration count under compensated dot products are
  validation-suite items, with an \fp{64}-grade reduction (the Kulisch
  route of \S\ref{sec:cqladder}) as the upgrade path on failure.
  \emph{Reduces
  to:} \fp{32}+Kahan (Grade~B; the bounded peripheral path).
\item \textbf{\textsc{axpy} updates $x,r,p$.}  Streaming
  $\mathrm{OI}\!\sim\!1/12$ vector updates (two reads, one write per
  $2$-flop update), trivially memory-bound on any
  pipe; no matrix work.
\end{enumerate}

Every arithmetic step is therefore either an Ozaki~II matrix product
bottoming out at the FP8 MMA, or a bounded \fp{32}/INT32 vector operation
that is memory-bound on its own pipe.  No step is compute-bound on the
collapsed \fp{64} units, and the dominant cost---the SpMM of step~1---is
exactly the case the sparse-primitive analysis of Appendix~\ref{app:sparse}
shows stays at $\beta\!\approx\!1$.  The block-CG variant strengthens the
conclusion, since widening to $s$ right-hand sides moves the SpMM further
into the regime where the FP8 tensor core is efficiently used.

\subsection{Exemplar 2: LOBPCG eigensolver}
\label{app:lobpcg}

Locally optimal block preconditioned conjugate gradient (LOBPCG) computes
the lowest $k$ eigenpairs of $Ax=\lambda x$ and is dominated, per
iteration, by (i) a block sparse product $AX$ with $X\in\mathbb{R}^{n\times
k}$, (ii) preconditioner applications on the block residual, and (iii) a
small dense Rayleigh--Ritz step on the $3k\times 3k$ projected problem.
The reduction:

\begin{enumerate}[leftmargin=2.2em,itemsep=2pt]
\item \textbf{$AX$ and $AW$, $AP$ (L3 $\rightarrow$ L2 SpMM).}  The
  block width is the eigencount $k$ (typically $16$--$64$), so these are
  squarely SpMM with a wide, tensor-core-friendly $n\!=\!k$ dimension---the
  most favourable sparse case in Appendix~\ref{app:sparse}, with
  intensity $\sim\!0.2k$ and the MMA $n$-lane fully utilised.
  \emph{Reduces to:} Ozaki~II SpMM (L1) $\rightarrow$ L0.
\item \textbf{Gram matrices $S^{\!\top}\!AS$, $S^{\!\top}\!S$ (L3
  $\rightarrow$ L2 dense GEMM).}  Here $S=[X,W,P]\in\mathbb{R}^{n\times
  3k}$, so the Gram products are tall-skinny dense GEMM ($n\times 3k$ by
  $3k\times n$ contracted over $n$), operational intensity growing with
  $n$ and well into the regime Ozaki~II dense GEMM handles at the memory
  roof or above.  \emph{Reduces to:} Ozaki~II GEMM (L1) $\rightarrow$ L0.
\item \textbf{Rayleigh--Ritz: dense $3k\times 3k$ eigensolve.}  This is a
  tiny dense symmetric eigenproblem ($3k\!\sim\!100$), solved once per
  iteration; its cost is $O((3k)^3)$ independent of $n$ and therefore
  asymptotically negligible against the $O(nk)$ block products as $n$
  grows at fixed $k$; its measured fraction depends on $n$, $k$, the
  preconditioner cost, and the implementation, so it is a profile item
  rather than a guarantee (small for $n \gtrsim 10^5$ at
  $k \sim 32$).  It is
  the panel-residual situation of \S\ref{sec:l3}(c): negligible at scale,
  and in any case executable on the FP32 pipe.
\end{enumerate}

Again the inner loop closes over the FP8 primitive: the two dominant costs
are SpMM and tall-skinny GEMM, both L2 dwarfs reducing to Ozaki~II and
thence to FP8, while the only dense-\fp{64} work---the $3k\times 3k$
projected eigensolve---is asymptotically negligible and off the critical
precision path.  LOBPCG is thus a faithful instance of the general claim:
a composite kernel whose every dominant operation is a matrix product that
bottoms out at the FP8 tensor-core op, with the residual non-matrix work
bounded and memory-bound.

\section{The Sparse Primitive: Why \texorpdfstring{$\beta$}{beta} Stays Near Unity Without a Large Matrix}
\label{app:sparse}

A recurring and reasonable objection to Ozaki-style emulation as a sparse
primitive is that the scheme is efficient only on \emph{large} dense
matrix products---large enough to amortise decomposition and
reconstruction and to drive the tensor core toward peak---whereas sparse
kernels present small, skinny matrix work after blocking, so the tensor
core cannot be driven to high efficiency, the overheads are not amortised,
and $\beta$ cannot reach unity.  This appendix shows, quantitatively, that
the objection conflates two distinct quantities and that the conclusion
does not follow for memory-bound kernels.  All figures use the B300
parameters of Table~\ref{tab:specs} ($P_{\fp{8}}=5$~PFLOP/s,
$B_{\text{mem}}=8$~TB/s, $r{=}12$, $\alpha=3r{+}1=37$).

\paragraph{Two efficiencies that must be separated.}  There are two
independent ``efficiency'' questions: \emph{(A)} the arithmetic
utilisation of the FP8 tensor core on the inner modular GEMM---are the MMA
tiles large enough, and is the right-hand-side dimension wide enough, to
approach the $5$~PFLOP/s peak?---and \emph{(B)} the bandwidth multiplier
$\beta$---does the residue and reconstruction traffic stay on-chip rather
than spilling to HBM?  Only \emph{(B)} decides whether a memory-bound
kernel is recovered to the memory roof.  The ``matrix too small''
objection is entirely about \emph{(A)}.  The central point of this
appendix is that \emph{for a memory-bound kernel, \emph{(A)} need not be
anywhere near unity for \emph{(B)} to hold}.

\paragraph{Ozaki does not change the intensity; it relocates the work.}
The useful \fp{64}-equivalent work in $y = Ax$ is the same two flops per
nonzero whether performed in native \fp{64} or emulated.  Ozaki multiplies
the \emph{on-chip} work by $\alpha=37$, not the HBM traffic: each element
of $A$ is streamed from memory once and reused across $3r{+}1$ FP8 MMAs
that never touch HBM.  The operational intensity of SpMV therefore remains
its structural $\sim\!0.2$--$0.5$~\flops{}/B, far below the emulated ridge
$P_{\fp{8}}/(\alpha B_{\text{mem}}) \approx 16.9$~\flops{}/B---a margin of
roughly $80\times$.  The kernel is memory-bound, decisively.

\paragraph{Keeping pace with HBM, not saturating the core.}  When a kernel
is memory-bound, the relevant question is not whether the FP8 unit is
saturated but whether it can keep pace with the memory stream.  Per
element of $A$ streamed in \fp{64} (eight bytes, the paper's canonical
storage), $\alpha=37$ FP8 MACs must
complete before the next element arrives; at 8~TB/s that is 1~Telem/s,
a required FP8 rate of $\approx 74$~TFLOP/s, or
\emph{$1.5\%$ of the FP8 peak}.  The tensor core may run at a small
fraction of peak utilisation and the compute still disappears beneath the
memory wall.  Charging the worst-case single-right-hand-side
tile-granularity cost of Eq.~\eqref{eq:wmma}---the $m16\,n8\,k32$ atom
uses $1/8$ of the MMA lanes at $B{=}1$, derating the
effective peak to $\sim\!625$~TFLOP/s---the keep-up requirement rises to
$\approx 12\%$; even storing $A$ compactly in \fp{16} (quadrupling the
element arrival rate to 4~Telem/s) leaves it at $47\%$---so the kernel
remains memory-bound on the MMA side in the least favourable geometry.
The objection would
bite only if the required utilisation exceeded $100\%$, leaving
substantial tensor-efficiency loss in hand before $\beta$ is
threatened.  We stress the scope of this argument: it disposes of the
\emph{tensor-side} keep-up question only.  The \emph{integer-side}
keep-up---the deconstruction of each streamed element into its residues,
charged by the fourth TME term of \S\ref{sec:model}---is an independent
and, at the reference constants of~\cite{bayraktar2026tme}, the binding
constraint for this kernel class; nothing in this appendix relaxes it.

\paragraph{The amortisation dimension is $r$ and the row length, not the
block size.}  The objection tacitly assumes Ozaki needs a large dense
block to amortise.  For SpMV it does not: the amortisation dimensions are
the moduli count $r$ (the residue planes of $x$ are reused across the
whole row) and the row length $k$ (over which the per-output Garner
reconstruction is spread), neither of which depends on block density.  A
$1{\times}1$ ``block''---plain CSR---still reuses the $r$ residue planes
across its row.  Under the ${\approx}r$-per-output reconstruction hosting
that the body carries as an undelivered codesign target
(\S\ref{sec:impl}; the conservative SIMT-Garner cost is $O(r^{2})$),
reconstruction would amortise once $k \gtrsim r \approx 12$, which
covers essentially all finite-difference and finite-element matrices (a
$27$-point stencil has $k=27$; general FEM, $k\sim 50$--$100$); at the
SIMT-Garner cost the thresholds are those of \S\ref{sec:impl}.  The
genuine edge case is an extremely short reduction combined with a single
right-hand side---a lone $7$-point stencil SpMV---and even there the
reconstruction overhead is on-chip, inflating compute \emph{time} rather
than HBM traffic, so $\beta$ is unaffected and only the (already
$80\times$) compute margin is reduced.

\paragraph{Why emulation is the \emph{better} sparse primitive.}  Native
scalar/vector \fp{64} SpMV has no amortisation dimension at all: every
$A_{ij}x_j$ is one scalar FMA on the collapsed $1.3$~TFLOP/s pipe, and the
\fp{64} vector unit is itself wasted at a single right-hand side exactly as
the tensor core is.  Ozaki~II \emph{manufactures} a reuse dimension---the
$r$ residue planes---that the FP8 tensor core feeds on.  The
single-right-hand-side tensor waste ($1/8$) is worse in ratio than the
\fp{64} vector waste, but it derates a peak that is some $3850\times$
higher: even at that pathological operating point, the emulated
\fp{64}-equivalent ceiling is $\sim\!13\times$ the native \fp{64} ceiling,
and with eight or more right-hand sides (the common case for block-Krylov,
LOBPCG, $s$-step, and ensemble methods, where the matrix work is in fact
SpMM, not lone SpMV) it is $\sim\!104\times$.  The moduli dimension turns
the geometry that the objection treats as a liability into the very source
of reuse that makes the FP8 tensor core effective on sparse work.

\paragraph{Summary.}  Across the canonical sparse cases the bandwidth
multiplier remains at or near unity, and exactly one corner---a shortest-
reduction kernel at a single right-hand side---requires care; it is
cheaply detected (short rows, single right-hand side) and has three clean
escapes (batch right-hand sides into SpMM, widen the block, or fall back
to the surviving native \fp{64} pipe).  This is an identifiable
second-stage corner case, expressible entirely as a statement about the
compute margin protecting $\beta$, not a failure of the primitive.
Batching right-hand sides also raises the useful flops per converted
element and hence the conversion-bound cap against native \fp{64}
(Eq.~\eqref{eq:oistar} scales with $\mathrm{OI} \propto B$)---though it
does \emph{not} reduce the deconstruction work per streamed byte, so the
distance to the memory roof at reference constants is $B$-independent
and is closed only by an engineered $c_q$ or hardware support
(\S\ref{sec:model}, \S\ref{sec:discussion}).

\section{Provenance of the Memo-Counted Reference Constants}
\label{app:memo}

The reference constants of \S\ref{sec:model}---$c_q = 16$ and the
$P_{\text{int}} \approx 75$-\tops{} normalisation---derive from a
private NVIDIA technical note by Bayraktar, Gunnels, and
Caday~\cite{bayraktar2026tme}, cited with permission.  To make the
anchor auditable we reproduce the non-confidential stage structure
here; this summary parallels the provenance appendix of the companion
paper~\cite{matsuoka2026ozaki25}, and the note's authors have been
asked to confirm it for publication.  Per streamed element and per
modulus on the shipping conversion path, as SASS-counted
\emph{estimates}---instruction counts, not throughput measurements,
on an architecture and toolchain marked preliminary: scale and
truncate-to-integer, ${\sim}4$ per element (amortised $4/r$ per
modulus); per-modulus byte-plane reduction, ${\sim}8$--$10$ (constant
division compiling to multiply-high sequences); final reduction to
the residue interval, ${\sim}2$; Karatsuba split into three \fp{8}
pieces, ${\sim}3$; totalling $c_q \approx 16$ under the stated
amortisation ($13$ counted from the shipping SASS plus $3$ for the
\fp{8} plane converts).  One \texttt{dp4a} counts as one issued SIMT
instruction throughout ($P_{\text{int}} = 75\cdot10^{12}$
instructions/s in the note's normalisation); \inteight{} tensor
capacity is accounted separately, in operations per second (two
operations per MAC), which is why the tensor-migrated rungs of
\S\ref{sec:cqladder} are charged against \inteight{}-tensor capacity
rather than against $P_{\text{int}}$.  The note also states the
two-sided $c_q$ decision criterion quoted in \S\ref{sec:model}.
Nothing else from the note is used in this paper.

\end{document}